Jefferson Science Associates, LLC

# Science Requirements and Conceptual Design for a Polarized Medium Energy Electron-Ion Collider at Jefferson Lab

2012

# Science Requirements and Conceptual Design for a Polarized Medium Energy Electron-Ion Collider at Jefferson Lab

S. Abeyratne[12], A. Accardi[1,7], S. Ahmed[1], D. Barber[6], J. Bisognano[18], A. Bogacz[1], A. Castilla[13,17], P. Chevtsov[14], S. Corneliussen[1], W. Deconinck[5], P. Degtiarenko[1], J. Delayen[13], Ya. Derbenev[1], S. DeSilva[13], D. Douglas[1], V. Dudnikov[11], R. Ent[1], B. Erdelyi[12], P. Evtushenko[1], Yu. Filatov[9,10], D. Gaskell[1], R. Geng[1], J. Grames[1], V. Guzey[1], T. Horn[4], A. Hutton[1], C. Hyde[13], R. Johnson[11], Y. Kim[8], F. Klein[4], A. Kondratenko[16], M. Kondratenko[16], G. Krafft[1,13], R. Li[1], F. Lin[1], S. Manikonda[2], F. Marhauser[11], R. McKeown[1], V. Morozov[1], P. Nadel-Turonski[1], E. Nissen[1], P. Ostroumov[2], M. Pivi[15], F. Pilat[1], M. Poelker[1], A. Prokudin[1], R. Rimmer[1], T. Satogata[1], H. Sayed[3,] M. Spata[1], M. Sullivan[15], C. Tennant[1], B. Terzić[1], M. Tiefenback[1], H. Wang[1], S. Wang[1], C. Weiss[1], B. Yunn[1], Y. Zhang[1]

[1] Thomas Jefferson National Accelerator Facility, Newport News, VA 23606, USA
[2] Argonne National Laboratory, Argonne, IL 60439, USA
[3] Brookhaven National Laboratory, Upton, NY 11973, USA
[4] Catholic University of America, Washington, DC 20064, USA
[5] College of William and Mary, Williamsburg, VA 23187, USA
[6] Deutsches Elektronen-Synchrotron (DESY), 22607 Hamburg, Germany
[7] Hampton University, Hampton, VA 23668
[8] Idaho State University, Pocatello, ID 83209, USA
[9] Joint Institute for Nuclear Research, Dubna, Russia
[10] Moscow Institute of Physics and Technology, Dolgoprydny, Russia
[11] Muons Inc., Batavia, IL 60510, USA
[12] Northern Illinois University, DeKalb, IL 60115, USA
[13] Old Dominion University, Norfolk, VA 23529, USA
[14] Paul Scherrer Institute, 5232 Villigen PSI, Switzerland
[15] SLAC National Accelerator Laboratory, Menlo Park, CA 94305, USA
[16] Science and Technique Laboratory Zaryad, Novosibirsk, Russia
[17] Universidad de Guanajuato, 36000 Guanajuato, Mexico
[18] University of Wisconsin-Madison, Madison, WI 53706, USA

**Editors: Y. Zhang and J. Bisognano**

# Table of Contents

# Executive Summary

Researchers have envisioned an electron-ion collider with ion species up to heavy ions, high polarization of electrons and light ions, and a well-matched center-of-mass energy range as an ideal gluon microscope to explore new frontiers of nuclear science. In its most recent Long Range Plan, the Nuclear Science Advisory Committee (NSAC) of the US Department of Energy and the National Science Foundation endorsed such a collider in the form of a "half-recommendation." As a response to this science need, Jefferson Lab and its user community have been engaged in feasibility studies of a medium energy polarized electron-ion collider (MEIC), cost-effectively utilizing Jefferson Lab's already existing Continuous Electron Beam Accelerator Facility (CEBAF). In close collaboration, this community of nuclear physicists and accelerator scientists has rigorously explored the science case and design concept for this envisioned grand instrument of science.

An electron-ion collider embodies the vision of reaching the next frontier in Quantum Chromodynamics—understanding the behavior of hadrons as complex bound states of quarks and gluons. Whereas the 12 GeV Upgrade of CEBAF will map the valence-quark components of the nucleon and nuclear wave functions in detail, an electron-ion collider will determine the largely unknown role sea quarks play and for the first time study the glue that binds all atomic nuclei. The MEIC will allow nuclear scientists to map the spin and spatial structure of quarks and gluons in nucleons, to discover the collective effects of gluons in nuclei, and to understand the emergence of hadrons from quarks and gluons.

The proposed electron-ion collider at Jefferson Lab will collide a highly polarized electron beam originating from the CEBAF recirculating superconducting radiofrequency (SRF) linear accelerator (linac) with highly polarized light-ion beams or unpolarized light- to heavy-ion beams from a new ion accelerator and storage complex. Since the very beginning, the design studies at Jefferson Lab have focused on achieving high collider performance, particularly ultra-high luminosities up to $10^{34}$ $cm^{-2}s^{-1}$ per detector with large acceptance, while maintaining high polarization for both the electron and light-ion beams. These are the two key performance requirements of a future electron-ion collider facility as articulated by the NSAC Long Range Plan. In MEIC, a new ion complex is designed specifically to deliver ion beams that match the high bunch repetition and highly polarized electron beam from CEBAF.

During the last two years, both development of the science case and optimization of the machine design point toward a medium-energy electron-ion collider as the topmost goal for Jefferson Lab. The MEIC, with relatively compact collider rings, can deliver a luminosity above $10^{34}$ $cm^{-2}s^{-1}$ at a center-of-mass energy up to 65 GeV. It offers an electron energy up to 11 GeV, a proton energy up to 100 GeV, and corresponding energies per nucleon for heavy ions with the same magnetic rigidity. This design choice balances the scope of the science program, collider capabilities, accelerator technology innovation, and total project cost. An energy upgrade could be implemented in the future by adding two large collider rings housed in another large tunnel to push the center-of-mass energy up to or exceeding 140 GeV. After careful consideration of an alternative electron energy recovery linac on ion storage ring approach, a ring-ring collider scenario at high bunch repetition frequency was found to offer fully competitive performance while eliminating the uncertainties of challenging R&D on ampere-class polarized electron sources and many-pass energy-recovery linacs (ERLs).

The essential new elements of an MEIC facility at Jefferson Lab are an electron storage ring and an entirely new, modern ion acceleration and storage complex. For the high-current electron collider ring, the upgraded 12 GeV CEBAF SRF linac will serve as a full-energy injector, and, if needed, provide top-off refilling. The CEBAF fixed-target nuclear physics program can be simultaneously operated since the filling time of the electron ring is very short. The ion complex for MEIC consists of sources for polarized light ions and unpolarized light to heavy ions, an SRF ion linac with proton energy up to 280 MeV, a 3 GeV prebooster synchrotron, a large booster synchrotron for proton energy up to 20 GeV, and a medium-energy collider ring with energy up to 100 GeV. The ion complex can accelerate other species of ions with corresponding energies at each accelerating stage. There are three collision points planned for MEIC. Two of them are for collisions with medium-energy ions; the third is for low energy ion beams stored in a dedicated low-energy compact storage ring, as a possible follow-on project.

The MEIC interaction region is designed with ultra-strong final focusing with a betatron function ($\beta^*$) value of 2 cm for a full-acceptance detector and with an even smaller value of 0.8 cm for a large-acceptance, high-luminosity detector. Such an interaction region design, enabled by the very short bunch lengths of colliding electron and ion beams together with a high collision frequency, will attain luminosities of $10^{34}$ $cm^{-2}s^{-1}$. Although relatively small, such values of $\beta^*$ have been achieved in B-factories with good dynamic aperture. A modern electron cooling facility, which provides the essential high phase space density, will utilize two key accelerator technologies—an energy-recovery SRF linac and a circulator ring—to deliver the required high-average-current, high-energy electron cooling beam. Crab crossing of colliding beams is used to accommodate the high bunch repetition rates without parasitic crossings. One important feature of the Jefferson Lab electron-ion collider design is the figure-8 shape of the collider rings and two booster rings. This unique design choice most effectively preserves the high polarization of light ions during their long and multi-stage acceleration. In particular, for polarized deuterons, due to their small value of (*g-2*), the figure-8 shaped ion booster and collider rings are the only viable solution. On the electron side, electron polarization can be preserved and enhanced by the Sokolov-Ternov effect in the storage ring, and a set of energy-independent spin rotators will align the electron spins in the longitudinal direction at collision points and in the vertical direction in the arcs.

This report summarizes the design concept of MEIC at Jefferson Lab. It also serves as a vision for guiding future design optimization and accelerator R&D. Aside from a brief discussion of the potential nuclear physics experiments possible at the collider, this report focuses on a description of the facility layout and types of accelerators needed to complete the collider at Jefferson Lab. We present preliminary designs of the accelerators and the interaction regions, and estimates of collider performance over a range of collision energies and ion species. Our conclusion is that, with development of promising ion beam cooling methods, it will be possible to support collision luminosity up to $10^{34}$ $cm^{-2}s^{-1}$ or above for a range of collision center-of-mass energy up to 65 GeV. Our plans support the eventual upgrade of the collider complex to higher collision energies should it be found scientifically compelling to do so.

# 1. Introduction

As the second century of nuclear physics begins, a new frontier has emerged—understanding how quarks and gluons are assembled to form nucleons. It is the studied, collective judgment of the nuclear physics community that exploration of this new frontier will require an electron-ion collider (EIC). In its most recent report, *The Frontiers of Nuclear Science: A Long-Range Plan for the Next Decade* [1], the DOE/NSF Nuclear Science Advisory Committee considered and proposed, in the form of a "half-recommendation," an electron-ion collider as the ideal gluon microscope for elucidating this domain of nuclear science. The report further articulated performance requirements for this facility in terms of the ion species and the range of center-of-mass energy, as well as the luminosity and beam polarization it must achieve. In addition, taking a longer-term view, the advisory committee recommended that US government funding agencies ensure "the allocation of resources to develop accelerator and detector technology necessary to lay the foundation for a polarized Electron-Ion Collider." [1]

Currently, within the United States, two groups are independently studying the possibilities for electron-ion colliders. Such an EIC could be built upon and significantly extend capabilities of existing nuclear physics facilities at either Jefferson Lab [2,3] or Brookhaven National Laboratory [4], utilizing either Jefferson Lab's CEBAF recirculating SRF linear accelerator [5] or BNL's Relativistic Heavy Ion Collider [6]. This report presents the design concept of the Jefferson Lab Medium-Energy Electron-Ion Collider (MEIC).

Jefferson Lab has been engaged in feasibility studies for a polarized electron-ion collider based on the CEBAF accelerator for over a decade [2,3]. The overall goal has been to collide a polarized electron beam at an energy of 3 to 11 GeV from the 12 GeV CEBAF with polarized light ions or nonpolarized light to heavy ions in an energy range from 20 to 250 GeV from a new ion complex.

The earliest design concept was, quite naturally, a linac-ring collider scenario in which a polarized electron or positron beam from the CEBAF recirculating linac collides with an ion beam in a storage ring [2]. The CEBAF facility would have been converted to an ERL in order to provide a high-average-current beam. Later, a circulator electron ring was added to the ERL-ring collider design to achieve factor-of-hundreds reduction of average current from the polarized source [2]. With a better understanding of performance tradeoffs, the design has now converged to a more straightforward ring-ring collider in which the colliding electron beam will also be stored in a ring while CEBAF will be used as a full-energy lepton injector [3]. No further upgrade of CEBAF beyond 12 GeV is required. The key realization in coming to this decision was that at the gigahertz level of repetition frequency offered by a CEBAF injector, the electron beam-beam effect in the storage ring is manageable and not a significant extra constraint. Overall performance has not been compromised while robustness has been increased with reduced technical risk. The design concept supports up to three interaction points to ensure high scientific productivity. The high-repetition-rate, ring-ring collider scenario reaches high luminosity while eliminating the burden of challenging R&D for a high-current polarized electron source and a many pass ERL.

The Jefferson Lab MEIC design takes full advantage of the existing CEBAF high-bunch-repetition-rate electron beam [5] and a new ion complex, which can be specially designed to deliver ion beams matching the time and phase-space structures of the colliding electron beam. The ultrahigh luminosity, exceeding $10^{34}$ $cm^{-2}s^{-1}$ subject to the requirements of the physics detectors, is achieved for MEIC based on short-bunch, high-repetition rate CW beam luminosity concepts experimentally proved at several lepton colliders including the two B-factories at KEK [7] and SLAC [8]. Polarization of light ions (p, d, $^3$He and possibly Li) is supported and maintained by the figure-8 shaped booster and collider rings, a simple and unique design choice providing great simplicity [9].

During the last two years, both evolution of the science program and optimization of the accelerator design have been driving the Jefferson Lab EIC design toward a medium energy electron-ion collider, MEIC, as the clear immediate goal, while maintaining opportunities for a future center of mass energy upgrade to larger rings should the science case be compelling. For the present baseline, MEIC has a center-of-mass energy range up to 65 GeV, covering 3 to 11 GeV for electrons, 20 to 100 GeV for protons and up to 40 GeV per nucleon for light to heavy ions. Such a staged approach is considered an optimized balance of the science program, accelerator R&D, and project cost. The following chapters present a design concept addressing all significant details in the MEIC and delineating a few key R&D items.

One important aspect of the EIC design approach at Jefferson Lab is a balance of technology innovation for maximum collider performance and overall design simplification for reduced technical challenges and required R&D. We have also imposed certain limits on several machine or beam parameters in order to minimize accelerator R&D and to improve robustness of the design. We have worked to stay, for the most part, within operational limits derived largely from previous lepton and hadron collider experience and the present state of the art of accelerator technology. Such a technical approach allows us to focus limited resources on a few critical accelerator R&D challenges such as electron cooling.

This report summarizes the present baseline design of the MEIC at Jefferson Lab. It is intended to be concise, focusing primarily on a laying out a self-consistent, manifestly feasible accelerator design concept. This report will also serve as a guideline for the next-level in-depth machine design calculations, simulations, and R&D plans. The report is organized as follows: in Chapter 2 we present a brief summary of the science case for the MEIC. In Chapter 3 we present an overview of the MEIC facility, with a schematic layout and main machine parameters. We discuss the MEIC high-luminosity concept and other key design features. A comparison of ERL-ring and ring-ring collider scenarios is also included in that chapter for supporting the design concept chosen. In the later chapters, we present detailed descriptions of each important component of the MEIC, namely, the electron facility, the ion complex, the electron cooler, and the interaction region. We close this report with a brief discussion of future R&D in chapter 8.

# 2. Nuclear Physics with MEIC

Understanding the internal structure of hadrons and nuclei on the basis of the fundamental theory of strong interactions, Quantum Chromodynamics (QCD), is one of the central problems of modern nuclear physics, as explained *e.g.* in the U.S. Department of Energy Office of Science Nuclear Science Advisory Committee's 2007 Long Range Plan [1]. It is the key to understanding the dynamical origin of mass in the visible universe and the behavior of matter at astrophysical temperatures and densities. It is an essential element in describing nuclear structure and reactions from first principles, a project with numerous potential applications to science and technology. Theoretical methods to apply QCD to hadronic and nuclear systems have made dramatic advances in the last two decades but rely crucially on new experimental information for further progress.

Electron scattering has been established as a powerful tool for exploring the structure of matter at the sub-femtometer level (<1 fm=$10^{-15}$ m). Historically, such experiments provided the first proof of the extended nature of the proton and revealed the presence of pointlike constituents, or quarks, at smaller scales, revolutionizing our understanding of strong interactions. Subsequent experiments established the validity of QCD and the presence of gluonic degrees of freedom at short distances and measured the basic number densities of quarks and gluons in the nucleon (proton, neutron). While much progress has been made, several key questions remain unanswered [1]:

I) What role do non-valence ("sea") quarks and gluons play in nucleon structure? What are their spatial distributions? How do they respond to polarization? What is their orbital motion, and how does it contribute to the nucleon spin? The answers to these questions will provide essential information on the effective degrees of freedom emerging from QCD at distances of the order of the hadronic size (~1 fm).

II) What are the properties of the fundamental QCD color fields in nuclei with nucleon number A > 1? What are the nuclear gluon and sea quark densities? To what extent are they modified by nuclear binding, quantum-mechanical interference, and other collective effects? These questions are the key to understanding the QCD origins of the nucleon-nucleon interaction at different energies, the role of non-nucleonic degrees of freedom, and the approach to a new regime of high gluon densities and saturation at high energies.

III) How do colorless hadrons emerge from the colored quarks and gluons of QCD? What dynamics governs color neutralization and hadron formation? By what mechanisms does the color charge of QCD interact with nuclear matter? We are still far from understanding the fundamental processes by which high-energy radiation converts into hadronic matter.

It is now widely accepted that a polarized *ep*/*eA* collider (Electron-Ion Collider, or EIC) with a variable *ep* center-of-mass (CM) energy in the range $\sqrt{s}$=20–70 GeV, and a luminosity of ~$10^{34}$ $cm^{-2}s^{-1}$ over most of this range, would offer a unique opportunity to address these questions [2]. Such a facility would provide the necessary combination of kinematic reach (momentum transfer viz. spatial resolution, energy span), luminosity (precision, multi-dimensional binning, rare processes), and detection capabilities (resolution, particle identification) to study nucleon and nuclear structure through scattering experiments with a variety of final states. It would represent the natural next step after the high-luminosity fixed-target *ep*/*eA* experiments (SLAC [3], JLab 6 and 12 GeV [4,5]) and the high-energy HERA *ep* collider [6,7]. It would be the first ever high-energy electron-nucleus collider and open up qualitatively new possibilities to study QCD in the nuclear environment. Finally, polarized beams would allow one to investigate proton and neutron spin structure with unprecedented accuracy and kinematic reach; such measurements were so far possible only in fixed-target experiments (EMC, SMC, SLAC, HERMES, COMPASS, JLab; for a review see Refs. [8,9,10]) or polarized *pp* collisions at RHIC [11]. In this chapter we briefly review what measurements with such a medium-energy EIC (or MEIC) could contribute to answering the above questions [12].

## 2.1 The Three Dimensional Structure of the Nucleon in QCD

The nucleon in QCD represents a dynamical system of fascinating complexity. In the rest frame it may be viewed as an ensemble of interacting color fields, coupled in an intricate way to the vacuum fluctuations that govern the effective dynamics at distances ~1 *fm*. In this formulation its properties can be studied through large-scale numerical simulations of the field theory on a discretized space-time (Lattice QCD) [13] as well as analytic methods. A complementary description emerges when one considers a nucleon that moves fast, with a momentum much larger than that of the typical vacuum fluctuations. In this limit the nucleon's color fields can be projected on elementary quanta with point-particle characteristics (partons), and the nucleon becomes a many-body system of quarks and gluons. As such it can be described by a wave function, in much the same way as many-body systems in nuclear or condensed matter physics (see Figure 2.1). In contrast to these non-relativistic systems, in QCD the number of pointlike constituents is not fixed, as they constantly undergo creation/annihilation processes mediated by QCD interactions, reflecting the essentially relativistic nature of the dynamics. A high-energy scattering process takes a "snapshot" of this fast-moving system with a spatial resolution given by the inverse invariant momentum transfer $1/Q$. The energy transfer, parameterized by the Bjorken variable $x \equiv Q^2/(2\, M_N\, \nu)$, defines the momentum fraction of the struck constituent and thus determines what particle configurations are intercepted in the scattering process ($\nu$ is the energy of the virtual photon in the nucleon rest frame). Scattering experiments at different $x$ and $Q^2$ can thus probe in detail the various components of the nucleon's wave function and map out their properties.

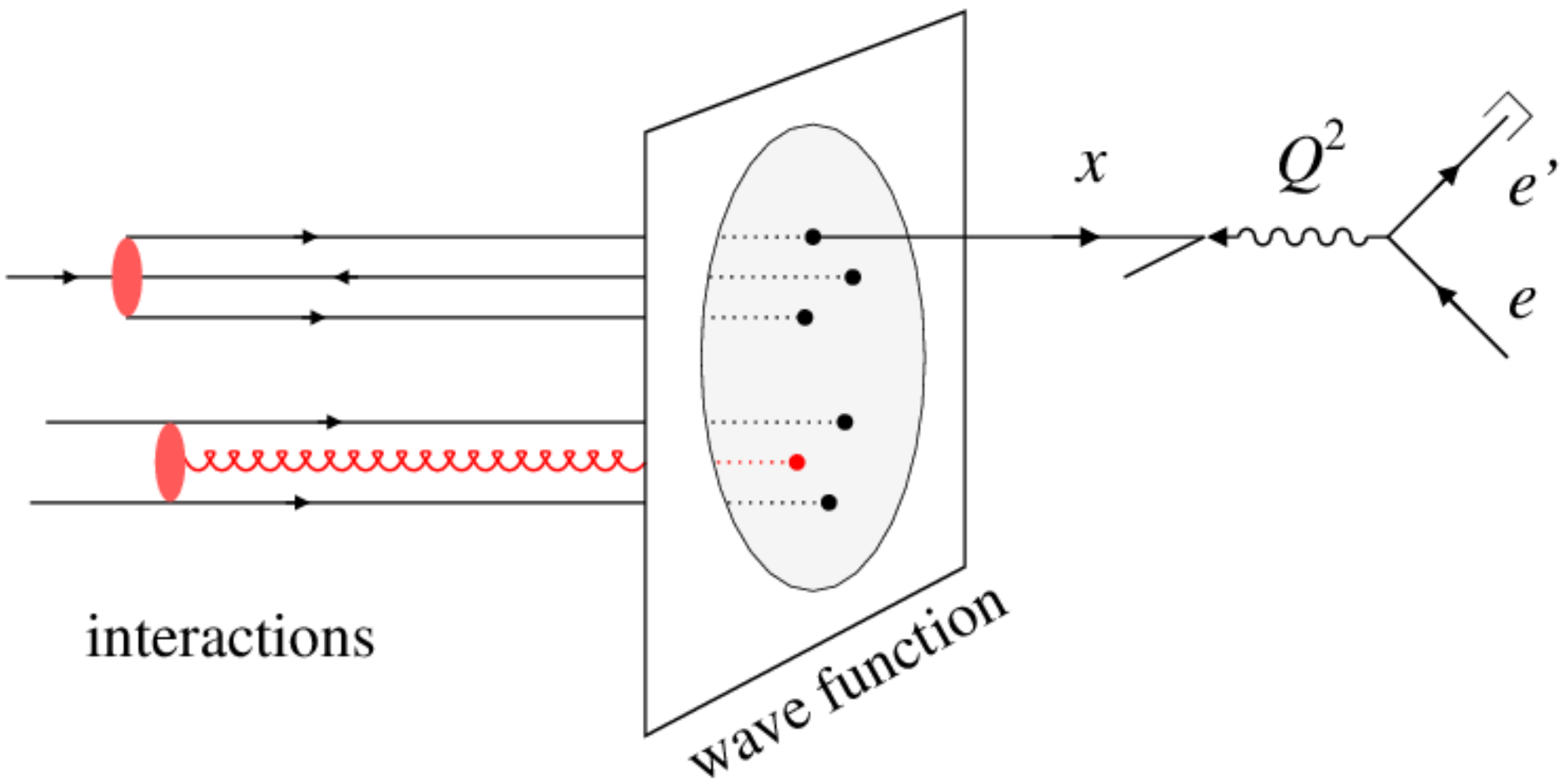

Figure 2.1: The fast-moving nucleon in QCD as a many-body system. A high-energy scattering process takes a "snapshot" of the interacting system with spatial resolution $1/Q$ and probes partons with a certain momentum fraction $x$.

The range of $x$-values attainable in electron scattering at given $Q^2$ is kinematically restricted to $x > Q^2/s$ and thus determined by the *ep* CM energy; the coverage in $x$ and $Q^2$ available at different facilities is shown in Figure 2.2a. Measurements with JLab 12 GeV [5] probe nucleon structure in the region dominated by the valence quark component of the nucleon wave function ($x > 0.1$, see right panel in Figure 2.2b), which carries most of the nucleon's spin and charge and acts as the "source" of the other components. These measurements will in particular explore the unknown $x \to 1$ region, corresponding to extreme configurations in which a single quark carries almost all of the nucleon's momentum [5].

In addition to the valence quarks, the nucleon contains a "sea" of quark--antiquark pairs that is created by non-perturbative QCD interactions and reflects the complex structure of the ground state (or vacuum) of the theory. The spin and flavor quantum numbers carried by the sea sit mainly in the region $0.01 \leq x < 0.1$ (see central panel in Figure 2.2b) and are poorly constrained by present data [14,15]. A medium--energy EIC could measure the distribution of sea quarks through semi-inclusive measurements, in which the charge and flavor of the struck quark/antiquark are "tagged" by detecting hadrons ($\pi^{\pm}$, $K^{\pm}$, $p, \bar{p}$, …) produced from its fragmentation. Compared to fixed-target experiments the energy available with the collider ensures that the hadronization of the struck quark proceeds independently from the target remnants and cleanly preserves the original spin-flavor information. The kinematic coverage and detection capabilities are uniquely suited to such measurements, allowing for a precise mapping of this largely unexplored component of the nucleon.

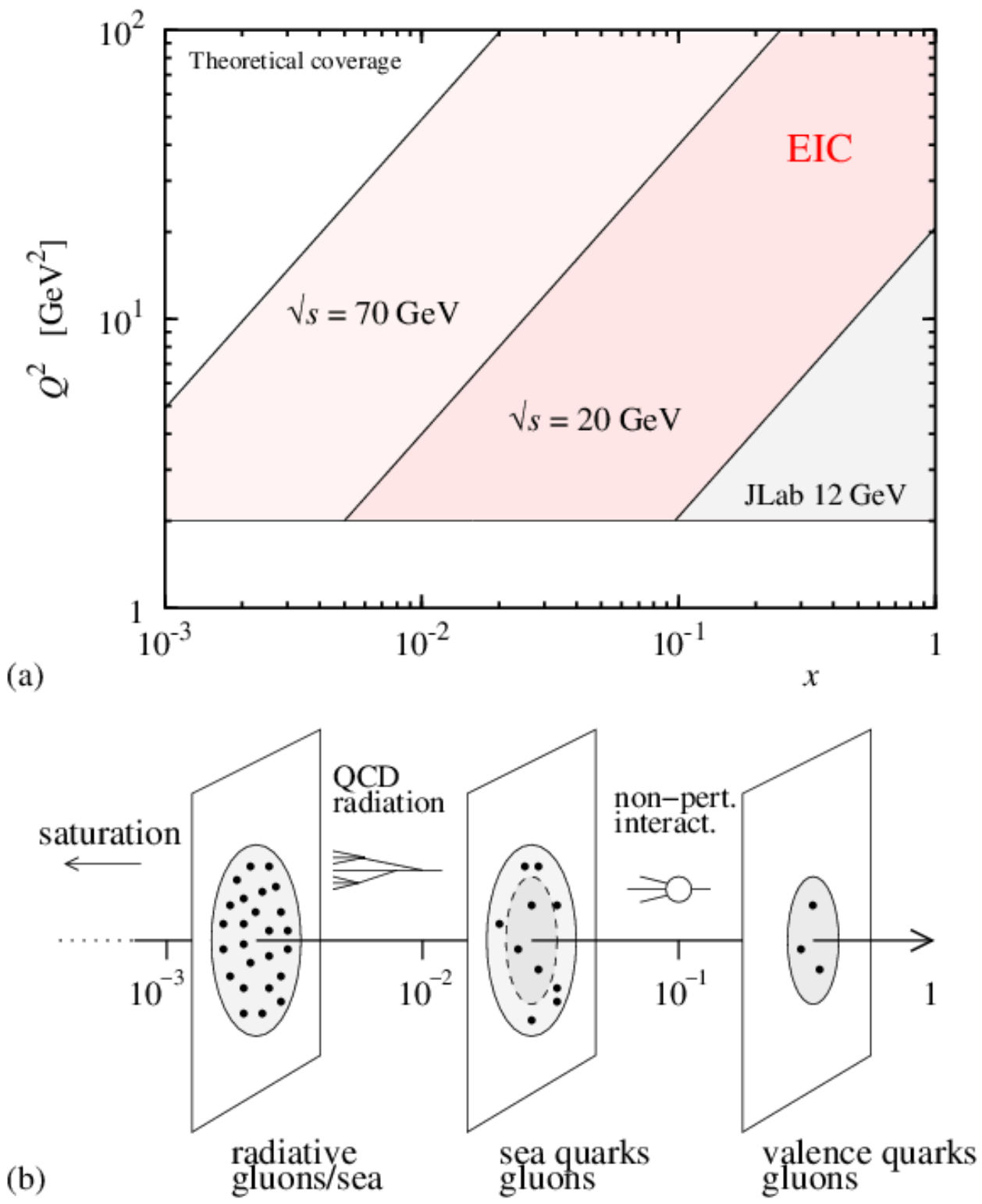

Figure 2.2: (a) Kinematic coverage in $x$ and $Q^2$ in $ep$ scattering experiments with JLab 12 GeV and a medium-energy EIC of CM energy $\sqrt{s}$=20 and 70 GeV. The minimum momentum transfer here was chosen as $Q^2_{min}$=2 GeV$^2$. (b) Components of the nucleon wave function were probed in scattering experiments at different $x$ (see text for details). Plots (a) and (b) share the same $x$-axis and should be read together.

Another distinctive component of the nucleon's sea is expected to arise from strange and charmed quark-antiquark pairs with large momentum fractions $x \geq 0.1$, which can be created by pairwise QCD interactions between valence quarks in the wave function; see Ref. [16] for a review. A medium-energy EIC would for the first time permit to study this component through the measurement of hadrons produced in the breakup up the proton remnant (target fragmentation), qualitatively improving the sensitivity compared to measurements of hadrons produced from the struck quark alone (current fragmentation) [17].

Equally important is the distribution of polarized gluons in the nucleon. Besides its intrinsic importance, its measurement is needed to solve the “puzzle” of the nucleon spin decomposition and quantify the role of orbital angular momentum in the nucleon wave function [8]. Since gluons carry no electric charge, electromagnetic scattering can probe them only indirectly, through the $Q^2$ dependence of the nucleon structure functions. Present $eN$ data ($N \equiv p$, $n$) and the data from polarized $pp$ collisions at RHIC practically do not constrain the polarized gluon density for $x \leq 0.05$ [18]. Inclusive measurements with a medium-energy EIC would dramatically extend the data set and determine the polarized nucleon structure function $g_1(x,Q^2)$ down to $x$~few × $10^{-3}$ with a substantial range in $Q^2$ (see Figure 2.2a), allowing one to extract the polarized gluon density from the $Q^2$ dependence [2].

Other fundamental characteristics of the nucleon are the transverse spatial distributions of quarks and gluons carrying a certain momentum fraction $x$ (see Figure 2.3). They define the basic size and "shape" of the nucleon in QCD and convert the one-dimensional picture conveyed by the longitudinal momentum densities into a full three-dimensional image of the fast-moving nucleon [19,20,21]. Information on the transverse distribution of quarks and gluons is obtained from exclusive scattering $e\,N \rightarrow e' + M + N$, ($M$ = meson, γ, heavy quarkonium). Such processes probe the generalized parton distributions (GPDs), which combine the concept of the quark/gluon momentum density with that of elastic nucleon form factors. Measurements of $J/\psi$ photo- and/or electroproduction with a medium-energy EIC would be able to map the transverse spatial distribution of gluons in the nucleon above $x \sim$ few $\times 10^{-3}$ in unprecedented detail [2]. In particular, these measurements would cover the unexplored gluons in the valence region at $x \geq 0.1$, whose presence is inferred from global fits to deep-inelastic scattering data but has proved difficult to confirm directly; their dynamical origins are one of the outstanding questions of nucleon structure in QCD. Information on the transverse spatial distribution of gluons is needed also to describe the final states in *pp* collisions at LHC (underlying event in hard processes, multiparton processes) and understand the approach to the regime of high gluon densities at small $x$ (initial conditions for non-linear QCD evolution equations) [22]. Measurements of real photon production (γ, deeply virtual Compton scattering) with an EIC would differentiate gluon and quark spatial distributions and study how the latter are deformed in a transversely polarized nucleon. Production of light mesons with charge/isospin (π, K, ρ, $K^*$) would map the transverse distributions of sea quarks and provide additional insight into their dynamical origins. This program of "quark/gluon imaging" requires differential measurements of low-rate processes and relies crucially on the high luminosity provided by the EIC in the envisaged energy range, and the possibility to longitudinally and transversely polarize the proton beam.

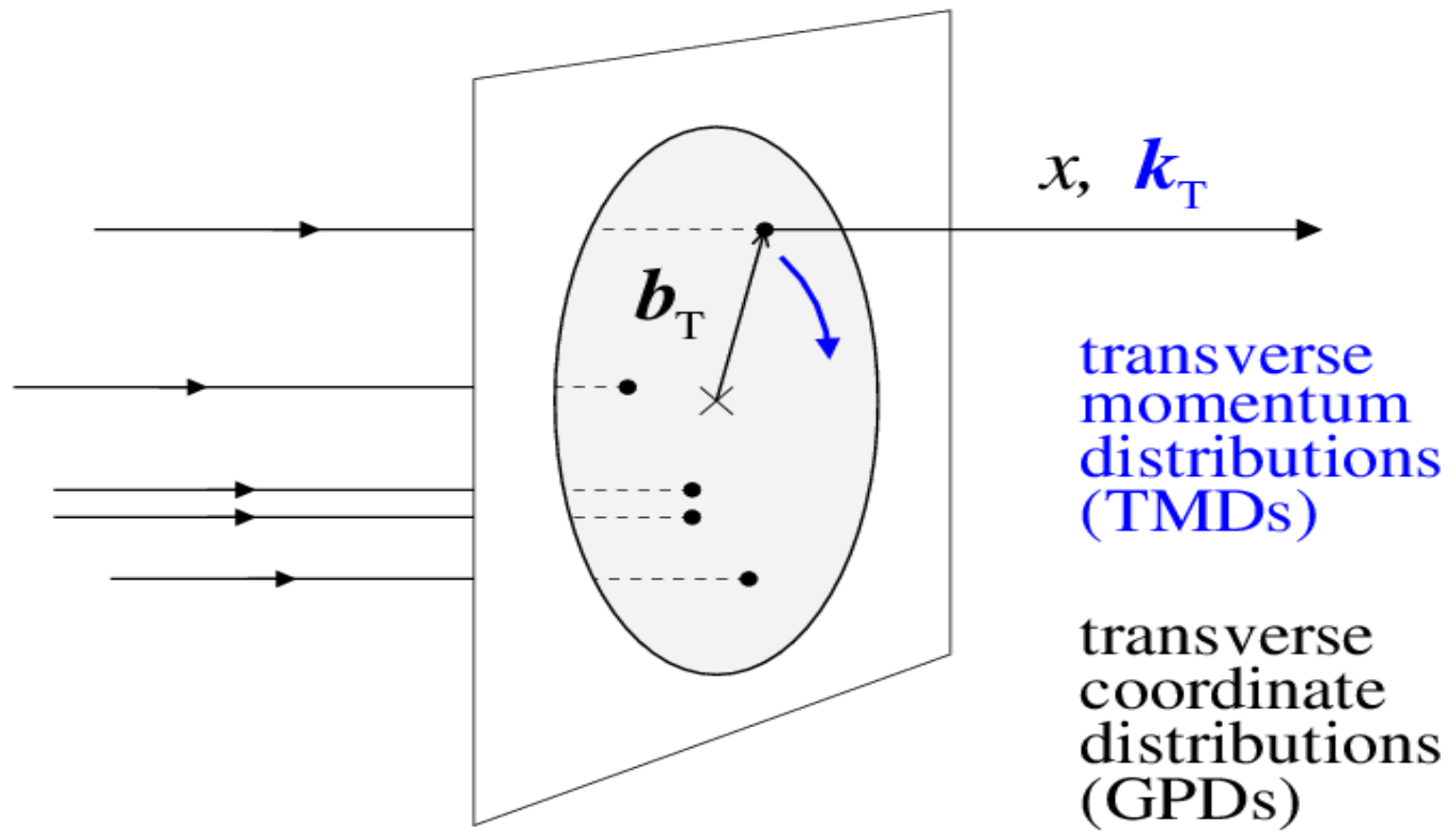

Figure 2.3: Three-dimensional structure of the fast-moving nucleon in QCD. The distribution of partons (quarks, gluons) is characterized by the longitudinal momentum fraction $x$ and the transverse spatial coordinate $\boldsymbol{b}_T$ (GPDs). In addition, the partons are distributed over transverse momenta $\boldsymbol{k}_T$, reflecting their orbital motion and interactions in the system (TMDs). Polarization distorts both the spatial and momentum distributions. Note that $\boldsymbol{b}_T$ and $\boldsymbol{k}_T$ are not Fourier conjugate; a joint description in both variables can be formulated in terms of a Wigner phase space density. Observables sensitive to either $\boldsymbol{b}_T$ or $\boldsymbol{k}_T$ help to establish a three-dimensional dynamical picture of the nucleon in QCD.

Closely related is the question of the orbital motion of quarks and gluons and its role in nucleon structure (see Figure 2.3). This information is encoded in the transverse momentum distributions (TMDs) and their response to nucleon and quark/gluon polarization [23,24]. They provide a three-dimensional representation of the nucleon in momentum space, complementing the spatial view offered by the GPDs. The TMDs can be measured in semi-inclusive scattering processes $\gamma^* N \rightarrow h + X$, where particles produced by fragmentation of the struck quark ($h = \pi$, $K$, $J/\psi$, open charm), as well as the nucleon fragments, can reveal the quark and gluon transverse momentum and its correlation with the nucleon spin. The various structure functions, each of which describes certain facets of nucleon structure (transverse motion and deformation, spin-orbit correlations, orbital angular momentum, final-state interactions of the struck quark with the color fields in the nucleon) can be separated by measurements with different combinations of beam and target polarizations, including the transverse nucleon polarization that is easily available with the collider. Measurements with a medium-energy EIC will be able to precisely determine, *e.g.*, the valence and sea quark Sivers function sensitive to spin-orbit interactions in the region $x > 0.01$, where it is expected to be sizable. They will also study for the first time the $Q^2$ evolution of TMDs and the region of large transverse momenta, $k_T$ »1 GeV, where TMDs can be related to multiparton correlations in the nucleon. Measurements with open charm and $J/\psi$ mesons in the final state can directly probe the gluon TMDs. All these studies require multi-dimensional binning in $x$, $Q^2$, and the energy fraction $z$ and transverse momentum $P_T$ of the produced meson, which can be performed only with high-statistics data as would become available with the planned EIC luminosity.

## 2.2 The Fundamental Color Fields in Nuclei

A basic quest of nuclear physics is to understand the structure and dynamics of the QCD color fields in nuclei with nucleon number $A$>1. Information on these fields is obtained by studying the scattering of small-size probes — *e.g.*, a virtual photon with $Q^2$»1 GeV$^2$ — from nuclei over a range of incident energies. Of particular interest is how the nuclear fields differ from the sum of the color fields of the individual nucleons. The lifetime of the probe in the target rest frame is defined by the coherence length, $l_{coh} \propto (x M_N)^{-1}$, where the coefficient depends on the size of the probe but is typically of order unity. If the coherence length is much smaller than the nuclear radius $R_A$ ~ few fm, the scattering involves only a single nucleon in the nucleus (see Figure 2.4b, right picture). In this regime the results can be interpreted in terms of a modification of single-nucleon structure through nuclear binding and reveal the QCD origins of the nucleon-nucleon interaction. If the coherence length becomes comparable to or larger than the nuclear radius, $l_{coh} \geq R_A$ (see Figure 2.4b, left picture), the color field seen by the probe is the quantum-mechanical superposition of the fields of the individual nucleons, resulting in a rich spectrum of coherence effects such as shadowing [24,25,26], diffraction, and eventually the approach to the unitarity limit (saturation) at high energies [27]. In addition, the fields change with the energy and the resolution scale $Q^2$ as a result of QCD radiation. An EIC with a CM energy in the range $\sqrt{s}$ ~ 20-70 GeV would for the first time provide the coverage in $l_{coh}$ and $Q^2$ necessary to observe these phenomena (see Figure 2.4a), opening up a whole new area of study.

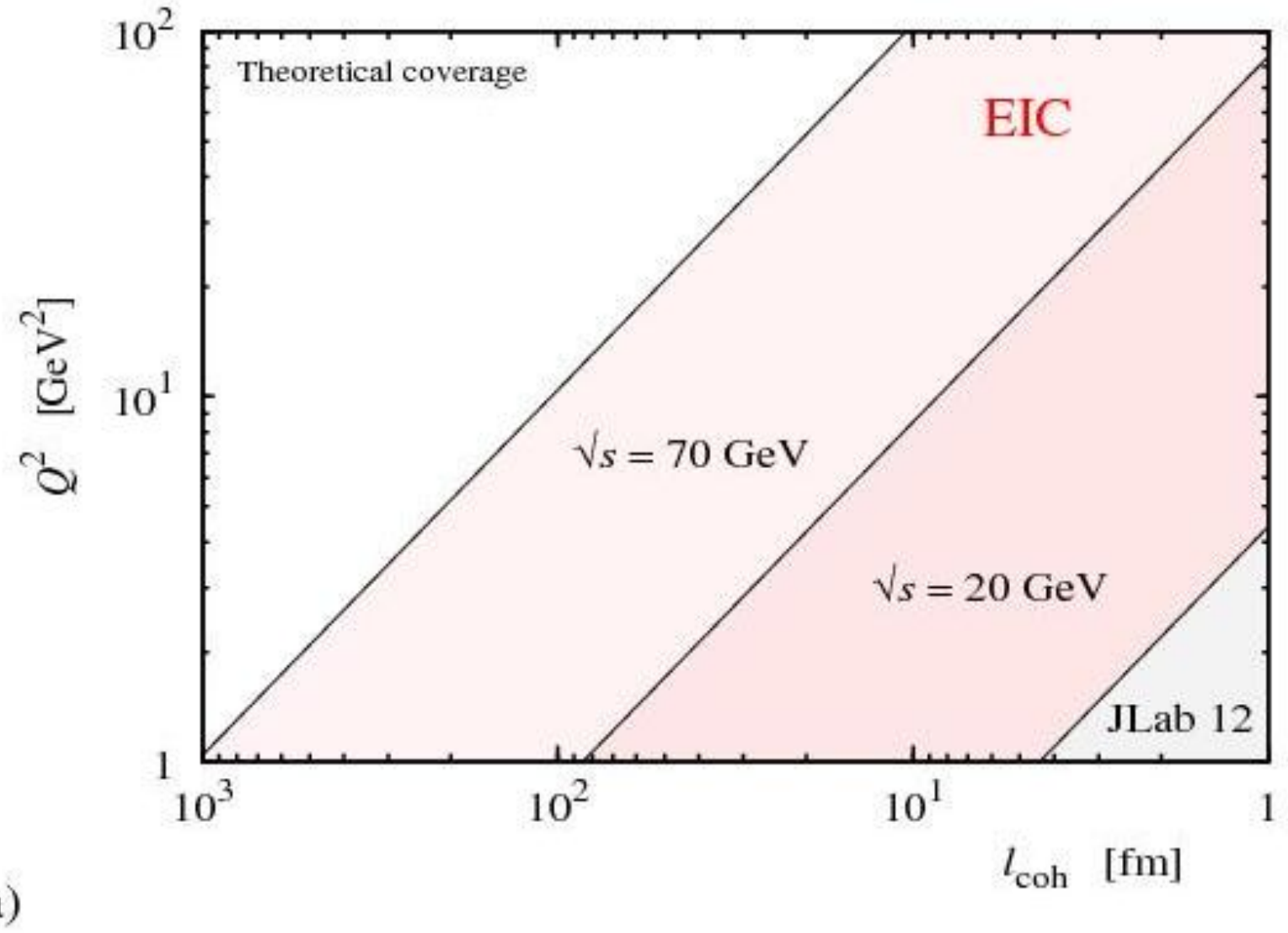

(a)

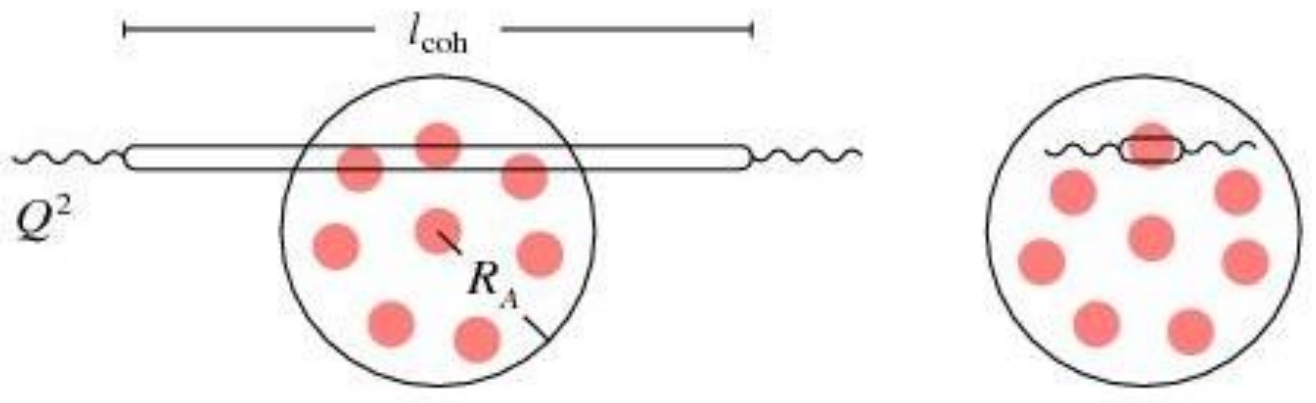

(b)

Figure 2.4: (a) Kinematic coverage in the typical coherence length $l_{\rm coh} = (M_{\rm N}\ x)^{-1}$ and the scale $Q^2$ in electron-nucleus scattering with a medium-energy EIC. (b) Interaction of a small-size probe with a nucleus. *Right picture:* $l_{\rm coh} \ll R_{\rm A}$. *Left picture:* $l_{\rm coh} \geq R_{\rm A}$ (see text for details).

Much interesting information already comes from the basic quark and gluon densities of nuclei. A peculiar pattern of nuclear modifications was observed in fixed-target experiments and caused much excitement; it shows suppression compared to the free nucleon for $0.2 < x < 0.8$ (the famous "EMC effect," to be explored further with JLab 12 GeV), some signs of enhancement for $0.05<x<0.2$, and significant suppression at smaller $x$ ("shadowing") [29,30,31]. However, such experiments were unable to reach deep into the shadowing region, distinguish valence and sea quarks, or probe gluons. Nuclear deep-inelastic scattering with a medium-energy EIC would for the first time allow one to determine the gluon and sea quark densities in a range of nuclei. Thanks to the wide kinematic coverage, the EIC will be able to penetrate deep into the shadowing region, while simultaneously having sufficient $Q^2$ range to extract the nuclear gluon densities through the $Q^2$ dependence of the structure function $F_2^{\rm A}$ at both small and large $x$. Measurements at different energies would isolate the longitudinal structure function $F_{\rm L}^{\rm A}$, which provides direct access to gluons. Using a combination of inclusive measurements and gluon tagging through charm production, an EIC will be able to explore nuclear gluons also in the antishadowing and EMC effect regions — a step that might prove as revolutionary for our understanding of nuclei as the discovery of the quark EMC effect 30 years ago.

Further information on the nuclear modification of the quark/gluon structure of the proton and the neutron can be gained from deep-inelastic measurements with detection of the spectator system of ($A$-1) nucleons in the final state. In particular, scattering from the deuteron with a tagged spectator proton can measure the structure functions of the bound neutron at controlled virtualities, from which the free neutron quantities can be obtained by extrapolation to the on-shell point. Measurements with a tagged spectator neutron, which are extremely difficult with a fixed target but feasible with a collider using a zero degree calorimeter, provide completely new information on the bound proton structure functions that constrains theoretical models of binding effects and the on-shell extrapolation (contrary to the neutron, the free proton structure function is known from independent measurements with a proton target). Tagged measurements on heavier nuclei could explore the effects of nucleon embedding in a complex nuclear environment. Measurements of coherent nuclear scattering, in which the nucleus remains intact and is detected with a small recoil momentum of $\leq$100 MeV in the final state, can map the transverse gluonic radius of nuclei and study shadowing as a function of the impact parameter — information essential for the analysis of high-energy *pA* and *AA* collisions.

Experiments with nuclear targets also provide qualitatively new insight into the short-distance dynamics of deep-inelastic processes. A fundamental prediction of QCD as a gauge theory is color transparency: the interaction of small-size colored configurations with hadronic matter is governed by their color dipole moment and vanishes proportionally to their transverse size. A medium-energy EIC would allow one to test this prediction through measurements of meson electroproduction on nuclei over a wide range of $l_{\text{coh}}$ and $Q^2$, controlling the longitudinal extent of the interaction region and the transverse size of the $q\bar{q}$ dipole. Previous fixed-target measurements (E665 [32], HERMES [33]) could not vary these parameters fully independently.

Detailed studies of color transparency and a reliable determination of the nuclear gluon density in the shadowing region $0.001 < x < 0.1$, as envisaged with a medium-energy EIC, are essential also for a quantitative assessment of the approach to the saturation regime at small $x$. In this regime the transverse density of gluons interacting with a high--energy probe becomes so large that it constitutes a new dynamical scale that enables systematic calculations of inclusive cross sections as well as final-state characteristics [28]. Saturation dynamics is expected to be important in *AA* and central *pp* collisions at the LHC and has been associated with phenomena observed in heavy-ion collisions at RHIC [28]. Nuclear shadowing, as would be established with a medium-energy EIC, may slow down the approach to gluon saturation at small $x$ [25,26,27]. The study of the saturation regime proper will be the object of high-energy colliders such as a high-energy EIC (*eA*) or the LHeC (*ep* and *eA*, see below). This program would involve measurements of inclusive and diffractive nuclear structure functions at small $x$, analysis of final states in which the saturation scale could manifest itself directly (*e.g.*, $p_T$ spectra of leading forward particles), and measurements of multiparticle correlations sensitive to the dynamics of the radiation processes generating the dense gluon medium.

## 2.3 The Emergence of Hadrons from Color Charge

The emergence of colorless hadrons from the elementary color charge produced by short-distance probes — the so-called hadronization process — is a principal aspect of QCD which still lacks a quantitative understanding from first principles [34]. Empirical fragmentation functions, which encode the probability that a quark or gluon decays into a hadron and a colored remnant, have been obtained by fitting experimental data, but knowledge of the underlying dynamics remains sketchy and model-dependent. Basic questions concern the characteristic time scales for the neutralization of color charge (sometimes referred to as pre-hadron formation) and the formation of physical hadrons (see Figure 2.5). Measuring these time scales would be the first step toward understanding how hadrons emerge dynamically from the color charge of QCD, complementing the information obtained from hadron structure and spectroscopy studies.

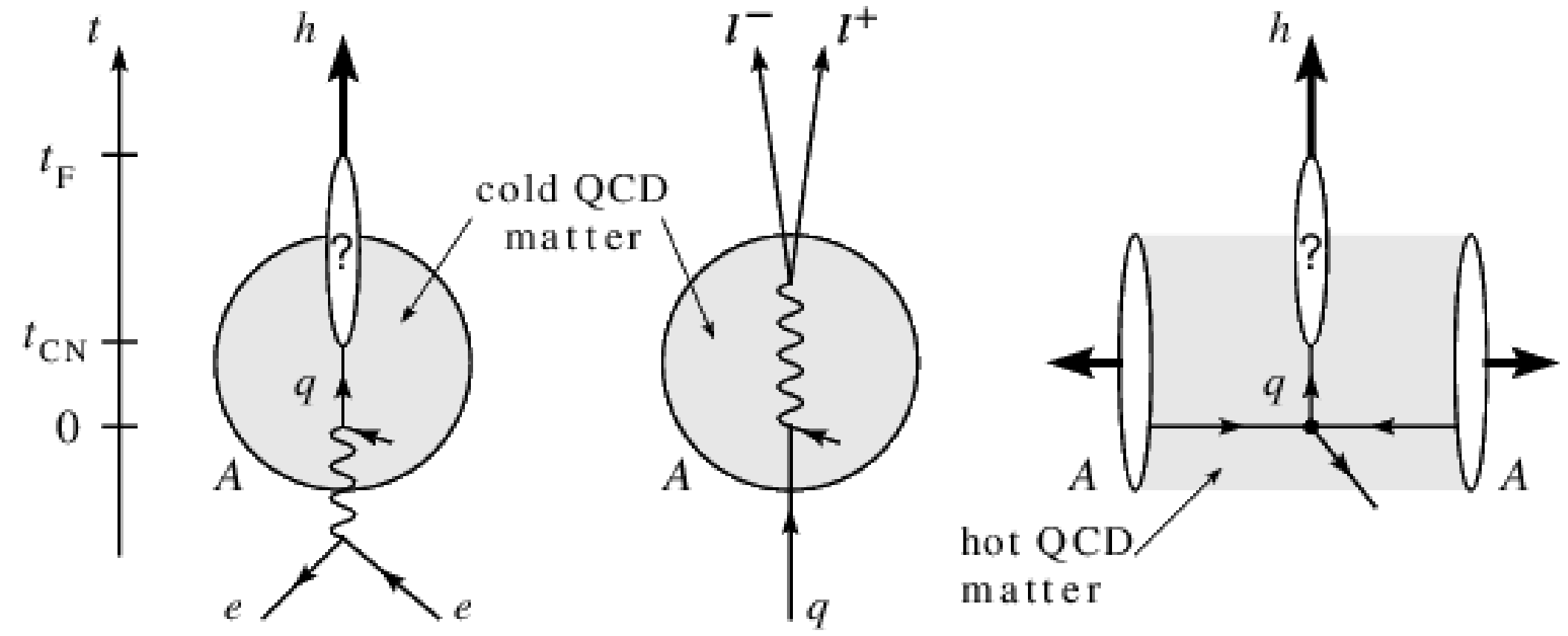

Figure 2.5: Parton propagation and hadronization in cold and hot nuclear matter. The color neutralization ($t_{CN}$) and hadron formation ($t_F$) time scales are indicated on the vertical time axis

Nuclear deep-inelastic scattering provides a known and stable nuclear medium ("cold QCD matter") and a final state with good experimental control of the kinematics of the hard scattering. This permits one to use nuclei as femtometer-scale detectors of the hadronization process (see Figure 2.5). By measuring the energy loss and transverse momentum broadening of leading hadrons induced by the nuclear environment one can discriminate the different dynamical processes (medium-induced gluon bremsstrahlung, pre-hadron reinteraction) and infer the space-time evolution of hadronization. Theoretical models of these processes can be calibrated in *eA* scattering and then applied to study, *e.g.*, the Quark-Gluon Plasma created in high-energy nucleus-nucleus collisions ("hot QCD matter," see Figure 2.5). The combination of high energy and luminosity offered by the EIC promises a truly qualitative advance in this field, compared with current and planned fixed target experiments. The large $Q^2$ range permits measurements in the fully calculable perturbative regime with enough leverage to determine nuclear modifications in the QCD evolution of fragmentation functions; the high luminosity permits the multidimensional binning necessary for separating the many competing effects and for detecting rare hadrons. The large range of virtual photon energies $\nu \approx 10$–1000 GeV allows one to experimentally boost the hadronization process in and out of the nuclear medium, and thus cleanly extract the color neutralization and hadron formation times (small $\nu$) and isolate in-medium parton propagation effects (large $\nu$). The quark and gluon in medium energy loss measured in this way is of major interest in its own right, as it addresses the fundamental question of the interaction of an energetic color charge with hadronic matter in QCD. With an EIC one will be able for the first time to study also the in-medium propagation and hadronization of heavy quarks (charm, bottom) in *eA* collisions, which is necessary to test predictions for their energy loss and confront puzzling measurements of heavy flavor suppression in the Quark-Gluon Plasma at RHIC [35].

Furthermore, an EIC with $\sqrt{s}\geq 30$ GeV will permit for the first time to measure jets and their substructure in *eA* collisions. The modifications compared to jets in *ep* scattering in the same kinematics can be related to the propagation of the colored parton shower in the nuclear medium and offers new insight into its space-time evolution. It can also be used to measure the cold nuclear matter transport coefficients which encode basic information on the non-perturbative gluon fields in nuclei.

Another interesting aspect of hadronization is the evolution of the system from which a color charge has been removed by the hard process. In deep-inelastic *ep* scattering with an EIC at $\sqrt{s}\sim 20$–70 GeV one would for the first time be able to cleanly separate the virtual photon (or current) and target fragmentation regions in the final state and study the properties of the latter using forward detectors. In this way one could follow the materialization of the "color hole" in the nucleon created by the hard process. Measurements of particle correlations between the current and target fragmentation regions (*e.g.*, particles originating from $s$ and $\bar{s}$ or $c$ and $\bar{c}$ quarks) would provide new insight into the nucleon's spin-flavor structure and could reveal dynamical pair correlations in the nucleon's partonic wave function, as are expected to be induced by the dynamical breaking of chiral symmetry in the QCD vacuum.

# 3. Baseline Design and Luminosity Concept

In this chapter, we present a comprehensive overview of the current design concept of MEIC. In the first section, a set of high-level accelerator design goals derived from the MEIC physics program is presented. It is followed by a general description of the baseline design of the accelerator facility in section 3.2 and main parameters of a representative point design of 60 GeV protons and 5 GeV electrons in section 3.3. Luminosities of electron-proton or electron-ion collisions for various energies and ion species are presented in section 3.4 and are followed by a discussion of the MEIC concept for maximizing luminosity. Some critical topics such as electron cooling, polarization and interaction region constraints are discussed in section 3.6. In the last section, tradeoffs of ERL-ring collider designs and technical challenges are presented.

## 3.1 Machine Design Goal

The nuclear physics programs outlined in the previous chapter provide a set of high-level requirements for MEIC at Jefferson Lab as follows:

1. ***Energy***
   The center-of-mass (CM) energy of this collider should be between 15 and 65 GeV. (The value of $s=(4E_eE_u)^{1/2}$ is from a few hundred to a few thousand GeV$^2$, where $E_e$ and $E_u$ are kinetic energies of electron and nucleon) Thus energies of the colliding beams should range
   - from 3 to 11 GeV for electrons,
   - from 20 to 100 GeV for protons, and
   - up to 40 GeV per nucleon for ions.

   Protons or ions with energies below 20 GeV per nucleon are also interesting to investigate certain potentially important physics processes.

2. ***Ion species***
   Ion species of interest include polarized protons, deuterons, and helium-3. Other polarized light ions are also desirable. Heavy ions up to lead do not have to be polarized. All ions are fully stripped at collision.

3. ***Multiple detectors***
   The facility should be able to accommodate up to three detectors with at least two of them available for collisions of electrons with medium energy ions. A third detector is desirable for collisions of electrons with ions whose energies are lower than 20 GeV/u.

4. ***Luminosity***
   The luminosity should be in the range of mid $10^{33}$ to above $10^{34}$ cm$^{-2}$s$^{-1}$ per interaction point over a broad energy range. Further, optimization of luminosity should be centered around 45 to 50 GeV CM energy (the value of $s$ is around 2000 to 2500 GeV$^2$).

5. ***Polarization***
   Longitudinal polarization for both electron and light-ion beams at the collision points should be achieved with greater than 70% polarization. Transverse polarization of the ions at the collision points and spin-flip of both beams are extremely desirable. High-precision (1–2%) ion polarimetry is required.

6. ***Positrons***
   Polarized positron beams colliding with ions are desirable, with a high luminosity similar to that of the electron-ion collisions.

In addition, an MEIC accelerator design should be flexible to allow an option of a future energy upgrade for reaching electron energy up to 20 GeV, proton energy up to 250 GeV, and ion energy up to 100 GeV per nucleon.

## 3.2 Baseline Design

The present MEIC baseline is a traditional ring-ring collider [1,2,3] in which the colliding electron and ion beams are stored in two collider rings. This choice, which has evolved from an ERL-ring collider of an earlier design stage [4], was adopted in 2007 [5] after it was realized that an electron-ERL/ion-ring design could not significantly improve luminosities beyond a ring-ring collider that utilized a high bunch repetition rate. Moreover, an ERL-ring collider scenario would add a substantial burden on the polarized electron source, and many-pass energy recovery is not well established. A brief comparison of two collider scenarios—ring-ring and ERL-ring—for the MEIC design is presented in section 3.7.

The central part of the proposed facility is a set of figure-8 shape electron and ion storage rings as shown in Figure 3.1. The electron ring is made of normal conducting magnets and will store an electron beam of 3 to 11 GeV. The CEBAF SRF linac [6] serves as a full-energy injector into the electron collider ring, requiring no further upgrade for energy, beam current, or polarization beyond the 12 GeV upgrade. The ion collider ring is made of high-field superconducting magnets and will store a beam with energy of 20 to 100 GeV for protons or up to 40 GeV per nucleon for light to heavy ions. The ion beams are generated and accelerated in a new ion injector complex that will be described below. The two collider rings are stacked vertically and housed in the same underground tunnel as shown in Figure 3.2. They have nearly identical circumferences of approximately 1.4 km, occupying a compact footprint of 500 m by 170 m, which is actually smaller than that of CEBAF, as shown in a Jefferson Lab site map in Figure 3.3. In addition, there is depicted in the figures a large figure-8 ring (in light grey color in Figure 3.1 and dashed red line in Figure 3.3) which represents two high energy collider rings (2.5 km or larger) for a future energy upgrade for reaching up to 20 GeV electrons, and up to 250 GeV protons or 100 GeV/u ions. The upgraded high-energy collider can use the same experimental halls and, possibly, the detectors of MEIC, and the medium-energy ion collider ring would then serve as the final booster in staged acceleration of ion beams.

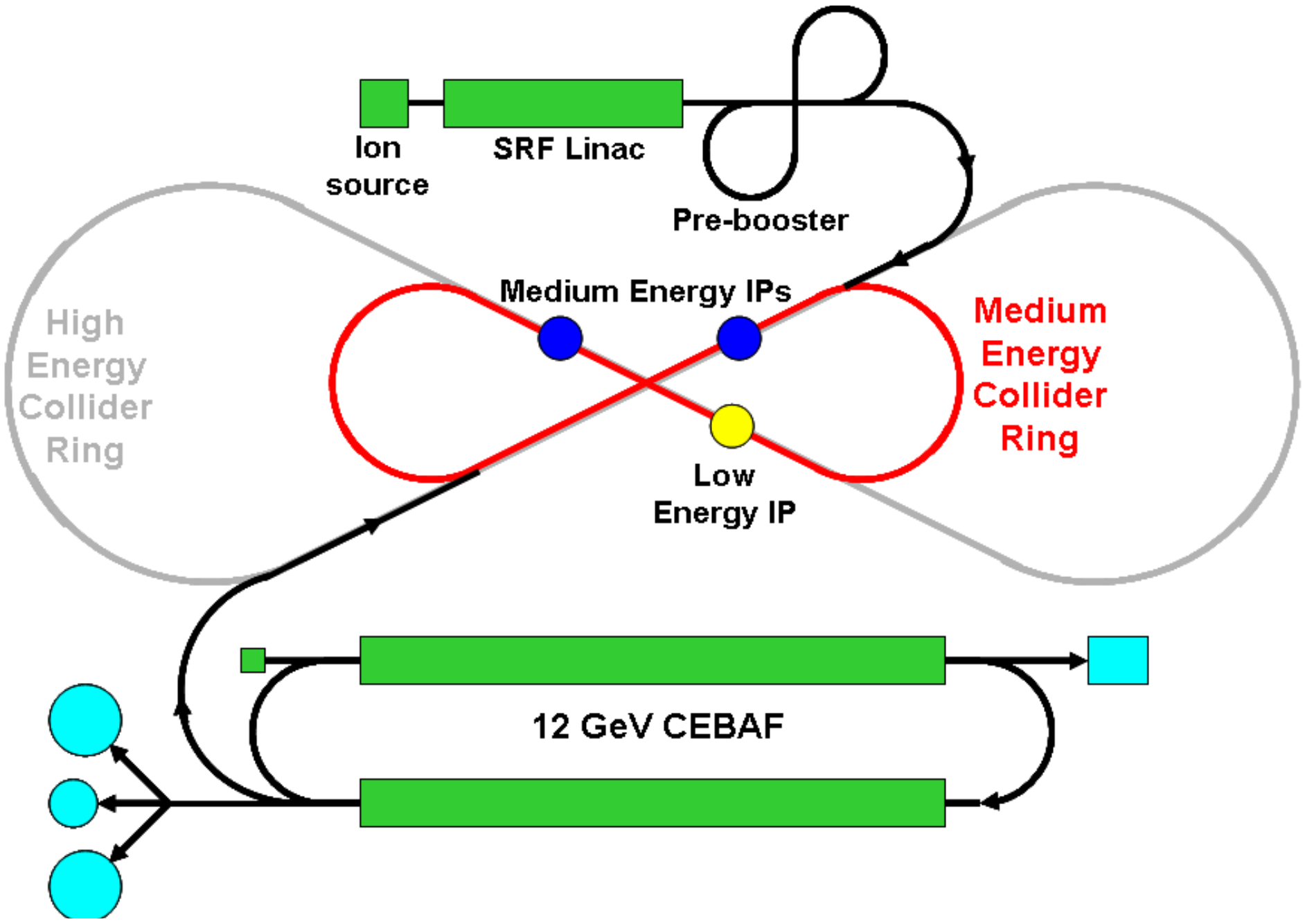

Figure 3.1: A schematic drawing of MEIC. The high energy ring is a future possibility.

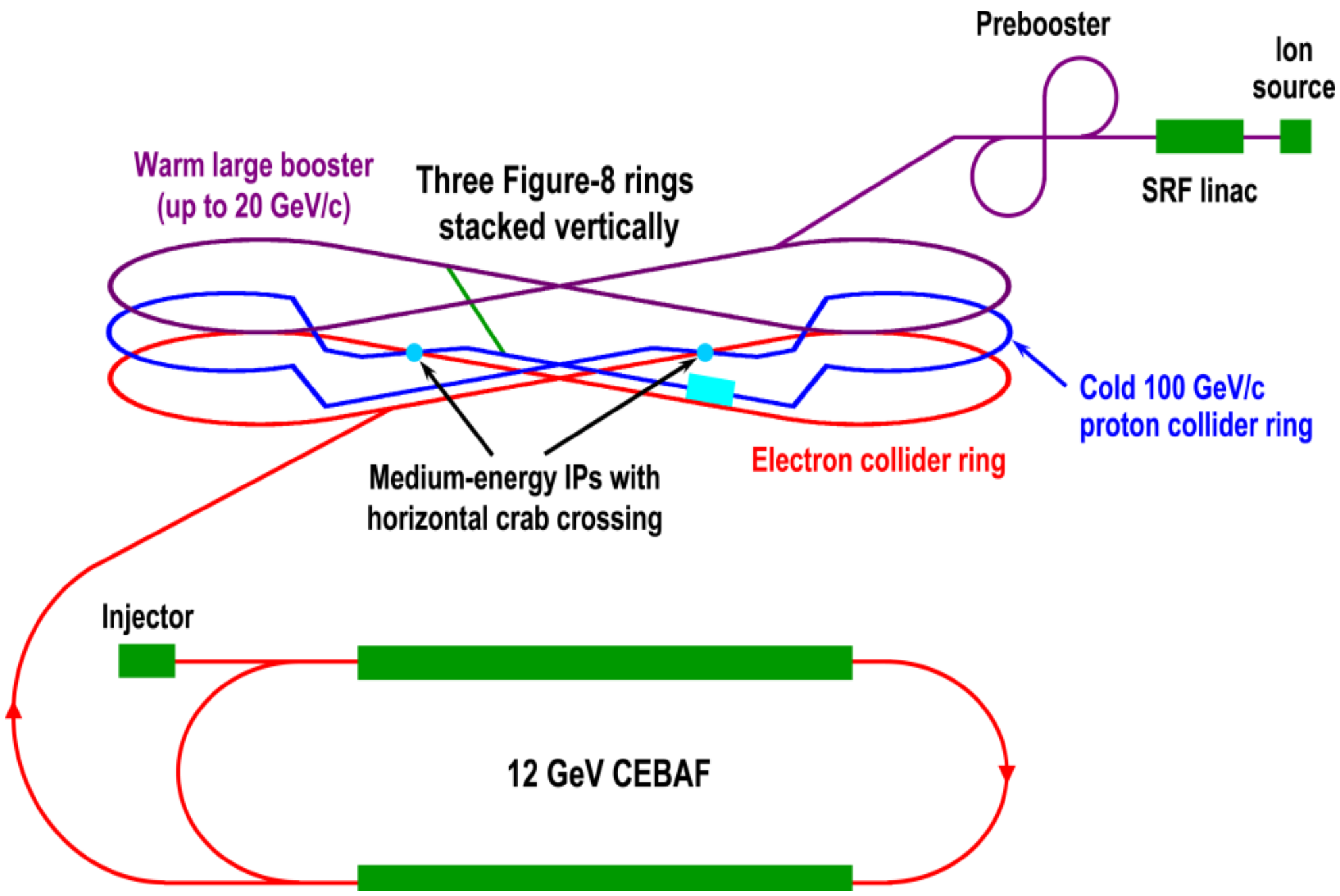

Figure 3.2: The MEIC electron and ion collider rings and the large ion booster rings are stacked vertically and housed in the same tunnel.

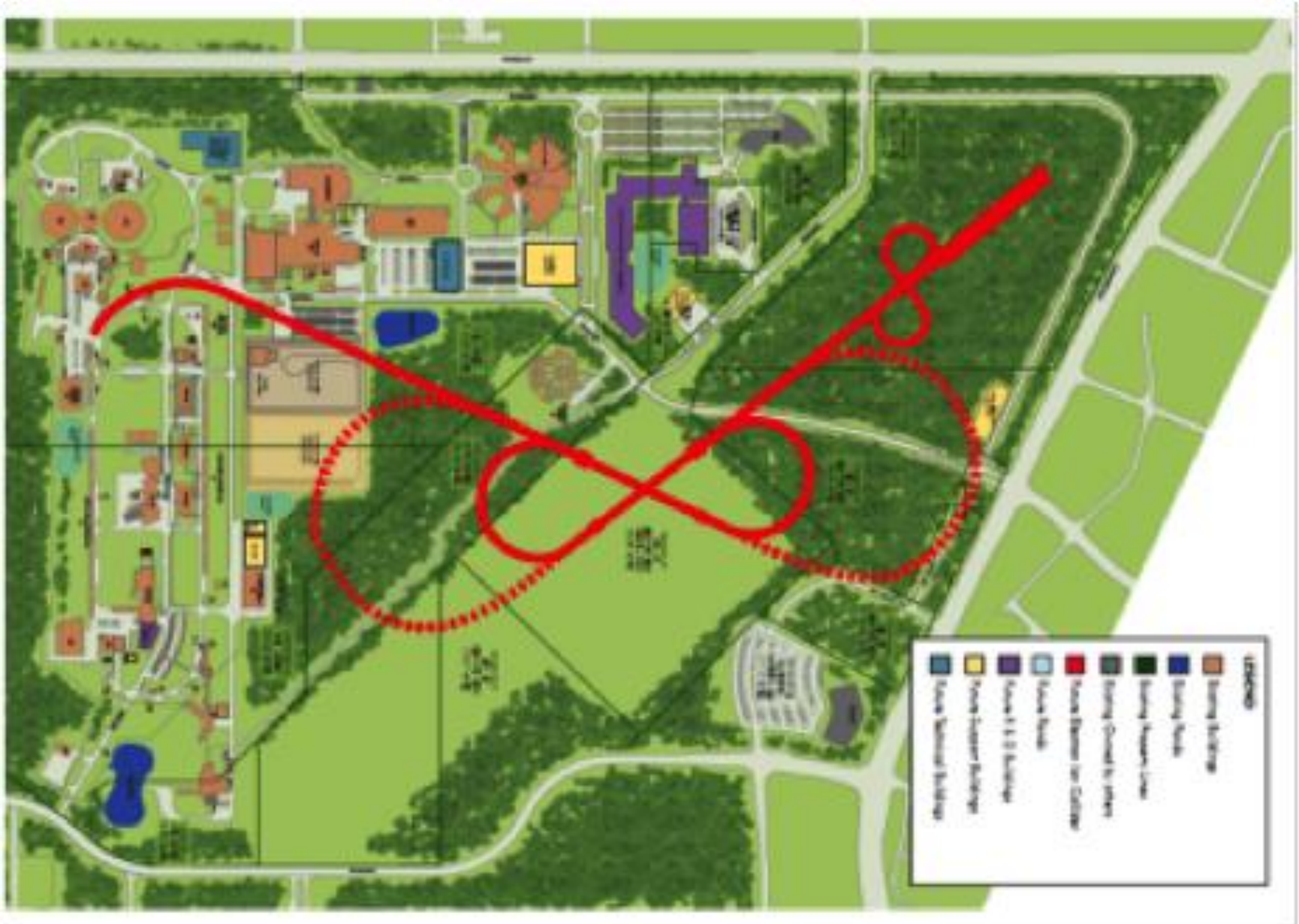

Figure 3.3: MEIC (solid red line) and its future upgrade (dashed red line) on the Jefferson Lab site. The ion injector (sources, linac and pre-booster) is in the upper right corner. CEBAF is on the left with a transfer line to the electron storage ring.

The unique figure-8 shape of the MEIC rings has been chosen for best preservation of ion polarization during acceleration in the multiple rings [7]. The crossing angle of this figure-8 shape is 60º, partitioning a collider ring roughly equally into two half arcs and two long straights. There are two additional short (20 m) straights in the middle of two half arcs for ion spin manipulation elements (Siberian snakes [8]). The electron and ion collider rings intersect at two symmetric points in two long straights as shown in Figure 3.2, thus accommodating two detectors. Presently, the ion beam executes a vertical excursion to the plane of the electron ring to realize a horizontal crossing for electron-ion collisions. The two long straights also accommodate other utility components of the collider rings, among them, injection/ejection, RF system, electron cooling, and polarimetry.

A third interaction point for collisions of electrons and low-energy ions may be added to one of two long straights as shown in Figure 3.1, if dictated by the nuclear physics. In that case, the electron beam participates in collisions at all three IPs while the medium-energy ion beam bypasses this third IP. The new ion beam, which has energy lower than 20 GeV/u, will be stored either in the large booster or in a dedicated third ring of compact size (a fraction of size of the large booster) to reduce the space-charge tune shift.

The MEIC accelerator design takes full advantage of two design elements for delivering high luminosities: an existing high-repetition rate CW electron beam from CEBAF and a new ion complex fully optimized for electron-ion collisions. More specifically, the new ion complex can be specially designed to deliver ion beams that match the phase-space structures (bunch length and transverse emittance) and bunch frequency of the colliding electron beam, enabling implementation of an advanced luminosity scheme based on high repetition rates. This luminosity scheme will be discussed in section 3.5.

In the present design, the MEIC ion complex consists of sources for polarized light ions and non-polarized light to heavy ions, a 280 MeV pulsed SRF ion linac, a 3 GeV pre-booster ring, a 20 GeV large booster, and the medium energy collider ring. All the above energy parameters are for the proton beam; they must be scaled appropriately for ion beams using the mass-to-charge ratio for the same magnetic rigidity. The large booster is currently designed with the identical footprint of the collider rings as shown in Figure 3.2 and thus will be housed in the same tunnel. Figure 3.4 depicts a cross section of the collider tunnel, including the configuration of the long straights if the ultimate energy upgrade is pursued. The large booster is placed above the medium-energy collider ring while the electron ring is below it. The detailed design of this ion complex will be presented in Chapter 5.

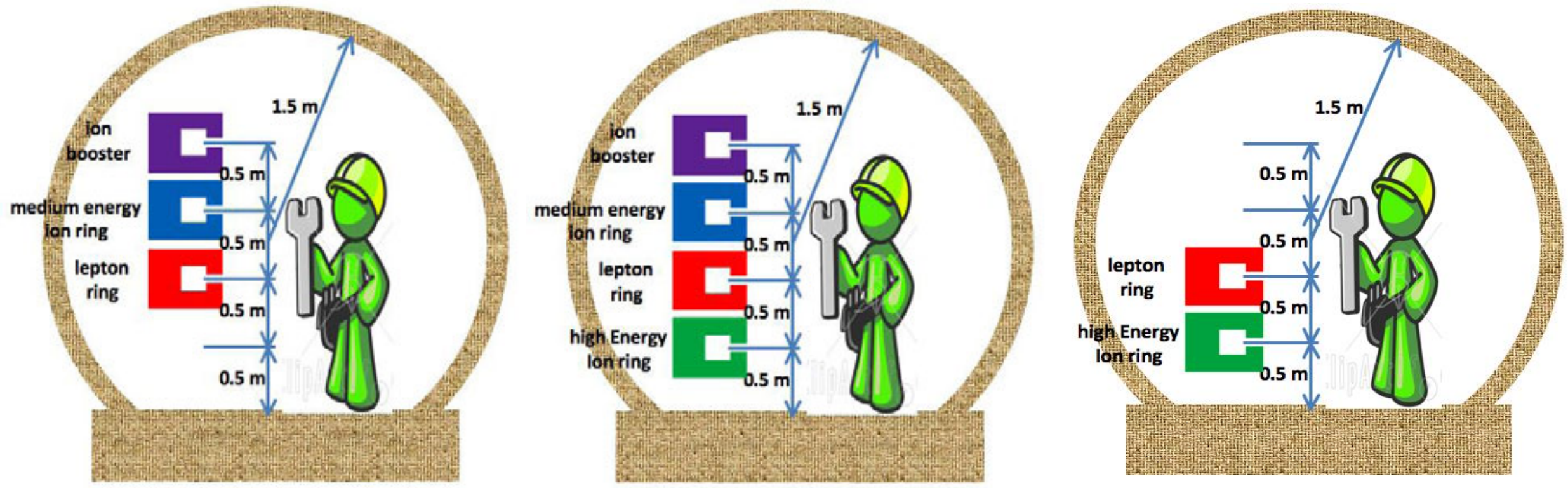

Figure 3.4: Cross sections of tunnels for MEIC and its future upgrade. The left one is for the medium-energy collider ring; the middle and right ones are for long straights and the large ring after the energy upgrade.

We close this section with a brief discussion of three additional MEIC design features. Firstly, since CEBAF is a full-energy injector for the electron collider ring, the filling time of this storage ring is short, about a few seconds. This leads to the following two consequences: the stored beam in the electron ring can be replaced quickly if the beam quality, emittance or polarization in particular, becomes unsatisfactory; CEBAF can operate in a top-off mode to inject electrons periodically into the storage ring. Secondly, the CEBAF fixed target program can be operated simultaneously along with the collider program, with essentially no loss of duty factor. Thirdly, the ring-ring collider design also enables collisions of polarized positrons to ions with luminosity similar to that of the polarized electron beam. The only new component required is a positron source since the CEBAF SRF linac can accelerate positrons as efficiently as electrons. Polarization of the stored positron beam can be achieved through the self-polarization effect of Sokolov-Ternov [9] due to synchrotron radiation if the positron source is not polarized.

## 3.3 Main Design Parameters

The MEIC nominal parameters at a design point of $5\times60$ GeV$^2$ electron-proton collision are presented in Table 3.1. The luminosity reaches $5.6\times10^{33}$ cm$^{-2}$s$^{-1}$ for a full-acceptance detector. To reach this exceptional acceptance, the detector space, i.e., the distance from an IP to the front face of the first final focusing quadrupole, must be at least 7 m for the ion beams; however, it can be shortened to 3.5 m for the electron beam. For the second detector, optimized for higher luminosity while still retaining a fairly large detector acceptance, the detector space can be reduced to 4.5 m so the luminosity is doubled to above $10^{34}$ $cm^{-2}s^{-1}$.

Table 3.1: MEIC main design parameters for a full-acceptance detector.

| | | | |
|---|---|---|---|
| Beam energy | GeV | 60 | 5 |
| Collision frequency | MHz | 748.5 | |
| Particles per bunch | $10^{10}$ | 0.416 | 2.5 |
| Beam current | A | 0.5 | 3 |
| Polarization | % | >70 | ~80 |
| RMS bunch length | cm | 1 | 0.75 |
| Normalized emittance ($\varepsilon^n_x$ / $\varepsilon^n_y$) | µm | 0.35 / 0.07 | 53.5 / 10.7 |
| Horizontal β* | cm | 10 (4) | |
| Vertical β* | cm | 2 (0.8) | |
| Vertical beam-beam tune-shift | | 0.015 | 0.03 |
| Laslett tune-shift | | 0.06 | small |
| Detector space | m | ±7 (4.5) | ±3.5 |
| Luminosity per IP | $10^{33}$ $cm^{-2}s^{-1}$ | 5.6 (14.2) | |

(Values for a high-luminosity detector with a 4.5 m ion detector space are given in parentheses.)

To derive this set of MEIC parameters, certain limits were imposed on several key machine or beam parameters in order to reduce accelerator R&D challenges and to improve robustness of the design. These limits are based largely on previous lepton and hadron collider experience and represent state-of-the-art accelerator technology and are listed below:

- The stored beam currents are up to 0.5 A for protons or ions and 3 A for electrons.
- Electron synchrotron radiation power density should not exceed 20 kW/m.
- Maximum bending field of ion superconducting dipole magnets is 6 T.
- Maximum betatron function at a beam extension area near an IP is 2.5 km.

In addition, the bunch repetition rate is reduced to 750 MHz, half of the nominal CEBAF RF frequency, in order to ease R&D and hardware cost for a high-current high-repetition RF system required by the electron ring. Further, it is a well-established fact that ultimately the collective beam effects pose fundamental limits to the parameters of a stored beam and the performance of a collider [10,11]. In MEIC, we have limited

- the direct space-charge tune-shift [12] of ion beams to 0.1,
- the total beam-beam tune-shift [13] to 0.03 for proton or hadron beams, and
- the total electron beam-beam tune-shift to 0.1 (due to its fast radiation damping).

## 3.4 Luminosity for Different Energies and Ion Species

MEIC is designed to achieve a high luminosity over a relatively broad CM energy region. Table 3.2 shows luminosities of electron-proton collisions for various representative design points. It should be pointed out that the same limits have been applied to the key design parameters such as maximum beam currents and beam-beam tune-shift as discussed in the previous section.

The equilibrium emittance of the electron beam in the storage ring is determined by synchrotron radiation [14]; its normalized value, varying as the third power of the beam energy, can have a factor of 50 variations over an energy range from 3 to 11 GeV. In the case of MEIC electron ring optics design, lattice type and betatron phase advance per base cell are utilized to greatly narrow this variation of emittance. On the other hand, the ion beam emittance is largely determined by the collective beam effects and the efficiency of beam cooling, and it is assumed to be relatively stable in MEIC. As a consequence, the emittances of the two colliding beams of MEIC are usually not matched. In order to match the beam spot sizes at an IP, the value of electron betatron function at the IP ($\beta^*$), which characterizes the strength of the final focus of the electron beam, is adjusted accordingly for different energies, as shown in Table 3.2. Such matching of spot sizes of colliding beams at a collision point is critical in minimizing the effects of the highly nonlinear beam-beam forces and maximizing the tolerable beam-beam tune-shift and ultimately the luminosity.

Table 3.2: MEIC luminosity for e-p collisions

| Energy (*p*/*e*) | GeV | 60/5 | 72/6 | 85/7 | 100/9 | 100/11 |
|---|---|---|---|---|---|---|
| CM energy | GeV | 34.7 | 41.6 | 48.8 | 60.0 | 66.3 |
| Current (*p*/*e*) | A | 0.5/3 | 0.5/2 | 0.5/1.1 | 0.5/0.4 | 0.5/0.18 |
| Particles/bunch (*p*/*e*) | $10^{10}$ | 0.42/2.5 | 0.42/1.7 | 0.42/0.92 | 0.42/0.33 | 0.42/0.15 |
| Proton normalized emittance (*x*/*y*) | mm | 0.35/0.07 | 0.37/0.05 | 0.4/0.03 | 0.45/0.03 | 0.45/0.03 |
| Electron normalized emittance (*x*/*y*) | mm | 53.5/10.7 | 65/8.67 | 90/7.5 | 150/10 | 225/15 |
| Proton β* (*x*/*y*) | cm | 10/2 (4/0.8) | 15/2 (6/0.8) | 24/2 (9.6/0.8) | 30/2 (15/1) | 30/2 (18.8/1.25) |
| Electron β* (*x*/*y*) | cm | 10/2 (4/0.8) | 13.1/1.75 (5.25/0.7) | 16.2/1.35 (6.6/0.55) | 15/1 (7.5/0.5) | 12/0.8 (7.5/0.5) |
| Vertical beam-beam tune-shift (*p*/*e*) | | 0.015 / 0.029 | 0.01 / 0.025 | 0.005 / 0.019 | 0.002 / 0.012 | 0.001 / 0.008 |
| Space-charge tune-shift, proton | | 0.059 | 0.051 | 0.045 | 0.033 | 0.033 |
| Luminosity ($10^{33}$) | $cm^{-2}s^{-1}$ | 5.6 (14.2) | 4.3 (10.7) | 2.6 (6.0) | 1.0 (2.0) | 0.44 (0.71) |

(Values for a high-luminosity detector with 4.5 m ion detector space are given in parentheses.)

In principle, collisions of electrons and ions can be arranged similarly to electron-proton collisions in MEIC, and luminosities of *e-A* collisions should be comparable to *e-p* collisions. Table 3.3 summarizes luminosities for different ions at their maximum energies in the MEIC ion ring and assumes the same electron beam parameters for all cases. The maximum ion energies are derived from the highest proton energy (100 GeV) in the ring by a scaling factor of *Z/A*, the charge-mass ratio, assuming the same maximum field strength of arc bending dipoles. The stored ion currents are capped again at 0.5 A, as in the case of protons in Tables 3.1 and 3.2.

Table 3.3: MEIC luminosities for different ion species.

| | | Proton | Deuteron | Helium | Carbon | Calcium | Lead |
|---|---|---|---|---|---|---|---|
| Ion species | | P | d | $^{3}He^{++}$ | $^{12}C^{6+}$ | $^{40}Ca^{20+}$ | $^{208}Pb^{82+}$ |
| Ion energy | GeV/u | 100 | 50 | 66.7 | 50 | 50 | 40 |
| Ion current | A | 0.5 | | | | | |
| Ions per bunch | $10^{9}$ | 4.2 | 4.2 | 2.1 | 0.7 | 0.2 | 0.05 |
| Ion β* (x/y) | | 6/2 (2.4/0.8) | | | | | |
| Ion beam-beam tune shift (vertical) | | 0.014 | 0.008 | 0.01 | 0.008 | 0.008 | 0.006 |
| Electron beam | | Energy: 6 GeV; Current: 3 A; Electrons per bunch: $2.5\times10^{10}$;<br>Vertical β*: 1.55 to 2.8 cm (0.61 to 1.1 cm)<br>Vertical beam-beam tune shift: 0.022 to 0.029 | | | | | |
| Luminosity/IP ($10^{33}$) | $cm^{-2}s^{-1}$ | 7.9 (19.8) | 5.5 (13.8) | 7.3 (18.6) | 5.5 (13.8) | 5.5 (13.8) | 4.4 (11.0) |

(Values for a high-luminosity detector with a 4.5 m ion detector space are given in parentheses.)

## 3.5 Luminosity Concept

The key to the MEIC high luminosity is utilizing high bunch repetition rate CW colliding beams. Briefly, both the electron and ion beams have very short bunch length and small transverse emittances such that a strong final focusing can be applied to reduce the beam spot sizes to as small as a few μm at collision points. This configuration combined with a high bunch repetition rate greatly boosts the collider luminosities. An ultrahigh bunch repetition rate ensures a very small bunch charge (hence relatively much weaker collective and inter beam effects) of the colliding beams, particularly the ion beams, while maintaining the same average beam current, hence a high luminosity. A detailed discussion of this concept can be found in reference [15]. It should be noted that this luminosity concept has been proved by several lepton-lepton collider B-factories worldwide and has led to the present world record of the highest achieved luminosity at the KEK-B factory [16].

To illustrate this concept, let us first examine the standard luminosity formula as a function of a beam-beam parameter for a head-on electron-proton collision [17]

$$L = \frac{\gamma_p N_p f_c \xi_{y,p}}{2 r_p \beta^*_{y,p}} \left(1 + \frac{\sigma^*_y}{\sigma^*_x}\right) \tag{3.1}$$

where the beam-beam parameter (tune shift) for the proton beam is

$$\xi_{y,p} = \frac{r_p N_e}{\gamma_p} \frac{\beta^*_{y,p}}{2\pi\sigma^*_y(\sigma^*_x + \sigma^*_y)} \tag{3.2}$$

The other parameters in these two formulas are the collision frequency (bunch repetition rate) $f_c$, numbers of electrons and protons per bunch $N_e$ and $N_p$ , and horizontal and vertical beam spot sizes (assuming matched) at the collision point $\sigma_x$* and $\sigma_y$*. Values of these parameters depend on the collider design and are usually limited by collective beam effects. The luminosity formula (3.1) can also be cast using the electron beam-beam parameters.

It is clear that a high luminosity could be achieved by increasing both the beam current $N_p f_c$ and the beam-beam tune shift $\xi_{y,p}$ as well as decreasing the vertical $\beta_y$*. Since the ion bunch charge is limited not only by the highly nonlinear beam-beam effects but also by additional collective beam effects such as space-charge tune shifts, increase of bunch repetition rate is a highly preferable way to boost the proton beam currents. On the other hand, one must also watch the electron beam-beam tune shift. However, at the low ion charge per bunch due to high repetition rates, the electron beam-beam tune shift can also be kept at reasonable, already achieved levels. This is the principal advantage of high repetition rates. Present experience from years of operation of existing ring-ring colliders shows that a value of 0.035 is a practical limit of the total beam-beam tune shifts for hadron beams; nevertheless the recent LHC operation has indicated that such a limit may be even higher [18]. It is roughly a factor of four to five times larger for lepton beams thanks to their synchrotron radiation damping. With these constraints, the luminosity of a ring-ring collider can be optimized by pushing up the bunch collision frequency and squeezing down the $\beta$* values.

For all ring-ring colliders involving hadron beams, the collision frequency, i.e., bunch repetition rates, is traditionally very small; therefore, there are very small numbers of bunches per beam, ranging from just a handful (9 for SPS) to several dozen (36 for Tevatron). It is worth to note that RHIC now operates at up to 112 bunches (about 8.9 MHz repetition rate). LHC indeed has more bunches however the circumference is much large, so the bunch repetition rate is 34 MHz, a significant increase from all previous hadron colliders. With the large bunch charges (up to $10^{11}$ and above) necessary to maintain even a modest hadron beam current, bunch lengths are usually very long (of the order of 1 m), mainly due to limits of collective effects and scattering processes. Long bunches prevent a strong final focusing (small $\beta$*) due to the so-called hourglass effect [19], and combined with large transverse emittance without beam cooling, lead to fairly large beam spot sizes at collision points. All this results in lower than optimal luminosities.

This high-repetition-rate luminosity concept have been validated at today's lepton collider B-factories [16], which use very large bunch collision frequency (storing tens to hundreds times more bunches compared to the hadron colliders) and hundreds times smaller $\beta$* (and corresponding spot sizes) at collision points through the strong final focusing enabled by short bunch lengths. The net effect, with the appropriate interaction region design discussed below, is a several orders of magnitude increase of luminosity in a storage ring on storage ring scenario.

To summarize, it has been a primary design strategy of MEIC to break away from the traditional approach of hadron colliders and to adopt this luminosity concept of short bunch and high repetition rate CW beams in a ring-ring collider involving ion beams. The MEIC luminosity concept can be summarized as follows:

- Very short bunches for both electron and ion beams
- Very small transverse emittance
- Ultrahigh collision frequency CW beams
- Staged electron cooling
- Very small final focusing $\beta^*$
- Large attainable beam-beam tune shift
- Crab crossing of colliding beams

The first three items above specify the design of MEIC colliding beams in terms of the phase space structure (bunch length and emittance) and time structure (bunch frequency and CW). The staged electron cooling is the essential ingredient for formation of these high performance ion beams. The last three specify design of MEIC interaction regions in order to take advantage of high bunch repetition CW colliding beams to achieve high luminosity. Crab crossing is required to eliminate parasitic collisions and the associated harmful long-range beam-beam effects. In the next section, we provide more detailed considerations for these subsystems.

## 3.6 Luminosity-Enabling Subsystems

CEBAF will deliver an extremely high quality CW electron beam at a 748.5 MHz bunch repetition rate. We will utilize this facility as a full-energy injector to the electron storage ring. On the ion beam side, we need to produce and store high-average-current beams with matched properties in terms of bunch length, emittance and repetition rate. Such ion beams have yet to be achieved in practice; however, they are technically feasible given advances in accelerator technologies over the last several decades—notably in ion sources, SRF linacs and the cooling of ion beams. Significant studies have been devoted over the last several years at Jefferson Lab to the accelerator physics of such a modern ion complex, and this work provides confidence that these capabilities are realizable.

It is quite clear that in the energy range of MEIC, unlike the electron beam, the synchrotron radiation damping of the ion beams essentially does not exist. A damping mechanism must be introduced into the MEIC ion collider ring for reduction of the 6D phase-space volume that the ion beam occupies to deliver high luminosity. The best candidate for such damping is staged electron cooling [20] of the ion beam. In addition to a conventional low energy DC electron cooling at the prebooster to assist accumulation of heavy ion beams, this scheme employs electron cooling first at the injection energy of the ion collider ring for initial 6D emittance reduction, and then at the top beam energy for conditioning the beam to the designed phase space density for collision. This process for formation of MEIC ion beams together with a detailed description of the ion complex in general will be presented in Chapter 6 of this report. Electron cooling also will be utilized continuously during collisions in order to suppress intra-beam scattering and other nonlinear collective effects. The cooling electron beam must have good quality at two to three times the current compared to the ion beam, and therefore an essential part of the MEIC is an ERL-based circulator electron cooler to deliver such an electron beam. This critical system is discussed in detail in Chapter 6.

The MEIC luminosity concept could not work without a proper design of the interaction regions, and several important issues warrant special attention. The first issue is elimination of a large number of parasitic collisions due to the ultrasmall (about 40 cm) bunch spacing that comes with the 748.5 MHz repetition rate. Current MEIC detector design requires a magnet-free space of ±4.5 to ±7 m near a collision point, and thus could generate more than 70 parasitic collisions in a head-on collision setup. It is well known that long-range beam-beam interactions at parasitic collisions are major sources of beam loss, detector background, and luminosity lifetime reduction. Following the KEK-B e+e- collider [21], the MEIC design adopts a finite crab crossing [22] angle with a crab-cavity scheme to mitigate this problem. A crossing angle of 50 mrad or larger is sufficient to separate the two MEIC beams to eliminate all parasitic collisions. The MEIC baseline design includes multi-cell SRF crab cavities on both sides of a collision point to recover luminosity loss by restoring head-on collisions. The second issue is correction of the large natural chromaticity [21,23,24,25] induced by strong final focusing. The value of the natural chromaticity, to first order, depends on the ratio of the final focal length and interaction region $\beta^*$. In the MEIC design with $\beta^*$ of a few centimeters, the natural chromaticity could be 10 times larger than that in existing hadron colliders. This issue is further amplified in MEIC compared to B-factories since the magnet-free space (which roughly equals the final focal length) in KEK-B and PEPII is much smaller (less than 1 m) [21,26]. A chromatic compensation must be carefully designed such that it not only provides adequate correction of chromaticity but leaves a large enough dynamic aperture for good beam lifetime.

An additional unique design feature of MEIC is a figure-8 shape of all ion rings. As pointed out above, it provides superior preservation of high polarization of ion beams. The basic idea is a complete cancellation of spin precession in the left and right half ring of the figure-8 ring; thus the net spin tune [27] is zero. The spin tune can be further made to be a half integer with insertion of Siberian snakes, leading to a stable spin polarization in the figure-8 ring. In both cases, the spin tunes are energy independent, thus eliminating crossing of many dangerous spin resonances [27] during acceleration of ion energies. The selection of a figure-8 configuration does not introduce any technical challenges or design complications, and the impact on the project cost is minimal.

## 3.7 Linac-Ring Tradeoffs and Technical Challenges

The standard argument for the advantage of the ERL-ring collider proceeds from the following logic [1]. In an ERL-ring collider, the electron beam-beam tune shift can be much higher than in a ring-ring collider because the electron beam does not have to be stored for long periods of time, merely transported to an electron dump. This increased flexibility in the choice of the electron beam-beam parameter should allow larger circulating ion current and hence larger overall luminosity. In practice, however, because of limitations on the electron beam current—given that even in an ERL, the electron current will not be at the multi-ampere levels that have been achieved in storage rings—a high disruption configuration must be established that limits this advantage. In the MEIC design, moreover, with gigahertz repetition rate ion and electron beams, the electron beam-beam tune shift can be maintained at reasonable levels with luminosities competitive with ERL-ring arrangements. So, indeed, no advantage remains.

The principal cost for an ERL-ring collider is that (polarized) electrons must be produced at a very high average current (0.1 A to 1 A), several orders of magnitude beyond current technical capabilities. Because of the technical difficulties in making the ERL electron source, Jefferson Lab's approach to electron-ion collider design has been to prefer a ring-ring collider as long as there is only a small luminosity penalty in doing so. In explorations of ring-ring designs for electron-ion collider applications in 2006 [2], considerations along the lines of the previous paragraph showed that the luminosity penalty by replacing the energy recovery linac for electrons by the storage ring was not as great as might have been earlier expected, *a priori*. In fact, for MEIC parameters, with the electron and ion interaction region optics identical between the two cases, the ultimate luminosity is identical. The simple luminosity scaling formulas can explain this somewhat surprising result.

The argument comparing the ring-ring and ERL-ring collider is presented in reference [28]. As a *gedanken* experiment, suppose that one has a stable ring-ring collider design. Because the ions must remain stable in the ring in going from a ring-ring collider to an ERL-ring collider, there isn't much possibility to change their beam-beam parameters up (in fact most of our parameter lists have assumed the ion beam-beam parameters are at the level of performance in the best present colliders). If we compare colliders at the same ion energy, $\gamma_i$ cannot change, and $r_i$ is a fundamental constant. Luminosity gains must arise from the possibility of increasing the ion current $fN_i$ circulating in the ring, or from decreasing the ion vertical beta-function at the interaction point. If the parameters in the ion ring are already chosen so that there is little possibility to further increase the current in going to the ERL-ring design—e.g., the ion current is already limited by some non-beam-beam effect such as the Lasslett tune shift or IBS heating that can no longer be electron cooled—there will not be any possibility to increase the luminosity in an ERL-ring design compared to a ring-ring design from the current factor. Likewise, a similar argument shows that unless there is some advantage to the ERL-ring design that allows one to design in smaller vertical beta-functions for the ions at collision, again, there will be no inherent advantage of the ERL-ring design.

One principal advantage of the ERL accelerator for the electrons is that the emerging emittance is expected to be lower than that of a similar current stored in a storage ring. Because in a collider the electron spots cannot be made arbitrarily smaller, to take advantage of the better emittance, due to the nonlinear beam-beam effect, and because, unfortunately, for collider applications where the ion beam-beam tune shift is the most important limit to satisfy in order to allow stable motion in the ion ring, the luminosity no longer depends directly on the electron emittance. The principal advantage of the ERL-ring collider has limited value to increase collider luminosity in many parameter regimes, including the MEIC parameter regimes as seen in Reference [29].

# 4. Electron Complex

## 4.1 Introduction

In this chapter, we will present the design concept for the MEIC electron complex. The central component of this complex is an electron collider ring that will be used to store a high repetition rate and high current polarized electron or positron beam for collisions at multiple interaction points. The main design requirements are:

- The collider ring should accommodate electrons with 3 to 11 GeV energy.
- The total current of the stored beam should be up to 3 A for energy up to 6 GeV; however, the current can be lower for higher electron energies.
- The stored beam should have a very short bunch length (1 cm or less RMS size) and small transverse emittances over a wide energy range to support the luminosity concept.
- The polarization of the stored beam should be at or above 70%, with a lifetime longer than 20 minutes.
- The collider ring should support three independent collision points/detectors.

On the accelerator technology side, we require that:

- The collider ring uses warm magnets with a maximum dipole field less than 1.7 T.
- The maximum linear density of synchrotron radiation power [1,2,3] should not exceed 20 kW/m and the total power should be less than 10 MW.

The basic feasibility of this electron collider is clear-cut given the several highly successful lepton storage rings with similar beam energies, currents, and bunch repetition frequencies in lepton-lepton colliders and modern light sources. The following are design choices we have taken to address some special needs of the MEIC electron collider ring as well as to optimize performance of the complex:

1. ***Full energy injection***
   It is considered an advantage to utilize the 12 GeV CEBAF as a full energy injector into the electron ring. No acceleration will be required in the ring in its baseline mode.

2. ***Top-off injection option***
   One advantage of a full energy injector is that the collider ring can be continuously or periodically refilled. This top-off injection capability may prove to be crucial for maintaining high current of the electron beam, and thus high luminosity of MEIC.

3. ***Figure-8 shape collider ring***
   Though the choice of a figure-8 shape of the collider ring is entirely for the benefit of ion polarization, the electron ring will adopt the same ring topology for the purpose of housing both colliding rings in the same tunnel to minimize costs. This choice should not compromise electron beam quality, nor harm the electron polarization.

4. ***Flat ring***
   The electron ring is completely flat to avoid emittance degradation and depolarization caused by synchrotron radiation [4] in the vertical bends, and also to minimize synchrotron radiation at the interaction regions. The ions will execute vertical excursion to the plane of the electron ring for collisions.

5. ***Arrangement of polarization in the collider ring***
   Electron spins are kept in a vertical direction and anti-parallel to the dipole magnetic field in the arcs in order to utilize the Sokolov-Ternov effect [5] for maintaining a high polarization. To achieve a longitudinal polarization at the collision points, a set of spin rotators [6] will be used to align spin to desired directions in the arcs and in the straights.

This chapter is organized as follows: a brief description of the CEBAF recirculating SRF linac and the injection scheme of the MEIC electron collider ring will be presented in section 4.2. It is followed by the optics design of the electron collider ring in section 4.3 and a short discussion of the RF system for this collider ring in section 4.4. Two important topics, electron polarization and beam instabilities, are covered in sections 4.5 and 4.6 respectively.

## 4.2 CEBAF as a Full Energy Electron Injector

The Continuous Electron Beam Accelerator Facility (CEBAF) at Jefferson Lab is the world's first high energy (above 1 GeV) recirculating SRF linear accelerator [7]. It can accelerate electrons up to 6 GeV, but an energy-doubling upgrade to 12 GeV is in progress and expected to be completed by 2015. The discussions and parameters provided in this section and in the rest of this report, unless otherwise specified, refer to the 12 GeV upgraded CEBAF.

As shown in Figure 4.1, the 12 GeV CEBAF consists of two 1.1 GeV SRF linacs connected by two sets of five vertically separated arc beam lines to form a mile-long racetrack. These ten arcs are needed to accommodate different energies of the electron beam during multiple passes through the two linacs in the racetrack. An electron beam is generated in a polarized photo-cathode DC injector and is accelerated in five passes of the two SRF linacs to 11 GeV for hitting the fixed targets at three currently existing experimental halls (A, B and C) or with an extra pass of one of the two SRF linacs to 12 GeV for experiments at a new hall (D). Electron energies below 11 or 12 GeV can also be made available by adjusting the number of passes and the accelerating gradients of the SRF linacs. One important feature of CEBAF is a high repetition rate CW beam allowed by the 1.497 GHz frequency of the SRF cavities. Beam can be split three ways by an RF beam switcher for delivery of beam to three halls simultaneously (Halls B and D taking beam alternatively), with each of the three beams at a 499 MHz bunch repetition frequency. The electron beam from the source reaches over 80% polarization, which can be nearly perfectly preserved during acceleration in the recirculating linac.

Table 4.1 summarizes the main parameters of the electron beam from the 12 GeV CEBAF. The total current of three beams is 90 μA CW, limited primarily by the maximum beam power acceptable to the dumps for used beams of the fixed targets. As an injector to the MEIC electron collider ring, the current from the CEBAF can be higher, up to 1 mA in a macro-pulse. Such beam current has already been demonstrated experimentally at Jefferson Lab and represents a readily achievable value for the MEIC design concept. The bunch length is extremely small, with an RMS bunch length of 1 ps or less.

It should be pointed out that the CEBAF SRF linac is able to accelerate a positron beam as effectively as the electron beam, provided there is a positron source and all magnets switch their polarizations. A positron source R&D program is currently underway at Jefferson Lab for the future CEBAF fixed target program. With a much smaller current from the CEBAF linac, the accumulation time of a stored positron beam in the collider ring could be much longer compared to an electron beam. Furthermore, the stored positron beam would most likely be un-polarized initially since the current of a polarized positron beam is too small to be practical for a storage ring. Ultimately, positron beam polarization must rely on the Sokolov-Ternov effect [5], provided the beam energy is sufficiently high (9 GeV and above) for attaining a reasonable self-polarization time. We discuss this polarization issue in section 4.7.7.

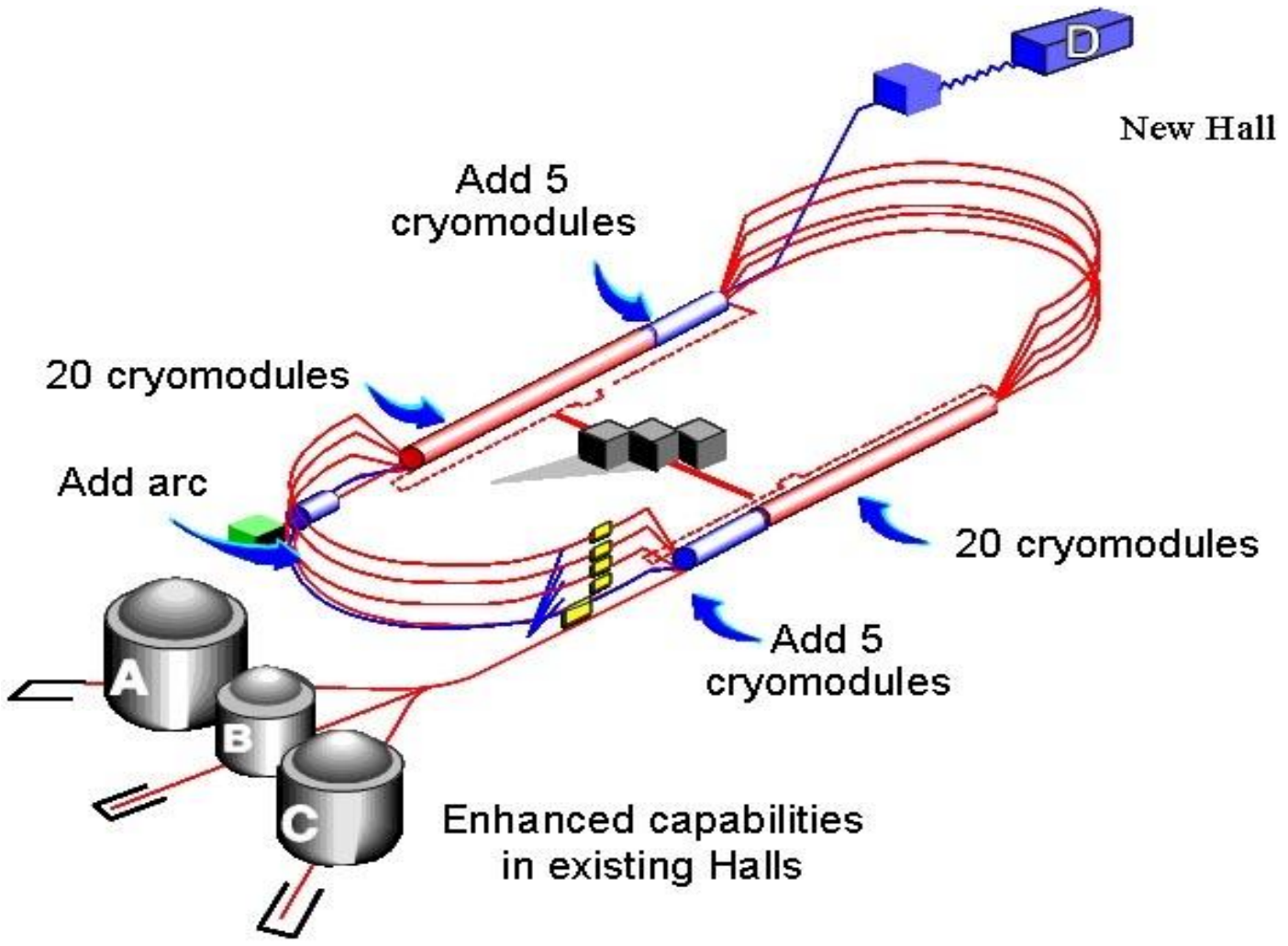

Figure 4.1: A schematic drawing of the CEBAF recirculating linac and experimental halls. It also illustrates the technical approach of the 12 GeV upgrade

Table 4.1 Nominal parameters of the 6 GeV and 12 GeV CEBAF

| 12 GeV Upgrade | | Before | After |
|---|---|---|---|
| Maximum energy to Halls A,B,C / D | GeV | 6 | 11 / 12 |
| Number of passes for Halls A,B,C / D | | 5 | 5 / 5.5 |
| Max. current to Halls A + C / B | μA | 200 / 5 | 85 / 5 |
| Emittance at maximum energy *x*/*y* | nm-rad | 1 / 1 | 10 / 2 |
| Energy spread at maximum energy | $10^{-5}$ | 2.5 | 50 @ 11 GeV<br>500 @12 GeV |
| Bunch length (RMS) | ps | 0.2 | ~1 |
| Polarization | | ~80% | ~80% |

To utilize CEBAF as a full energy injector, an RF deflector will be needed to divert the beam from the CEBAF racetrack and send it to the collider ring through a transport beam line. CEBAF will be operated at a 748.5 MHz bunch repetition rate (filling every other bucket [8] in the bunch train) by adjusting the photo-cathode driving laser pulse repetition rate. A stored beam of 3 A average current will be accumulated by stacking the CEBAF linac beam with 1 mA current. The synchrotron radiation damping [9] will assist phase space equilibration of the accumulated beam. The injection scheme is simple: a 10-turn injection of the CEBAF linac beam (about 45 μs, assuming 1.35 km nominal circumference of the electron collider ring) followed by a 50 ms pause (more than 5 radiation damping times) to allow phase space equilibration, then repeating this cycle of injection-equilibration approximately 300 times. The total injection/accumulation time is about 15 s, very short compared to the design lifetime (minimum 20 min) of the beam current or polarization, and therefore presents an opportunity for continuing the fixed target program of CEBAF. Figure 4.2 illustrates structures of both micro and macro bunch trains from CEBAF, and their parameters are listed in Table 4.2.

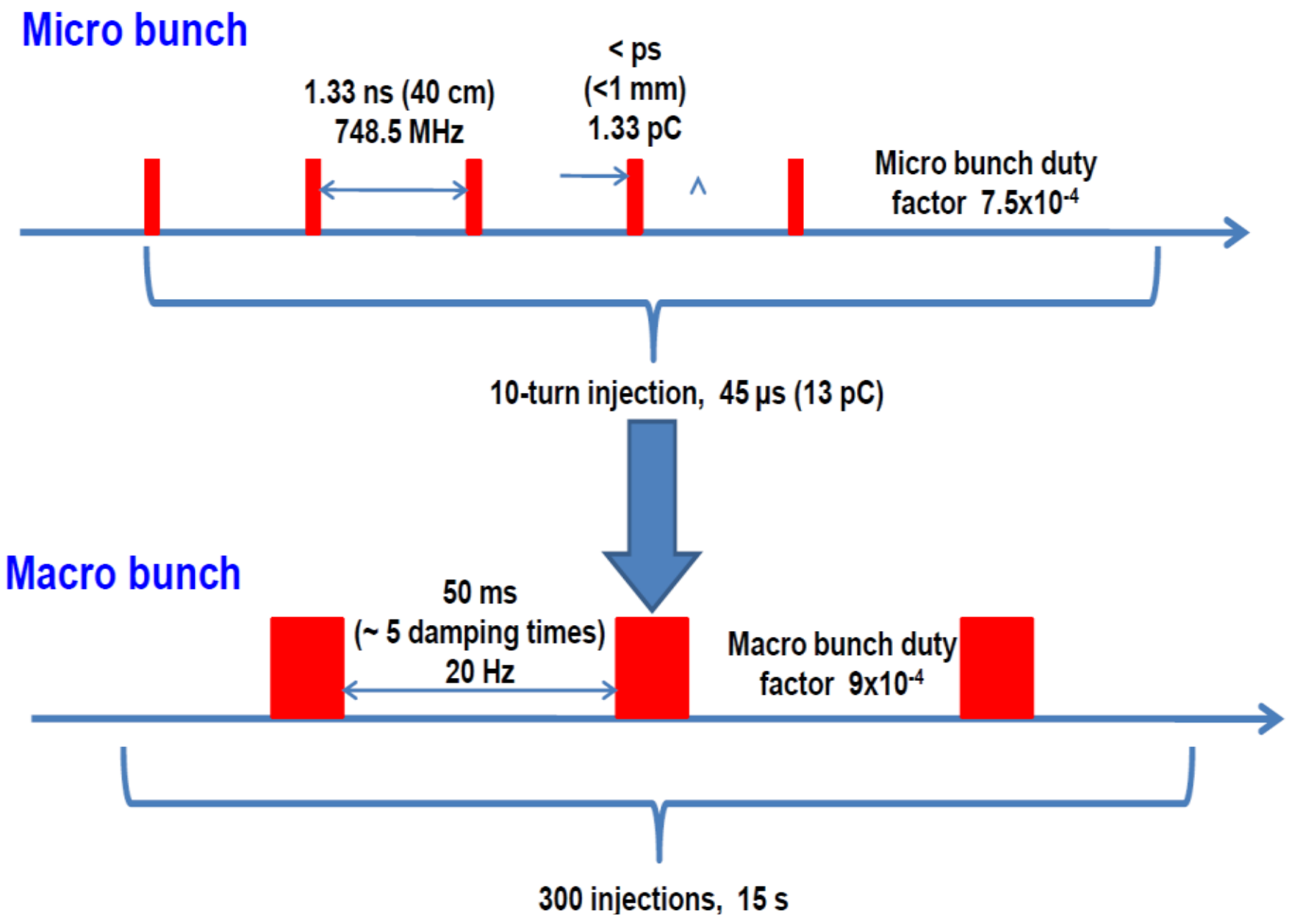

Figure 4.2: Macro and micro bunches of injected beam from CEBAF

Table 4.2 Parameters for injection of electron beam into the MEIC collider ring

| Bunch train | | Micro bunch | Macro bunch |
|---|---|---|---|
| Injection current | mA | 1 | |
| Number of turns in each injection | | 10 | |
| Bunch train length | μs | 5 | |
| Injection beam repetition rate | MHz | 750 | |
| Injection bunch charge | pC | 1.33 | |
| Micro bunch spacing | ns | 1.33 | |
| Macro bunch repetition rate | Hz | | 20 |
| Macro bunch spacing | ms | | 50 |
| Duty factor | $10^{-4}$ | 7.5 | 9 |

## 4.3 Electron Collider Ring

### 4.3.1 Layout

While a compact design of the MEIC electron collider ring is naturally favored as it would be most cost effective, one practical constraint of the ring design is that, because both electron and ion collider rings will follow the same footprint, size and dimension of these rings should be sufficiently large to be able to accommodate all machine components of both electron and ion rings. The components and their dimensions and locations are listed as follows.

- Universal Spin Rotators [10] for the electron beam, each consisting of two solenoids and two sets of arc dipoles of 8.8° and 4.4° bending angles respectively, 47.6 m in length and located on each end of two arcs.
- Short straights for Siberian snakes [11] for polarized ions, 26 m in length and located in the middle of two arcs.
- Three interaction regions [12], each 125 m long, and placed in long straights of the figure-8 ring.
- An electron cooling channel in each of two long straights, 30 m in length and placed near the vertex of the figure-8 ion ring.
- Electron and ion polarimeters.
- Usual storage ring elements for injection, ejection, RF systems, etc.

From a design point view, these components come into the collider ring optics design as geometric constraints. For example, the electron ring arcs must contain short straight sections matching those in the middle of the ion ring arcs where Siberian snakes are placed, and similarly, the ion ring arcs must conform to the geometry of the electron spin rotators located at both ends of each electron arc. As a guide for optics design, Figure 4.3 presents a possible layout of the MEIC electron collider ring that could accommodate all the above machine components. Figure 4.4 shows a footprint of the baseline ring optics design with main parameters listed in Table 4.3. The crossing angle of the figure-8 is 60°; it should provide sufficient separation of the two collision points on two separated straights so adequate sized experimental halls can be constructed. This crossing angle also partitions the circumference of the ring roughly equally into two half arcs and two straights, providing sufficiently long arcs (of 240° per half arc given the figure-8 reverse bends on each side of the crossing) to accommodate the higher number of bending dipoles required by the highest ion energies and for reduction of electron synchrotron radiation power.

It should be noted that a half arc of the figure-8 electron ring, either on the left or right side, comprises a beam line section consisting of two Universal Spin Rotators, two quarter arc sections and a short straight section in a sequence as shown in Figure 4.3. The name of quarter arc is just for convenience since that section of arc actually only bends electrons 106.8° rather than 120° because the spin rotators also contain arc bending dipoles.

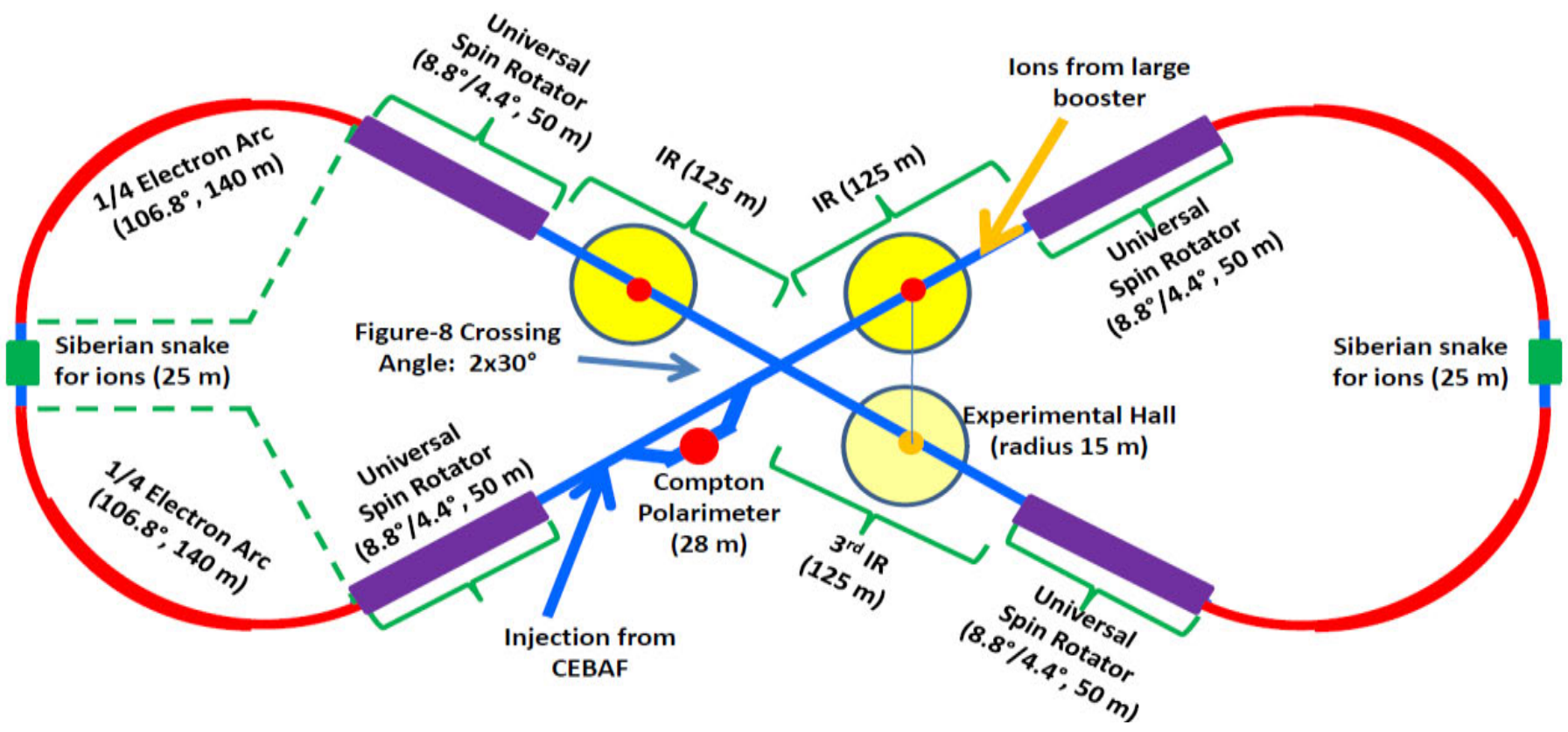

Figure 4.3: A layout of the electron ring which accommodates all machine components.

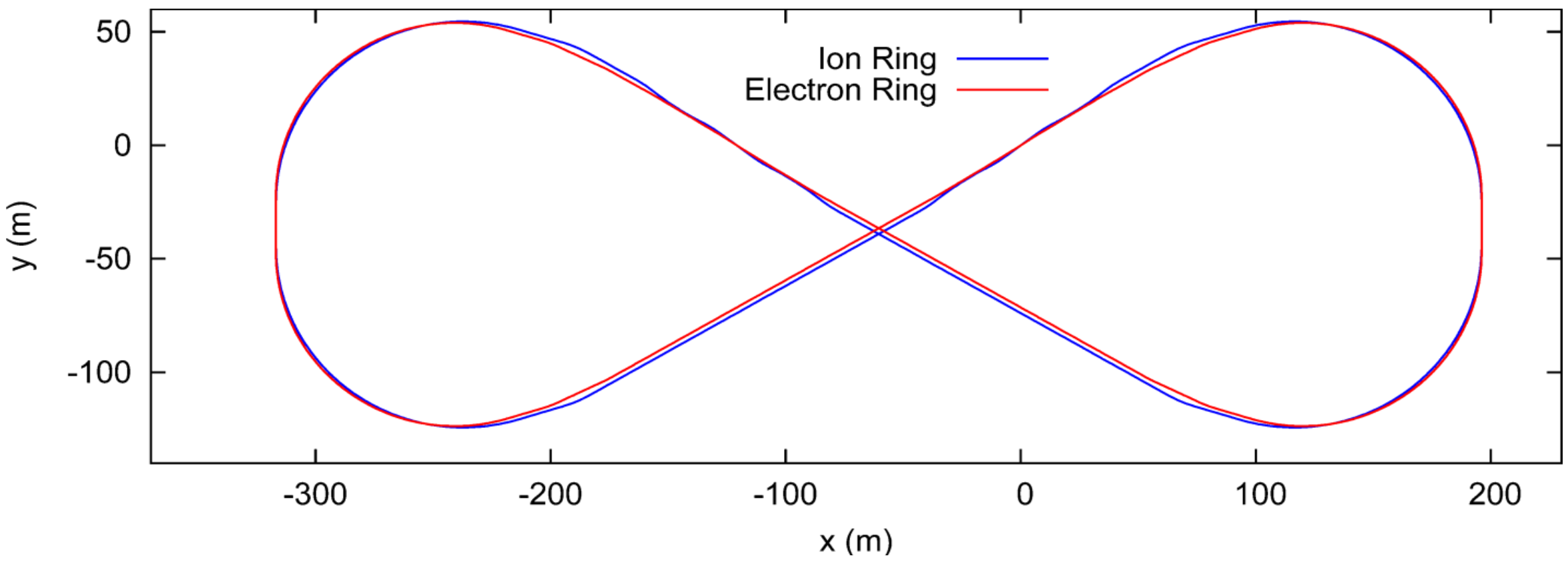

Figure 4.4: Scaled layout of the electron and ion collider rings using MAD-X survey output.

Table 4.3: Main parameters of the electron ring layout

| | | |
|---|---|---|
| Circumference | m | 1340.41 |
| Figure-8 crossing angle | deg | 60 |
| Total bending angle per (half) arc | deg | 240 |
| Length of one half arc* / arc radius | m | 405.75 / 96.86 |
| Universal Spin Rotator | m | 47.6 |
| Length of long/short straight | m | 264.46 / 25.9 |

* One arc also includes one short straight section and two Universal Spin Rotators

### 4.3.2 Lattice Design

The electron ring is currently designed as a FODO lattice [13] in both arc and straights. There are functional blocks such as spin rotators [10,14,15] and interaction regions [12] inserted into the base lattices, and optics matching is required. In this section, we first describe the arc lattice, followed by the lattice of long straights; at the end we present some global properties of the whole ring. An earlier version of the linear optics design can be found in reference [16].

The arc lattice is designed mirror symmetrically about the short straight in the center. Each quarter arc contains 27 FODO cells [13]. Among them 21 are regular cells, 2½ for dispersion suppression [13] near the short straight, and the remaining 3½ cells for matching optical functions to those of the spin rotator. Each arc cell is 5.25 m long and filled with two 1.5 m long normal conducting dipole magnets. The filling factor is 57%. The 54 dipoles of one quarter arc provide a total of 106.8º bending angle. The other 13.2º bending angle will be provided by two sets of dipoles in one Universal Spin Rotator (see section 4.7.4).

The optical functions of a regular arc cell are plotted [17] in Figure 4.5 (left) and the magnet parameters are listed in Table 4.4. For simplicity, currently the betatron phase advance per cell is 120° in both transverse planes. While this value is fairly close to 135°, the optimal value [18] associated with the minimum equilibrium horizontal emittance of a FODO lattice, it is nevertheless an important tuning knob for adjusting electron equilibrium emittance for matching spot sizes of colliding electron and ion beams at an IP over a wide beam energy range.

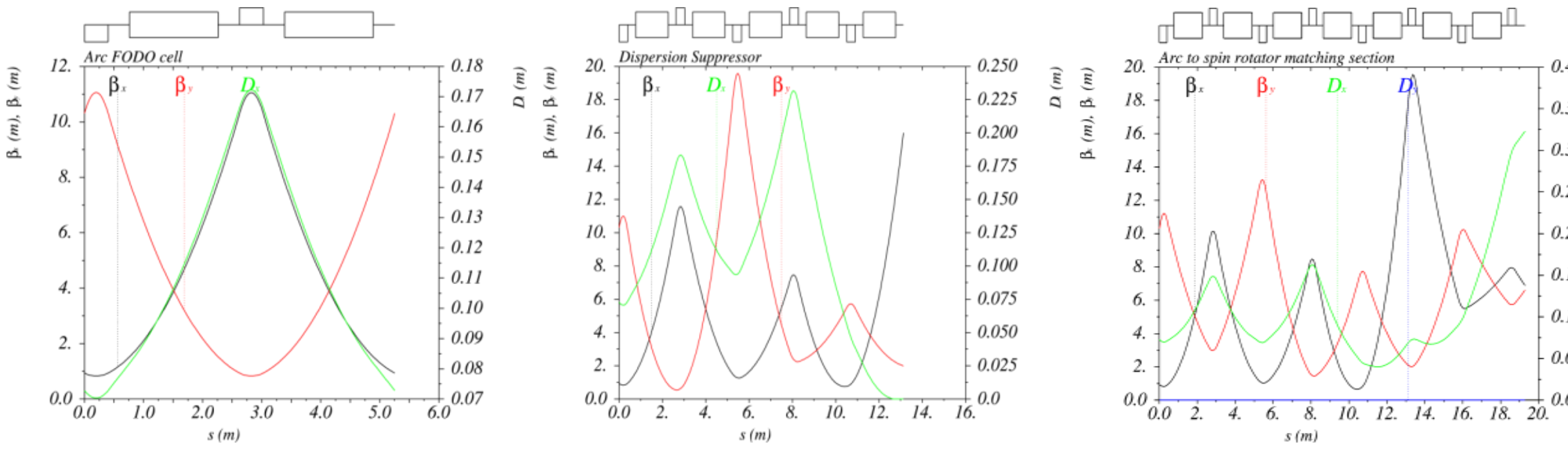

Figure 4.5: Optics of building blocks of a MEIC electron arc: a regular arc FODO cell (left), a dispersion suppressing section (center) and a matching block (right) for connecting the electron arc and the Universal Spin Rotator

A short straight between two quarter arcs consists of 5½ FODO cells. Currently it is allocated for part of the electron ring RF system, which provides compensation of synchrotron radiation energy loss and longitudinal focusing. This requires zero dispersion. To achieve this goal, a dispersion suppression scheme based on adjusting quadrupole field gradient [13] has been adopted at the end of the arc adjacent to the short straight. Figure 4.5 (center) shows optical functions of a dispersion suppression block consisting of 2½ FODO cells. The key advantage of this scheme compared to the more commonly used missing-dipole dispersion compressor is preservation of uniformity of the arc geometry and the avoidance of the inter-dependence of geometry and horizontal betatron phase advance in the suppressor.

Table 4.4: Electron ring magnet parameters at 5 GeV energy

| | | |
|---|---|---|
| Dipole length | m | 1.5 |
| Dipole bending radius | m | 43.5 |
| Dipole bending angle | deg | 1.98 |
| Dipole bending field | T | 0.384 |
| Quadrupole length | m | 0.4 |
| Quadrupole strength in the arc / straight | T/m | 29.0 / 27.2 |
| Maximum quadrupole strength in the ring | T/m | 31.4 |

Lastly, to complete a half arc, an optics matching block [13] is introduced to connect the regular arc FODO lattice to a Universal Spin Rotator. The matching block consists of three and a half FODO cells and its optics is plotted in Figure 4.5 (right). The optics of a whole arc of the figure-8 electron collider ring is shown in Figure 4.6.

Many machine components are placed in the long straights of the electron ring. They include interaction regions [13], more RF units, electron polarimetry, injection and ejection. The following approach has been taken: the long straights are designed with a FODO lattice and all the machine components come into the ring as insertions; as the design work progresses, it goes to more and more details. This modular design approach improves not only efficiency as alterations proceed for optimization but also operational reliability of the ring.

There are total 48.5 FODO cells in one long straight; 43.5 of them are regular ones and 5 are in two optics matching blocks at both ends of the long straight. All cells have the same length of 5.58 m, and parameters of the magnets are listed in Table 4.4. The optical functions of a regular FODO cell, a matching block and an entire long straight are plotted in Figure 4.7. Descriptions of a few important insertion blocks are presented in remaining parts of this report: the Universal Spin Rotator in section 4.7 and the low-beta interaction region in chapter 7.

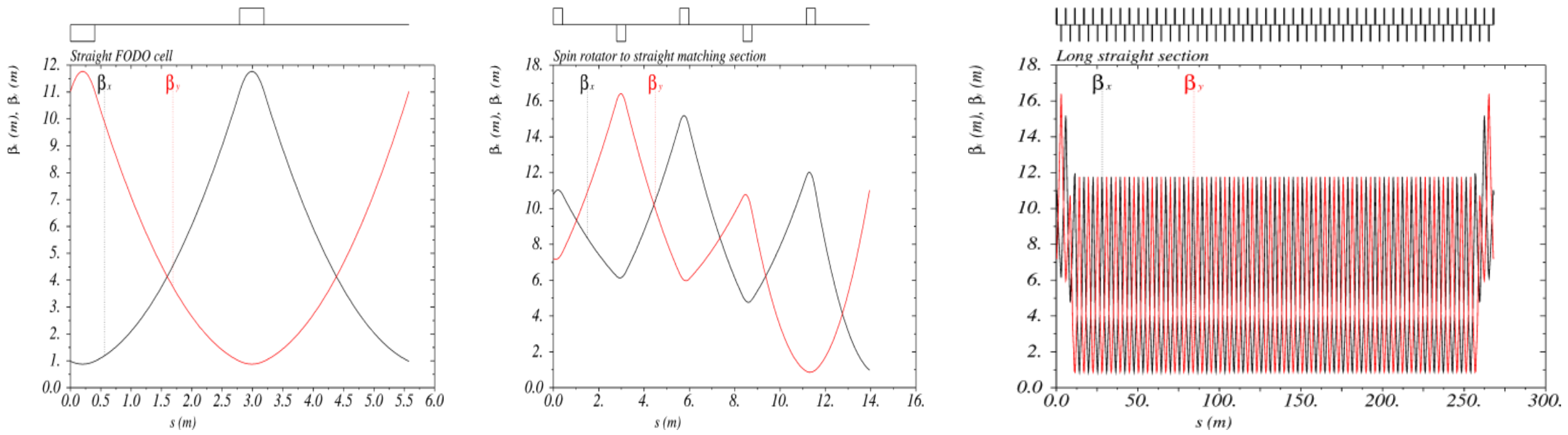

Figure 4.7: Optical functions of a long straight in the electron collider ring: a regular FODO cell (left), an optical matching block (center) and an entire long straight (right).

For the designed energy range of MEIC, electrons are already in the ultra-relativistic regime; however, the speed of ions varies slightly for different energies from 20 to 100 GeV per nucleon. This results in a small but nontrivial energy dependent difference of revolution times between electrons and ions, and an issue of synchronization of beams at collision points. The ramifications of this important issue will be addressed in the next chapter. Corrections of chromatic aberrations [19] are treated in Chapter 7 for interaction region design, since the low-beta insertions [20] are the dominant contributors. Dynamic aperture is also discussed there.

Before closing this subsection, we list some global optics parameters in Table 4.5. Optics of the complete electron collider ring is shown in Fig. 4.8. The momentum compaction factor [21] of this ring is $5.6 \cdot 10^{-4}$, leading to a relatively large transition gamma [22] of 42.3 or a transition energy of 21.6 GeV.

Table 4.5: Electron ring optics parameters.

| | | |
|---|---|---|
| Maximum beta functions, horizontal/vertical $\beta_{x,y}^{max}$ | m | 19.0 / 18.6 |
| Maximum dispersions, horizontal/vertical $D_{x,y}^{max}$ | m | 0.745 / 0.179 |
| Betatron tunes, horizontal/vertical, $\nu_{x,y}$ | | 71.25 / 73.50 |
| Natural chromaticities, horizontal/vertical $\xi_{x,y}$ | | -119 / -121 |
| Momentum compaction factor α | $10^{-4}$ | 5.6 |
| Transition gamma $\gamma_{tr}$ | | 42.3 |

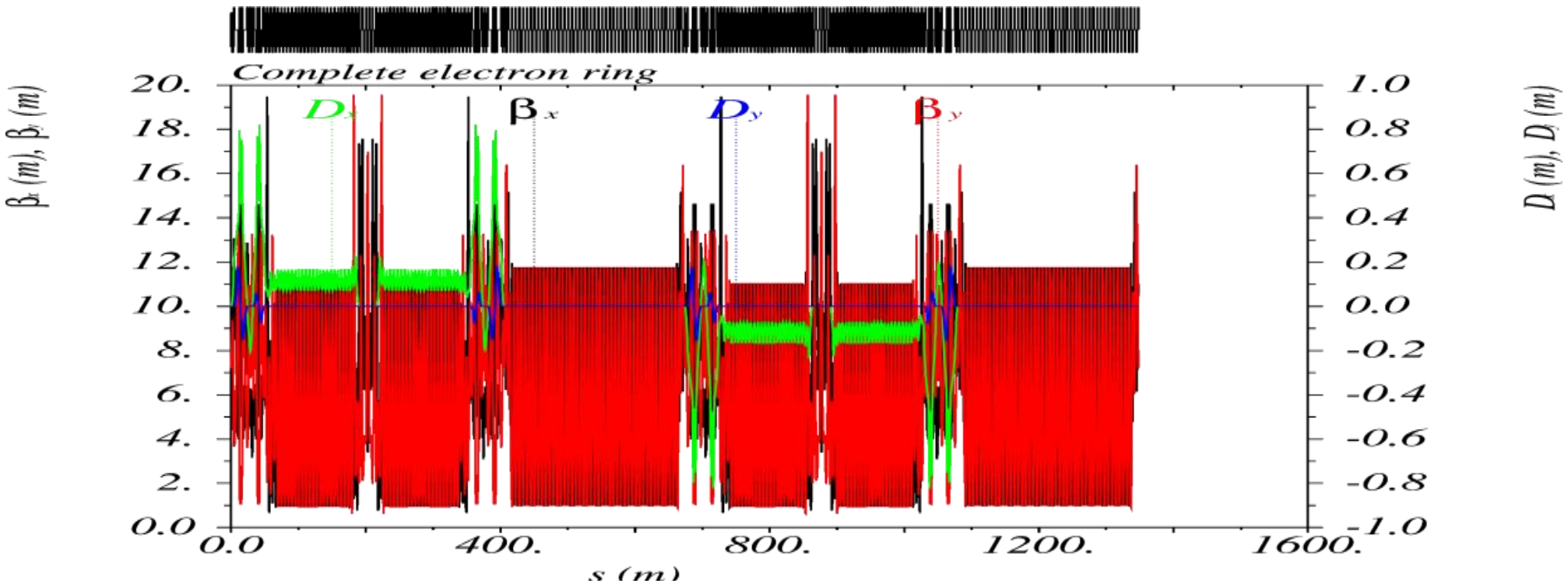

Figure 4.8: Optics of the complete figure-8 electron collider ring.

## 4.4 Synchrotron Radiation and Beam Parameters

In the MEIC electron ring, an electron beam emits strong synchrotron radiation. This leads to several important consequences: (1) electrons lose significant energy on each revolution around the ring, thus requiring accelerating RF cavities to restore and maintain constant energy; (2) the radiation power is very significant for a high current beam; it not only demands a high RF power to compensate the lost beam power but also brings a serious challenge to dissipate this radiation power and maintain vacuum inside the beam chambers; and (3) the electron beam is in a radiation dominated regime such that its phase space parameters are determined entirely by an equilibrium of radiation damping [9], quantum excitation [23] and accelerating forces of RF cavities, while its initial state from the linac is quickly erased.

### *Synchrotron Radiation Power*

The total radiation power of an electron beam circulating in the figure-8 collider ring is

$$P_{tot} = C_\gamma \frac{I}{e}\frac{E^4}{\rho}\left(1 + \frac{\varphi}{\pi}\right) \tag{4.1}$$

where $C_\gamma$=8.8575×$10^{-5}$ m/GeV$^3$, $e$ is charge of electron, $E$ and $I$ are energy and current of the electron beam, ρ is the bending radius assuming it is an iso-magnetic ring. The factor (1+φ/π) takes into account the fact that a figure-8 ring has a total bending angle greater than 2π, and φ is the crossing angle. The linear density of the synchrotron radiation can be derived by dividing the total radiation power by the total length of arc dipoles, yielding

$$P_{linear} = C_\gamma \frac{I}{e}\frac{E^4}{\rho}\frac{1}{2\pi\rho} = 14.085\ \frac{I(A)E^4(GeV)}{\rho^2(m)}\ \ (kW/m) \tag{4.2}$$

where the last part casts the formula in some practical units. The following table 4.6 presents the radiation power and electron energy loss per turn in the MEIC collider ring for several electron energies. The energy loss per turn equals total radiation power divided by the beam current.

The high radiation power dumped on the interior surface of the beam chamber needs to be carefully dissipated. We have set a limit of 20 kW/m for the maximum power density; thus, currents of electron beam at high energies have been adjusted to meet this limit. Although this limit is almost two times higher than what B-factories (KEK-B [24]) have achieved, it is well below the capability of new beam chamber designs developed for the KEK-B upgrade.

Table 4.6: Parameters of colliding electron beam of MEIC

| | | | | | | | |
|---|---|---|---|---|---|---|---|
| Beam energy | GeV | 3 | 5 | 6 | 7 | 9 | 11 |
| Beam current | A | 3 | 3 | 2.0 | 1.1 | 0.4 | 0.18 |
| Total SR power | MW | 0.67 | 5.15 | 7.1 | 7.1 | 7.1 | 7.1 |
| Linear SR power density | kW/m | 1.85 | 14.3 | 20 | 20 | 20 | 20 |
| Energy loss per turn | MV | 0.22 | 1.7 | 3.6 | 6.5 | 18 | 40 |
| Energy spread | $10^{-3}$ | 0.4 | 0.66 | 0.8 | 0.93 | 1.19 | 1.46 |
| Longitudinal damping time | ms | 58.4 | 12.6 | 7.3 | 4.6 | 2.2 | 1.2 |

### *Horizontal Emittance*

The equilibrium emittance [25] of an electron beam in a storage ring depends on the lattice design of the ring. For an iso-magnetic FODO lattice, the horizontal geometric emittance is given by the following formula under a thin-lens approximation where $C_q$=3.84•10$^{-13}$ m,

$$\epsilon_{FODO} = C_q \gamma^2 \theta^3 \frac{l_b}{l_{b,0}} \frac{<H>}{\rho\theta^3} \tag{4.3}$$

$\gamma$ is the relativistic factor, $\theta$ is the bending angle of dipoles, $l_{b,0}$ is the actual effective length of one bending magnet and $l_b$ is half of the length of a FODO cell. $<H>$ is the well known H-function [8] for a lattice. The lattice dependent factor $<H>/\rho\theta^3$ is a function of horizontal betatron phase advance per cell and has a minimum at approximately 135°, corresponding to a minimum emittance.

For the case of the MEIC electron collider ring, the equilibrium value of normalized horizontal emittance at 5 GeV is 36 µm, approximately 33% below the nominal design value (53.5 µm) in the parameter table (Table 3.1). This gives a healthy margin for this baseline design since it is foreseen that several insertions such as spin rotators and chromatic compensation blocks will likely have negative impact on the emittance due to the dipoles in these insertions.

The equilibrium horizontal emittance strongly depends on the beam energy, leading to a large (a factor of 13) change from 3 to 11 GeV under one type of lattice. To achieve a match of spot sizes of the colliding electron and ion beams at collision points, the variation of electron beam emittance must be narrowed significantly. The betatron phase advance per FODO cell is one turning knob to adjust the emittance values. Different types of optics such as a Theoretical Minimum Emittance (TME) [26] lattice may also be adopted when the collider is operated at higher energies. In practice, a hybrid lattice has been conceptually developed for the MEIC electron collider ring such that, at low electron energies, the ring sees a FODO lattice; however, at higher energies, by changing settings of a few quadrupoles, the ring uses a different, lower emittance lattice such as a TME lattice [27].

### *Vertical Emittance and Aspect Ratio*

Theoretically, the intrinsic value of the vertical emittance of an electron beam in a completely flat and optically uncoupled storage ring is a negligible fraction of the horizontal emittance. In practice, the vertical emittance depends on the coupling of optics of the horizontal and vertical degrees of freedom, and this will be utilized in MEIC. Since the ion beam emittances depend on the efficiency of the electron cooling [28], its emittance aspect ratio varies greatly from low to medium energies (see Chapter 6). The MEIC design strategy is to match aspect ratios of the colliding electron and ion beams over the whole energy range for optimization of the interaction region design and luminosity. Therefore, electron beam emittance aspect ratio will be varied using the electron storage ring coupling as the tuning knob, invoking certain special magnetic lattice elements such as solenoids and skew quadrupoles, to achieve an adjustment ranging from 1 (round beam) to 7 (modestly oval) for different operating ion energies.

### *Bunch Length and Energy Spread*

The energy spread of an electron beam in the electron storage ring is

$$\frac{\sigma_E^2}{E_0^2} = C_q \frac{\gamma^2}{J_s} \frac{<\frac{1}{\rho^3}>_s}{<\frac{1}{\rho^2}>_s} \tag{4.4}$$

and the resulting bunch length is

$$\sigma_l = C_q \frac{\sqrt{2\pi}c}{\omega_{rev}} \sqrt{\frac{\eta_s E}{heV cos\psi_s}} \frac{\sigma_E}{E} \tag{4.5}$$

where $\eta_s$ is the momentum compaction factor of the ring, $\omega_{rev}$ is the revolution frequency of the electrons in the ring, $V$ and $\psi_0$ are the amplitude and synchronous phase of the RF accelerating field, and $h$ is the harmonic number. The damping partition number $J_s$ approximately equals 2. The values of energy spread are listed in Table 4.7. The equilibrium bunch length depends on the synchronous phase. For the design point of MEIC, we have the following table

Table 4.7: MEIC electron beam bunch length, RF voltage and synchronous phase

| Energy | Current | Energy loss per turn | Synchronous phase | Integrated RF voltage | Bunch length |
|---|---|---|---|---|---|
| GeV | A | MeV | deg | MV | Mm |
| 5 | 3 | 1.7 | 25 | 4 | 7.5 |

### *Synchrotron Radiation Damping Time*

The synchrotron radiation damping times in an electron storage ring can be expressed in the following simple analytic formula

$$\tau_\alpha = \frac{2}{J_\alpha}\frac{E}{\Delta E}T_0 \qquad (4.6)$$

where index $\alpha$ runs over $x$, $z$ and $s$, corresponding to the horizontal, vertical and longitudinal directions respectively, $\Delta E$ is the electron energy loss per turn in the ring and $J_\alpha$ (stands for $J_x$, $J_z$ and $J_s$) are the damping partition numbers. $T_0$ is the revolution of the ring. There is an identity $J_x+J_z+J_s=4$, or explicitly, $J_x=1-D$, $J_z=1$ and $J_s=2+D$, and usually $D$ is very small, thus leading to $\tau_x \approx \tau_z \approx \tau_s/2$. Values of these damping times are given in Table 4.6.

## 4.5 RF System

The electron collider ring RF frequency has been chosen as 748.5MHz based on a reasonable extrapolation from the existing RF technology at lower frequencies either by normal conducting cavities (PEP-II, 476MHz [29]; BESSY, 499MHz [30]) or by superconducting cavities (CESR-III, 499.765MHz [31]; KEKB-HER, 509MHz [32]). A higher bunch frequency such as 1497 MHz would be favorable for high luminosity, but the technology would be much more challenging due to the high average current (up to 3A). More R&D would be needed before such a choice could be viable. The smaller frequency increase requires less modification to the existing coupler and loads designs for damping the HOM power on the already reduced cavity volume. Both normal conducting and superconducting options are available for this technology choice. At the 5 GeV energy, the electron ring does not require a high acceleration gradient, so a normal conducting cavity would be possible, but at 11 GeV, a high gradient of 20 MV/m is required and this necessitates an SRF solution. Gradients in this range have been demonstrated in SRF single cell cavities at this frequency. To handle RF power deposition [33] on the cavity wall and also the high HOM power damping [34] through the waveguide couplers, the preference is for a superconducting cavity system operated at 4.2K with liquid helium bath cooled directly on the cavity surface and active cooling of coupler intersections. Given the availability of 2K helium supply at CEBAF, operation at superfluid temperature could also be considered. High power RF sources and couplers are available in this frequency range. In Table 4.8, we list the RF design parameters at both 5 GeV and 11 GeV, scaled from the CESR-III SRF system illustrated in Figure 4.9.

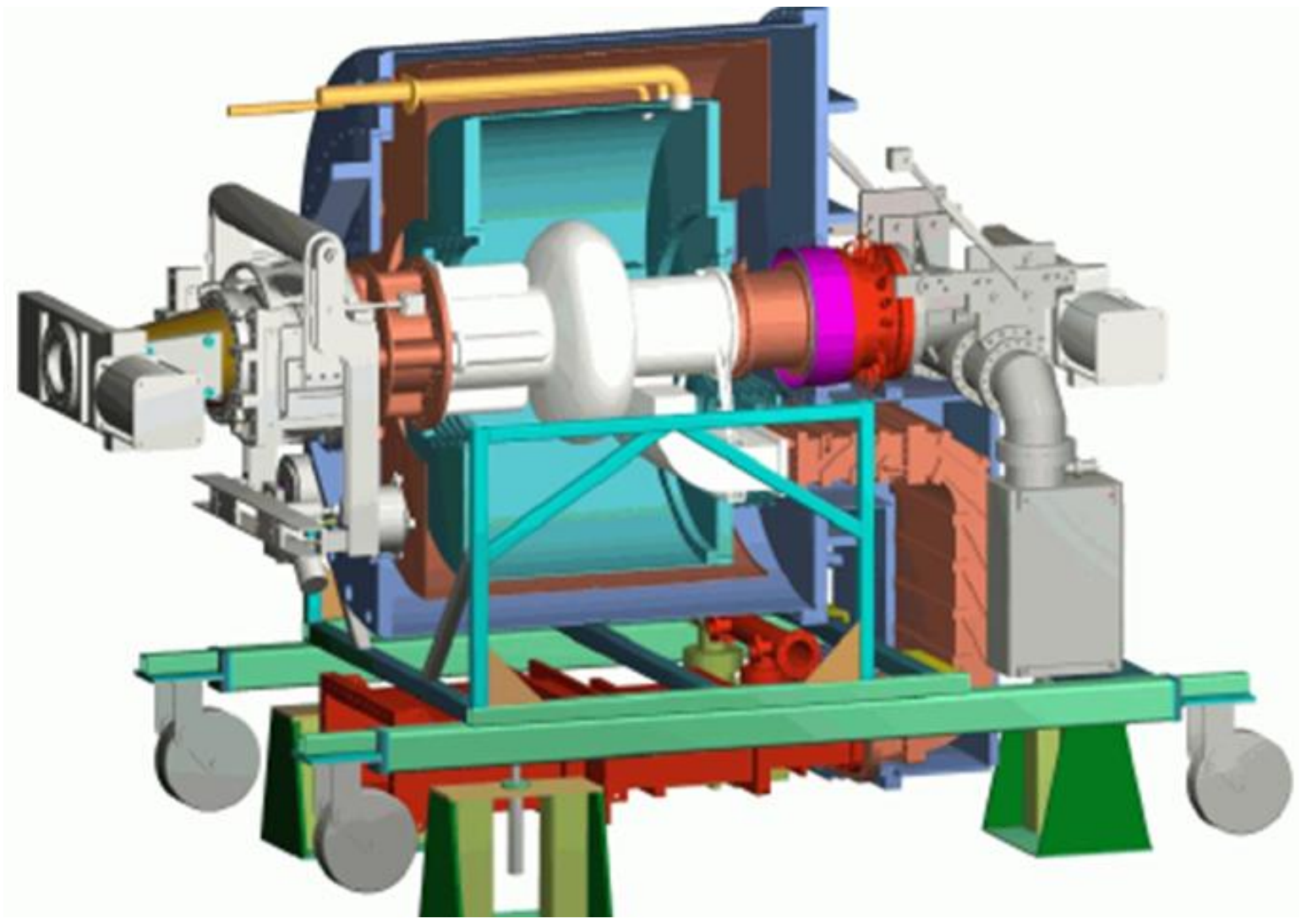

Figure 4.9: CESR-III SRF cavity system (500 MHz) at Cornell University.

Table 4.8: RF system design parameters for SRF technology

| | | | |
|---|---|---|---|
| Beam energy | GeV | 5 | 11 |
| Frequency | MHz | 748.5 | 748.5 |
| Number of cells | | 1 | 1 |
| $R/Q=V_{eff}^2/\omega U$ | Ω | 89 | 89 |
| $R/Q$/cell | Ω | 89.0 | 89.0 |
| Material independent geometry factor $G=R_s*Q_0$ | Ω | 270.0 | 270.0 |
| $R/QG$ | Ω | 24030 | 24030 |
| Active length | m | 0.2 | 0.2 |
| Insertion length | m | 1.91 | 1.91 |
| Operating temperature | K | 4.2 | 4.2 |
| $Q_0$ | $10^9$ | 1.17 | 1.17 |
| Shunt impedance ($R=V_{\rm eff}^2/P$) | MΩ | $1.04\times10^5$ | $1.04\times10^5$ |
| Input power (total losses) | kW | 544.8 | 513.7 |
| $P_{\rm cavity}$ (surface losses) | W | 0.26 | 123.23 |
| $P_{\rm beam}$ (beam loading) | kW | 493.78 | 465.43 |
| $P_{\rm beam}$ (beam loading on crest) | kW | 544.8 | 513.5 |
| Average beam current | A | 3 | 0.13 |
| Minimum gap voltage required | kV | 164.6 | 3580.3 |
| Accelerating gradient | MV/m | 0.91 | 19.73 |
| $Q_{\rm ext}$ matched $Q_0/(1+P_{\rm beam}/P_{\rm cavity})$ | | $5.59\times10^2$ | $2.80\times10^5$ |
| Coupling factor ($Q_0/Q_{\rm ext}$) | | $2.09\times10^6$ | $4.17\times10^3$ |
| Total radiated power | MW | 6.42 | 6.52 |
| Energy loss per turn | MeV | 2.14 | 50.12 |
| RF effective accelerating voltage | MV | 2.14 | 50.12 |
| Synchronous phase, 0 is on crest | deg | 25 | 25 |
| RF peak voltage required | MV | 2.361 | 55.305 |
| Number of cavities needed | | 13 | 14 |
| Insertion length | m | 24.825 | 26.734 |
| Straight section length in storage ring | m | 20 | 20 |

## 4.6 Collective Effects and Beam Stabilities

Collective effects are usually the dominant phenomena limiting beam current in the storage ring in the parameter regime of interest, and require careful analysis and mitigation. The MEIC electron beam is designed in a parameter regime (energy, current, and bunch frequency) in which *e+-e-* colliders of B-factories have been operated successfully for many years. It is expected that collective effects should be very similar in these two types of electron rings and therefore manageable in MEIC. Experience and knowledge learned in the B-factories should be directly applicable in developing and refining the MEIC design in this regard.

Like B-factories, the electron beam of MEIC has a high beam current at lower beam energy (3 to 5 GeV) but a relatively low average current at higher energy (9 GeV and above). Bunch repetition frequencies of both MEIC and B-factories are very high, leading to moderate bunch intensities but very small bunch spacing (~40 cm for MEIC and ~120 cm for KEK-B). As a consequence, one would expect moderate single bunch instabilities but fairly strong multiple bunch coupling in MEIC.

### 4.6.1 Impedance Budget

A beam circulating in a ring excites electromagnetic fields through interaction with the vacuum chamber and other structures. The induced electromagnetic fields in turn act on the beam, closing a feedback loop that can lead to beam instabilities. The strength of this self interaction is described by the notion of machine impedance [35].

A ring contains various impedance generating machine components. The total impedance of a ring is usually estimated by summing the impedances of individual components. Table 4.9 presents an impedance budget for the electron collider ring of MEIC, and the total low frequency (inductive) impedance from hardware parts not including RF cavities is about 0.2 Ω, including a value of 0.05 Ω for those elements not included in the table, such as masks, collimators, etc. It should be pointed out that the estimation of individual components is based on impedance calculations from the PEPII design. The MEIC e-ring is about half in size compared to PEPII rings [36].

Table 4.9: The MEIC inductive impedance budget

| Components | Quantities | Inductive impedance $Z/n$ ($10^{-2}$ Ω) |
|---|---|---|
| Bellows | 128 | 1.3 |
| Flanges | 1024 | 0.2 |
| Valves | 10 | 1.0 |
| BPMs | 256 | 0.046 |
| Vacuum ports | 256 | 0.009 |
| Tapers | 8 | 2.0 |
| DIP screens | | 6.0 |
| IR, injection, crotches | 2 | 2.2 |
| Feedback pickups, kickers | 5 | 1.2 |

We consider a beam pipe configuration similar to PEPII. Thus, the MEIC beam pipe will have an elliptical cross section with half-axes of 4.5×2.5 cm in the arcs and have a round pipe with a radius 3 cm in the straight sections. The pipe will be made of copper in the arcs and stainless steel in the straight sections. Our estimation for the average resistive wall impedances of the MEIC e-ring is

$$\left(\frac{Z}{n}\right)_{resistive} = (1-i)\frac{2.4}{\sqrt{n}}\,(\Omega) \tag{4.7}$$

$$(Z_T)_{resistant} = (1-i)\frac{0.85}{\sqrt{n}}\;\;(M\Omega/m) \tag{4.8}$$

where $n$ is the harmonic number.

The CEBAF cavity originally consisted of five cells and more recently was upgraded to seven cells. Consequently, there exist 5 or 7 frequencies for each resonant mode. At frequencies below 3.2 GHz, which is the cutoff frequency of the 3.5 cm radius beam pipe, the impedance of the CEBAF cavity consists of very high-$Q$ resonant modes typical of a superconducting cavity. The cavity impedance consists of a narrow band impedance and high frequency impedance tail which can be represented by

$$\left(\frac{Z}{n}\right)_{cavity} = (1+i)\frac{Zave}{\sqrt[3]{n}} \tag{4.9}$$

The low frequency limit of cavity impedance is inductive and one can compute the RF-equivalent broadband impedance of a cavity from the cavity parameters. A single 5-cell CEBAF cavity contributes to the broadband impedance about 17 mΩ of *Z/n*.

In MEIC four 5-cell CEBAF cavities are required to supplement radiation energy loss and maintain the short bunch length. Cavity contribution to the broadband impedance is expected to be 0.07 Ω [37]. The total broadband impedance is therefore about 0.3 Ω. We will adopt somewhat conservatively $(Z/n)^0{}_{BB}$=0.5 Ω for the MEIC ring for the overall broadband impedance of the ring leaving a room for including unexpected impedance generating machine elements.

The HOM loss is given by

$$P_{HOM} = k_{loss} I_{avg}^2 / k_B f_R \tag{4.10}$$

where $k_B$ is the number of bunches and $f_R$ is the revolution frequency. Note that HOM loss will be less with more bunches for a fixed total charge in a ring. The total loss factor is estimated to be 25 *V/pC* of which 80 percent coming from cavities. The total HOM loss for 5 GeV electron beam at the design current of 3 A is rather modest 150 kW considering the very high beam current.

### 4.6.2 Instabilities

The longitudinal microwave instability limits the bunch current [38]. We find the longitudinal microwave instability threshold bunch current at 1.8 mA using the broad band impedance. The design value for the bunch current is 0.9 mA. The SPEAR scaling reduction in the impedance for a short bunch is not used even though we expect the effective impedance the bunch sees would be reduced from the 0.5 Ω value. The SPEAR scaling may not be valid in MEIC as the machine impedance is not completely dominated by RF cavities.

Transversely the strong head-tail instability is known to limit the single bunch current in a large ring [39]. We find that the threshold current for this transverse mode coupling instability is about 5.6 mA. We have assumed the average beta function of 20 m and the transverse impedance of 0.2 MΩ.

In Table 4.10, we have also compared MEIC design with that of a few other electron facilities (built or proposed) with comparable beam parameters and ring sizes [40,41]. Maximum allowed impedances in each case have been calculated by setting design beam current equal to instability thresholds.

Table 4.10: Impedance requirements based on single bunch instability consideration

| | | MEIC | eRHIC | KEKB | PEPII |
|---|---|---|---|---|---|
| Energy | GeV | 5 | 5 | 3.5 | 3.1 |
| Circumference | m | 1340 | 1277.95 | 3016.26 | 2200 |
| Bunch intensity ($N_e$) | $10^{10}$ | 2.5 | 10 | 3.3 | 5.911 |
| RMS Bunch length ($\sigma_l$) | mm | 7.5 | 16.0 | 4 | 10 |
| Bunch current ($I_b$) | mA | 0.9 | 3.75 | 0.5 | 1.3 |
| Current | A | 3 | 0.45 | 2.6 | 2.14 |
| Number of bunches ($k_B$) | | 3350 | 120 | 5120 | 1658 |
| Harmonic number ($h$) | | 3350 | 2040 | 5120 | 3492 |
| Integrated RF Voltage ($V$) | MV | 4.8 | 5 | 5 – 10 | 6 |
| Synchrotron tune ($v_s$) | | 0.016 | 0.05 | 0.01-0.02 | 0.0371 |
| Momentum compaction (α) | $10^{-3}$ | 5.6 | 9.1 | 0.1 - 0.2 | 1.8 |
| Energy spread (δp/p) | $10^{-4}$ | 6.6 | 4.8 | 7.1 | 8 |
| Synchronous phase ($\psi_s$) | Deg | 25 | 9.2 | | 13 |
| Broadband impedance $(Z/n)_{BB}$ | Ω | 0.28 | 0.5 | 0.012 | 0.15 |
| $(\|Z_T\|_{BB}\beta_{av}/b)$ | GΩ/m | 1.7 | 1.3 | 0.6 | 1 |
| $(Im(Z_T)_{BB}\beta_{av})$ | MΩ | 25 | 42 | 4.7 | 20 |

Note: All data quoted are original design values; eRHIC parameters are for the ring-ring design.

It is clear that broadband impedance requirements at the MEIC electron ring are much less demanding (a few times to an order of magnitude) compared to other existing and proposed rings of comparable beam parameters as a result of designing a high luminosity collider by filling every bucket at 750 MHz RF frequency.

Narrowband impedance in a storage ring, typically from the higher order modes of RF cavities, can induce coupled bunch instabilities—the wakefields generated by a bunch ring long enough to interact with the following bunches. Superconducting cavities exacerbate the problem, since natural *Q*s are high; HOM damping is critical. Many modes can be excited, with some modes counteracted by Landau damping from the synchrotron frequency spread from the nonlinearity of the RF bucket. In addition, the natural synchrotron radiation damping will also stabilize the beam if mode growth rates are slower than the damping. However, from experience with B-factories, it is expected that many modes will neither be Landau nor synchrotron radiation damped, and a broadband feedback will be required to provide stability. The art of such feedback systems has been well developed for B-factories, where amperes of current are stored. Since the bunch repetition frequency of MEIC is only somewhat higher than today's B-factories (and lower than some proposed super-B-factories), the technology for these feedback systems is at hand. It is critically important to damp cavity higher order modes to the level of cavities of B-factories.

### 4.6.3 Scattering Effects

Multiple small-angle Coulomb scattering causes diffusion in both longitudinal and transverse phase space, resulting in a degradation of beam emittances in both spaces [42]. Compared to PEPII, the transverse phase space volume is about 10 times smaller, but the number of electrons in a bunch is smaller by a factor of 5. Consequently, IBS growth rate is expected to be worse by a factor of 2 for the same energy electron. Since no significant emittance growth from intra-beam scattering was to be expected even for the 2.5 GeV beam operation in the low energy ring of PEPII, it is our estimate that intra-beam scattering will not cause a problem at MEIC for the 5 GeV beam.

Large-angle single scattering events during intra-beam collisions can change the momentum sufficiently to make it fall outside the momentum acceptance of a ring (Touschek Scattering [43]). The momentum acceptance may come from the RF system or by the dynamic and physical aperture of the accelerator. For the 5 GeV electron beam the momentum acceptance from the RF system is 0.42%. With this RF aperture as limiting we have estimated the half-lifetime of MEIC electron beam to be 12 hours. From experience with B-factories, we expect that the momentum aperture from dynamic aperture considerations, including an aggressive interaction region, can be managed with careful design to exceed this value considerably.

Electron beam-gas scattering with residual gas nuclei results in the loss of beam particles either from the excitation of betatron oscillation or from a momentum change exceeding the dynamic and/or the momentum acceptance of the ring [44]. Two processes of particular interest are elastic scattering on nuclei and the bremsstrahlung on nuclei, which is likely to be dominant for the 5 GeV electron beam in MEIC. The lifetime is driven by the gas pressure $P$ ($N_2$ equivalent). We found $30/P$ (in nTorr) hours for the MEIC electron beam. The beam lifetime from beam-gas scattering is about 6 hours at a gas pressure of 5 nTorr. The MEIC vacuum system will be designed to produce this preliminary value of 5 nTorr pressure of background gas. The overall beam lifetime from gas and Touschek scattering is about 4 hours. The MEIC electron ring can be operated in top-off mode as full energy injection is envisioned. The beam lifetime of 4 hours is quite acceptable.

#### 4.6.4 Ion Trapping and Electron Clouds

Trapped ions resulting from beam interactions with residual gas molecules in the vacuum chamber can degrade the performance of electron storage rings [45]. Critical masses are much less than one with MEIC design beam parameters. All ion species will be trapped and stable with a uniform bunch spacing at 750 MHz. In order to avoid ion trapping there will be a gap (or several gaps) in the MEIC electron bunch train and the total length of the gap will be about 5 to 10% of the ring circumference.

Single turn ion accumulation, though unstable, can still affect the beam. It produces a betatron tune spread from bunch to bunch and the fast beam-ion instability, for example [46]. The large number of bunches with a short bunch spacing seems to provide a favorable environment for the fast beam-ion instability. The strength of this instability is comparable to that of B-factories, and MEIC will utilize a feedback system like those found at B-factories.

For positron beams, the very short bunch spacing in MEIC interferes with the formation of electron clouds generated by the multipacting mechanism. An electron generated at the chamber wall has to cross to the other side of the wall while interacting with several bunches along the way, which makes it difficult to satisfy the multipacting resonant condition. However, electron clouds once formed can produce both single bunch and multi-bunch instability in the positron beam. A single bunch transverse mode coupling instability is possible when the density of electron clouds is greater than the threshold density of $5\times10^{12}\,/\mathrm{m}^3$ for MEIC electron ring [4]. Also, there may be an electron cloud generated coupled bunch instability. The strengths of these instabilities are similar to those of B-factories (smaller number of particles per bunch is compensated by shorter bunch spacing). A rough estimate gives the growth time of about 0.2 ms for the instability. MEIC will use various measures (solenoid coils, coating the vacuum chamber with TiN or NEG, etc.) that have been taken at B-factories to control electron cloud effects.

In summary, with the design of the vacuum chamber following the example of B-factory ring colliders, MEIC will be safe from the single bunch instabilities. No bunch lengthening and widening due to the longitudinal microwave instability and no current limitations from the transverse mode coupling instability are expected. The performance of the MEIC e-ring is most likely to be limited by multi-bunch instabilities that can be handled by feedback systems well within the state-of-the-art.

## 4.7 Polarization

The physics program of MEIC requires that

- Polarization of the colliding electron beam should be 70% or above.
- Electron polarization must be longitudinal at collision points.
- Polarization of electron bunches can be flipped at a high frequency.

These requirements have guided the MEIC electron beam polarization design, and in this section we will show, in principle, that they all can be met. Our discussion will focus on the transport, manipulation and preservation of the electron polarization. In particular, the following issues are considered: polarized source; self-polarization and depolarization; polarization lifetime; and spin rotators. We will also briefly touch on the topics of spin flip, polarimetry and the positron beam.

### 4.7.1 An Overview of Polarization in a Lepton Storage Ring

It is useful to start with a brief review of spin dynamics in electromagnetic fields. The evolution of a spin vector $\vec{S}$ of a relativistic charged particle such as an electron is given by the following Thomas-BMT equation [48,49,50,51]:

$$\frac{d\vec{S}}{dt} = \vec{\Omega} \times \vec{S}, \tag{4.11}$$

$$\vec{\Omega} = -\frac{e}{\gamma m_e}\left[(1+a\gamma)\vec{B} \;+ (1+a)\vec{B}_{||} + \left(a\gamma + \frac{\gamma}{\gamma+1}\right)\frac{1}{c^2}\vec{E}\times\vec{v}\right] \tag{4.12}$$

where $e$ and $m_e$ are the charge and rest mass of the electron, $a$ is the gyromagnetic anomaly of electrons with a value of 0.00115965, $\vec{B}$ and $\vec{B}_{\|}$ are the magnetic fields perpendicular and parallel to the momentum respectively, and $\vec{E}$ is the electric field. It should be noted the Thomas-BMT equation is in a mixed reference frame in the sense that the electromagnetic fields are in a laboratory frame while the spin is in the rest frame of the electron. On the design orbit in a storage ring, the electric field of an accelerating RF cavity is always parallel to the electron momentum; thus, $\vec{E}\times\vec{v} = 0$. Therefore, on the design orbit of a perfectly aligned ring, only precession of spins around the direction of magnetic fields needs to be considered. In a magnetic dipole, a spin on the design orbit rotates around its magnetic field by an angle $\alpha = a\gamma\theta$ with respect to the orbit, where $\theta$ is the bending angle of the electron orbit. In a solenoid, a spin on the design orbit rotates around the longitudinal field by an angle $\psi = (1+a)\theta_s$ , where $\theta_s = \frac{e}{mc\beta\gamma}\int B_{\|}dl$ and $\beta = v/c$. Such rotations, from sets of dipoles and solenoids, can be combined to give special required spin rotations. Such magnet systems are called spin rotators. For example, they can be designed to rotate vertical polarization into the longitudinal direction independent of the electron energy as in the Universal Spin Rotator, which is discussed in subsection 4.7.4. Other types of spin rotators such as Siberian Snakes will be utilized in the ion rings and discussed in the next chapter.

One important consequence of synchrotron radiation from an electron beam in a storage ring is the Sokolov-Ternov self-polarization effect. It is a pure quantum effect associated with disparity of transition rates for spins to flip from along the direction of the dipole field to against the direction of the field and vice versa, leading to a maximum equilibrium polarization of 92.38% in a simple flat ring. The self-polarization time [52] is given by

$$\tau_{sp}^{-1} = \frac{5\sqrt{3}}{8}\frac{r_e\gamma^5}{m_e|\rho|^3} \approx \frac{2\pi}{99}\frac{E[GeV]^5}{C[m]|\rho|[m]^2}[s^{-1}] \tag{4.13}$$

where $r_e$ is the classical radius of the electron and  is the Planck constant. The last part of the above equation casts the self-polarization time in some practical units.

The polarization is also affected by the stochastic nature of emission of synchrotron radiation photons which imparts stochastic components to particle trajectories. These, together with the non-uniform fields of some magnets (e.g. quadrupoles) and the spin-orbit coupling due to the Thomas-BMT equation, cause so-called radiative depolarization [53] of the electron beam. Taking this effect into account, the polarization of an electron beam in a storage ring can be viewed as a dynamic equilibrium of radiative self-polarization and depolarization. Furthermore, the basic Sokolov-Ternov formula needs modification in a ring where spin rotators are used to manipulate the direction of the polarization. Then, when this heuristic picture of equilibrium is replaced by a unified rigorous treatment, the equilibrium polarization is given by the Derbenev-Kondratenko formula [52,54,55,56]

$$P_{dk} = \frac{8}{5\sqrt{3}}\frac{\oint ds\,\langle\frac{1}{|\rho|^3}\hat{b}\cdot\left(\hat{n}-\frac{\partial\hat{n}}{\partial\delta}\right)\rangle_s}{\oint ds\langle\frac{1}{|\rho|^3}\left(1-\frac{2}{9}(\hat{n}\cdot\hat{s})^2+\frac{11}{18}\left(\frac{\partial\hat{n}}{\partial\delta}\right)^2\right)\rangle_s} \tag{4.14}$$

where $\langle\ \rangle_s$ denotes an average over phase space at azimuth $s$, with integration over the whole ring. Here $\hat{b}$ is a unit vector along the magnetic field, $\hat{n}$ is a one-turn periodic unit 3-vector field over the phase space satisfying the Thomas-BMT equation along the particle trajectory and $\delta = \Delta E/E$ is the relative energy deviation. Further, in the presence of radiative depolarization, the formula for a self-polarization time in Eq. (4.13) becomes

$$\tau_{sp}^{-1} = \frac{5\sqrt{3}}{8}\frac{r_e\gamma^5}{m_e}\frac{1}{C}\oint ds\,\langle\frac{1}{|\rho|^3}\left(1-\frac{2}{9}(\hat{n}\cdot\hat{s})^2+\frac{11}{18}\left(\frac{\partial\hat{n}}{\partial\delta}\right)^2\right)\rangle_s \tag{4.15}$$

where the first two terms in the parentheses represent the Sokolov-Ternov self-polarization effect generalized for an arbitrary magnetic field and polarization direction along the closed orbit, and the third term quantifies radiative depolarization.

The equilibrium polarization is one-turn periodic and aligns itself along the unit vector $\hat{n}_0$ which is the vector $\hat{n}$ evaluated on the closed orbit. Designing a layout to give a required orientation of the polarization at each point in a ring amounts to setting up the layout so that with perfect alignment $\hat{n}_0$ has the desired orientation at each point around the ring. The other key quantity at this stage of a design is the closed-orbit spin tune $\nu_0$ [57]. This is the number of precessions of a spin around $\hat{n}_0$ that a spin makes for one traversal of the closed orbit.

Misalignments and the resulting distortion of the closed orbit can cause $\hat{n}_0$ to deviate from its ideal orientation. Likewise, misalignments can cause $\nu_0$ to deviate from its ideal value. The third term in parenthesis in Eq. (4.15) which quantifies depolarization depends on the orientation of $\hat{n}_0$ and on the optical state of the ring. Depolarization tends to be particularly strong at so-called spin-orbit resonances defined by the condition

$$\nu_0 = k_0 + k_x\nu_x + k_y\nu_y + k_z\nu_z \tag{4.16}$$

where $k_0$, $k_x$, $k_y$, $k_z$ are integers and $\nu_x$, $\nu_y$, $\nu_z$ are orbital tunes.

Depolarization can often be reduced by a technique called “spin matching”. These involve careful attention to the optical state of the ring and to the adjustment of the closed orbit. Once high polarization has been attained, the vectors $\hat{n}$ are typically within a few milliradians of $\hat{n}_0$. Then the product $< \hat{b}.\hat{n} >$in the numerator of Eq. (4.14), which represents the “driving term” for the Sokolov-Ternov effect, can be replaced by $< \hat{b}.\hat{n}_0 >$ without significant loss of precision. The derivative in the numerator of (4.14), resulting from the rigorous treatment, represents so-called “kinetic polarization”. For the MEIC this is insignificant compared to the driving term. More background material on these topics can be found in [52].

Depolarization can also be caused by the beam-beam interactions in collision. It has been observed that such beam-beam depolarization effects could be significant and will require careful measures to mitigate it.

### 4.7.2 Polarized Electron Source of CEBAF

In spite of the above effects, it is intended that the main source of polarization of the MEIC electron beam be the polarized source of CEBAF. Electrons with longitudinal spins are emitted from a cathode illuminated by a polarized laser beam. Electrons circulate for five turns in the CEBAF racetrack while being accelerated up to 11 GeV, and are then injected into the MEIC electron ring. Electron spins precess along five passes of the racetrack and in the CEBAF-to-MEIC beam transport line, reaching a final polarization orientation at the injection point of the MEIC electron ring. This final orientation is energy dependent since rotation of spin by dipoles or solenoids varies with the electron energy. Currently, two Wien filters, a special type of spin rotator which does not alter the electron orbit, have been installed in the CEBAF photo-injector. These elements carefully set up the initial polarization direction at entry to the CEBAF linac to ensure that the final polarization orientation is always in the longitudinal direction at the point of the MEIC collider ring (where $\hat{n}_0$ is longitudinal), regardless of the beam energy. The sign of the polarization should be chosen so that it becomes anti-parallel to the magnetic fields of the dipoles in the arcs. Then it is already compatible with the natural polarization direction caused by the Sokolov-Ternov effect. In past CEBAF operation, polarization at 6 GeV was above 85%. It is expected that a similar high polarization will be achieved after completion of the 12 GeV upgrade of CEBAF.

### 4.7.3 Electron Spin in the Collider Ring

The MEIC design strategy is to utilize the Sokolov-Ternov effect to preserve the electron polarization and increase its lifetime. This requires aligning $\hat{n}_0$ in the vertical direction, i.e., along the magnetic field of the arc bending dipoles, as shown in Figure 4.10. This, in turn, requires four spin rotators on each end of two half arcs of the figure-8 ring to rotate the vertical $\hat{n}_0$ in the arcs to longitudinal at the collision points. The first spin rotator rotates a downward spin to the longitudinal direction in one long straight; the second spin rotator then rotates the spin another 90° to upward orientation in the other half ring. This spin manipulation is repeated for the second long straight, and the electron will finally return to the original state of downward spin in the original half ring. This arrangement ensures that the driving term for the Sokolov-Ternov effect has the same sign in both half rings, so that the half rings give identical contributions. Since $\nu_0$ depends on the beam energy, one or more extra solenoids will be needed at positions where $\hat{n}_0$ is longitudinal in order to keep $\nu_0$ away from spin-orbit resonances.

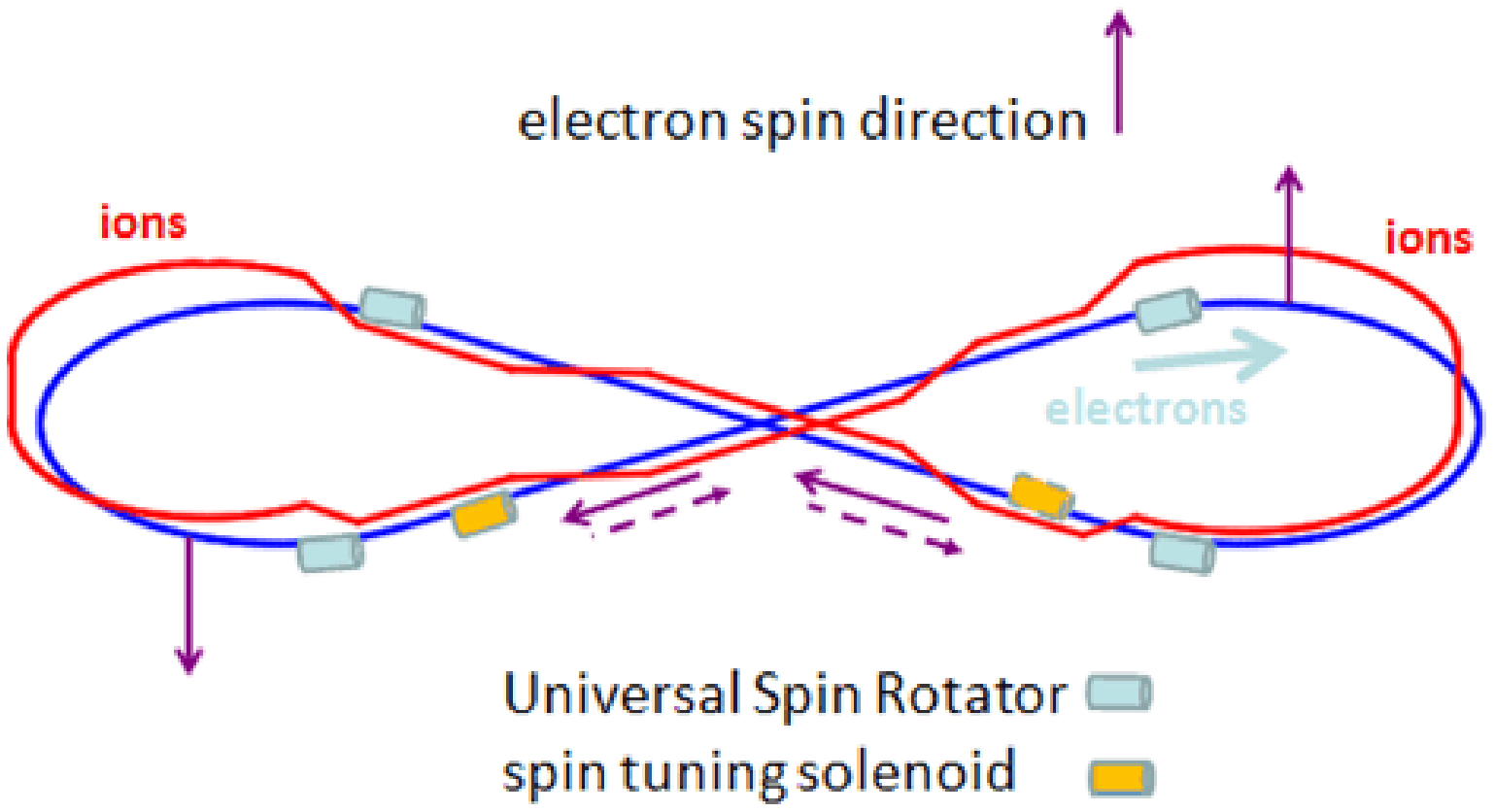

Figure 4.10: An illustration of polarization orientation in the MEIC electron collider ring.

### 4.7.4 Universal Spin Rotator

As indicated above, a 90° spin rotator is needed to rotate the polarization from the vertical direction to the longitudinal direction in MEIC. Such a spin rotation must be energy independent since the MEIC electron beam covers an energy range from 3 to 11 GeV. A Universal Spin Rotator (USR) [10] has been adopted to fulfill this need. As shown in Figure 4.11, a USR is made of two solenoids interleaved with two dipoles (or two sets of dipoles) to form one combined magnetic device. Figure 4.12 illustrates how a USR works by showing a step-by-step rotation of an electron spin. It should be noted that a USR will not change the design orbit over the entire range of electron beam energy of MEIC.

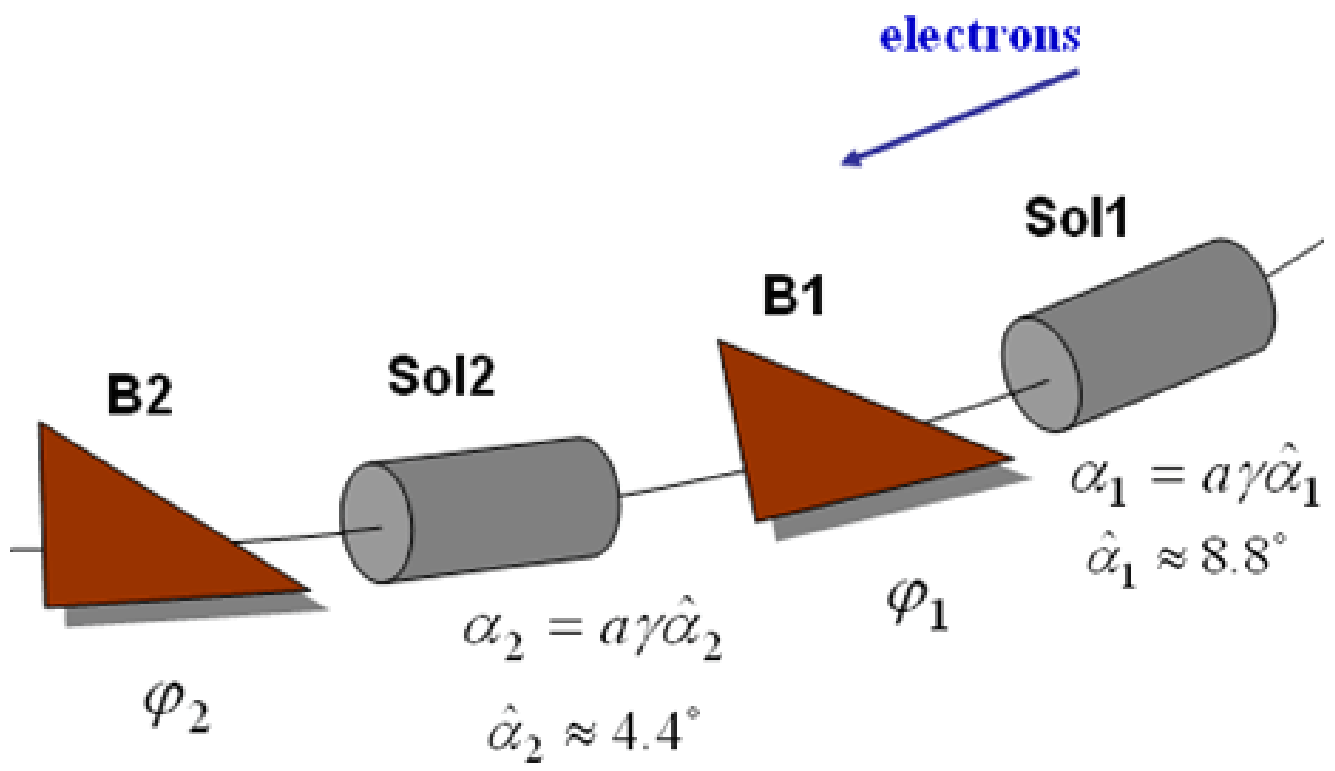

Figure 4.11: A schematic drawing of a universal spin rotator. B1 and B2 are the arc bends rotating spins by $\alpha_1$ and $\alpha_2$. Sol1 and Sol2 are solenoids with spin rotation angles $\varphi_1$ and $\varphi_2$.

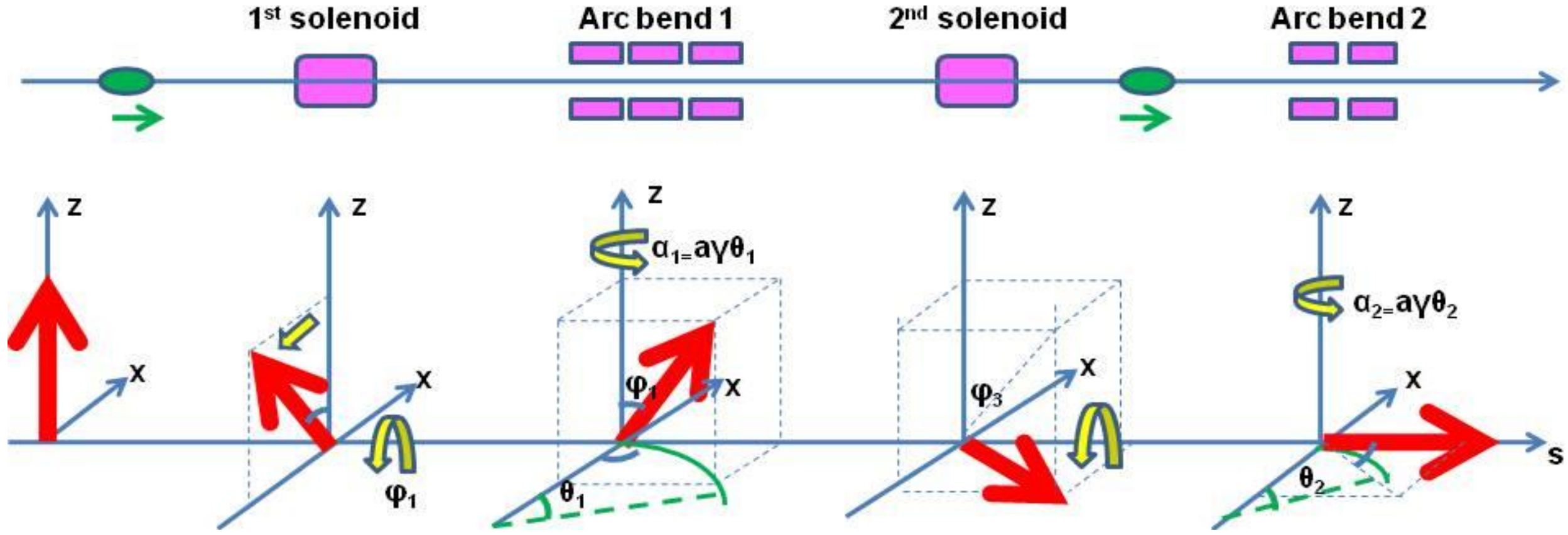

Figure 4.12: An illustration of step-by-step spin rotation by a Universal Spin Rotator

In principle, a USR could rotate a spin from the vertical direction to an arbitrary direction in the *s-z* plane; for the case of the MEIC electron ring, a 90° rotation to the longitudinal direction is desired. Implementing this concept requires allocating two arc dipoles of bending angles $\theta_1$ and $\theta_2$ to the spin rotator (see Figure 4.11). Each of two arc dipoles provides a spin rotation of $\alpha = a\gamma\theta$ which depends on the electron energy. Thus, the values of solenoid magnetic fields must vary to make the net rotation of a USR always equal to a fixed value (90° for the MEIC) regardless the beam energy; hence the name "universal." The complete treatment of a USR can be found in reference [10]; here we present the final analytic results in the following two equations:

$$\sin\varphi_1 \sin\alpha_1 = \cos\alpha_2 \tag{4.17}$$

$$\sin\varphi_1 \sin\varphi_2 \cos\alpha_1 = \cos\varphi_1 \cos\varphi_2 \tag{4.18}$$

where $\varphi_1$, $\varphi_2$, $\alpha_1$ and $\alpha_2$ are angles of spin rotations in the two solenoids and the two bending dipoles respectively, as shown in Figure 4.11.

From Eq. (4.17) and (4.18), if we allocate two sets of arc dipoles with 8.8° and 4.4° bending angles respectively into a USR, it can be shown that, with proper solenoid magnetic fields, i.e., the solenoid spin rotation angles $\psi_1$ and $\psi_2$, the required electron spin rotation will be achieved for all MEIC energies. Table 4.11 lists the angles and magnetic field strengths (BDL) of two solenoids for several representative energies of the MEIC electron beam. Since the bending angles of the dipoles are fixed, it should be noted that with a USR, the design orbit remains the same at the planned MEIC electron beam energies.

Table 4.11: Parameters of Universal Spin Rotators for MEIC

| E | Solenoid 1 | | Arc Dipole 1 | Solenoid 2 | | Arc Dipole 2 |
|---|---|---|---|---|---|---|
| | Spin Rotation | BDL | Spin Rotation | Spin Rotation | BDL | Spin Rotation |
| GeV | rad | T·m | rad | rad | T·m | rad |
| 3 | π/2 | 15.7 | π/3 | 0 | 0 | π/6 |
| 4.5 | π/4 | 11.8 | π/2 | π/2 | 23.6 | π/4 |
| 6 | 0.62 | 12.3 | 2π/3 | 1.91 | 38.2 | π/3 |
| 9 | π/6 | 15.7 | π | 2π/3 | 62.8 | π/2 |
| 12 | 0.62 | 24.6 | 4π/3 | 1.91 | 76.4 | 2π/3 |

### 4.7.5 Compensation of Optics Coupling of Solenoids

A solenoid introduces transverse coupling which is usually undesirable and therefore should be corrected. A compensation technique has been developed by Litvinenko and Zholents [58]. The basic idea is the following: a solenoid is divided into two equal parts and a decoupling insert consisting of several quadrupoles is placed between them, as shown in Figure 4.13. With a proper choice of quadruple strengths, the 4×4 transport matrix for the whole block (solenoid and insert) can be block diagonalized as in the following equation

$$\mathbf{M}_{\mathbf{sol/2}} \cdot \begin{pmatrix} \mathbf{T} & \mathbf{0} \\ \mathbf{0} & -\mathbf{T} \end{pmatrix} \cdot \mathbf{M}_{\mathbf{sol/2}} = \begin{pmatrix} \mathbf{A}_{\mathbf{L/2}}\mathbf{T}\mathbf{A}_{\mathbf{L/2}} & \mathbf{0} \\ \mathbf{0} & -\mathbf{A}_{\mathbf{L/2}}\mathbf{T}\mathbf{A}_{\mathbf{L/2}} \end{pmatrix} \tag{4.19}$$

where $\mathbf{M}_{\mathbf{sol/2}}$ is the 4×4 transport matrix of a solenoid defined as follows

$$\mathbf{M}_{\mathbf{sol/2}} = \begin{pmatrix} \mathbf{A}_{\mathbf{L/2}} & \mathbf{0} \\ \mathbf{0} & \mathbf{A}_{\mathbf{L/2}} \end{pmatrix} \begin{pmatrix} \mathbf{I}_{2\times2}\cos\varphi & \mathbf{I}_{2\times2}\sin\varphi \\ -\mathbf{I}_{2\times2}\sin\varphi & \mathbf{I}_{2\times2}\cos\varphi \end{pmatrix} \tag{4.20}$$

and $\mathbf{T}$ is a 2×2 matrix and $\mathbf{I}_{2x2}$ is a 2×2 identity matrix.

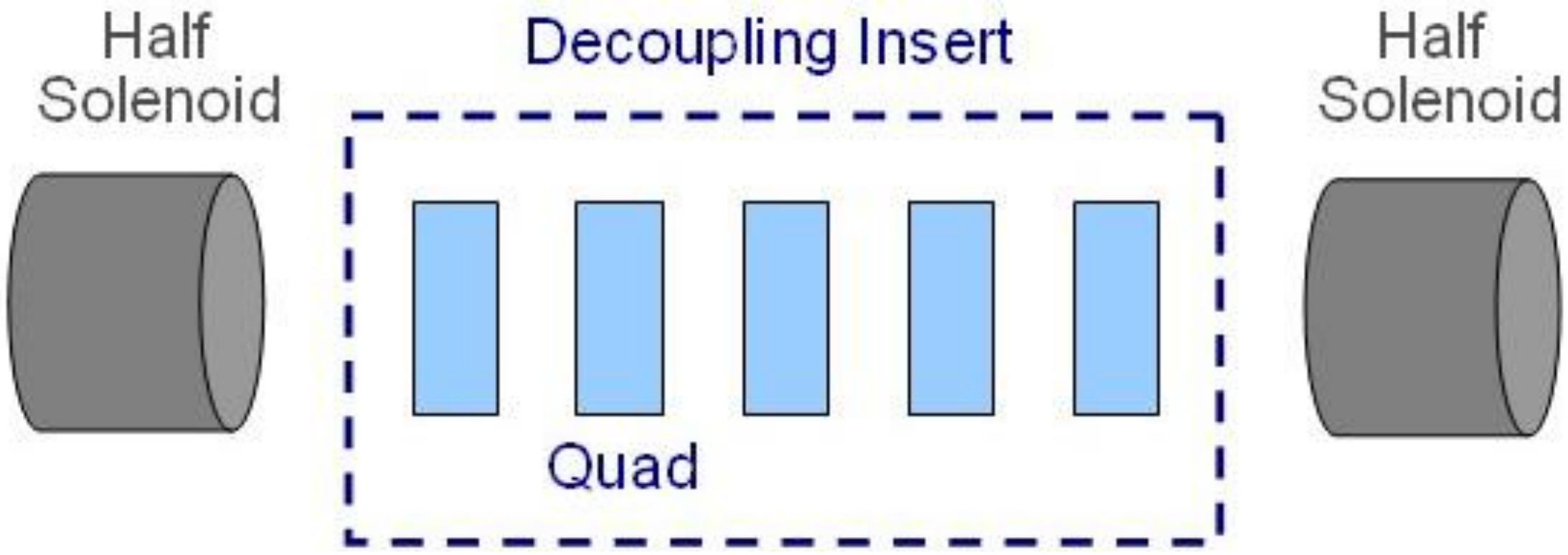

Figure 4.13: A schematic drawing of the scheme for cancelling the *x-z* coupling of a solenoid

Various implementations of the coupling compensation scheme can be devised. The chosen one [59,60] utilizes five quadrupoles and arranges them in a symmetric way: two symmetric doublets separated by one singlet quadrupole. Such an insert provides three independently adjustable parameters for satisfying three conditions of the above matrix equation. The compactness of such a system was incorporated in the optimization process yielding relatively short drifts between the quadrupoles. The quadrupole parameters are summarized in Table 4.12. The total length of the insert is 9.21 m. The optical functions of a solenoid and its optics compensation block are shown in Figure 4.14.

Table 4.12: Parameters of quadrupoles in a solenoid-coupling compensation block

| | Length (m) | Strength K ($m^{-2}$) |
|---|---|---|
| Drift | 1 | |
| Quadrupole 1 | 0.25 | 2.4 |
| Drift | 1 | |
| Quadrupole 2 | 0.51 | -1.25 |
| Drift | 1.5 | |
| Quadrupole 3 | 0.70 | 1.24 |

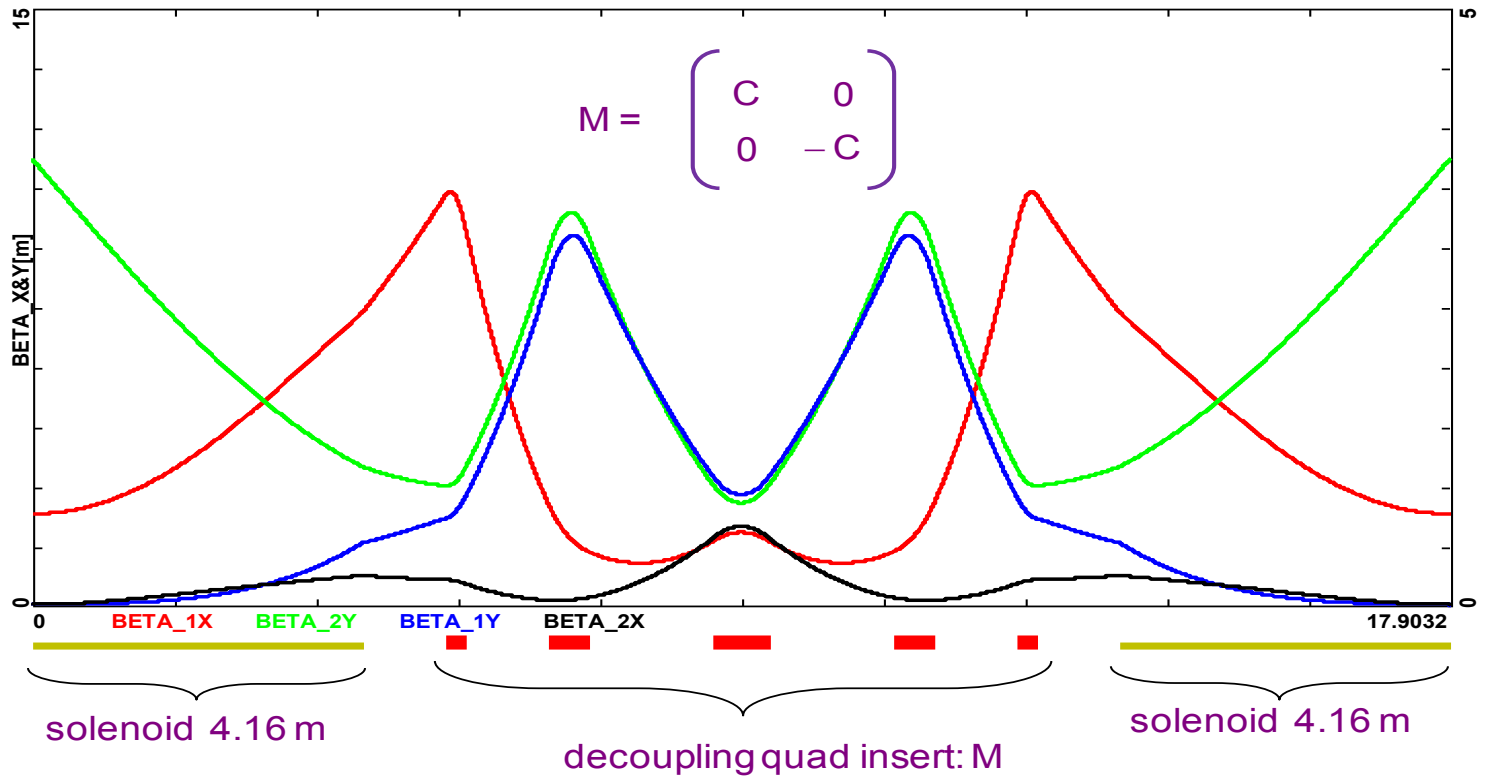

Figure 4.14: The optics functions of local decoupling of a solenoid via a mirror symmetric 5-quadrupole insert.

### 4.7.6 Equilibrium Polarization and Polarization Time

In ideal circumstances, the Sokolov-Ternov effect delivers an equilibrium polarization of 92.38%. However, for a perfectly aligned MEIC electron ring, and before depolarization is included, the equilibrium polarization is about 85% owing to the regions around the collision points where $\hat{b}.\hat{n}_0$ is zero in some dipoles. It is also nearly independent of energy. Once radiative depolarization is taken into account, the equilibrium polarization is in general lower than the above 85%. Studies show that the depolarization is driven mainly by the stochastic component of the synchro-betatron motion in the quadrupoles and solenoids around the collision points. Therefore, the spin matching must be focused on those regions.

The polarization lifetime in the absence of depolarization can be estimated using Eq. (4.13) for various energies of MEIC electron beams, and their values are given in Table 4.13.

Table 4.13: Polarization time of lepton beams in MEIC collider ring

| Energy | GeV | 3 | 4 | 5 | 6 | 9 | 11 |
|---|---|---|---|---|---|---|---|
| Polarization time | Hours | 14.6 | 3.5 | 1.1 | 0.46 | 0.06 | 0.02 |

### 4.7.7 Positron Beam Polarization

With the present source technologies, the current of a polarized positron beam is too low (at the nA level) to be practical for storage ring operation. Only a non-polarized positron source will be able to fill the collider ring of MEIC on a reasonable time scale (a couple of minutes or less). The stored positron beam, not polarized initially, will start to self-polarize due to the Sokolov-Ternov effect with polarization times given in Table 4.13. It is clear that only for positron beam energies of 9 GeV and above will self polarization time be small enough for acceptable MEIC operation.

When the desired collision energy of the positron beam is lower than 9 GeV, the self-polarization time could be comparable or even much larger than the beam lifetime in the MEIC collider ring. To reduce this polarization time, one can first accelerate the positron beam in the CEBAF to very high energy (for example, 11 GeV) in order to take advantage of small polarization time. After the beam has reached the equilibrium polarization, one can adiabatically decelerate it to the desired low collision energy while avoiding spin-orbit resonances in order to preserve the high polarization.

### 4.7.8 Spin Flip

Frequent and quick spin flip is important for reducing systematic uncertainties in polarized beam experiments. Specifically:

- For single spin asymmetry experiments, one will need to
  - flip electron spins for single electron spin asymmetry measurements, such as parity violation experiments;
  - flip ion spins for single nucleon spin asymmetry measurements, such as transverse spin structure study.
- For double spin experiments such as longitudinal spin structure experiments, fast spin flipping of either electrons or ions is required.

For a typical measurement for transverse-momentum-dependent distributions or generalized parton distributions, a flip rate of 1-10 per minute is required while a higher flip rate will help further reduce the target-spin correlated systematic effects. On the other hand, experience of the parity-violating electron scattering program at Jefferson Lab shows that a flip rate of ~100 Hz is needed to control typical helicity-correlated systematic effects. This is very demanding. In the case of longitudinal spin structure studies utilizing double spin asymmetry measurements, requirements for the flipping rate should be similar to those for the transverse spin study.

There are two ways to provide required spin flipping of an electron beam in MEIC. The first way is to change the electron polarization at the source by changing the polarization of the photo-injector driver laser. The bunch trains will have alternate polarization at a required frequency, as illustrated in Figure 4.15. Nevertheless, two problems could arise with this method. Firstly, since the polarization state alternates from bunch to bunch, the polarization state with anti-parallel direction to the magnetic field of the arc dipoles will be supported by the Sokolov-Ternov effect while the polarization of the other polarization state will be driven down through zero and up to the expected equilibrium value anti-parallel to the dipole fields. This problem can be avoided by periodically replacing the electron beam in the storage ring. Secondly, there may be a systematic difference between the bunches of different polarization states coming out of the source due to, for instance, movement of the spot of the driving laser beam on the photo-cathode. This systematic effect on the polarization can be mitigated by adiabatic damping during acceleration and can be further reduced if needed by a feedback loop in the electron source.

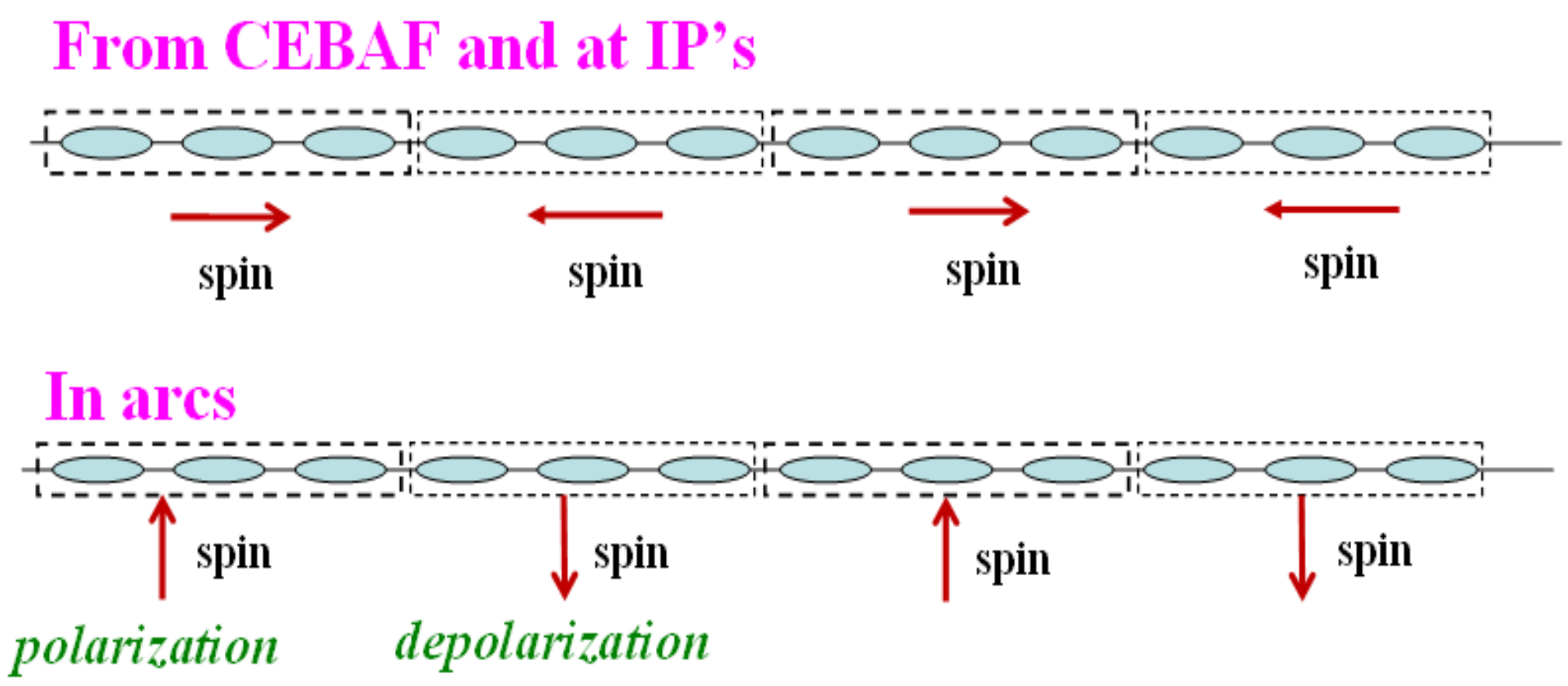

Figure 4.15: Alternating polarization of bunches of electron beams.

The second way to provide the flipping is to flip polarization of a stored beam using RF magnetic fields [61,62,63,64]. An adiabatic spin reversal could be achieved by sweeping the frequency of an RF magnet through a harmonic of the spin precession frequency [61,62]. This ensures that the different polarization states have the same polarization magnitudes. It can also provide better control over the polarization of the colliding beams stored for extended periods.

### 4.7.9 Electron Polarimetry

Measurement of the electron beam polarization for the MEIC is envisioned to include rapid, high precision Mott and Compton polarimeters. Mott polarimetry at low energy (5 MeV) presently exists at the CEBAF nuclear physics facility [65] and can be used to both absolutely measure the electron polarization (1-2% precision) of the polarized source and to assist aligning the orientation of the electron polarization to the transverse direction, prior to acceleration to full energy and injection into the MEIC storage ring. Compton polarimetry can be used to make rapid, non-destructive measurements of the electron beam polarization in the MEIC storage ring. This non-destructive aspect is crucial in order to maintain the electron beam lifetime in the storage ring. In addition, Compton polarimetry can be used to measure the transverse beam polarization as well as the longitudinal polarization, although the experimental configurations for the measurement are typically somewhat different. Longitudinal polarimeters tend to yield higher precision results since they are based on a counting, or energy-weighted asymmetry, and are the preferred type of device for accurate polarization measurements. Transverse polarimeters rely on a spatial asymmetry, which is more difficult to control systematically. Despite the likely lower precision, a transverse polarimeter would be useful as an independent monitor of the electron beam polarization in the storage ring proper. In particular, it would serve as a very useful diagnostic in the case of gross mistuning of the longitudinal spin rotators resulting in anomalously low longitudinal polarizations in the interaction region. In the ideal situation, the MEIC would accommodate both a longitudinal and transverse Compton polarimeter.

A Compton polarimeter uses a laser, typically providing IR to green light, colliding with a high-energy electron beam. The ~180° backscattered photons are boosted to high energies and the asymmetry for this process is well known from quantum electrodynamics. For a low energy photon colliding head-on with an electron, the (unpolarized) Compton scattering cross section is given by (see for example [66]),

$$\frac{d\sigma}{d\rho} = 2\pi r_0^2 a \left[ \frac{\rho^2(1-a)^2}{1-\rho(1-a)} + 1 + \left( \frac{1-\rho(1+a)}{1-\rho(1-a)} \right)^2 \right], \tag{4.21}$$

where $\rho = E_\gamma / E_\gamma^{max}$ with $E_\gamma$ the backscattered photon energy and $E_\gamma^{max}$ the maximum backscattered photon energy (corresponding to 180° backscattered photons), $r_0$ is the classical electron radius, and $a$ is a kinematic factor. This cross section is shown in Figure 4.16 for 532 nm (green) laser photons colliding with 3 and 11 GeV electrons. Note that the integrated cross section only weakly depends on electron beam energy.

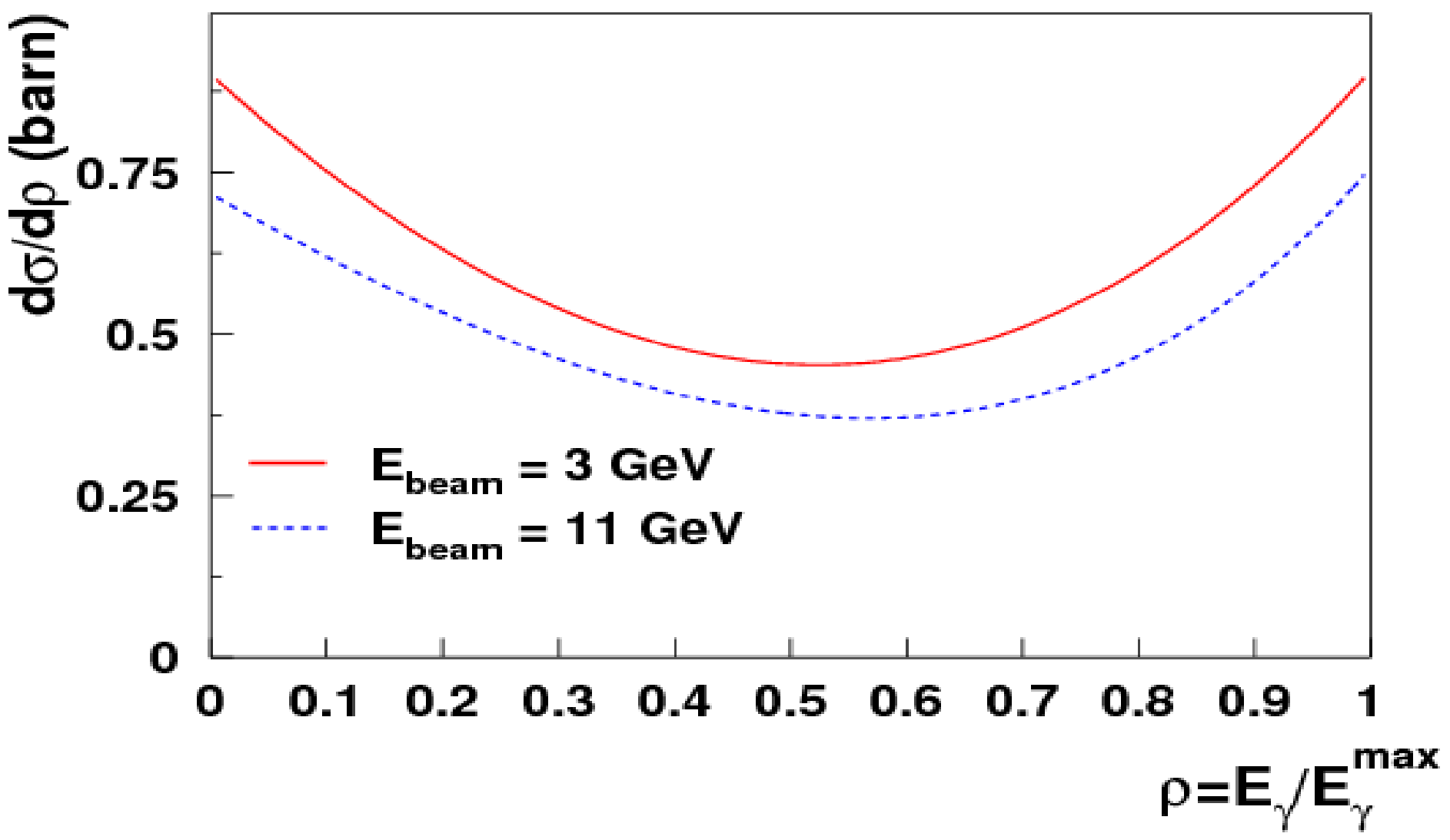

Figure 4.16: Unpolarized Compton scattering cross section for a green laser colliding with 3 GeV (red solid curve) and 11 GeV (blue dashed curve) electrons. Note that for 3 (11) GeV incident electrons, the maximum backscattered photon energy is 290 MeV (3.1 GeV).

### *Longitudinal Compton Polarimeter*

The asymmetry for circularly polarized light incident on longitudinally polarized electrons is shown in Figure 4.17. The asymmetry is maximized at the endpoint (180° scattering). Also note, however, that the asymmetry goes to zero and changes sign at low backscattered photon energies. For methods that integrate over the full energy spectrum, this tends to reduce the total figure of merit.

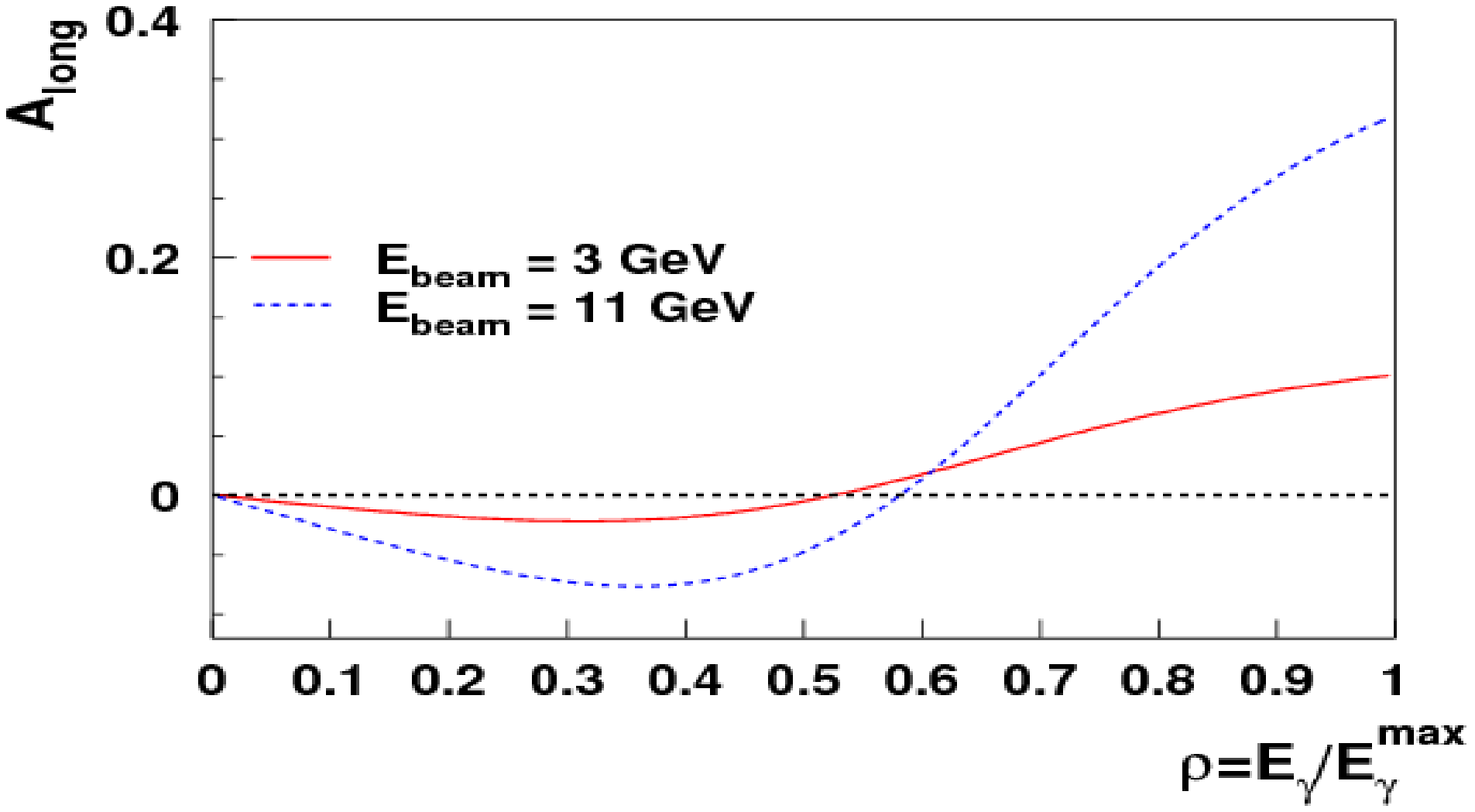

Figure 4.17: Asymmetry for circularly polarized photons incident on longitudinally polarized electrons. The asymmetry is maximized at the largest backscattered photon energy (180° scattering).

A longitudinal Compton polarimeter must be located near the IP in the region between the spin rotators (or alternatively could be located in its own IP). The luminosity of the interaction is maximized for small crossing angles between the laser and the electron beam; in addition, dedicated chicane magnets are useful for allowing ample space for photon detection and providing momentum analysis of the scattered electron, so space on the order of 10s of meters is desired. Additional electron beam focusing to achieve beam sizes on the order of 100–200 μm is also desired to maximize rate. A Compton polarimeter can detect either the backscattered photon or the scattered electron, or both, as is done in Halls A [67] and C at Jefferson Lab. In either case, the relevant detector must be located after a dipole downstream of the laser-electron beam interaction. A schematic of a conceptual design of a longitudinal Compton polarimeter is shown in Figure 4.18.

This proposed polarimeter in Fig. 4.18 [68] uses a dipole chicane to offset the electron beam from its nominal trajectory by about 1 m, where it interacts with a high power laser system (either a CW Fabry-Perot cavity or an RF-pulsed, single-pass laser system). The backscattered photons can be detected using either an $e^+$-$e^-$ pair spectrometer (for event mode reconstruction and analysis) or a photon calorimeter (for integration mode analysis). The scattered electrons are momentum-analyzed in the 3rd dipole of the chicane and detected in a silicon or diamond strip detector just before the 4th dipole. This polarimeter has the advantage of three quasi-independent measurements of the electron polarization, each with different and complementary systematic errors. On the other hand, the space required for this polarimeter is large; the chicane (assuming four 3 m long dipoles) is 28 m long, while the 2nd (integrating) photon detector is about 54 m from the laser-electron interaction point. A shorter device could be made by sacrificing one or more detectors; however, this would likely be at the expense of somewhat larger systematic errors.

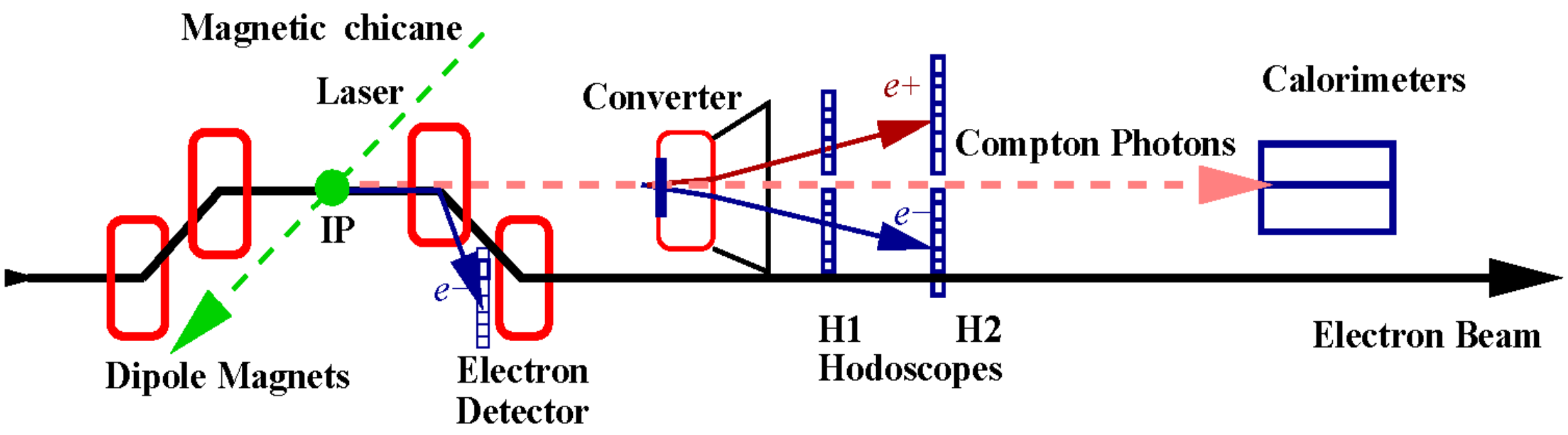

Figure 4.18: Schematic of conceptual design for a Compton polarimeter to measure the longitudinal polarization of the electron beam. Key components include a 4-dipole chicane, high-power laser system, scattered electron detector, and a sampling calorimeter and pair spectrometer for photon detection.

### *Transverse Compton Polarimeter*

Measurement of the transverse polarization of the electron beam is a bit more complicated than the technique used to measure the longitudinal polarization. In this case, the asymmetry depends on the direction of the scattered photon with respect to the electron spin direction. For electrons polarized vertically, in the $y$-direction, the asymmetry for circularly polarized (left-handed) photons is

$$A(y,E_\gamma)=\frac{N_L(y,E_\gamma)-N_L(-y,E_\gamma)}{N_L(y,E_\gamma)+N_L(-y,E_\gamma)}. \tag{4.22}$$

A key aspect of this measurement is that the asymmetry depends on the vertical position of the backscattered photon. If one were to integrate the signal over all transverse dimensions, the asymmetry would vanish. This asymmetry is shown in Figure 4.19. Alternatively, one can flip either the laser or electron helicity rather than calculate the up-down asymmetry; however, the measurement must be made at some non-zero average value of $y$.

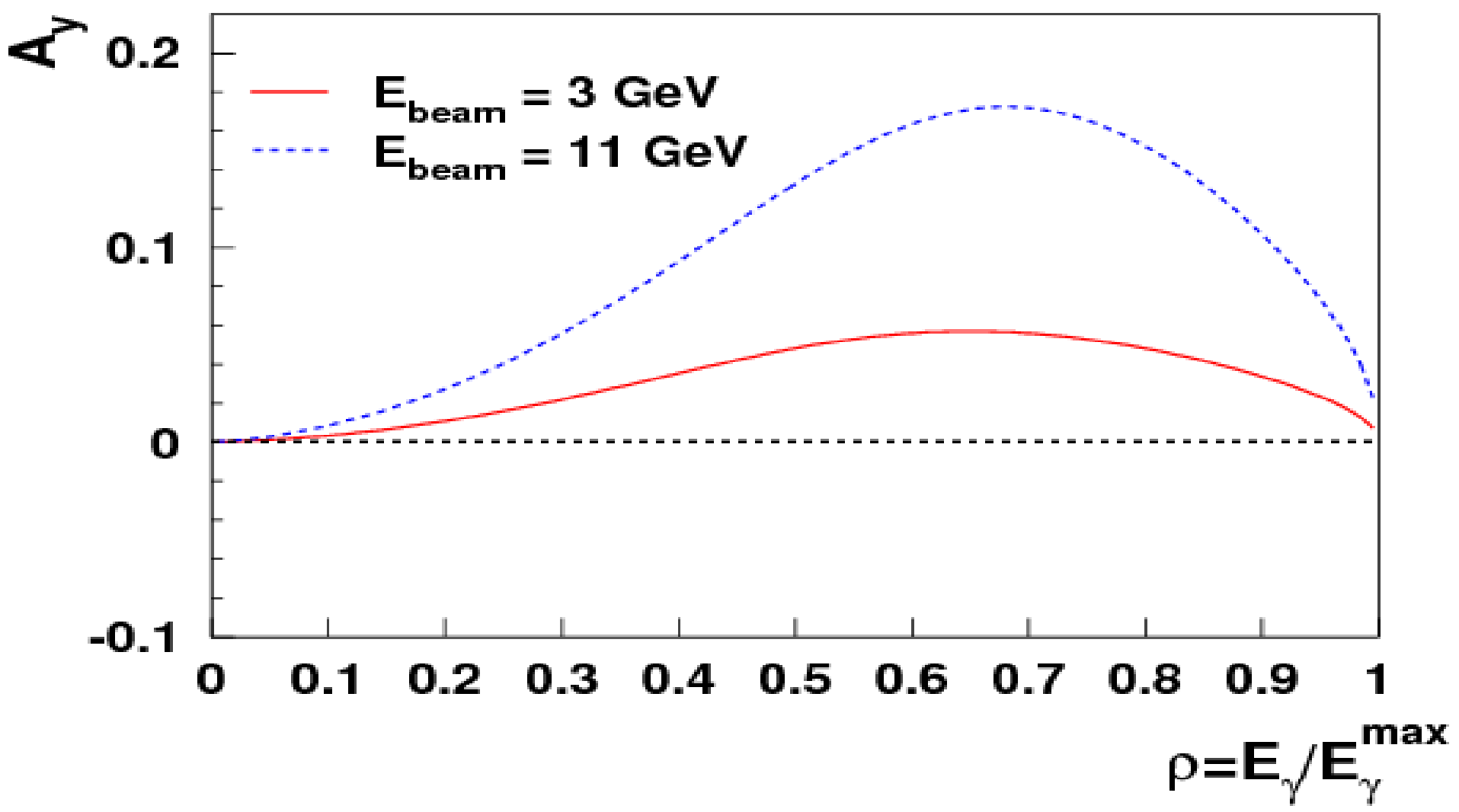

Figure 4.19: Up-down asymmetry for circularly polarized photons from a vertically polarized electron beam. In this case, the asymmetry is not maximized at the endpoint, but around $E_\gamma \approx 0.7E_\gamma^{max}$.

An additional complication in this measurement is that the high-energy backscattered photons are emitted in a very narrow cone about 180 degrees. This is shown in Figure 4.20 – here the vertical distance of the backscattered photon from the electron beam plane is shown at a theoretical detector plane 50 m away. The majority of backscattered photons are contained in a vertical region on the order of 1 cm tall. Because of these small displacements, accurate knowledge and/or fine segmentation of the backscattered photon detector will be key to minimize systematic errors. A transverse polarimeter of this nature has been used at HERA and regularly achieves systematic errors on the order of 2-4% [69].

It is worth reiterating that control of the systematic errors in a transverse Compton Polarimeter is a significant challenge due to the narrow backscattered photon cone. Minimization of these systematic errors will require a great deal of effort if high precision is required for the transverse polarimeter.

Given the high electron beam currents that will be used in the MEIC, 1% (statistics) measurements should be able to be achieved very rapidly in either the longitudinal or transverse Compton polarimeters – on the scale of minutes or less depending on the laser system and detector being analyzed.

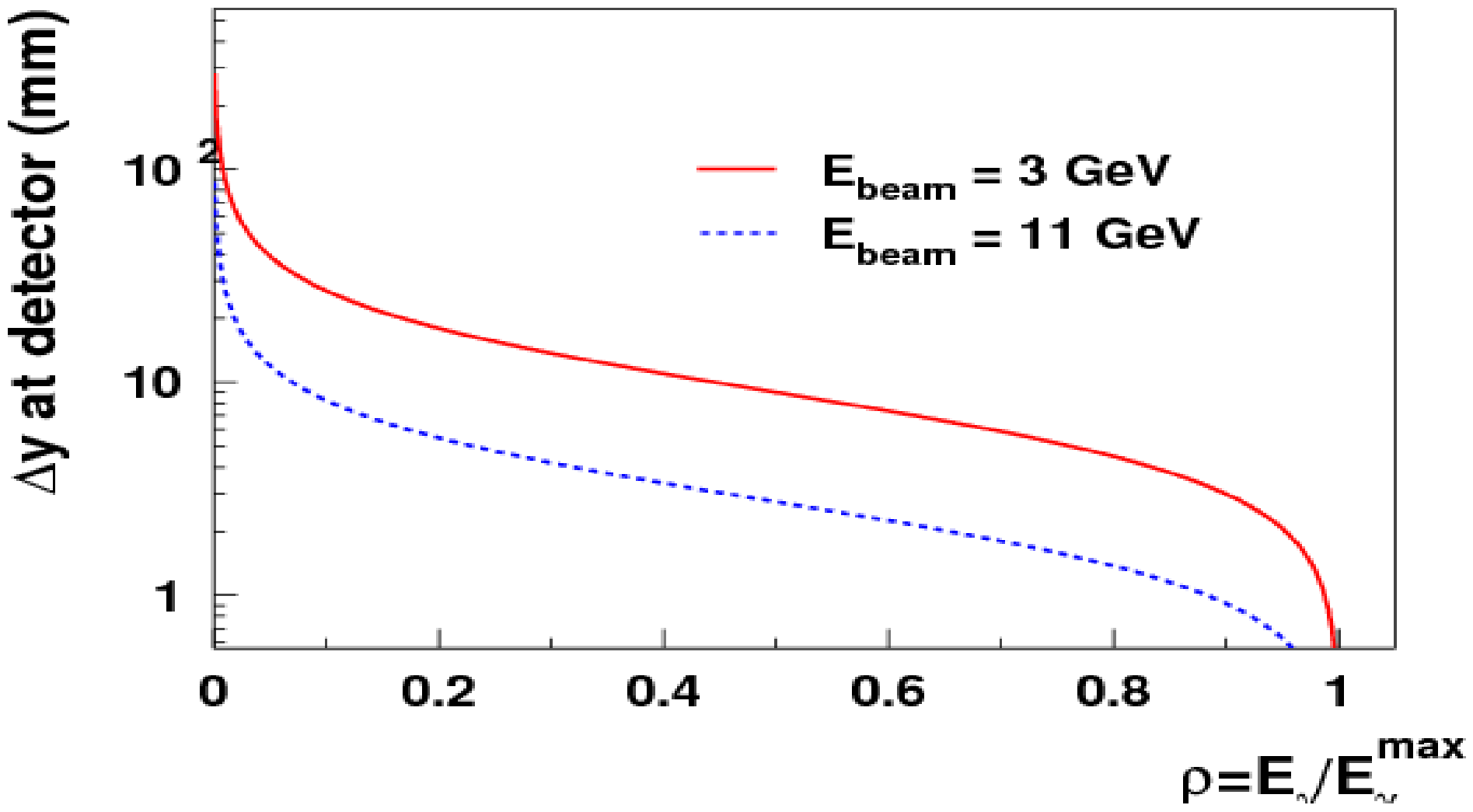

Figure 4.20 Vertical position of backscattered photon at a theoretical detector plane about 50 m from the interaction point.

# 5. Ion Complex

## 5.1 Introduction

At present, Jefferson Lab has no proton or ion beam; thus, an ion acceleration and storage complex will be a major new facility for the proposed MEIC. This chapter presents a conceptual design with design goals as follows:

- The facility can deliver polarized proton, deuteron and helium-3 beams, as well as un-polarized ion beams up to lead for collisions, with all electrons stripped.
- The collider ring can accommodate a proton beam with 20 to 100 GeV energy, and ion beams with energy per nucleon of the same magnetic rigidity.
- Current of the stored ion beam in the collider ring should reach up to 0.5 A.
- The colliding ion beams have a 748.5 MHz bunch repetition rate, a short bunch length (2 cm or less RMS) and small emittances to support the luminosity concept.
- Polarization of the light ion beams exceeds 70%, and can be arranged in both longitudinal and transverse directions at all collision points, with lifetimes comparable to that of the beam current.
- The ion collider ring supports at least two independent collision points/detectors.

Thorough investigations have resulted in the following design choices to achieve these goals:

- ***Ion linac***
  An ion linear accelerator consisting of both normal and superconducting RF cavities will provide the first stage fast acceleration of ion beams, as a way to minimize space charge effect and improve current and emittance of the injected beam into the pre-booster.

- ***Booster synchrotrons and no transition energy crossing***
  A set of two booster synchrotrons will provide the next stage acceleration; these rings will be carefully designed such that crossing of a transition energy will not be allowed for any ion species during acceleration, so that particle loss will be minimized.

- ***Superconducting collider ring***
  High field superconducting magnets will be used for the collider ring to maximally extend ion energy ranges with a compact ring size.

- ***Figure-8 rings***
  A figure-8 shape is adopted for all booster and collider rings as an optimized solution to preserve ion polarization by avoiding spin resonances during acceleration and to ensure energy independence of spin tune. In addition, a figure-8 shape ring is the only practical way for accelerating and storing polarized deuterons at a medium energy range.

- ***Multi-phased electron cooling***
  Cooling will be an essential part of the ion complex and utilized at low energy for assisting accumulation of ions, and at high energies for suppressing IBS, maintaining beam phase space density and reaching high luminosity and long lifetime.

This chapter is organized as follows: a general description of formation of MEIC ion beams and a schematic layout of the ion complex is presented in the next section, followed by conceptual designs of major components including ion sources, linac, pre-booster, large booster and collider rings in section 5.3 to 5.7 respectively. Section 5.8 presents a scheme to address the beam synchronization issue, followed by an investigation of ion beam instabilities in section 5.9. The last section of this chapter is devoted to polarization of light ion beams. The MEIC electron cooling scheme and the design concept of the cooling facility, a critically important component of this ion complex, are presented in the next chapter.

A note should be added here regarding the parameter presentation of this design report. It should be understood that the MEIC ion complex is designed to accelerate and store all ion beams. We usually use protons, sometimes also lead ions, as examples for description of this facility in this report. Often, the design parameters, especially the beam energy, are given only for protons to avoid repetitions. However, unless otherwise explicitly explained, it is assumed that the machine components and the design parameters should be equally applied to other ion species, provided an appropriate scaling adjustment is applied. As an example, the beam energy in the linac, booster and collider rings should be scaled by a factor of mass-to-charge ratio of the ion species, to satisfy the constraint of the same magnetic rigidity.

## 5.2 Formation of MEIC Ion Beams

The MEIC design concept calls for superior ion beams to support high luminosity. The several factors crucial to the process of formation and acceleration of MEIC ion beams include in particular:

- Accumulating of low current beams from sources to reach full design current.
- Accelerating ions from the sources to the colliding energies.
- Increasing and preserving phase space density of the beams.

The conceptual design of the MEIC ion complex is geared to address these factors with special care taken to optimize the quality of the proton and ion beams. A high level description of the formation and acceleration of ion beams is presented first in this section. We start with a schematic layout of the ion complex, then proceed with discussions of several important aspects of this process, including accumulation of ion beams in the pre-booster, acceleration in the pre-booster and large booster, debunching and rebunching, and staged electron cooling.

### 5.2.1 Ion Complex Layout

The schematic layout of the MEIC ion complex shown in Figure 5.1 illustrates the flow of protons or ions within the complex. The complex consists of the following key components:

- ***Ion sources***
  Depending on their species, ions will be generated in two sources, namely, an Atomic Beam Polarized Ion Source (ABPIS) for polarized or non-polarized light ions and an Electron Cyclotron Resonance ion source (ECR) for un-polarized heavy ions up to lead. A third type, an Electron Beam Ion Source (EBIS), will also be a candidate for heavy ion production, due to the advantage of its relatively low cost. All ions will be extracted in pulses of many micro bunches from the sources (see section 5.3).

- ***Ion linac***
  The ion pulses will be accelerated first by a linac consisting of both warm and cold RF cavities to 280 MeV for protons and 112 MeV per nucleon for lead ions. Depending on ion species, a certain amount of electrons will be stripped out in the linac (see section 5.4).

- ***Pre-booster/accumulator ring***
  The linac pulses will be injected and accumulated in the pre-booster, then accelerated to 3 GeV energy for protons and corresponding energies per nucleon for ions with the same magnetic rigidity; for $H^-$ and $D^-$ ions, strip injection will be employed; for other positive ions, low energy DC electron cooling will be utilized to assist accumulation (see section 5.5).

- ***Large-booster***
  Acceleration will continue in the large booster to reach beam energy up to 20 GeV for protons and 8 GeV/u for heavy ions; there will be multiple fillings since the circumference of the large booster is five times of that of the pre-booster (see section 5.6).

- ***Collider ring***
  The beam will finally be injected into the medium energy collider ring, accelerated to 100 GeV for protons and up to 50 GeV/u for ions, re-bunched to high repetition frequency, and brought into collision with the electron beam (see section 5.7).

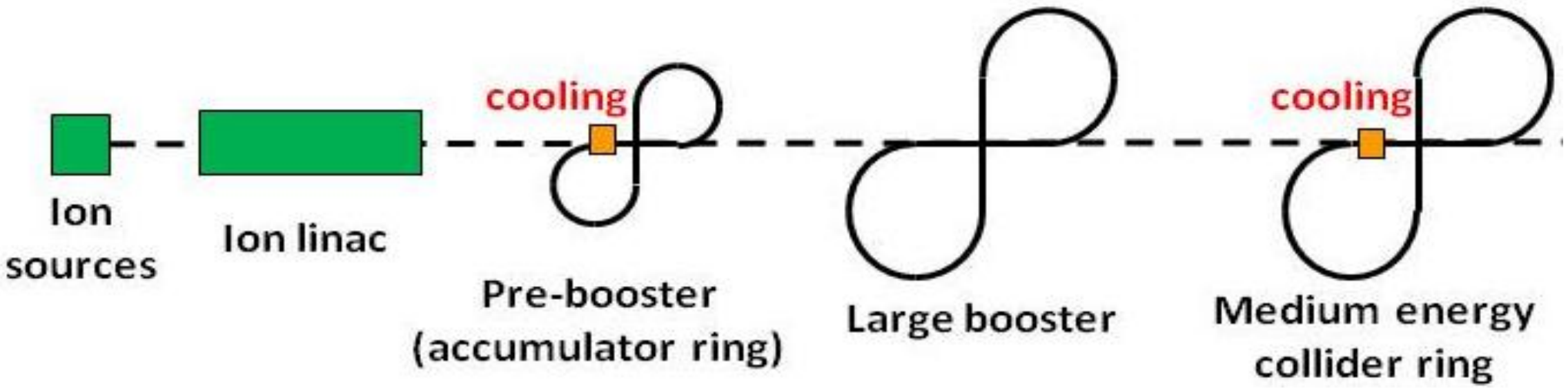

Figure 5.1: A schematic drawing of the MEIC ion complex.

### 5.2.2 Accumulation of Ion Beams

Ions are extracted from ion sources in pulses. The pulse currents are usually low, of order of several mA except non-polarized $H^-$ and $D^-$; therefore accumulation of pulsed beams from the ion linac is the way to store a high current beam in a storage ring. In the MEIC ion complex, accumulation of ion beams takes place in the pre-booster, combined with stripping of additional electrons from ions in the linac and pre-booster [1]. The accumulated ion beams will fill the large booster ring in several injections from the pre-booster. In order to achieve the designed beam parameters, the ions need to go through several phases of acceleration, accumulation, de-bunching and re-bunching. In this section, we use protons and lead ions as two examples to explain our design concepts. The evolution of the beam parameters according to this scheme is shown in Table 5.1 for protons and Table 5.2 for lead ions.

Table 5.1: Evolution of polarized proton beam current in MEIC ion complex

| | | ABPIS Source | Linac | Pre-booster | | Large Booster | Collider Ring |
|---|---|---|---|---|---|---|---|
| | | | At exit | At injection | After boost | After boost | After boost |
| Charge status | | $H^-$ | $H^-$ | $H^+$ | $H^+$ | $H^+$ | $H^+$ |
| Kinetic energy | MeV/u | ~0 | 13.2 | 280 | 3000 | 20000 | 100000 |
| γ | | | | 1.3 | 4.2 | 22.3 | 107.6 |
| Β | | | | 0.64 | 0.97 | 1 | 1 |
| Pulse current | mA | 2 | 2 | 2 | | | |
| Pulse length | ms | 0.5 | 0.5 | 0.22 | | | |
| Charge per pulse | μC | 1 | 1 | 0.44 | | | |
| Ions per pulse | $10^{12}$ | 3.05 | 3.05 | 2.75 | | | |
| Pulses | | | | 1 | | | |
| Efficiency | | | | 0.9 | | | |
| Total stored ions | $10^{12}$ | | | 2.52 | 2.52 | 2.52× 5 | 2.52×5 |
| Stored current | A | | | 0.33 | 0.5 | 0.5 | 0.5 |

Table 5.2: Evolution of lead beam current in MEIC ion complex

| | | ECR source | Linac | Pre-booster | | Large booster | | Collider ring |
|---|---|---|---|---|---|---|---|---|
| | | | After stripper | At injection | After boost | Before injection | After boost | After boost |
| Charge status | | $^{208}Pb^{30+}$ | $^{208}Pb^{67+}$ | | | $^{208}Pb^{82+}$ | | |
| Kinetic energy | MeV/u | ~0 | 13.2 | 100 | 670 | 670 | 7885 | 39423 |
| γ | | | | 1.11 | 1.71 | 1.71 | 9.4 | 43.0 |
| Β | | | | 0.43 | 0.81 | 0.81 | 0.99 | 1 |
| Pulse current | mA | 0.5 | 0.1 | | | | | |
| Pulse length | Ms | 0.25 | 0.25 | | | | | |
| Charge/pulse | μC | 0.125 | 0.025 | | | | | |
| Ions per pulse | $10^{10}$ | 1.664 | 0.332 | | | | | |
| No. of pulses | | | | 28 | | | | |
| Efficiency | | | 0.2 | 0.7 | | 0.75 | | |
| Total ions | $10^{10}$ | | | 4.5 | | 3.375×5 | | |
| Stored current | A | | | 0.26 | 0.5 | 0.447 | 0.54 | 0.54 |

### *Proton Beam*

For the case of polarized protons, the top part of Figure 5.2 (left) corresponds to micro bunch structure of proton pulses in the linac, namely, 25300 micro bunches in a 115 MHz repetition rate are combined into a macro pulse of 0.22 ms long. Such a pulse will fill up the pre-booster with the required number of protons, shown as the bottom part of Figure 5.2 (left). The proton bunch at the end of the acceleration ramp in the pre-booster is extracted and transferred to the large booster by a bunch-to-bucket transfer method. We assume that the large booster will operate with harmonic number 5 (equal to the ratio of circumferences of the large booster and the pre-booster), and hence five such transfers will be performed in a time period of around 0.8 s.

### *Lead Ion Beam*

Accumulation of a lead ion beam is very similar to the case of a proton beam; the only difference is in the beginning part of the acceleration cycle. With the present state-of-the-art source technologies, a pulse current up to 0.5 mA over a 0.25 ms time duration for lead ions could be expected from the heavy ion sources, but the pulse repetition rate could be relatively high, up to 10 Hz. Taking into account the particle losses during stripping in the ion linac, one fill-up of the pre-booster to the required intensity will need about 28 pulses of such a linac beam. Figure 5.2 (right) illustrates the bunch structures of lead ions in the MEIC ion complex. This pre-booster stacking is performed with the help of repeated multi-turn painting injection and electron cooling. Since the cooling time for the contemplated beam parameters is shorter than the time between the pulses of the maximum anticipated repetition rate, it is estimated that this accumulation process will take approximately 2.8 s. As discussed below, acceleration of lead ions in the pre-booster can be done in about 0.144 s; this leads to approximately 3 s cycle time in the pre-booster. Finally, we conclude that the large booster can be filled in about 15 s.

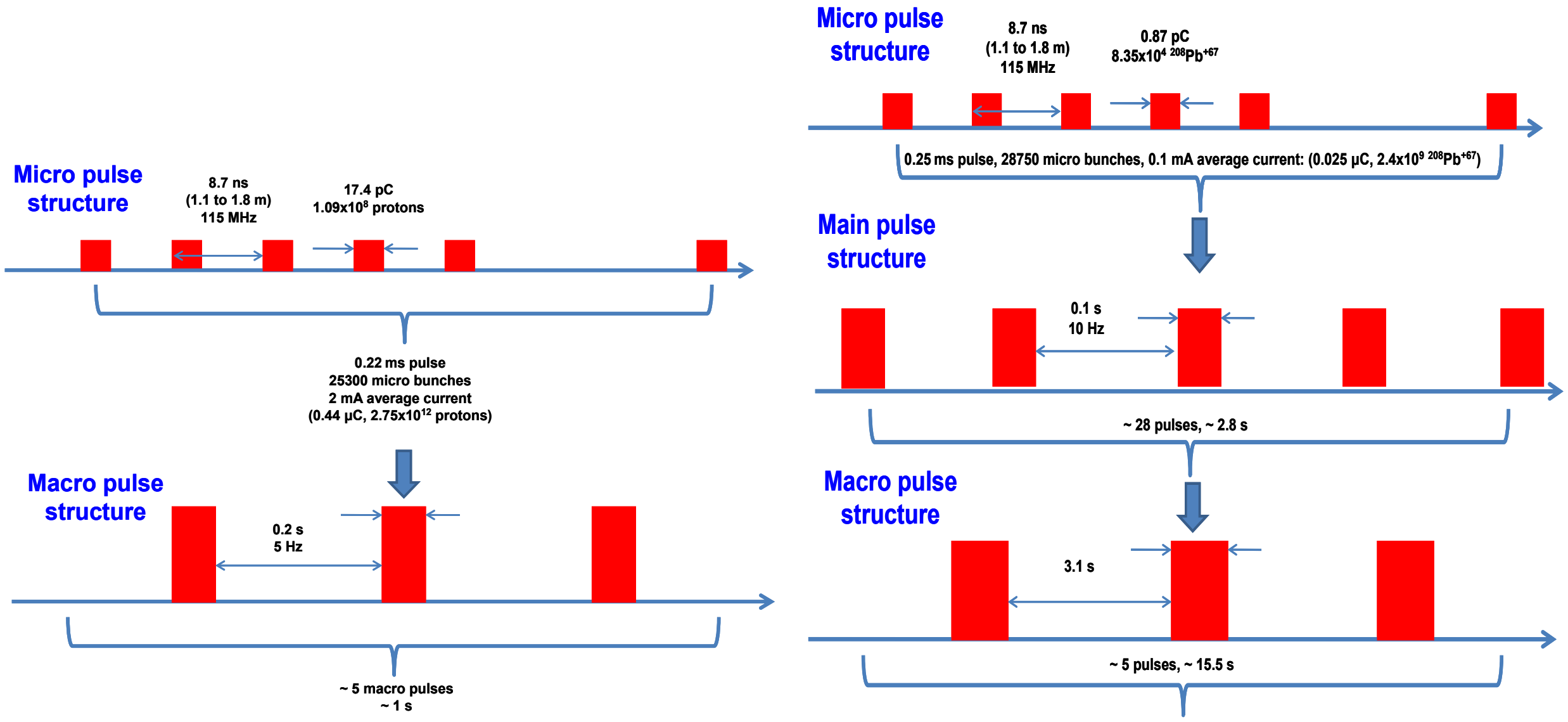

Figure 5.2: Proton (left) and lead ion (right) beams in the linac, pre-booster and large booster.

### 5.2.3 Acceleration of Ion Beams

#### *In Pre-booster*

The proton or lead ion pulse, once painting-injected into the pre-booster, becomes a coasting beam. It will be RF captured and accelerated utilizing LEIR-type FINEMET cavities [2]. These cavities are broadband, capable of covering the full frequency range required to accelerate protons to lead ions in the pre-booster, up to the third harmonic, with parameters listed in Table 5.3. Acceleration is achieved according to the voltage-synchronous-phase curves in Figure 5.3. The longitudinal phase space particle distributions at the beginning, middle and end of the acceleration ramp for protons and lead ions are presented in Figure 5.4.

Table 5.3. Acceleration and RF parameters in the MEIC pre-booster

| | | Proton | Lead |
|---|---|---|---|
| Initial & final kinetic energy | MeV/u | 280 / 3000 | 100 / 670 |
| Initial momentum spread | % | ±0.3 | ±0.054 |
| Initial beam emittance (RMS) | eV·s | 0.295 | 6.48 |
| RF harmonic number | | 1 | 1 |
| RF frequency | MHz | [0.82,1.25] | [0.55,1.04] |
| Maximum RF voltage | kV | 51.2 | 48.8 |
| Maximum phase | deg | 50 | 50 |
| Momentum compaction | | 0.039 | 0.039 |
| Final capture efficiency factor | | 0.999 | 0.95 |
| Final momentum spread | % | ±0.27 | ±0.11 |
| Final beam emittance (RMS) | eV·s | 0.19 | 8.00 |
| Total turns | 1000 | 130 | 115 |
| Acceleration period | ms | 117 | 144 |
| Bunch length | ms | 0.166 | 0.386 |

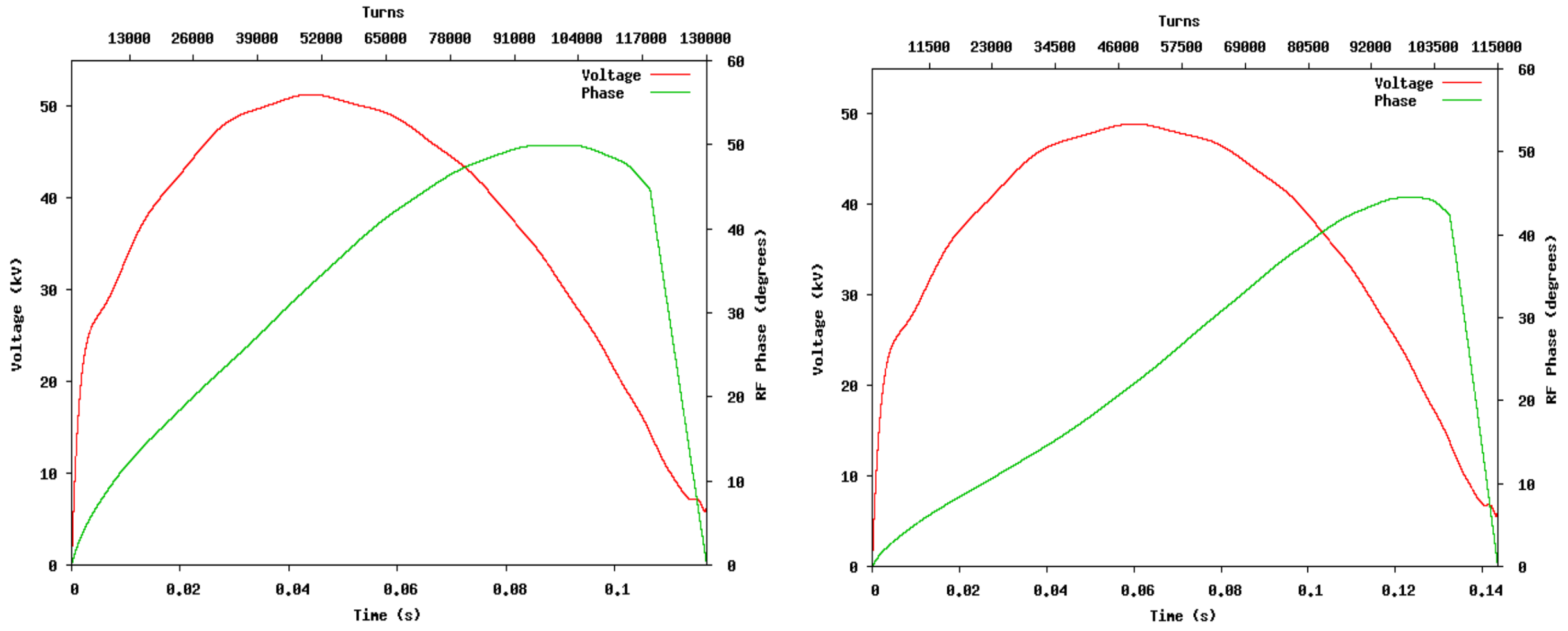

Figure 5.3: Acceleration in the MEIC pre-booster for protons (left) and lead ions (right).

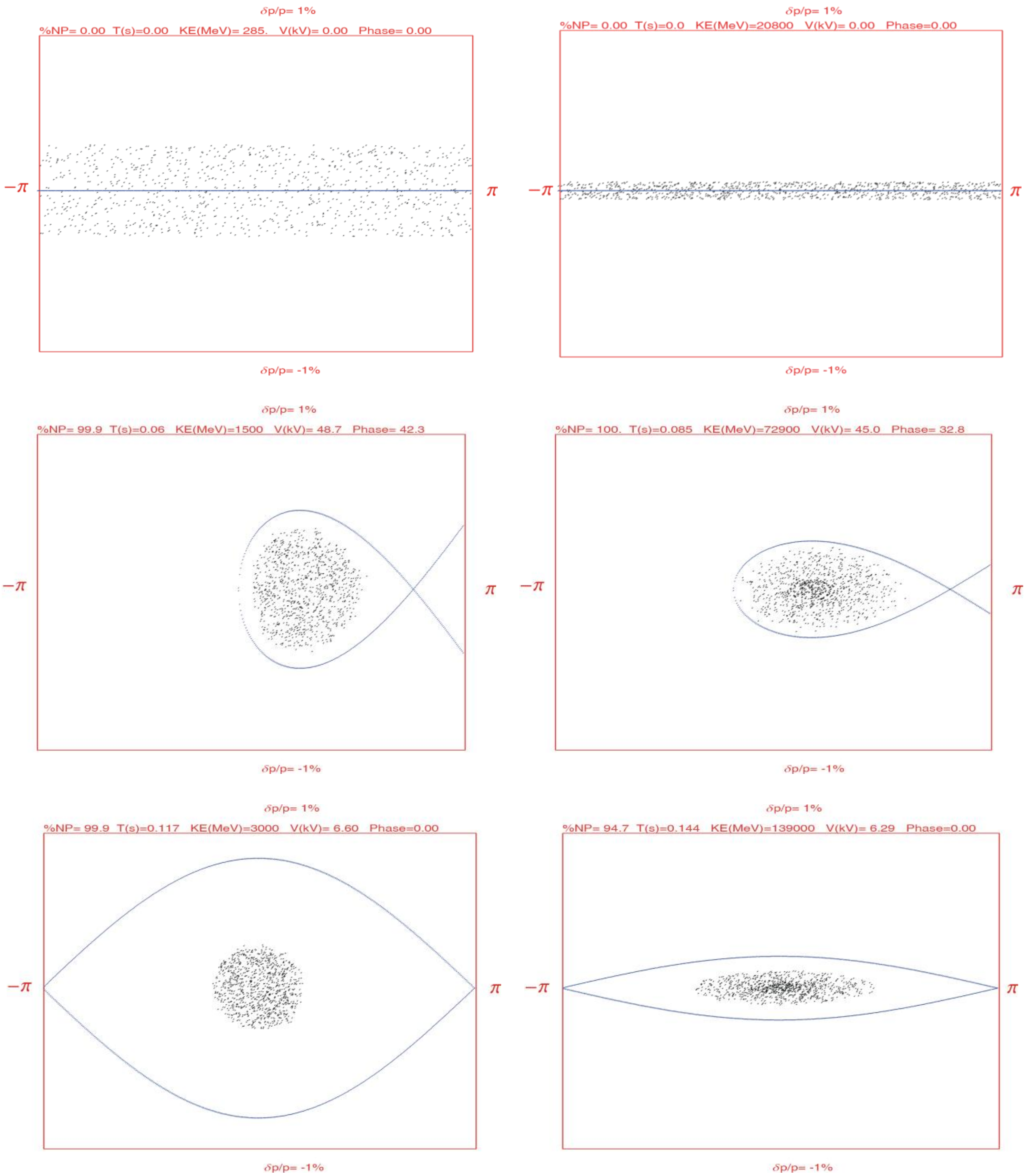

Figure 5.4: Longitudinal dynamics of proton (left column) and lead (right column) beams in the pre-booster at the beginning, middle, and end of the acceleration ramp.

### *In Large Booster*

The acceleration in the large booster will use voltage-phase curves appropriately scaled from the pre-booster case. The parameters for accelerating proton beams are listed in Table 5.4. The longitudinal phase space of a proton beam at the beginning and the end of the ramp can be seen in Figures 5.5.

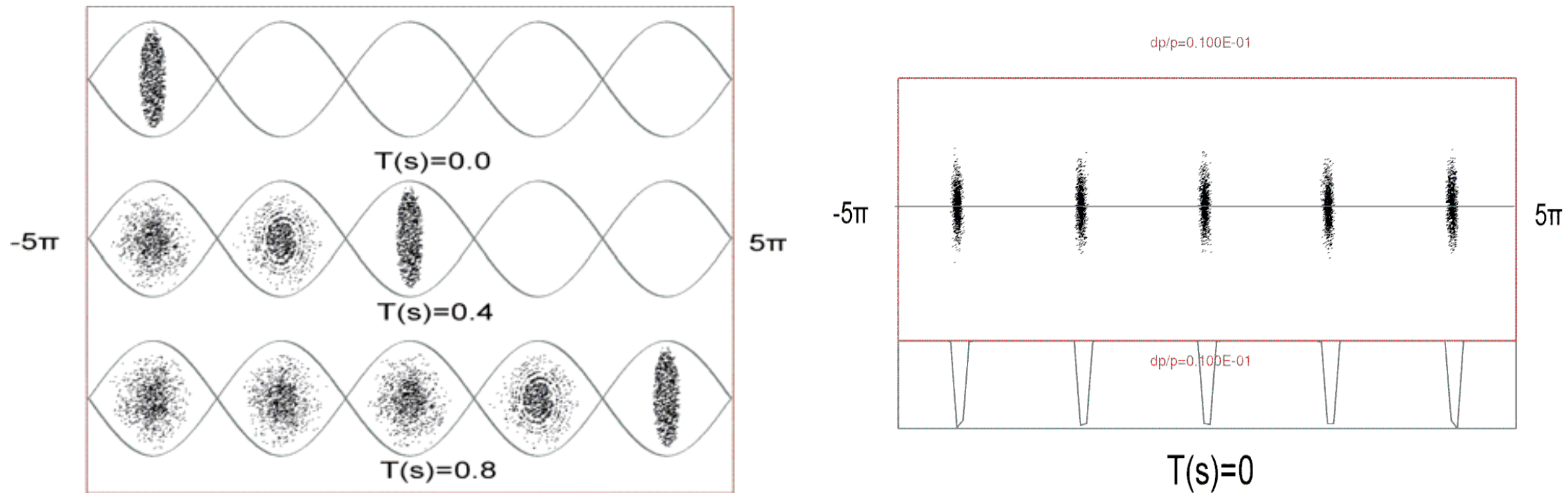

Figure 5.5: Proton beam at the start, middle and end of the accumulation process in the MEIC large booster ring (left) and after acceleration to 20 GeV (right).

Table 5.4: Accumulation and acceleration parameters of protons in the large booster

| | | |
|---|---|---|
| Initial & final kinetic energy | GeV/u | 3 / 20 |
| Assumed momentum compaction | | 0.039 |
| **Accumulation stage** | | |
| Initial momentum spread | % | ±0.27 |
| Initial beam emittance (RMS) | eV·s | 0.19 |
| RF harmonic number | | 5 |
| RF frequency | MHz | 1.21 |
| Maximum RF voltage | kV | 15 |
| Final capture efficiency factor | | 1 |
| Total turns | 1000 | 194.1 |
| Accumulation time | s | 0.8 |
| Minimum bunch separation | ms | 0.25 |
| Maximum bunch separation | ms | 0.48 |
| **Acceleration stage** | | |
| RF harmonic number | | 5 |
| RF frequency | MHz | [1.21,1.24] |
| Maximum RF voltage | kV | 208 |
| Final capture efficiency factor | | 1 |
| Final momentum spread | % | ±0.4 |
| Final beam emittance (RMS) | eV·s | 0.38 |
| Total turns | | 200000 |
| Acceleration period | ms | 806 |
| Average bunch separation | ms | 0.71 |

Space charge is obviously a limiting effect in the boosters. Preliminary studies indicate that ion current of 0.5 A is still manageable. In case the space charge in the pre-booster is too strong at injection energy, mitigation is required. There are two options for such mitigation: controlled emittance blow-up in the pre-booster and/or further beam accumulation in the large booster. Assuming acceleration parameters from this section and a kicker rise/fall time of approximately 100 ns, the maximum harmonic number possible in the large booster is 24 for protons and 10 for lead ions. Therefore, accumulation in the large booster with h=10 would allow a reduction of the required intensities in the pre-booster by a factor of two, appropriately reducing the space charge effects. In addition, mitigation of space charge effects would also allow reduction of the final energy of the ion linac, leading to cost savings. All these factors will be the subjects of future beam dynamics studies and cost optimizations.

### 5.2.4 Coasting Beam, De-bunching and Re-bunching

The beam transported into the collider ring still has a 1.24 MHz bunch structure from the big booster accelerating RF. It needs to be redistributed evenly into 750 MHz buckets for further acceleration and collisions [3]. There are two common approaches to accomplishing this RF gymnastic: debunch/rebunch, and adiabatic bunch splitting. Both require bunch-to-bucket transfer to maintain high transfer efficiency, so both booster rings should have matched 1.24 MHz RF buckets and transfer bunches either individually or in full-booster bunch trains.

#### *Debunch/Rebunch Scheme*

In this scheme, only 1.24 MHz and 750 MHz RF systems are required. After beam transfer into the collider ring, the 1.24 MHz RF voltage is adiabatically lowered to a value near zero before turned off. The longitudinally unfocused beam shears through the phase space until it uniformly fills the circumference of the collider ring except for an abort gap. A barrier bucket will be used to create and maintain the abort gap for machine protection. Initial studies show that the proton beam debunching time is about 4 s or 0.89 million turns.

After the beam is debunched, a 750 MHz RF voltage is adiabatically applied to recapture the beam in higher frequency RF buckets. Proton debunch/rebunch gymnastics with a barrier bucket was demonstrated at the Fermilab main injector in 2004, including intensity stacking by varying the barrier bucket length. The total time from source to start of re-bunching to high frequency for collisions in the collider ring will take approximately 6 s in the case of proton beams.

One advantage of a debunch/rebunch scheme is that it does not depend deeply on the 750 MHz RF harmonic number. This flexibility is important for the ability to provide collisions over a large energy range. There is some beam loss and longitudinal emittance growth during rebucketing, though such side effects will be somewhat ameliorated for MEIC by cooling. Longitudinal coasting beam instabilities must also be carefully controlled to create equal 750 MHz bunch intensities needed by the experiments.

#### *Adiabatic Bunch Splitting*

In this scheme, bunches are never completely debunched, but instead are split through a process of successive adiabatic bunch manipulations. This technique was developed in the CERN PS in 1989 [4] and is in routine use at several other facilities such as the Brookhaven booster. It requires more complicated bunch gymnastics than debunch/rebunch, but provides better longitudinal emittance control and instability suppression via Landau damping. Figure 5.6 shows a similar adiabatic bunch merging process of a gold ion beam in the Brookhaven booster.

The primary disadvantages of adiabatic bunch splitting in the context of the MEIC application are the need for broad-frequency (hence warm ferrite-loaded) RF in the collider ring, and the unprecedented ~600:1 splitting necessary to move from 1.24 MHz booster RF bunches to 750 MHz collider RF bunches.

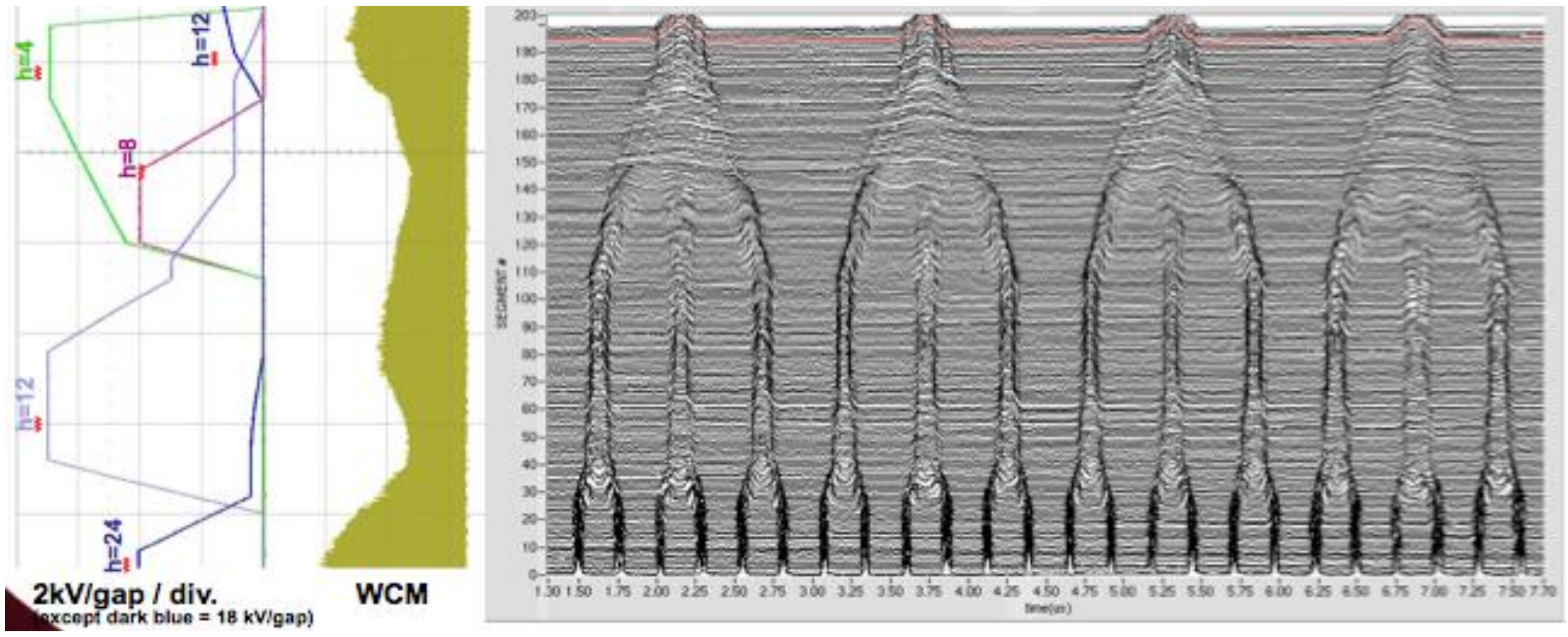

Figure 5.6: Gold beam adiabatic bunch merging in the Brookhaven Booster. Time flows from bottom to top. Four RF harmonics (*h*=4, 8, 12, 24) are used to perform successive 2-to-1 and 3-to-1 bunch merges for a final effective 6-1 merge.

### *Crabbing and Rebucketing*

Once the collider beam has been captured and accelerated with the 750 MHz accelerating RF, the bunches can be shortened by being rotated in phase space and captured by 1.5 GHz storage RF for crabbing and collision. This mechanism is called rebucketing, and has been in common use at RHIC since 2003 where 28 MHz bunches are recaptured into 7$^{th}$ harmonic 196 MHz RF buckets. Rebucketing in MEIC would be straightforward and transform a single 750 MHz bunch into a single 1.5 GHz bunch at the same intensity but with nearly twice the momentum spread. Given suitable RF, it may also be possible to adiabatically 1-to-2 split the 750 MHz bunches into separate 1.5 GHz bunches. We will evaluate the tuning capabilities of the MEIC SRF design to determine whether this adiabatic split is feasible.

## 5.2.5 Cooling of Ion Beams

Cooling of the ion beams is essential in the MEIC ion beam formation. It will be utilized in the pre-booster for assisting accumulation and in the collider ring for final preparation and preservation of phase space density during collision. At the pre-booster, a low energy DC electron beam is needed, and the electron energy is ranged from 155 keV for cooling 280 MeV protons to 54 keV for cooling 100 MeV/u lead ions. Such a DC electron cooler based on an electrostatic accelerator has been well developed [5]. At the collider ring, cooling will be performed at an energy range order of magnitude beyond state-of-the-art. The electron energy is from 4.3 MeV (for cooling lead ions at injection energy) to 54 MeV (for cooling protons at top collision energy). A conceptual design of an ERL based circulator electron cooler has been developed to meet this need. Chapter 6 provides a full description of this cooler design and discussions of key accelerator technologies to support this design.

## 5.3 Ion Sources

Three types of ion sources have been considered for the ion complex. They are an Atomic Beam Polarized Ion Source (ABPIS), an Electron Beam Ion Source (EBIS) and an Electron Cyclotron Resonance ion source (ECR), shown in Figure 5.7. Table 5.5 provides parameters based on the present state-of-the-art as well as performance extrapolations (in parentheses).

Table 5.5: Beams parameters after Low Energy Beam Transport (LEBT)

| Ions | Source Type | Pulse Width | Rep. Rate | Pulsed Current | Ions/pulse | Polarization | Emittance (90%) |
|---|---|---|---|---|---|---|---|
| | | μs | Hz | mA | $10^{10}$ | | π·μm·rad |
| $H^-/D^-$ | ABPIS | 500 | 5 | 4 (10) | 1000 | >90% | 1.0 / 1.8 |
| $H^-/D^-$ | ABPIS | 500 | 5 | 150 / 60 | 40000/15000 | 0 | 1.8 |
| $^3He^{++}$ | ABPIS | 500 | 5 | 1 | 200 | 70% | 1 |
| $^3He^{++}$ | EBIS | 10 to 40 | 5 | 1 | 5 (10) | 70% | 1 |
| $^6Li^{+++}$ | ABPIS | 500 | 5 | 0.1 | 20 | 70% | 1 |
| $^3Pb^{30+}$ | EBIS | 10 | 5 | 1.3 (1.6) | 0.3 (0.5) | 0 | 1 |
| $Au^{32+}$ | EBIS | 10 to 40 | 5 | 1.4 (1.7) | 0.27 (0.34) | 0 | 1 |
| $^3He^{++}$ | ECR | 500 | 5 | 0.5 | 0.5 (1) | 0 | |
| $Au^{32+}$ | ECR | 500 | 5 | 10.5 | 0.4 (0.6) | 0 | |

# Numbers in parentheses represent future performance extrapolations.

### *Atomic Beam Polarized Ion Source (ABPIS)*

This type of ion source can be used for production of polarized and un-polarized $H^-$, $D^-$, $H^+$, $D^+$, $^3He^+$ and $Li^+$ ions [6]. The main components of this type of H or D ion source include a source of polarized H or D atomic beams, a surface plasma source of cold negative $H^-$ and $D^-$ ions with arc discharge and surface plasma ionizer, a charge exchange solenoid with a grid extraction system, a bending magnet for separation of polarized and non-polarized $H^-$ and $D^-$ beams, and a polarimeter.

ABPIS with a selective resonant charge exchange ionization has a good potential for production of a $H^-/D^-$ ion beam with a high polarization. Low energy non-polarized $H^-/D^-$ transfers electrons only to H or D atoms but not molecules. Such an ionization scheme helps to enhance polarization of $H^-$ or $D^-$ beam more than that of hydrogen in atomic beam, because it excludes non-polarized molecules from the beam. It is possible to have up to 12% efficiency ratio of transforming polarized atoms to polarized $H^-$ ions. The high selectivity of polarized atom ionization can lead to 90% of polarization

For polarized $^3He^{++}$ production, it is possible to use the same source with a pulsed injection of nuclear polarized $^3He$ atoms (polarized by optical pumping) into an arc discharge plasma source. For polarized $Li^+$ production, the same ion source can be utilized, provided the hydrogen dissociator is replaced with an Lithium oven for Lithium atomic beam generation and a resonance frequency of the RF transition is changed. A hydrogen or helium plasma flux from the arc discharge plasma source can be used for polarized Li atom ionization with high efficiency.

## *Electron Beam Ion Source (EBIS)*

For production of heavy multi-charged ions, an EBIS source similar to the one used for RHIC injection at BNL is a cost effective candidate for MEIC [7]. In the afterglow mode it is possible to extract a high intensity beam with higher charge. Total charge of the accumulated ion pulse is limited by charge of the electron beam, according to the relation $N_z=N_e/Z$ where $N_e$, $N_z$ are numbers of electrons and ions respectively, and $Z$ is the charge state. As an example, the electron beam of the BNL EBIS, with a 2 m long pulse and 10 A pulse current, has $10^{11}$ electrons per pulse.

## *Electron Cyclotron Resonance Ion Source (ECR)*

An ECR source can also be used for production of heavy multiply-charged ions as at CERN [8]. It generates heavy multi-charged ions by ionization of electrons heated up to keV temperature by resonance RF fields. A complex magnetic system with minimum magnetic field is used for stable electron confinement. High resonance frequency is used in SC coils.

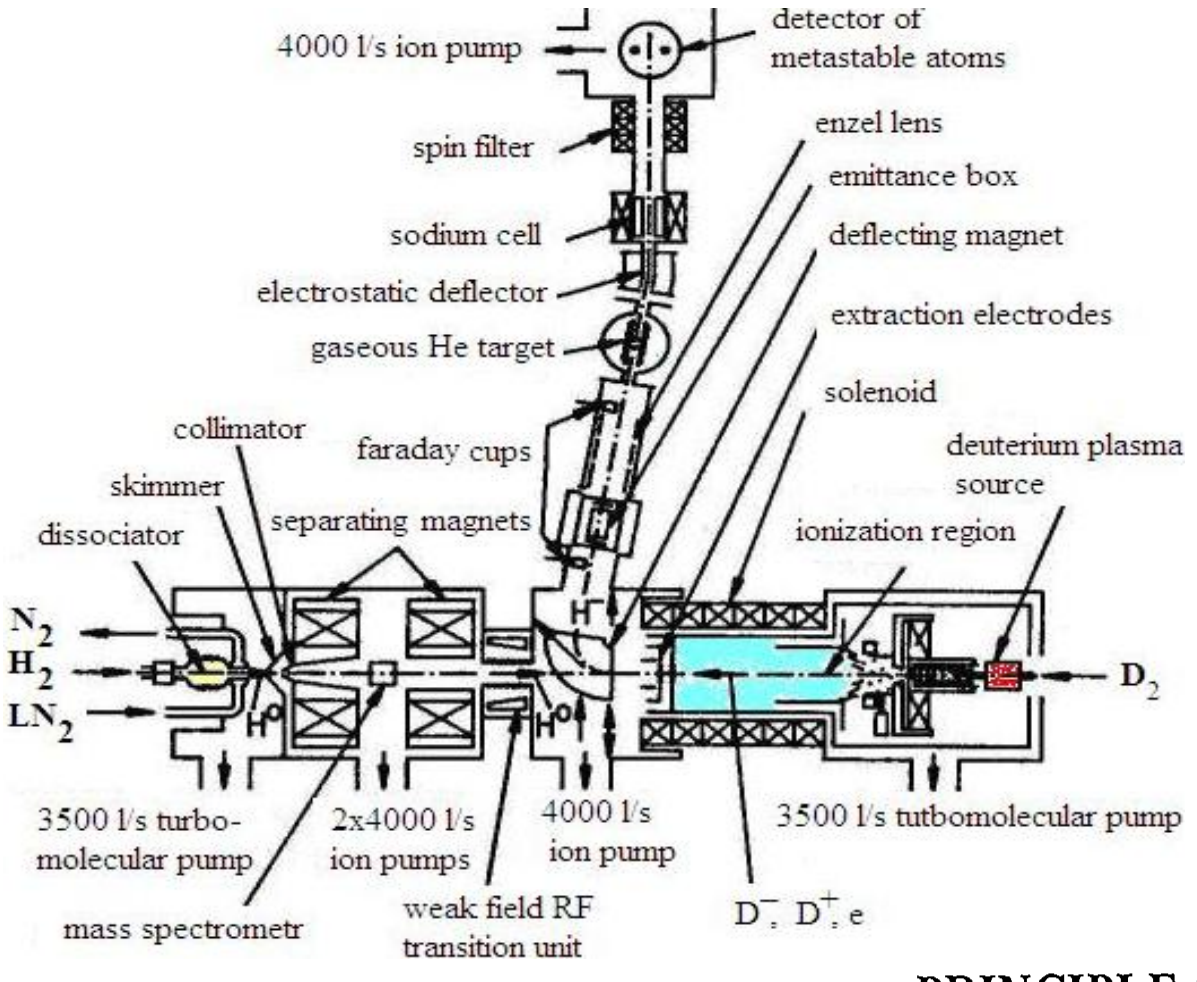

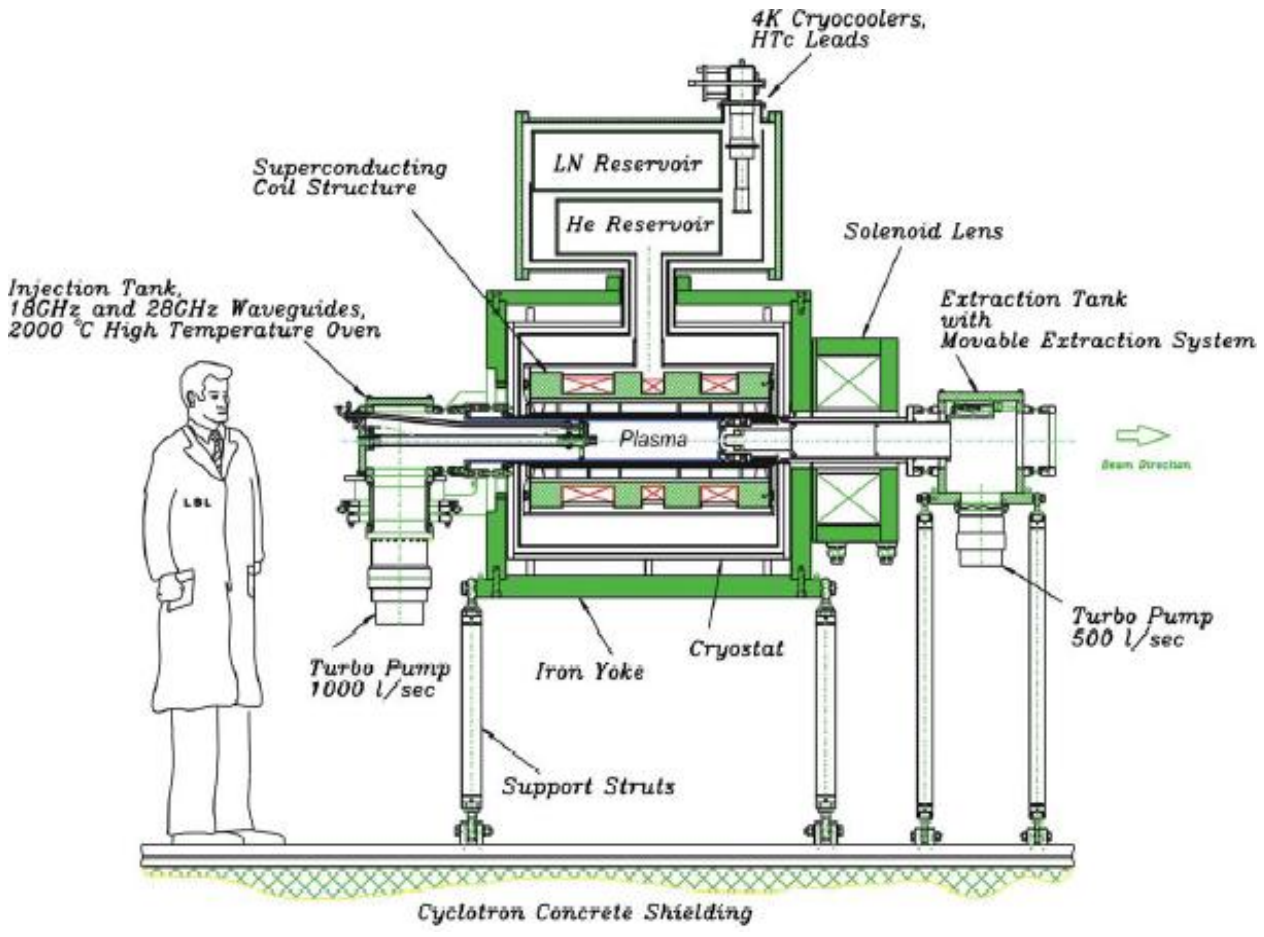

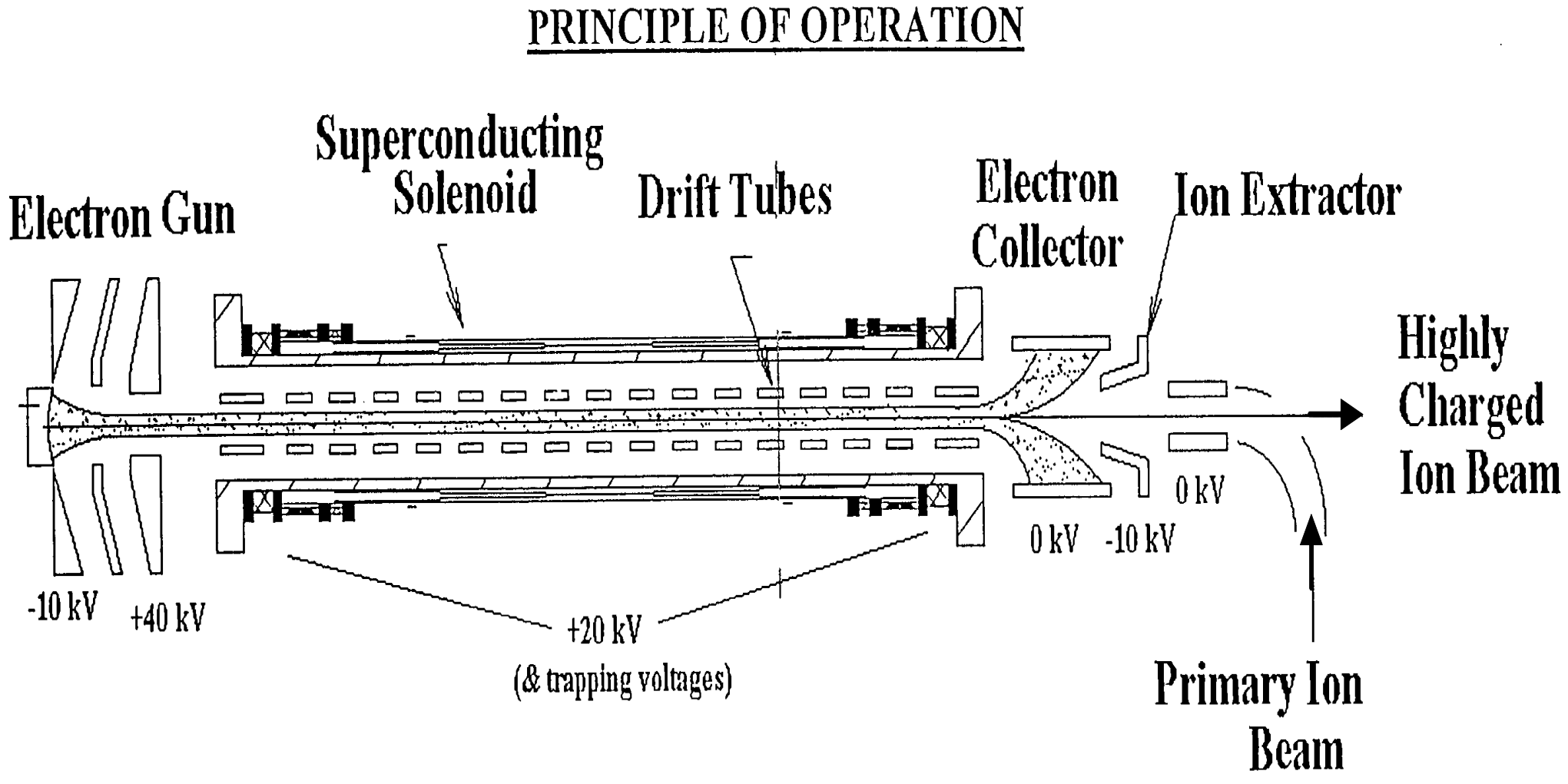

Figure 5.7: An ABPIS with resonant charge exchange ionization (top left), a 28 GHz ECR heavy ion source (top right) and an EBIS ion source (bottom).

Yield of ions in charge state q:

$$N_q = \frac{I_e \times L}{q \times \sqrt{V_e}} \times K_1 \times K_2$$

## 5.4 Ion Linac

A technical design of a SRF ion linac had been developed at Argonne National Laboratory originally as a heavy ion linear accelerator for a proposal of Facility of Rear Isotope Beam [9]. This 130-m long linac is very effective in accelerating a wide variety of ions from proton $H^-$ (280 MeV) to lead ion $^{208}Pb^{67+}$ (100 MeV/u), thus can be readily adopt with minimum modification as the MEIC driver linac. The basic design parameters of this linac are listed in Table 5.6 and a block-diagram is given in Figure 5.8 (left). Economic acceleration of lead ions to 100 MeV/u energy requires a stripper in the linac with an optimum stripping energy of 13 MeV/u, obtained from a plot of total accelerating voltage as a function of the stripping energy in Figure 5.8 (right). The stripping efficiency of lead ions to the most abundant charge state 67+ is 21%.

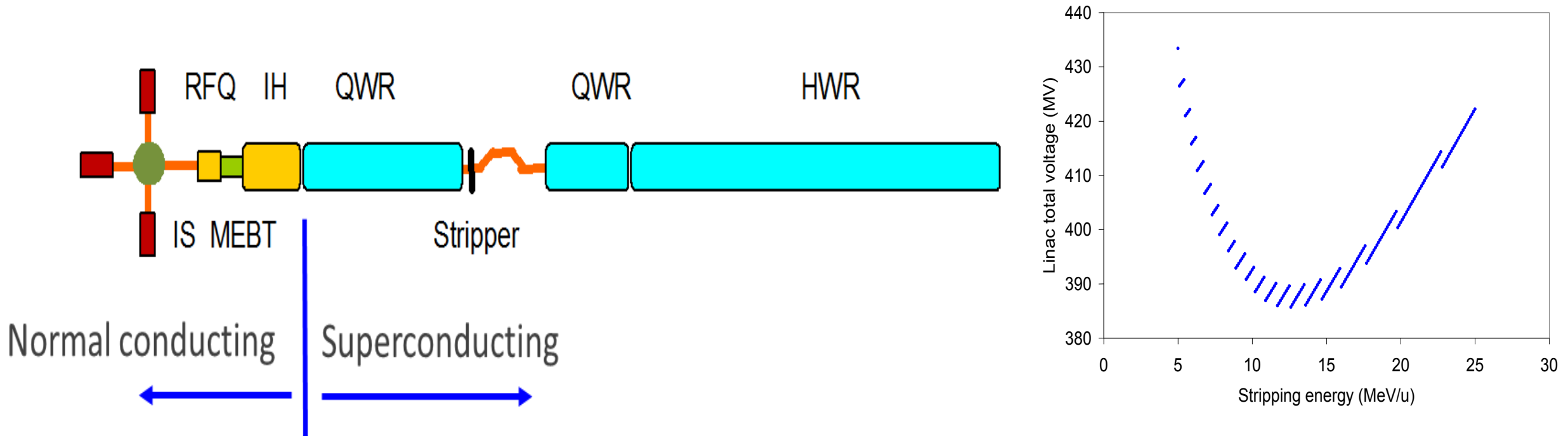

Figure 5.8: A schematic drawing of the MEIC ion linac (left) and effective linac voltage as a function of stripping energy of lead ions (right).

Table 5.6. Basic parameters of the linac.

| | | | |
|---|---|---|---|
| Ion species | | | proton to lead |
| Ion species for the reference design | | | $^{208}Pb$ |
| Kinetic energy of protons & lead ions | | MeV/u | 280 & 100 |
| Maximum pulse current | Light ions (A/q≤3) | mA | 2 |
| | Heavy ions (A/q>3) | mA | 0.5 |
| Pulse repetition rate | | Hz | up to 10 |
| Pulse length | Light ions (A/q≤3) | ms | 0.5 |
| | Heavy ions (A/q>3) | ms | 0.25 |
| Maximum beam pulsed power | | kW | 680 |
| Fundamental frequency | | MHz | 115 |
| Total length | | m | 121 |

The MEIC ion linac includes room temperature RFQ and inter-digital IH structure operating at a fixed velocity profile, similar to the CERN lead-ion linac [10] and to the BNL pulsed heavy-ion injector [7]. This linac section provides 4.8 MeV/u energy for all type of ions, and is considered highly effective, especially for pulsed machines. The ion beams will be subsequently accelerated by the SRF section of the linac which comprises two different types of accelerating cavities to cover velocity range from 0.1 to 0.5 of speed of light. The quarter wave resonator (QWR) and half wave resonator (HWR) have been developed for the application in next generation heavy ion driver linacs [9]. A photo of QWR and 3D view of the HWR are shown in Figure 5.9. All these cavities have been built and tested, providing excellent performance [11]. For the present MEIC application, 80 mT and 40 MV/m surface peak fields in all these cavities is proposed as design parameters. The SC cavities will be combined into cryostats with the length of about 6 m together with SC focusing quadrupoles. Table 5.7 shows dimensions of each linac section, and the total linac length excluding the injector and LEBT is 130 m. The performance of SC cavities in the velocity range required for the MEIC injector is continuously improving due to the development work for many projects [12,13]. To the time when the MEIC project starts, higher performance SC cavities will be available. This can result in less cavities, shorter linac and reduced cost. The fundamental frequency of the linac does not impact any parameter of the downstream chain of accelerators in MEIC. While fundamental frequency of the linac will be close to 100 MHz, the exact number can be different from 115 MHz and depends on availability of high-performance SC cavities developed by the beginning of the MEIC project.

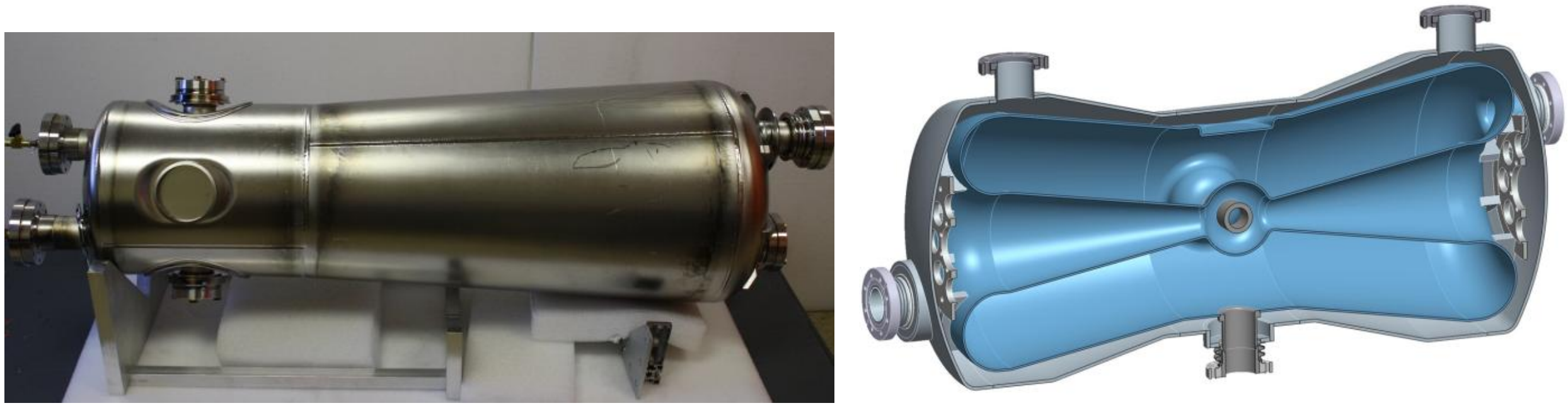

**QWR** **HWR**

Figure 5.9: Superconducting cavities for the MEIC linac.

Table 5.7. MEIC ion linac parameters

| | Length | Frequency | cryostats |
|---|---|---|---|
| | m | MHz | |
| RFQ | 3 | 115 | - |
| MEBT | 3 | | - |
| Room Temperature IH structure | 9 | 115 | - |
| QWR | 24 | 115 | 4 |
| Stripper and chicane | 10 | | |
| QWR | 12 | 115 | 2 |
| HWR | 60 | 230 | 10 |
| Total length (excluding injector, LEBT) | 121 | | |

The MEIC ion linac comprises 122 SC cavities, with basic parameters listed in Table 5.8. A dog leg magnet system is inserted after the stripping foil to dump unwanted charge states in a designated area. The linac can be re-phased to accelerate any ions from proton to lead, Table 5.9 shows energies of several selected ion beams. Though this linac was originally designed to provide optimal voltage gain for lead ions, due to the wide velocity acceptance of the proposed cavities, lighter ions can also be accelerated to higher velocities as is shown in Table 5.9. Figure 5.10 shows the voltage gain for leads and deuterons.

Before ending this section, it is worth to note that all technical systems of this linac are based on well-developed technologies and some of them are commercially available. The phase parameters of the ion beam at exit of the linac are as follows: 1 π·mm·mrad transverse normalized emittance, 10 π·keV/u·nsec longitudinal emittance and 0.05% momentum spread.

Table 5.8: Design parameters of SRF resonators in MEIC ion linac

| B | Length (βλ) | $E_{PEAK}$ (at $E_{ACC}$=1 MV/m) | $B_{PEAK}$ | R/Q | G | $E_{PEAK}$ | $E_{ACC}$ | Phase | Number of Cavities |
|---|---|---|---|---|---|---|---|---|---|
| | Cm | MV/m | mT | Ω | Ω | MV/m | MV/m | deg | |
| 0.15 | 39.1 | 4.7 | 57 | 509 | 42 | 30 | 6.5 | 20 | 42 |
| 0.3 | 39.1 | 4.5 | 78 | 250 | 58 | 40 | 9.0 | 20 | 80 |

Table 5.9. Ion beam energies in the MEIC ion linac.

| | Charge at source | Energy at stripper | Charge after stripper | Total Energy |
|---|---|---|---|---|
| | | MeV/u | | MeV/u |
| Proton | 1 | 55 | 1 | 280 |
| Deuteron | 1 | 32.8 | 1 | 162 |
| $^{40}$Ar | 12 | 22.4 | 18 | 143 |
| $^{132}$Xe | 26 | 16.5 | 48 | 117 |
| $^{208}$Pb | 30 | 13.2 | 67 | 105 |

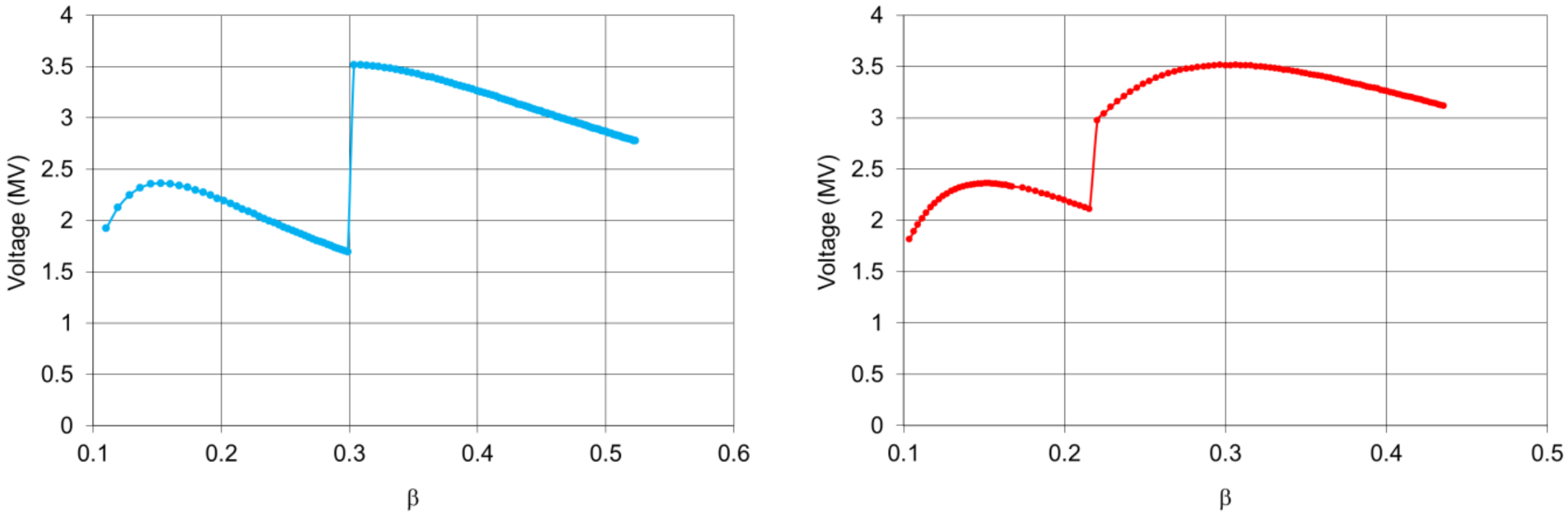

Figure 5.10: Voltage per resonator as a function of velocity of deuteron (left) and lead ion (right).

## 5.5 Pre-booster

The MEIC pre-booster synchrotron accepts ion pulses from the linac, and after accumulation and acceleration, transfers the beam to the subsequent large booster for further acceleration. Exact mechanisms of pre-booster operation will depend on the ion species. The two extreme cases are polarized protons and un-polarized lead ions. In the following we briefly describe the similarities and distinctions of these two scenarios.

### *Protons*

A 0.22 ms long pulse of 280 MeV protons is injected into the ring utilizing multi-turn injection with combined longitudinal and transverse painting and the charge exchange injection mechanism. The resulting coasting beam is then captured and accelerated with the first harmonic up to 3 GeV and fast-extracted to the large booster.

***Lead Ions***

A number of 0.25 ms long pulses of 100 MeV/u lead ions are injected into the pre-booster by a similar multi-turn painting scheme as in the case of protons. However, the painted phase space region is smaller (about half the acceptance) in order to leave room for stacking. Therefore, stacking/accumulation of lead ions must be performed under drag-and-cool conventional DC electron cooling. This process is repeated to eventually reach the required beam intensity. The cooled stack is accelerated to 0.67 GeV/u on first harmonic and finally fast-extracted to the large booster.

The pre-booster was designed by taking into account the following desired properties:

- Figure-8 shape for ease of spin transport, manipulation and preservation.
- Adequate space and optical properties for all the inserts including electron cooling, RF system, snakes, collimation, injection and ejection.
- Modular design, with (quasi)-independent module design optimization.

Since the pre-booster will be housed in a separate tunnel instead, there is no restriction on dimension and shape of the footprint except it must have a figure-8 ring; however, we prefer a compact design for cost efficiency. For the optics design, it has been decided that there be:

- FODO arcs for simplicity, compactness and ease optics correction schemes.
- No dispersion suppressors on ends of arcs
- Triplet straights for long dispersion-free drift and round beams.
- Matching/tuning modules between arcs and straights.
- Matched injection insertion with a large constant normalized dispersion region.
- Sizable (normalized) dispersion for/at injection.

In addition, it is desired that the pre-booster should have:

- Momentum compaction smaller than 1/25 so the transition gamma is large than 5.
- Maximum full beam size less than 2.5 cm, and 1 cm vertically in dipoles.
- Betatron tune working point chosen such that the tune footprint does not cross low order resonances (tunability requirement).

Further the pre-booster will be made of warm magnets with the following limits:

- Maximum dipole bending field is 1.5 T.

- Maximum quadrupole gradient is 20 T/m.

The footprint of the pre-booster is shown in Figure 5.11. It consists of two uneven half-rings connected by two long straights. The circumference is 234 m and the figure-8 crossing angle is 75 degrees. The left half ring consists of 9 standard FODO cells which provide a total 255° bending angle. The total phase advance over this arc section is a multiple of 2π, therefore making dispersion suppressors unnecessary at either end of the arc. The right half ring is divided into two identical arc sections with an injection insert in between. Each of two identical right arc sections consists of three standard FODO cells. The matched injection insertion, designed optically as an identity transfer matrix, contains a long straight with large constant normalized dispersion for efficient injection painting in all three phase planes simultaneously. The two long straights are composed of triplet cells which allow long drifts with round beam. The long drifts provide space for several necessary components such as electron cooling devices, RF cavities, collimation, extraction, and possibly further solenoids and other elements for horizontal-vertical decoupling. There are modules of five quadrupoles at the ends of long straights for matching optics between straights and arcs as well as providing knobs for tune adjustment. Since the phase space ellipses of the ion beams are upright by design at the ends of the matching sections, only two sets of independent adjusting knobs are needed; therefore, the other two sets will be arranged as a mirror symmetry of the first two [14].

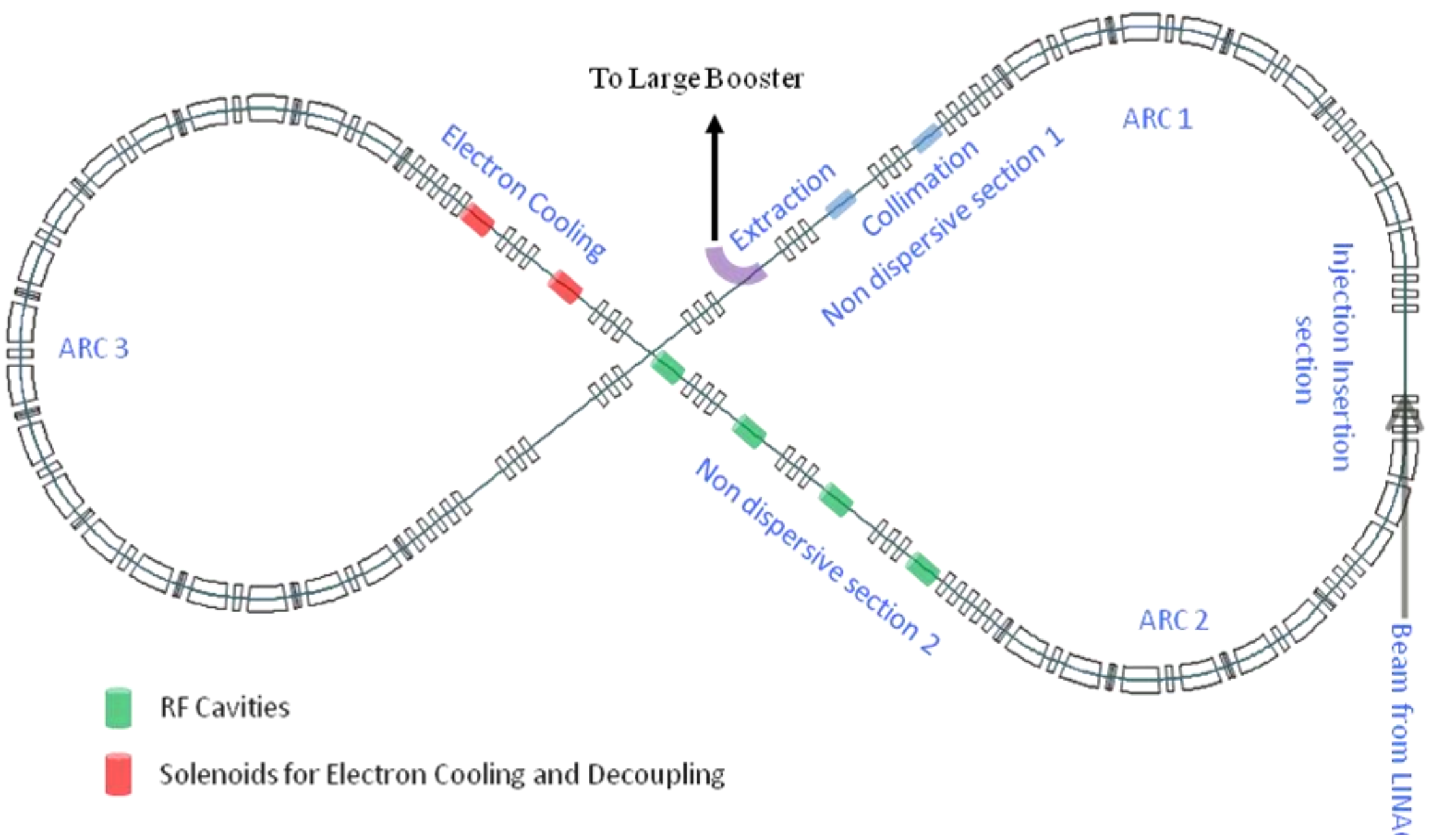

Figure 5.11: Footprint of the MEIC pre-booster.

Figure 5.12 shows optical functions of the pre-booster. Maximum beta functions are below 30 m, and the absolute value of dispersion is less than 3.4 m. The horizontal and vertical tunes are currently set to 7.96 and 6.79 respectively, with no working point optimization performed yet. Other optics parameters can be found in Table 5.10. The magnet parameters are summarized in Table 5.11. The ring will always operate below the transition energy.

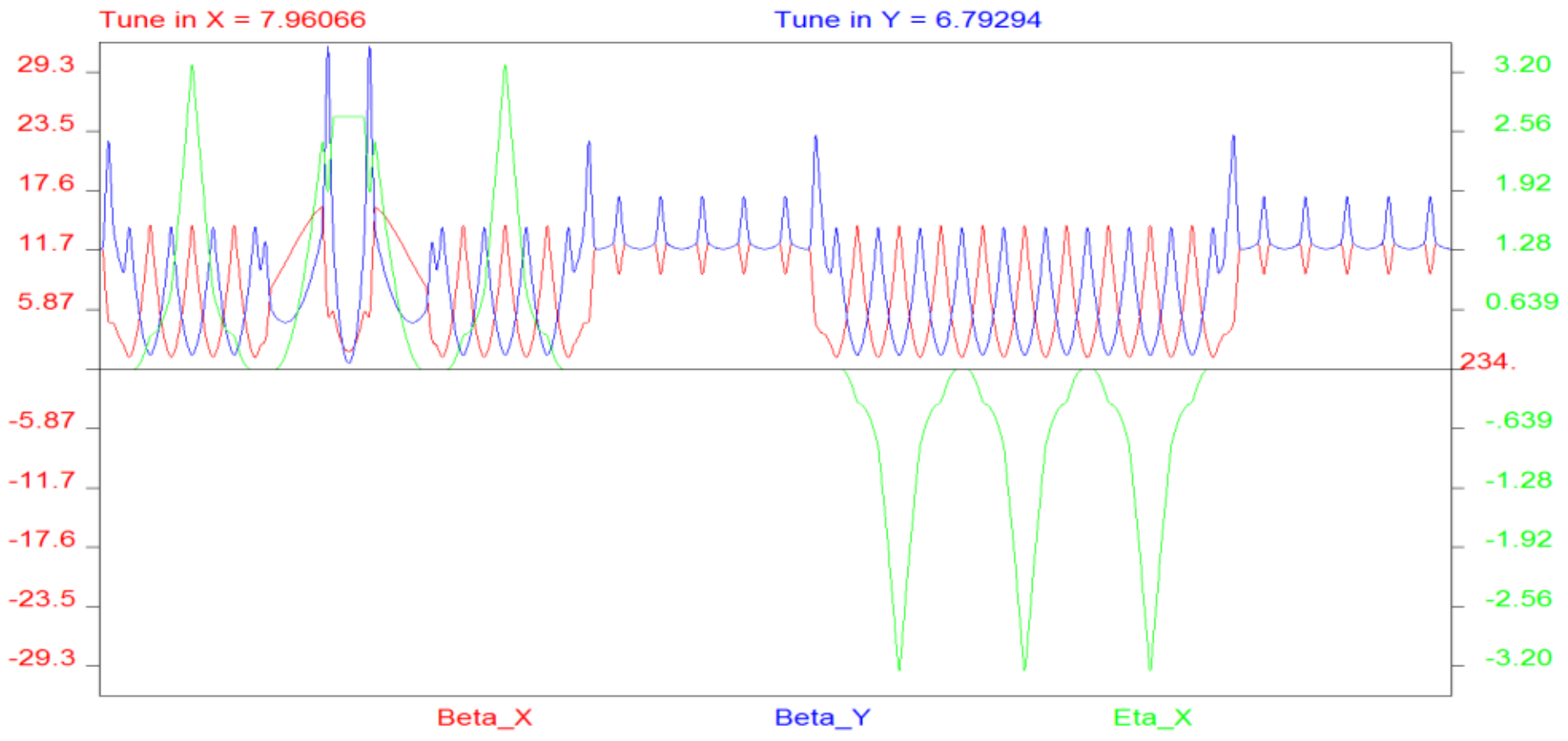

Figure 5.12: Optical function of the footprint of the MEIC pre-booster.

Table 5.10: Lattice parameters of the MEIC pre-booster

| | | |
|---|---|---|
| Circumference | m | 234 |
| Figure-8 crossing angle | deg | 75 |
| FODO cells in left arc and two right arcs | | 6 & 9 |
| Numbers of triplet cells and matching cells | | 10 & 4 |
| Minimum drift length between magnets | cm | 50 |
| Drift lengths in the insertion region | m | 5 |
| Drift between triplets (RF, collimation, cooling) | m | 5 |
| Maximum betatron function in $x$ and $y$ | m | 16 & 32 |
| Maximum beam size (and inside dipole) | cm | 2.3 (0.5) |
| Maximum dispersion | m | 3.36 |
| Normalized dispersion at injection | | 2.53 |
| Horizontal & vertical betatron tune | | 7.96 & 6.79 |
| Momentum compaction & transition gamma | | 0.04 & 5 |

Table 5.11: Pre-booster magnet parameters.

| | Quantity | Parameters | Units | Value |
|---|---|---|---|---|
| Quadrupole | 95 | Length | cm | 40 |
| | | Half aperture | cm | 5 |
| | | Maximum pole tip field | T | 1.53 |
| Dipole | 36 | Length | m | 2.19 |
| | | Vertical full gap | cm | 3 |
| | | Maximum strength | T | 1.41 |
| | | Radius | m | 9 |
| | | Angle | deg | ≈14 |

Finally, Table 5.12 summarizes parameters of proton and ion beams in the pre-booster as well as the RF cavities used to accelerate the beams. Emittances of the lead beam after cooling are set such that the maximum Laslett tune shift is below 0.3. This might lead to partial blow-up of transverse emittances after cooling, and will require further study.

Table 5.12: Main parameters of proton and lead beams in MEIC pre-booster

| | | Proton | Lead |
|---|---|---|---|
| Injection & extraction energy | GeV/u | 0.280 & 3 | 0.1 & 0.67 |
| Charge state in pre-booster | | +1 | +67 |
| Current at extraction | A | 0.5 | 0.5 |
| Total number of ions in the ring | $10^{10}$ | 252 | 4.5 |
| **Beam from linac** | | | |
| Pulse length | ms | 0.22 | 0.25 |
| Frequency | MHz | 115 | 115 |
| Number of micro bunches in pulse | | 25300 | 28750 |
| Average pulse current | mA | 2 | 0.1 |
| Charge per pulse | μC | 0.44 | 0.025 |
| Number of ions per pulse | $10^{10}$ | 275 | 0.24 |
| Injection efficiency | | 0.9 | 0.7 |
| Number of pulses | | 1 | 28 |
| **Timing** | | | |
| Cooling electron energy | MeV | | 0.56 to 0.88 |
| Electron cooler current | A | | 0.3 |
| Length of cooling section | m | | 3 |
| Cooling time, transverse & longitudinal | ms | | 16 & 55 |
| RF acceleration time | ms | 120 | 144 |
| Pre-booster cycle time | s | 2 | 3.1 |
| **Beam profile at injection** | | | |
| Acceptance in $x$ and $y$ (uniform, normalized) | π mm·mrad | 92 & 50 | 92 & 50 |
| Momentum acceptance ($\Delta p/p$) | % | ±0.3 | ±1 |
| Emittance in $x$ & $y$ ($\varepsilon_x$, $\varepsilon_y$) (uniform, normalized) | π mm·mrad | | 50 & 25 |
| Momentum spread ($\Delta p/p$) | % | | ±0.3 |
| **Beam profile after cooling/accumulation** | | | |
| Emittance in $x$ & $y$ ($\varepsilon_x$, $\varepsilon_y$) (full, uniform, normalized) | π mm·mrad | | 20 & 10 |
| Momentum spread ($\Delta p/p$) | % | | ±0.054 |
| **Beam profile at extraction** | | | |
| Emittance in $x$ & $y$ ($\varepsilon_x$, $\varepsilon_y$) (full, uniform, normalized) | π mm·mrad | | 20 & 10 |
| Momentum spread ($\Delta p/p$) (95% capture efficiency) | % | ±0.27 | ±0.21 |
| Bunch length | ms | 0.166 | 0.386 |
| **Space charge** | | | |
| Laslett tune shift (at injection) | | -0.025 | -0.16 |
| Max. Laslett tune shift (beginning of acceleration) | | -0.038 | -0.3 |
| Laslett tune shift (at extraction) | | -0.006 | -0.09 |

## 5.6 Large Booster

The MEIC large booster accepts ions from the pre-booster and accelerates them to 20 GeV (proton energy or equivalent ion energy with a same magnetic rigidity) before sending them to the collider ring. This booster synchrotron, which is made of normal conducting magnets, is placed in the same tunnel and stacked vertically above the ion collider rings [15]. This geometric constraint, along with a requirement of no crossing of the transition energy during acceleration for any ion species, drives much of the booster design, some parameters of which are detailed in Table 5.13. The no-crossing transition energy requirement is satisfied by keeping the transition energy above the highest energy attained in the synchrotron.

Table 5.13. Design parameters for the MEIC large booster

| | | |
|---|---|---|
| Proton kinetic energy | GeV | 3 to 20 |
| Circumference | M | 1343.9 |
| Figure-8 crossing angle | deg. | 60 |
| Quarter arc length | M | 189.1 |
| Length of long and short straight | M | 267.8 / 25.9 |
| Lattice base cell | | FODO |
| Arc/straight cell length | M | 5.91 / 6.1 |
| Phase advance per cell arc/straight | deg. | 85.7 / 90 |
| FODO cells in arcs/straights | | 32 / 43 |
| Dispersion suppression method | | Phase advance |
| Max. horizontal/vertical beta function | M | 15.0 / 19.0 |
| Maximum horizontal dispersion | M | 0.73 |
| Horizontal/vertical betatron tunes | | 53.078 / 52.54 |
| Natural horizontal/vertical chromaticity | | -66.9 / -67.1 |
| Transition Energy factor $\gamma_{tr}$ | | 25.0 |
| Horizontal/vertical normalized emittance | mm·mrad | 4 |
| Max. horizontal/vertical rms beam size | Mm | 7.7 / 8.8 |

The large booster has a figure-8 shape ring with a footprint matching the collider rings. The figure-8 shape also helps preserve the ion polarization in the large booster. The booster comprises four identical quarter arc sections and two short and two long straight sections. A set of six arc dipoles are placed about 30 m away from end of one of the four quarter arcs in order to follow the footprint of the universal spin rotators in the electron ring. The present booster design does not include a few details such as a required horizontal chicane to bypass the detectors in the machine. A maximum magnetic field of 1.7 T is assumed for the warm magnets.

The large booster optics is presently designed as a FODO lattice, with sextupoles placed in the arcs for chromaticity correction. The optical functions for the arc part are shown in Figure 5.13. The two short straight sections are currently filled with a FODO lattice that creates an identity matrix as placeholders for Siberian snakes.

The two long straights of the figure-8 each include one set of dipoles to fit around the local chromatic compensation blocks of collider rings in the IRs. This double bend geometry can ensure dispersion suppression without ruinously high phase advances. To improve tunability, the cells in a long straight can be adjusted in groups of seven to alter the overall phase advance of the booster ring. There are four such tuning groups in the large booster; therefore, even if a change of 180° phase advance is required, increase in the beta function is only about 1 m. This makes tunability of the machine a function of the electrical busses more than the physical layout. While an analysis of the ideal working point has not been done yet, we can reach any that is needed. The optical functions of a long straight section are shown in Figure 5.14.

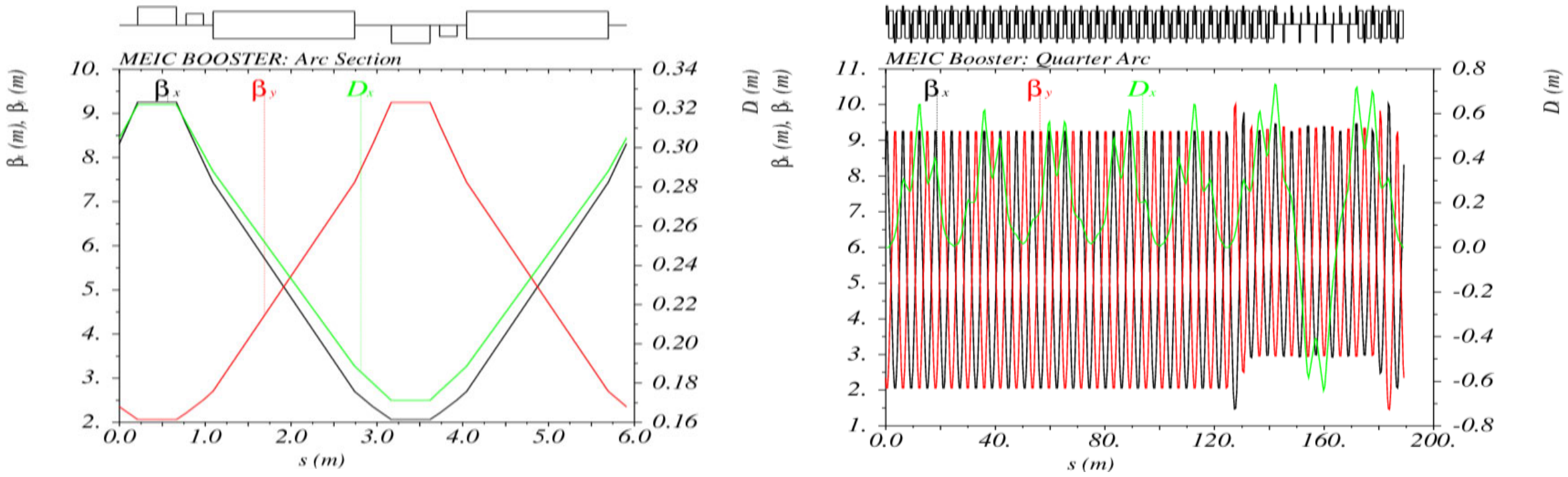

Figure 5.13. Optical functions of a single arc FODO cell (left) and of an entire quarter arc (right).

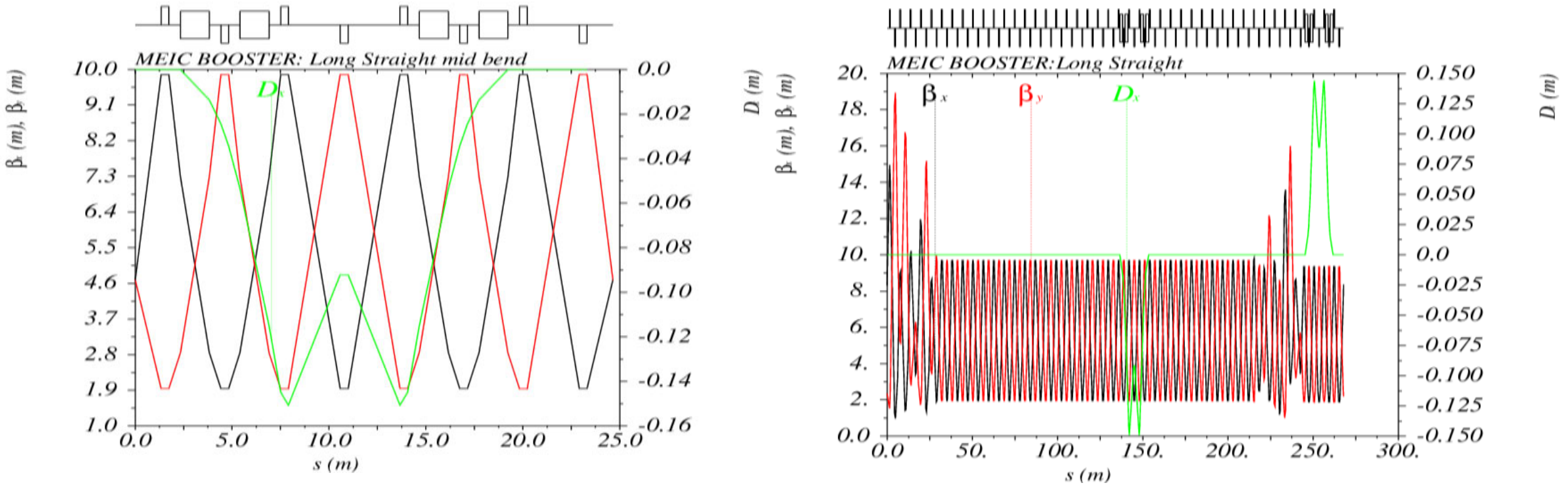

Figure 5.14. Optical functions of one bending section in a long straight (left) and of an entire long straight (right).

The injection and extraction systems have not been included in the present design of the large booster. The location of the injection line is likely in the area before the final bending of a quarter arc section, and the extraction is in one long straight. An emergency dump line will also be added to the design. The Laslett tune shift at the lowest energy should not exceed 0.1.

A rough estimation of inductance and resistance of the dipoles gives the ramping time as approximately 0.95 seconds. This leads to an acceleration requirement of about 150 keV per turn. The pre-booster is approximately 1/5 the size of the large booster, and a period of approximately 0.8 seconds is required to fill the large booster. The numbers and values for the various magnets in the design are shown in Table 5.14.

Table 5.14. The large booster magnet parameters

| | Quantity | Parameter | Units | Value |
|---|---|---|---|---|
| Dipoles arc/straight | 204/16 | Length | m | 1.65 |
| | | Field strength | T | 1.64 / 0.769 |
| | | Bending radius | m | 42.49/90.65 |
| | | Bending angle | deg. | 2.225 / 1.043 |
| Quadrupoles arc/straight | 192/286 | Length | m | 0.45 |
| | | Gradient arc straight | T/m | 75.22 / 75.59 |
| Sextupoles | 256 | Length | m | 0.2 |
| | | K2 type1/type 2 | $T/m^2$ | 20 / -33 |

## 5.7 Ion Collider Ring

The ion collider ring is stacked vertically above the electron ring and housed in the same tunnel [16]; therefore, the two footprints must be nearly identical. This means deviation of the two rings should not be larger than a few meters, roughly the width of the tunnel, though an exception could be made at isolated locations if required, with specially widened sections of provided. The layout considerations and geometric constraints concerning the two collider rings are discussed in Section 4.3.1, from which the optics design has followed. The main parameters of this collider ring are summarized in Table 5.15. Several simplifications have been introduced in this version of the conceptual design. For example, the vertical chicanes which bring the ion beams to the plane of the electron ring for collision are not included, nor are the other required elements such as Siberian snakes [17], injection and beam dump.

Table 5.15: Main design parameters of the MEIC ion collider ring

| | | |
|---|---|---|
| Proton kinetic energy | GeV | 20 to 100 |
| Circumference | m | 1340.92 |
| Figure-8 crossing angle | deg | 60 |
| Arc length and average radius | m | 391.0 / 93.34 |
| Length of long and short straight | m | 279.46 / 20 |
| Lattice base cell | | FODO |
| Arc / straight cell length | m | 9 / 9.3 |
| Phase advance per cell, hori. / vert. | deg | 60 / 60 |
| FODO cells in arcs / straight | | 52 / 20 |
| Dispersion suppression | | Adjusting quadrupole strength |

We start by showing details of the arc design. The four quarter arc sections, two on each side of the figure-8 ring and separated by two long and two short straights, are designed identically, and a FODO lattice [18] is adopted. The base unit is 9 m long and filled with two 3 m long superconducting dipoles for a 3.236° bending angle. There are a total 33 such SC dipoles in one quarter arc section that provides 106.8° bending. Three more SC dipoles inside additional FODO cells are placed 21 m away from the regular arc FODO cells, described above, in order to match the geometry of a spin rotator [19,20] in the electron ring. These SC dipoles have the same length of 3 m and provide a 4.4° bending angle per dipole, and three of them yield 13.2°. These additional FODO cells are further connected to the long straight of the figure-8 ring. The magnet parameters are listed in Table 5.16. The arc SC dipole strength is 6.2 T for 100 GeV protons, close to the 6 T limit in the design guideline. Dispersion suppression at the end of a quarter arc connecting to the short straight is accomplished by adjusting field gradients of three quadrupoles [18]. The optical functions of a quarter arc are plotted in Figure 5.15 [21].

Table 5.16: Ion collider ring magnet parameters at 60 GeV

| | | | | |
|---|---|---|---|---|
| Dipole | 144 | Length | m | 3 |
| | | Field strength | T | 3.77 |
| | | Bending radius | m | 53.1 |
| | | Bending angle | deg | 3.236 |
| Quadrupole | 298 | Length | m | 0.5 |
| | | Field gradient in arc / straight | T/m | 92.2 / 89.4 |

The betatron phase advances per arc FODO cell were chosen to be 60° in each plane. The horizontal phase advance also determines the transition energy [22,23] of the ion collider ring; the value associated to this lattice is 13.1 GeV, which is below the injection energies of protons and helium-3 from the large booster. Therefore, protons and helium-3 ions are always above this transition energy in the ion collider ring. However, the optics needs further optimization to lower the transition energy in order to achieve the no-crossing design goal for other ions.

The two short straight sections in the middle of left or right half rings where two Siberian snakes will be placed are filled with symmetric FODO cells. They are optically matched to the arc lattice on both ends and also dispersion free due to the dispersion suppressors on the end of an arc adjacent to the short straight. Figure 5.16 (left) shows Twiss functions of a short straight.

The two long straights of the figure-8 ring will accommodate several important collider components including two interaction regions [24], electron cooling, RF system and spin-tune SC solenoid or Siberian snakes. The optics design of the two long straights has followed the modular design principle, namely, the baseline has a FODO lattice, and special machine components will come in as insertions as the design work progresses. Some of these insertions and their integration to the baseline are covered in other sections of this report. On each end of the long straights, a matching block consisting of five quadrupoles is utilized to match the optics between the arc and the long straight. An example of the long straight's FODO lattice is shown in Figure 5.16 (right). The nominal phase advance per cell in the long straight is also 60° in both transverse planes; however, they are adjusted along with optics of the matching blocks for attaining the desired fractional parts of the betatron tunes. The optical functions of the complete figure-8 ion collider ring are shown in Figure 5.17, with global parameters listed in Table 5.17.

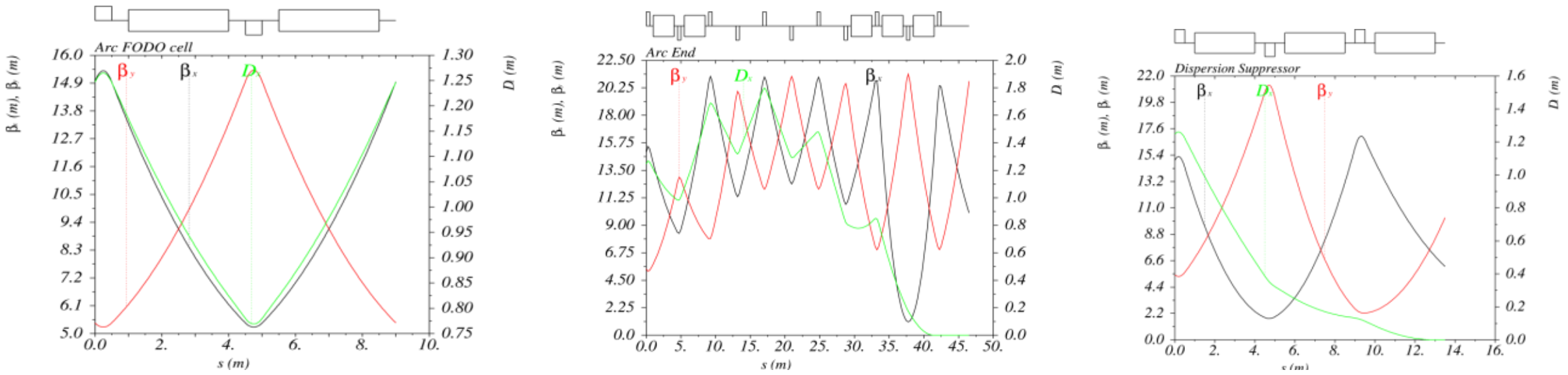

Figure 5.15: Optical functions of an arc FODO cell (left), an end section that matches geometry of electron spin rotator (middle) and dispersion suppression block (right).

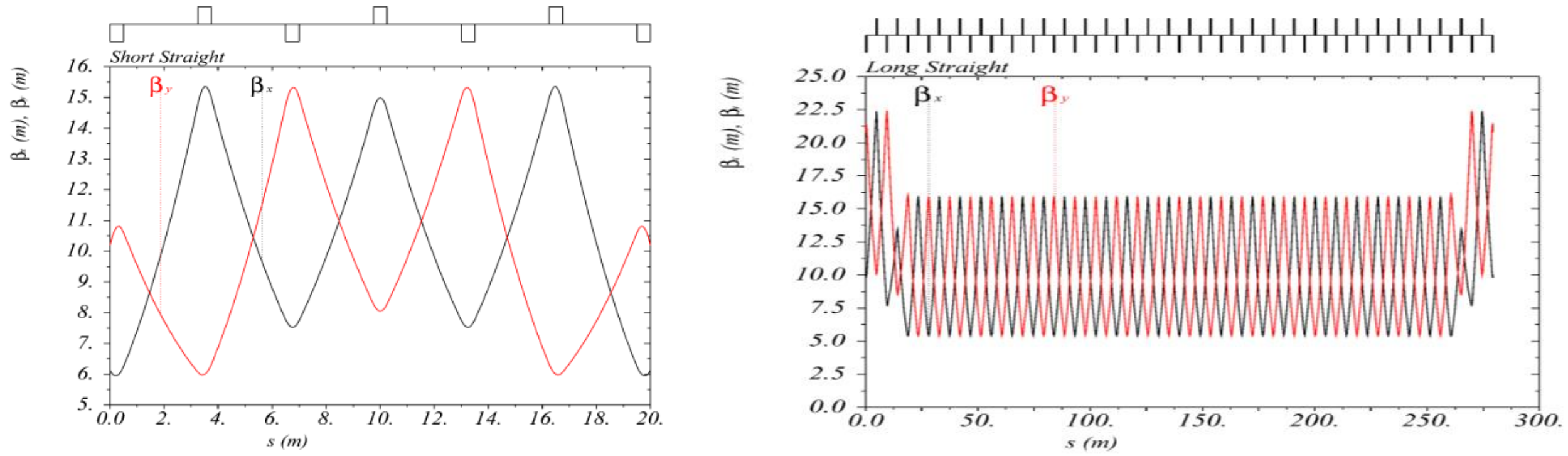

Figure 5.16: Optical functions of a short straight (left) and a long straight (right) of the ion collider ring.

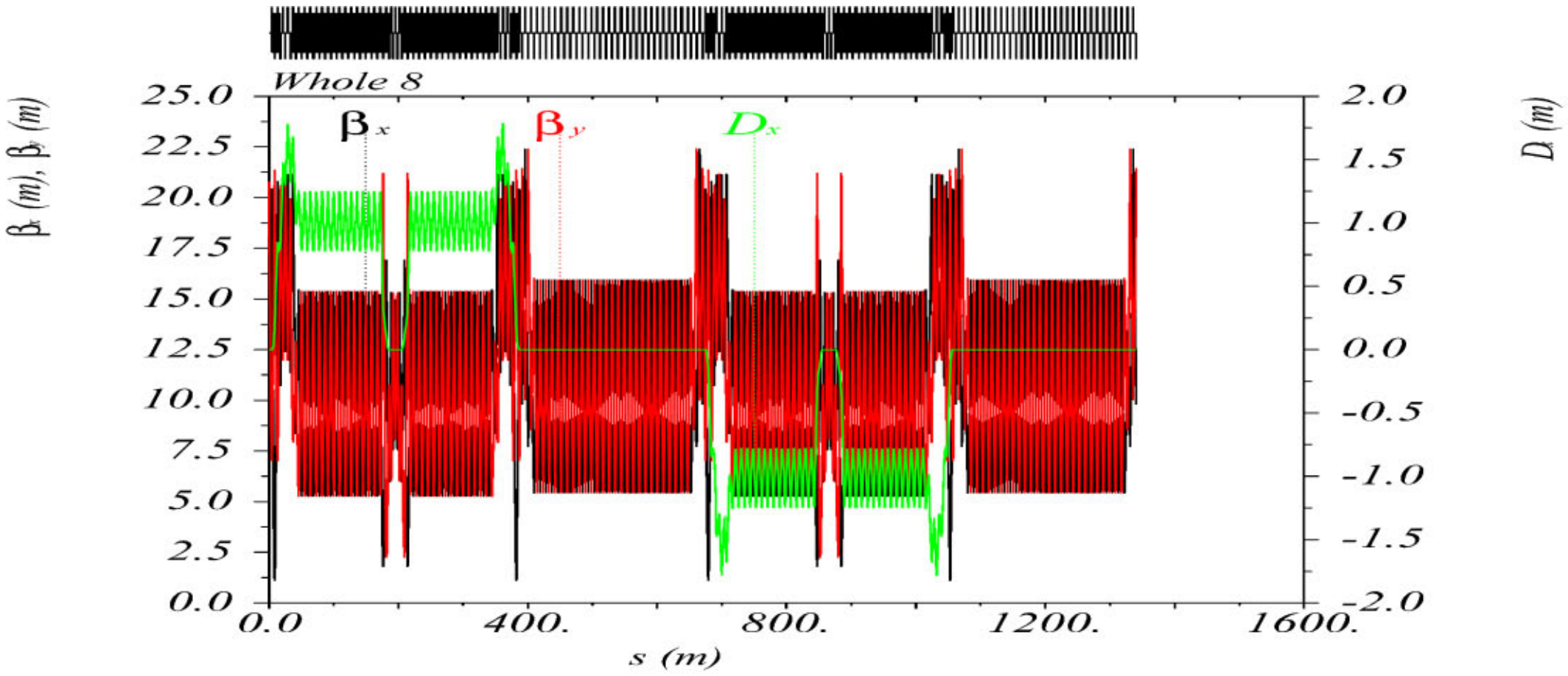

Figure 5.17: Optics of the complete figure-8 ion collider ring of MEIC.

Table 5.17: Global optics parameters of the MEIC ion collider ring

| | | |
|---|---|---|
| Maximum horizontal / vertical beta function | m | 22 / 22 |
| Maximum horizontal dispersion | m | 1.78 |
| Horizontal / vertical betatron tunes | | 24.53 / 23.47 |
| Uncompensated horizontal / vertical chromaticity | | -28.2 / -26.73 |
| Momentum compaction factor | | $5.52\times10^{-3}$ |
| Transition energy factor $\gamma_{tr}$ | | 13.46 |
| Horizontal / vertical normalized emittance<br>at 20 GeV/*c* proton injection<br>at 60 GeV/*c* proton collider-mod | mm·mrad | <br>4 / 4<br>0.35 / 0.07 |
| Maximum horizontal / vertical RMS beam size<br>at 20 GeV/*c* proton injection<br>at 60 GeV/*c* proton collider-mode | mm | <br>2 / 2<br>0.35 / 0.16 |

## 5.8 RF System in the Collider Ring

Two RF systems are needed in the ion collider ring for fulfilling two distinct tasks. One is a low frequency RF system for boosting the ions from the injection energy to the collision energy up to 100 GeV for protons or 40 GeV/u for lead ions. The other is a high frequency RF system which is used primarily for rebunching a coasting beam to a very high frequency (748.5 MHz) bunched beam. The process of debunching and rebunching is discussed in section 5.2.4. The process of debunching and rebunching is discussed in section 5.2.4. The two systems will be adiabatically switched over simultaneously after completion of the energy boost. Below we present a brief discussion of technologies for these two systems.

### *Low Frequency RF System*

The following table (5.18) shows main requirements for the low frequency RF system in the ion collider ring, assuming a 1350 m nominal circumference. The RF frequency is close to that of the ion large booster. However, it varies a small amount to accommodate increase of the ion velocity during the energy ramp up.

Table 5.18. Ions acceleration in collider ring

| Ion species | Injection energy | Injection frequency | Flat-top energy | Flat-top frequency |
|---|---|---|---|---|
| | GeV/u | MHz | GeV/u | MHz |
| Proton (p) | 20 | 1.109 | 100 | 1.110 |
| Lead ($^{208}Pb^{82+}$) | 8 | 1.104 | 40 | 1.110 |

At such a low frequency, most RF systems with a variable frequency are made of inductance-loaded co-axial cavities with the ferrites commonly used for inductance. The magnetic field ramping time of superconducting magnets, limited by the ramping rate, is about 1 minute; therefore the required energy gain per turn is about 6 keV. For a 0.5 A average current, the beam loading is about 3 kW; assuming the synchronous phase is 75$^{o}$ from crest, the total RF voltage should be 23.2 kV.

Two double-gapped cavities can be used to provide the voltage required. The gap voltage of an inductance loaded co-axial cavity is given by

$$V_{cavity} = 2\pi f_{rf} B_{rf} L r_i \ln(r_o/r_i) \tag{5.1}$$

where $f_{rf}$ is the RF frequency, $r_i$ and $r_o$ are the inner and outer radii of the ferrite toroid stack, $L$ is the ferrite toroid stack length and $B_{rf}$ is the average magnetic field inside ferrite toroids. The ferrite toroid inner and outer radii can be 15 and 25 cm respectively. If we use 40 ferrite toroids and each toroid is 2.5 cm thick, the length of the ferrite stack for a single cavity is 1 m. Using the above equation, $B_{rf}$ should be 219 G at 1.1 MHz frequency for 11.6 kV/cavity. The corresponding power dissipation [25] in ferrite can be found in Figure 5.18; its value is about 0.15 W/cm$^3$. Then for each cavity, the dissipated power $P_d$ in ferrite is about 19 kW. From this, we can further obtain the shunt impedance,

$$R_{shunt} = \frac{V_{cavity}^2}{P_d} = 7.1 \text{ k}\Omega \text{ per cavity} \tag{5.2}$$

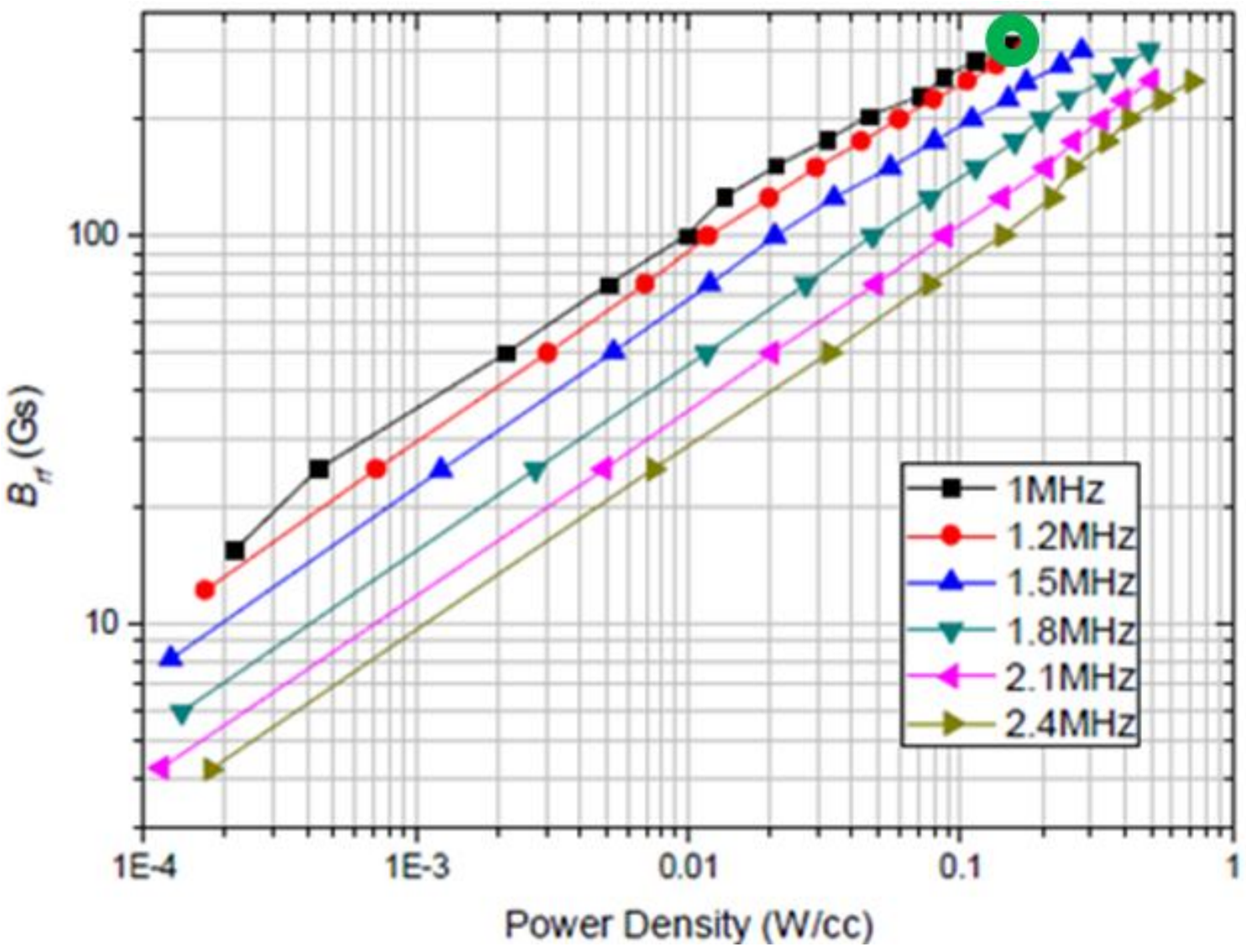

Figure 5.18. Power dissipation in Ferroxcube 4M2. The green circle corresponds to a data point of 11.6 kV/cavity.

The cavity design parameters are summarized in Table 5.19. The total power consumed by cavity and beam is 41 kW. Because of the low $Q$ of such kind RF cavity, the required forward RF power is not much larger. For $Q_0$=40 and synchronous phase 75° from crest, the total optimized forward RF power is 42 kW. If higher field gradient is desired because of limited available space, magnetic-alloy-loaded cavity can be used to replace the ferrite-loaded cavity.

Table 5.19. Parameters of the ferrite-loaded RF cavity

| | | |
|---|---|---|
| Cavity core material | | Ferroxcube 4M2 |
| Cavity number | | 2 |
| Cavity length | m | 1.5 |
| Gaps per cavity | | 2 |
| RF frequency range | MHz | 1.10 to 1.11 |
| RF voltage per gap | kV | 5.6 |
| Power dissipation | kW/cavity | 19 |
| Beam loading power | kW | 3 |
| Ferrite toroid inner radius | cm | 15 |
| Ferrite toroid outer radius | cm | 25 |
| Ferrite stack length | cm | 100 |
| Shunt impedance | kΩ/cavity | 7.1 |

### *High Frequency RF System*

The main function of the high frequency RF cavities is to re-bunch the coasting ion beams. The total accelerating voltage is determined by the desired bunch length as follows [26]

$$V_c = \frac{c^2}{2\pi f_{rev}^2}\frac{\eta E}{\sigma_l^2 He\cos\psi_s}\left(\frac{\delta E}{E}\right)^2 \tag{5.3}$$

where $\sigma_l$=1 cm is the RMS bunch length, $\delta E/E$=0.05% is ion energy spread, $\eta$ is the phase slip factor of the collider ring, and $H$ is the harmonic number. The synchronous phase $\psi_s$ is set at zero crossing (90° from the crest) for maximum bunching effect. Using the design parameters of the MEIC collider ring, it is found that the total required RF cavity voltage is 116.9 MV. Obviously, superconducting RF cavities are needed for the CW operation. The cavity is detuned to meet both the 90° synchronous phase requirement and stability requirement (the maximum stable current is 5% larger than beam current) [27,28,29]. The following table 5.20 provides main design parameters.

Table 5.20. Parameters for typical situation in MEIC ion ring high frequency RF system

| | | |
|---|---|---|
| Frequency | MHz | 748.5 |
| Number of cell per cavity | | 5 |
| R/Q per cavity | Ω | 525 |
| Material independent geometry factor G = $R_s$*$Q_0$ | Ω | 276.0 |
| Active length per cavity | m | 1.0 |
| Average insertion length per cavity | m | 1.50 |
| Operating temperature | K | 2.1 |
| BCS surface resistance $R_{BCS}$ | nΩ | 7.9 |
| Residual surface resistance assumed $R_{res}$ | nΩ | 10 |
| Quality factor $Q_0$ | | $1.54\times10^{10}$ |
| Shunt impedance per cavity ($R=V_{eff}^2/P$) | MΩ | $8.09\times10^6$ |
| Input power | kW | 233.5 |
| Cavity dissipation power $P_{cavity}$ | W | 47 |
| Average beam current | A | 0.5 |
| RF voltage per cavity | MV | 19.5 |
| Number of cavities needed | | 6 |
| Total insertion length | m | 9.0 |
| RF effective accelerating voltage | MV | 116.9 |
| Acceleration gradient | MV/m | 19.5 |
| Synchronous phase, 0 is on crest | deg | 90 |
| Detuning frequency | kHz | -5.285 |
| Detuning angle | deg | 87.38 |
| Optimum coupling $Q_{ext}$ | | $1.55\times10^6$ |

## 5.9 Beam Synchronization

### 5.9.1 The issue

The MEIC ion collider ring accommodates a wide variation in energies of ion beams, from 20 to 100 GeV for protons or 8 to 40 GeV per nucleon for lead ions. In this medium energy range, ions are not fully relativistic, which means values of their beta ($\beta=v/c$) are slightly below 1, leading to an energy dependence of revolution time of the collider ring. On the other hand, electrons with energy 3 GeV and above are already ultra-relativistic such that their speeds are effectively equal to the speed of light. The differences in speeds of colliding electrons and ions of MEIC are actually huge in terms of revolution times or path-lengths in the collider rings. While the circumferences of MEIC collider rings can be adjusted such that the revolution times of electrons and ions are matched (or identical) for one particular ion energy—such that an electron bunch that collides with an ion bunch at a collision point will collide with the same or another bunch of the ion beam after one revolution of the collider ring—this matched condition could not be maintained for the whole ion energy range. At other energies, the colliding bunches could miss each other due to different revolution times. This is the beam synchronization issue. Multiple collision points could further complicate the situation and make beam synchronization even more difficult.

To provide a measure of this beam synchronization issue in MEIC, it can be shown that the difference of revolution times of 3 GeV electrons and 20 GeV protons is approximately 4.3 ns, equivalently 1.3 m path-length difference, assuming two collider rings are matched at a 100 GeV proton energy with a nominal ring circumference of 1350 m. With approximately 1.3 ns (40 cm) bunch space associated with a 748.5 MHz repetition rate, the above path-length difference is large than 3 bunch spacings. The situation becomes much worse for the case of a lead ion beam at 8 GeV/u minimum operational energy; the path-length difference is now widened to 7.44 m, or roughly 18 times the bunch spacing.

It is clear that the conventional path-length adjustment schemes based on a chicane or a dog-leg type magnetic system are not feasible to handle such a large path-length difference. In this section, we present a new scheme for mitigating this issue. For simplicity, we assume two collider rings have identical circumference. Nevertheless the scheme and results presented in the following can be generalized straightforwardly to the cases in which circumferences of collider rings are different by a small amount, such as a few meters.

### 5.9.2 Variation of Ion Bunch Numbers

We start with a nominal circumference of 1350 m for the MEIC collider rings. There are 3370 electron or ion bunches stored in each of two rings when the MEIC facility is operated at 748.5 MHz RF frequency and a CW beam mode. Assuming the two collider rings are matched for 100 GeV protons, then the circumference of the ion ring is slightly shorter (by 6 cm) than that of the electron ring. Under this matched condition, the electron and proton revolution times are identical; therefore, an electron bunch will continuously collide with one particular proton bunch at the IP while both bunches are circulating in opposite direction in their own rings. This synchronization condition should be still valid when there are two or more IPs in the collider rings as long as these IPs are distributed uniformly in the ring; for an example of two IPs, the distance from IP1 to IP2 equals the distance from IP2 back to IP1. More general cases of multiple IPs with unequal distance between them can also be considered; however, they are generally more complicated. In the following, we discuss the case of one IP only, though occasional comments for cases of two or more IPs have been included.

When the proton beam in the MEIC varies its energy from 20 to 100 GeV, the proton velocity also varies, and so does the proton revolution time in the collider ring. As a consequence, the synchronization condition is no longer valid. However, at certain discrete energies, the difference of revolution times happens to equal an integer multiplier of the bunch spacing; thus the synchronization condition is automatically restored if the ion collider ring stores additional bunches. In these cases, an electron bunch colliding with one particular proton bunch at an IP will collide, after one complete revolution, with the $n$-th bunch down in the proton bunch train. Such a set of proton energies are called harmonic energies and the number of bunches in the ring is called the harmonic number. They satisfy the following relation

$$N_0\beta_0 = N\beta \tag{5.4}$$

where $N$ is number of bunches in the ion ring and $\beta=v/c$. The subscript 0 in the above equation denotes the values at the matched energy of the baseline; for our case, that is 100 GeV protons. The number of additional ion bunches required to be stored in the collider is $n=N-N_0$. Table 5.21 lists the first several harmonic energies of ion beams in the MEIC collider ring.

Table 5.21 Ion beam harmonic energies in MEIC collider ring

| Harmonic number | β | γ | Ion energy (GeV/u) |
|---|---|---|---|
| 3370 | 0.99996 | 106.6 | 100 |
| 3371 | 0.99966 | 38.32 | 35.95 |
| 3372 | 0.99936 | 28.02 | 26.23 |
| 3373 | 0.99907 | 23.15 | 21.72 |
| 3374 | 0.99877 | 20.17 | 18.93 |
| 3375 | 0.99848 | 18.11 | 16.99 |

It can be seen from the table that the harmonic energies are distributed more densely at low ion energies, particularly below 20 GeV/u, which means they provide abundant choices of ion energies with synchronization of beams at one IP for experiments. In the ion energy range between 20 to 36 GeV/u, there are still three harmonic energies. From a user point view, there should be enough choices of energies in the range of interest to underlie EIC physics research. Thus, one could conclude that this scheme is indeed a viable solution to the beam synchronization issue at ion energy below 36 GeV/u. There is a particularly important implication of the above statement: for heavy ions like lead, the energy range aimed at by this medium energy EIC is only up to 40 GeV/u; therefore, the scheme of variation of number of stored bunches in the ion collider alone already provides a working solution to the beam synchronization issue if there is only one IP.

This scheme can also be applied to the MEIC with multiple IPs; however, the number of harmonic energies that meet the beam synchronization condition at all IPs simultaneously is reduced. As an example, if the MEIC has two IPs symmetrically distributed in the collider ring, only the even harmonic numbers in the above table 5.21 meet the beam synchronization requirements. Such reduction should not affect the low energy end, since there are still abundant harmonic energies available for experiments.

### 5.9.3 Variation of Ion Path Length

For those energies lying between two consecutive harmonic energies, additional adjustments of collider rings will be needed. One possible solution is adjusting the circumference of the ion collider ring by up to half of the bunch spacing, which is 20 cm for a 748.5 MHz bunch repetition rate, for only one IP. For a figure-8 ring with a 60° crossing angle, this represents a change of arc radius by up to 1.2 cm. With two IPs, the amount of collider ring circumference adjustment is doubled in order to satisfy the synchronization condition at both IPs simultaneously. While this could be the simplest solution, it will be quite a technical challenge in implementation since the MEIC ion collider ring is made of SC magnets. Apertures of SC magnets are usually more difficult to be enlarged to accommodate a shift of the magnetic centers. Alternatively, the SC magnets must be mounted on movers which are indeed workable—however, with high cost.

### 5.9.4 Variation of Electron Path Length and RF Frequency

Beam synchronization can also be achieved at IPs by variation of circumference of the electron ring instead; however, the bunch repetition rate (i.e., RF frequency) must be adjusted accordingly. The main advantage of this scheme is that, being a warm magnet ring, change of the MEIC electron ring circumference is far easier since apertures of the magnets could be made large enough for a shift of the magnetic center up to 1.2 cm for one IP or 2.4 cm for two IPs. It could be easily shown that, for the case of one IP, changes of electron ring circumference and RF frequency must satisfy the following two simple equations

$$f/f_0=(N_p/N_{p0})(\beta/\beta_0) \qquad L_e/L_{e0}=(N_e/N_{e0})(f_0/f) \qquad (5.5)$$

where $f$ is the RF frequency, $\beta=v/c$ is the proton relativistic factor assuming the electron relativistic factor is one, and $N_p$ and $N_e$ are harmonic numbers in the ion and electron collider rings respectively. The subscript 0 again denotes values at the matched proton energy, which is 100 GeV for MEIC. We also take the simplest case of fixed number of bunches in the electron ring, namely, $N_e=N_{e0}$; thus the second equation in (5.5) can be further simplified. Table 5.22 provides an illustration of this scheme for MEIC with one IP.

From an accelerator technology point view, change of frequency of an RF system is quite conceivable, especially when the relative change is very small. When combined with variation of bunch numbers in the ion collider ring, the maximum change of RF frequency, as shown in Table 5.22, is less than 0.015%. The maximum variation of the electron ring circumference is less than 20 cm, which could be handled with no trouble.

Table 5.22: Change of ion harmonic number, electron ring circumference and RF frequency

| Energy (GeV/u) | | | | | Collider Ring | | | | |
|---|---|---|---|---|---|---|---|---|---|
| Proton | Deut. | Lead | $\gamma$ | B | Harmonic number | f (MHz) | $\delta f/f_0$ ($10^{-4}$) | $\delta L_e$ (cm) | $\delta R_e$ (cm) |
| 100 | | | 106.6 | 0.99996 | 3370 | 748.5 | 0 | 0 | 0 |
| 90 | | | 95.92 | 0.99995 | 3370 | 748.49 | -0.10 | 1.4 | 0.09 |
| 80 | | | 85.26 | 0.99993 | 3370 | 748.48 | -0.25 | 3.3 | 0.20 |
| 70 | | | 74.61 | 0.99991 | 3370 | 748.47 | -0.46 | 6.2 | 0.38 |
| 60 | | | 63.95 | 0.99988 | 3370 | 748.44 | -0.78 | 10.6 | 0.65 |
| 50 | 50 | | 53.29 | 0.99982 | 3370 | 748.40 | -1.32 | 17.8 | 1.09 |
| 40 | 40 | 40 | 42.63 | 0.99972 | 3371 | 748.54 | 0.66 | -8.8 | -0.54 |
| 38 | 38 | 38 | 40.5 | 0.99970 | 3371 | 748.53 | 0.36 | -4.8 | -0.29 |
| 35.9 | 35.9 | 35.9 | 38.32 | 0.99966 | 3371 | 748.5 | 0 | 0 | 0 |
| 34 | 34 | 34 | 29.84 | 0.99962 | 3371 | 748.47 | -0.40 | 5.4 | 0.33 |
| 32 | 32 | 32 | 27.71 | 0.99957 | 3371 | 748.43 | -0.89 | 12.1 | 0.74 |
| 30 | 30 | 30 | 26.65 | 0.99951 | 3372 | 748.61 | 1.48 | -20.0 | -1.22 |
| 28 | 28 | 28 | 25.58 | 0.99944 | 3372 | 748.56 | 0.76 | -10.2 | -0.62 |
| 26.3 | 26.3 | 26.3 | 28.02 | 0.99936 | 3372 | 748.5 | 0 | 0 | 0 |
| 26 | 26 | 26 | 27.71 | 0.99935 | 3372 | 748.49 | -0.14 | 1.9 | 0.12 |
| 24 | 24 | 24 | 25.58 | 0.99924 | 3372 | 748.18 | -1.27 | 17.2 | 1.05 |
| 22 | 22 | 22 | 23.45 | 0.99909 | 3373 | 748.52 | 0.24 | -3.2 | -0.20 |
| 21.7 | 21.7 | 21.7 | 23.15 | 0.99907 | 3373 | 748.5 | 0 | 0 | 0 |
| 20 | 20 | 20 | 21.32 | 0.99890 | 3374 | 748.60 | 1.29 | -17.4 | -1.06 |
| | 19 | 19 | 20.25 | 0.99878 | 3374 | 748.51 | 0.10 | -1.3 | -0.08 |
| | 18.9 | 18.3 | 20.17 | 0.99877 | 3375 | 748.5 | 0 | 0 | 0 |
| | 17.0 | 16.0 | 18.11 | 0.99818 | 3376 | 748.5 | 0 | 0 | 0 |

Note: Cells in orange colors are parameters for harmonic energies.

To summarize discussions of this section, we conclude that schemes have been found for ensuring beam synchronization in the MEIC collider rings. At sufficiently low proton energy and over the whole range of heavy ions, variation of the number of ion bunches in the collider ring provides a working solution. For proton energy of 40 to 100 GeV, variation of electron ring circumference has an advantage; however, tests are needed for required variation of frequency of SRF modules by up to 0.012%, which is believed achievable.

## 5.10 Collective Effects and Beam Stability

Nominal beam parameters of a proton bunch in the MEIC collider ring have been chosen to deliver a high luminosity in collision with electron bunches by filling the ring with many bunches possible. The number of protons in an MEIC single bunch is $4.16 \cdot 10^9$. The emittance and bunch length of the beam are 0.35 microns and 1 cm, respectively. These values are both roughly an order of magnitude smaller compared to presently achieved values for collider proton bunches [30,31]. Single bunch effects in the MEIC collider ring are mostly determined by the ring impedance at high frequencies of the wavelength comparable to or less than the bunch length on the order of 10 GHz.

Beam parameters in the various machines in the injection chain of the ion complex are yet to be specified. Most relevant collective effects in various injectors are the Laslett tune shift and microwave instabilities. The mutual repulsive forces of the charges of the particles as they ride along together transformed into the laboratory frame and including the image forces in the vacuum chamber produce a defocusing effect which shifts the betatron tunes, $\nu$. This will not cause a problem, if the shift is same for all particles. However, the change in $\nu$ depends on where the particle is located in the beam. Therefore, the beam attains a finite spread in $\nu$ proportional to intensity as beam particles execute synchrotron and betatron oscillations in the ring. This spread in the betatron tunes dictates the choice of working point which must be located safely for dangerous resonances. It also limits the intensity at low energy in each of the machines which form the injection chain for protons.

Potentially dangerous longitudinal instabilities in the MEIC collider ring are the microwave instabilities and the coupled-bunch instability [32,33]. The microwave instability is induced by the broadband coupling impedance of the ring. The instability is not fatal but produces a fast increase of the momentum spread when the peak beam current exceeds a threshold value which is proportional to the square of the momentum spread. The change in the momentum spread in turn leads to the bunch lengthening so that the peak intensity remains below the threshold value. The instability growth time is typically much shorter than one synchrotron period. We find that the broadband impedance seen by the bunch requires $|Z/n|_{\mathrm{bb}} \leq 9\ \Omega$ in order for proton bunches to stay away from the microwave instability during the storage. This is a rather large impedance requirement which should be attainable in present day storage ring design. The collider ring may have to satisfy a more stringent impedance requirement coming from the beam stability at the injection. However, beam parameters at the injection are yet to be established and such study cannot be performed now.

We noted earlier, in section 4.4, that the $\nu_s$ of higher order modes of RF cavities need to be damped below 10 to avoid neighboring bunch couplings. Consequently, multi-bunch instabilities are practically unavoidable in the MEIC proton ring. The resistive wall impedance, though large at low frequency, usually does not contribute significantly in driving longitudinal coupled bunch instabilities. The broadband coupling impedance does have an influence on the instability despite the impedance generating a short range wake cannot induce any growth of the oscillations. It can suppress Landau damping by shifting the frequency of the coherent modes out of the band of the single particle frequencies. A growing mode is expected to be Landau damped if the bunch mode spectrum averaged impedance $|Z/n|_{\text{eff}}$ is about 10 mΩ. Most longitudinal modes of RF cavities will be able to drive coupled bunch instabilities, and growing modes not Landau damped have to be suppressed by feedback systems.

The dominant single bunch instability in the case of transverse oscillations is the mode coupling instability, which is driven by the reactive component of the broadband impedance. Two adjacent head-tail modes of the bunch oscillate at frequencies that differ by an amount equal to the synchrotron frequency at low intensity. When the intensity is high enough, the interaction induces a betatron frequency shift comparable to $\omega_s$ and the instability occurs. MEIC proton bunches at the nominal energy of 60 GeV are safe from this instability if the impedance meets the condition $\text{Im}(Z_T)_{BB}\,\beta_{av} \leq 1.9$ GΩ. This limit should be met easily.

Transverse coupled bunch instabilities can be induced by transverse modes of RF cavities and also by the resistive wall wake fields. The resistive wall instability is usually most dangerous at low frequency since the impedance decreases as $\omega^{-1/2}$

Transverse coupled bunch modes except the transverse rigid are Landau damped as in longitudinal instabilities when the mode frequency shift is less than the effective spread of the bunch due to the natural synchrotron frequency spread comes from a sinusoidal RF bucket. For the rigid dipole mode, betatron tune spread on the order of a few times $10^{-3}$ is needed to obtain Landau damping. This spread may be naturally present in the beam due to magnet non-linearity of mainly IR quadrupoles and also from chromaticity sextupoles. If this is not the case, it may have to be provided by the system of octupolar lenses. As in the longitudinal case, a few transverse modes of RF cavities may induce instabilities in spite of Landau damping. These have to be suppressed by active feedback systems.

The intra-beam scattering which produces a growth of both longitudinal and transverse emittances grows at a rate which is proportional to the six dimensional phase space density of the beam [34]. The proton phase space density in the MEIC collider ring is very high as a result of selected beam parameters capable of delivering high luminosity. A rough estimate provides intra-beam scattering growth rates on the order of a few seconds. Ultimately equilibrium emittances are determined by the electron cooling rates counteracted by the intra-beam scattering growth rates.

Touschek scattering is related to the intra-beam scattering [35]. The main difference is that one is concerned now with large-angle, single scattering events that change the scattered particle's momentum sufficiently to make it fall outside the momentum acceptance of the accelerator. At 60 GeV, Touschek lifetime is estimated on the order of 15 hours with the assumed momentum acceptance of 0.4%.

Ion trapping and fast beam-ion instability are not of much concern with positively charged beams. On the other hand, electron trapping and electron cloud effects are problems to be studied in detail. Positively charged beams preferably interact with electrons. Electrons can be trapped in the potential well of the proton beam just as ions are trapped by the electron beam. Electrons are mostly generated through the ionization of gas and also from the beam loss in proton rings. If the condition is right, seed electrons can multiply by the multipacting mechanism. The electron buildup saturates when the attractive beam field is on the average compensated by the field of the electrons, and the result is electron clouds in the ring. A simulation study is necessary as the multipacting condition depends strongly on various factors like the energy gain from the beam field, the chamber dimension, the bunch spacing, etc. Electrons can more easily induce instabilities as the oscillation frequency in the beam is much larger than that for ions. As noted in section 4.4, we find from a crude estimate that beam parameters and the chamber dimension are not favorable for forming electron clouds in the MEIC rings. We are in the process of studying the electron clouds problem with several numerical codes obtained from CERN and KEK [36,37].

## 5.11 Ion Polarization

The MEIC ion beam polarization design requirements are:

- High polarization (over 70%) for proton or light ions (d, $^3He^{++}$, and possibly $^6Li^{+++}$).
- Both longitudinal and transverse polarization at all IPs.
- Sufficiently long lifetime to maintain high beam polarization.
- Spin flipping at a high frequency.

Spin flipping is required for adequate suppression of systematic effects in spin asymmetry measurements.

In principle, it is a challenge to provide highly polarized ion beams in MEIC because, unlike the electron beam, there is no synchrotron radiation at the medium energy range of MEIC, so ion beams cannot acquire polarization through the Sokolov-Ternev [38] self-polarization effect. As a consequence, polarized ion sources must be provided. It also takes a long process to accelerate ions to the collision energy in multiple rings while passing over many spin resonances [39,40], all of them posing dangers of polarization degradation [41,42,43,44]. Special care has been taken in the MEIC design to minimize such depolarization.

The primary design choice of MEIC is a figure-8 shape for all ion booster and collider rings. It is a revolutionary concept which enables energy independent spin tune and thus effectively bypasses all the spin resonances. With Siberian snakes for proton and helium-3 or a special magnetic insert for deuterons, stable spin motions in the ion collider rings can be achieved and desired polarization can be realized at collision points. This section presents the MEIC ion spin design scheme and also demonstrates advantages of the figure-8 design.

### 5.11.1 Polarized Ion Sources

If there is no depolarization during acceleration and storage, the final beam polarization is determined by the initial polarization at extraction from the ion source. The present state-of-the-art polarized ion source technology is adequate for MEIC needs. The most common sources for polarized ions [45] are the Atomic Beam Polarized Ion Sources (ABPIS) [46,47] with resonant charge-exchange ionization of polarized atoms by positive and negative ions and the Optically-Pumped Polarized $H^-$ Sources (OPPIS) [48,49]. Relevant parameters of some of the existing ion sources are listed in Table 5.23 [46,47,48,49]. There are also proposals to develop a universal ABPIS for production of polarized and unpolarized $H^-/H^+$, $D^-/D^+$, $T^-$, $^3He^{++}$ and $^6Li^{+++}$ beams with improved intensity, polarization and emittance [50].

Acceleration of polarized ions in a linac does not present a depolarization problem since there is not enough time for spin resonances to be developed on a single pass. When transferring a polarized beam from one accelerating stage into another, special care should be taken for matching polarization of the injected beam to the stable spin direction at the injection point. This can be accomplished straightforwardly using appropriate spin rotators in the transfer beam lines.

Table 5.23 Main parameters of existing ion sources

| | | OPPIS@BNL | CIPIOS@IUCF/INR | INR Moscow |
|---|---|---|---|---|
| Status | | in operation | shut down 8/2002 | test bed |
| Species | | $H^-$ | $H^-$ / $D^-$ | $H^+$ / $H^-$ |
| Pulse width | μs | 500 up to DC | Up to 500 | > 100 |
| Bunch current | mA | 1.6 | 2.0 / 2.2 | 11 / 4 |
| Max $P_z$ | % | 86 | 80 to > 90 | 80 / 95 |
| Emittance (90%) | π·mm·mrad | 2.0 | 1.2 | 1.0 / 1.8 |

### 5.11.2 Advantage of Figure-8 Ring

Acceleration of polarized ions in a conventional circular or racetrack synchrotron ring is complicated by a large number of depolarizing spin resonances [39,40,41,42,43,44]. The two major types [39,40] are imperfection resonance caused by the field errors and element misalignments in an accelerator while the spin tune equals to an integer, and intrinsic resonances caused by the betatron motion while the spin tune equals to an integer-multiple combination of the fractional parts of the betatron tunes. The spin tune of a conventional synchrotron depends linearly on the beam energy as $G\gamma$ [27], where $\gamma$ is the relativistic factor and $G$ is the anomalous magnetic moment of the particle. This means that numerous spin resonances may be crossed during acceleration.

One powerful technique for preserving the beam polarization during acceleration in a circular or racetrack synchrotron is stabilizing the spin motion with Siberian snakes [17], a magnet insertion device rotating the spin by 180° about an axis in the horizontal plane. This makes the spin tune equal to ½, thus independent of the beam energy, and effectively eliminating all imperfection and most intrinsic spin resonances.

A Siberian snake can be constructed using either longitudinal (solenoidal) [50] or transverse (dipole) [51] magnetic fields. Solenoidal snakes are preferable at low energies since they do not cause an orbit distortion; however, they become unpractical at high energies due to the requirement of unrealistically large field strengths. On the other hand, dipole snakes are the choice at high energies since the spin effect of transverse magnetic fields is nearly energy independent for relativistic particles and the orbit distortion caused by dipole snakes decreases inversely with energy. Medium energy from a few GeV to a few tens of GeV is most challenging for preserving beam polarization because solenoidal snakes are already impractical while orbit distortions caused by dipole snakes are still too large.

The figure-8 geometry [52,53] of the MEIC is an elegant way to preserve ion polarization during acceleration and storing; it significantly simplifies the spin control tasks. In this "twisted" ring design, two half arcs are connected by two crossing straights, making the spin motion degenerate. Precession of a spin in one half arc is exactly cancelled in the other half arc so the net spin rotation (and the spin tune) is zero, independent of beam energy. This means that there is no preferred direction of the polarization; any spin orientation in the ring is periodic and is repeated from turn to turn [54]. In such a case, it becomes possible to control and stabilize the beam polarization in any desired orientation by using relatively small magnetic fields [54].

In a circular or racetrack type synchrotron, one has to resort to employing strong magnetic fields, such as full Siberian snakes, to control spin dynamics [40]. But there is conceptually no need to introduce such large magnetic fields in a figure-8 ring [52,54]. The spin motion can be fully controlled by relatively weak magnetic fields such as a partial Siberian snake [55], as long as the spin rotation provided by the partial snake exceeds the strength of the zero imperfection resonance by a few times. Such partial snakes can make the spin tune energy independent. This is especially advantageous in the pre-booster where accommodation of full Siberian snakes may turn out to be difficult; thus stabilization of the polarization has to be achieved by using a partial solenoidal or dipole snake or a device described in Section 5.11.4 which rotates the polarization by a small angle around the vertical axis [56]. Even in the collider ring, where one could utilize full Siberian snakes for proton or $^3He^{++}$ beams, the figure-8 ring design makes its polarization schemes much more flexible.

Another important advantage of the figure-8 ring design is the possibility of accelerating polarized deuterons in a synchrotron in this medium energy range. Since full Siberian snakes are not feasible for deuterons due to their tiny anomalous magnetic moment, a figure-8 ring is presently the only practical way to accelerate polarized deuterons up to the MEIC collision energies. Like protons and $^3He^{++}$, the deuteron polarization can be stabilized in any desired orientation in the MEIC pre-booster, large booster and collider ring using a partial Siberian snake or a magnetic device that rotates the polarization by a small angle around the vertical axis [44]. For instance, a solenoid installed in a figure-8 straight stabilizes the longitudinal polarization in that straight. The only requirement is that the spin rotation provided by the solenoid or another device has to be a few times greater than the zero-frequency Fourier harmonic of the spin perturbation, i.e. the strength of the zero imperfection resonance caused by magnet misalignments and magnetic field errors [54].

### 5.11.3 Polarized Protons and Helium-3 in a Figure-8 Ring

Accelerating polarized protons in MEIC is conceptually very similar to accelerating $^3He^{++}$ ions. The spin resonances are generally stronger for $^3He^{++}$ than for protons due to the larger anomalous magnetic moment of $^3He^{++}$. However, the $^3He^{++}$ ion energy is lower due to the charge-mass ratio. Since the strengths of the spin resonances are generally proportional to energy [33], this makes the problems of preserving the proton and $^3He^{++}$ polarizations rather comparable. Therefore, the discussion below focuses on the proton polarization scheme only.

In the ion collider ring of MEIC, three full Siberian snakes are required to align the proton spin in the longitudinal direction at multiple IPs in both straights of the figure-8 simultaneously. As shown in Figure 5.19 (upper left), a full Siberian snake placed in the middle of one long straight rotates the spin 180° about the longitudinal axis; thus the spin tune equals ½. However, the direction of the stable polarization in the other long straight still depends on the energy. To make the polarization in the second straight also longitudinal and energy independent, two additional Siberian snakes with longitudinal axes are placed in the middle of the arcs. In this configuration, it can be seen that the stable spin direction is longitudinal in both straights at all energies.

A similar configuration can be arranged to orient the proton polarization in the transverse direction in both straights of the MEIC figure-8 collider ring as shown in Figure 5.19 (upper right). In this case, only two full Siberian snakes are needed and placed in the middle of the arcs with the snake axes perpendicular to each other. The stable polarization is in the vertical direction and the spin tune also equals ½. In fact, this scheme can be implemented using the same hardware as in the longitudinal polarization scheme described above. To go from the longitudinal to transverse polarization, the snake located in the long straight must be switched off and the axes of the arc snakes are configured to form a 90° angle. As discussed in the next subsection, for a helical Siberian snake [51,57], one can manipulate the axis orientation simply by adjusting currents in the snake windings. Thus, no physical movement or hardware replacement is required.

The figure-8 ring is flexible enough to provide other polarization configurations at IPs to meet physics needs. For instance, by inserting two Siberian-snake-type devices with certain spin rotation angles and axis orientations, as shown in Figure. 5.19 (bottom), one can orient proton polarization along the vertical direction in one straight and the longitudinal direction in the other [47]. Parameters of the spin rotators can be chosen such that the spin tune of the figure-8 ring always equals ½.

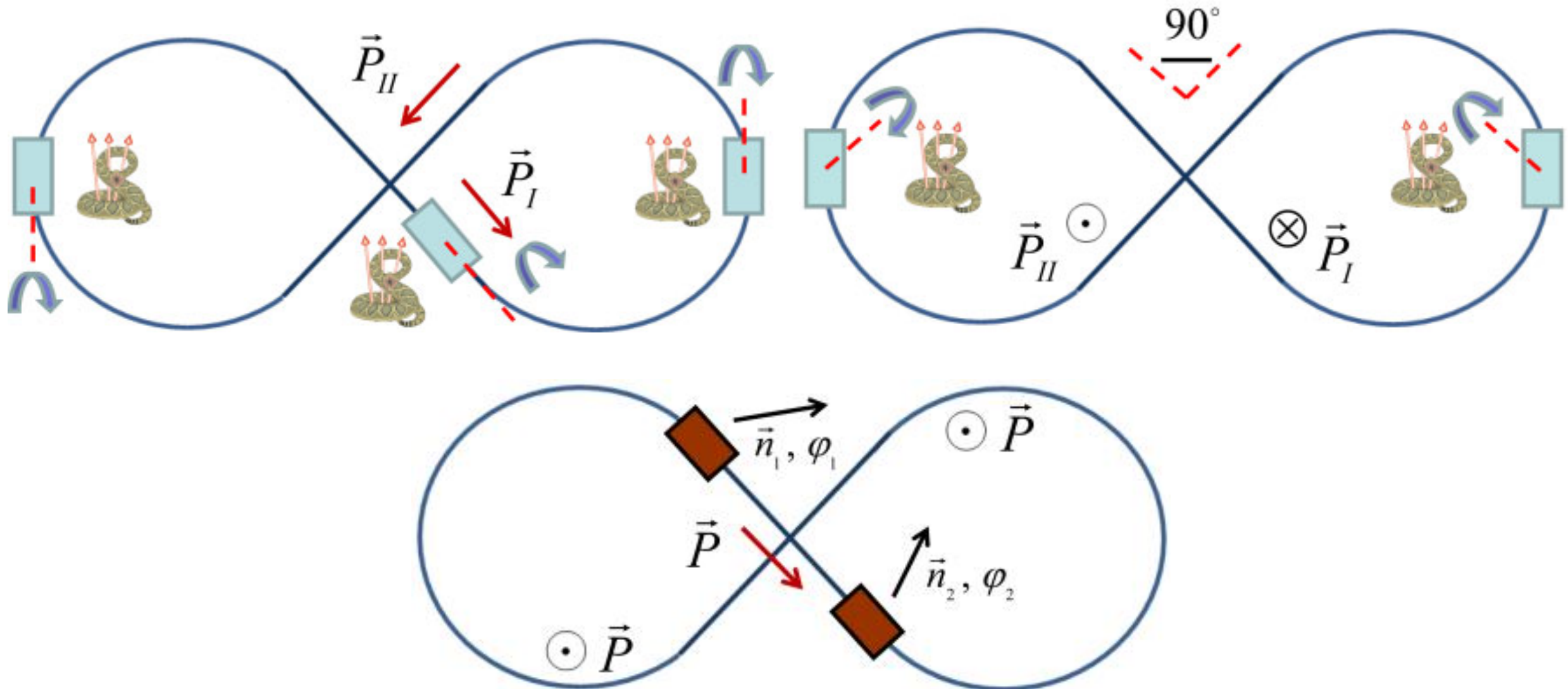

Figure 5.19: Proton and $^3He^{++}$ polarization configurations in the MEIC figure-8 collider ring. They provide stable longitudinal (upper left) and transverse (upper right) polarization at multiple IPs in both straights simultaneously. The bottom drawing shows a stable transverse polarization in one straight and a stable longitudinal polarization in the other.

### 5.11.4 Polarized Deuterons in a Figure-8 Ring

A polarized deuteron beam offers new and important opportunities in the electron-ion collider proposal. The nuclear science programs require a capability of providing both the longitudinal and transverse deuteron polarizations in collisions. As discussed above, with the present technology, full Siberian snakes are not feasible for deuterons due to their small anomalous magnetic moment. The MEIC figure-8 collider ring design, however, allows one to control the stable spin orientation with a small spin rotation around a certain axis using magnetic inserts [54]. The deuteron polarization is then stable and points along the rotation axis at the insert's location, as long the spin rotation angle exceeds the Fourier harmonics of the spin perturbing fields in the figure-8 structure.

The longitudinal deuteron polarization can be stabilized in one straight of the MEIC figure-8 collider ring by inserting a solenoid in that straight, as shown in Figure 5.20 (left). The spin tune in this case is $\phi/2\pi$ where $\phi$ is the spin rotation angle in the solenoid [54]. The stable polarization of deuterons in the other straight lies in the horizontal plane with its orientation depending on the proton or ion energy, and therefore can be along either the longitudinal or transverse direction for a discrete set of energies.

A vertical deuteron polarization can be achieved in both straights of the MEIC collider ring by placing one or more magnetic inserts [58], which rotate the spin by a small angle around the vertical axis, in one or both straights, as illustrated in the right drawing of Figure 5.20.

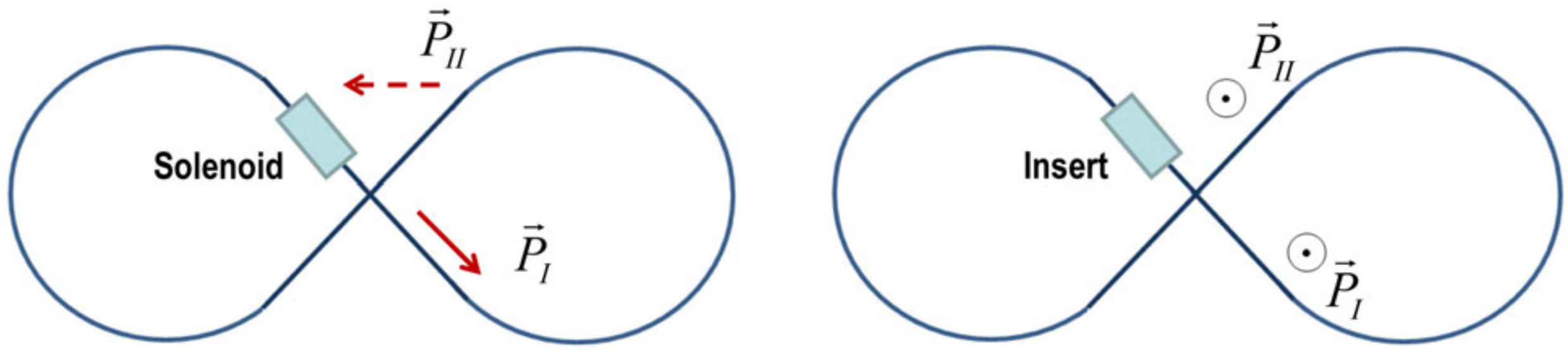

Figure 5.20: Configurations of deuteron polarization in the MEIC figure-8 ion collider ring. The schemes provide stable longitudinal (left) and transverse (right) deuteron polarization at IPs in two straights of the figure-8 ring.

### 5.11.5 Comments on Siberian snakes and SC Solenoids

It appears that the RHIC Siberian snakes [51,57] will work well for the MEIC ion ring. Each RHIC Siberian snake consists of four full-twist helical magnets, about 2.4 m long, combined in a special symmetric structure as illustrated in Figure 5.21, which completely compensates transverse beam orbit excursions introduced by helical magnetic fields. The helical magnets form two pairs, namely, the inner and outer ones. The magnets in each pair are wired to have equal values but opposite vertical directions of the magnetic field at the entrance, which guarantees that the spin rotation axis of the whole snake always lies in the horizontal plane. By operating helical magnets at different values of the magnetic fields $B_{\text{outer}}$ and $B_{\text{inner}}$ (two parameters of the snake) one can manipulate both the spin rotation angle and the direction of the spin rotation axis with respect to the beam orbit.

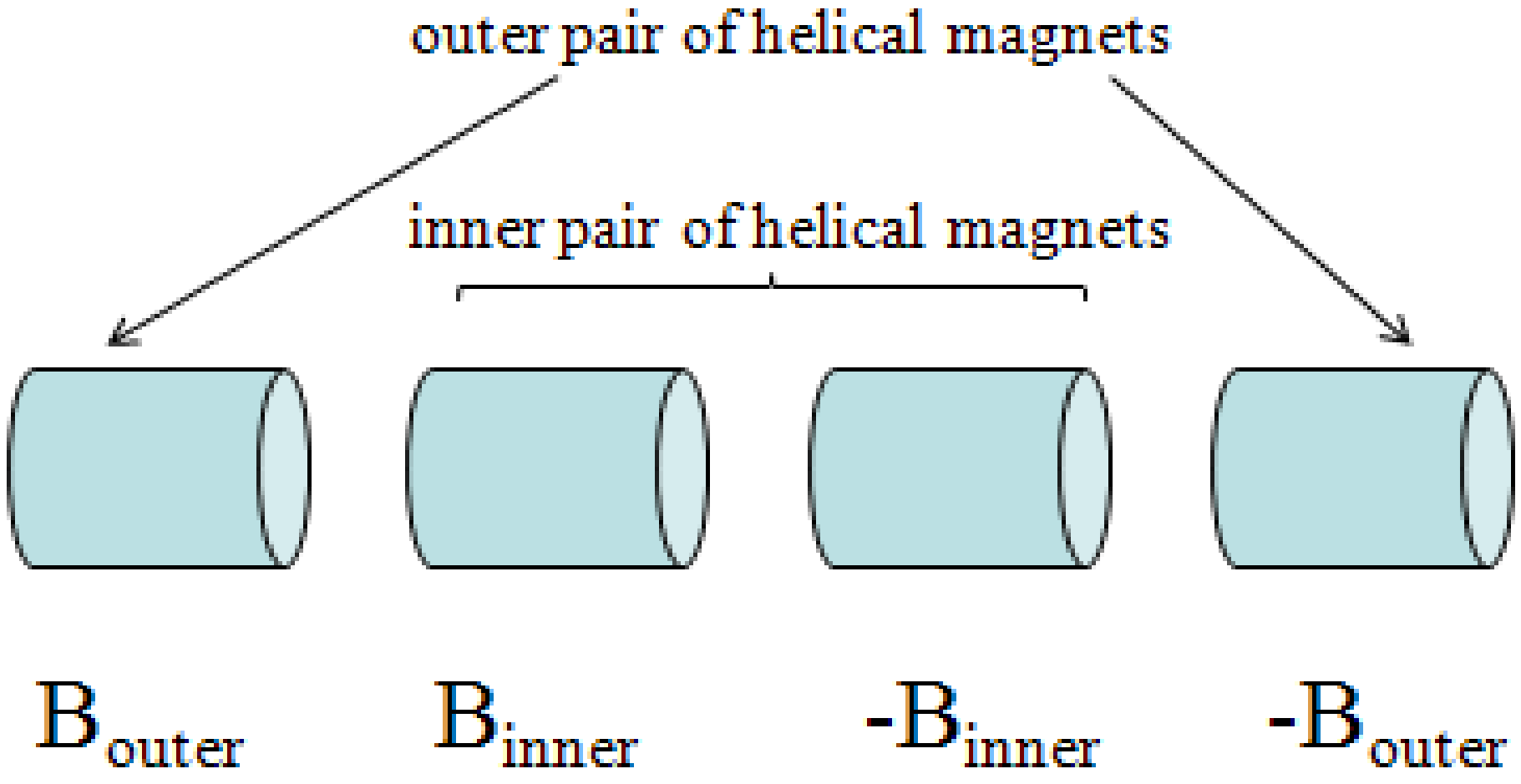

Figure 5.21: A schematic drawing of a RHIC Siberian snake.

The values of the magnetic fields of the helical magnets required for providing the longitudinal proton polarization in the interaction regions are listed in Table 5.24. It can be seen that the magnetic fields of the helical magnets are very close to each other. The maximum proton orbit excursion in the snake at 20 GeV is 32 mm in vertical and 15 mm in horizontal planes. It gets smaller when the energy grows.

Table 5.24: Fields of helical magnets in a Siberian snake with a longitudinal axis

| $E$ | GeV | 20 | 30 | 40 | 50 | 60 | 100 |
|---|---|---|---|---|---|---|---|
| $B_{\text{outer}}$ | T | -2.134 | -2.153 | -2.162 | -2.168 | -2.171 | -2.179 |
| $B_{\text{inner}}$ | T | 2.834 | 2.858 | 2.871 | 2.878 | 2.883 | 2.893 |

The transverse proton polarization scheme requires only two snakes in the middle of the arcs with their axes perpendicular to each other. If the snakes are configured to have their axes pointing at +45° and -45° with respect to the beam direction, we get a configuration similar to RHIC's. The parameters of the Siberian snakes in this case can be found in RHIC publications and are given in Table 5.25. One can see that the inner helical magnets work much harder than the outer ones. The maximum proton orbit excursion in the snake at 25 GeV is 33 mm in vertical and 15 mm in horizontal planes.

Table 5.25: Fields of helical magnets in a Siberian snake with an axis at 45° to the beam

| $E$ | GeV | 20 | 30 | 40 | 50 | 60 | 100 |
|---|---|---|---|---|---|---|---|
| $B_{\text{outer}}$ | T | -1.225 | -1.236 | -1.241 | -1.244 | -1.247 | -1.253 |
| $B_{\text{inner}}$ | T | 3.943 | 3.977 | 3.994 | 4.005 | 4.012 | 4.033 |

Implementation of superconducting solenoids to stabilize the longitudinal polarization in one of the figure-8 straights is rather straightforward. This is especially important for protons and $^3He^{++}$ in the pre-booster and large booster and for deuterons in all accelerating stages including the collider ring. When inserting a strong solenoid into a lattice, one has to account for its focusing and coupling effects. A rather straightforward approach to compensation of coupling is described in Section 4.5.5.

A special magnetic insert [58] for rotating ion spin is shown in Figure 5.22. It is made of a sequence of alternating dipole and solenoid, and generates a spin tune shift of

$$\nu_s = \frac{\varphi_x \varphi_y}{2\pi} \tag{5.5}$$

where $\varphi_x$ and $\varphi_y$ denote the spin rotation angles around the radial and longitudinal axes. Summation of contributions of all individual inserts will give the total spin tune of the figure-8 ring. The maximum orbit excursion in an insert is given by

$$|\Delta z_{\max}| = \frac{|\varphi_x|}{4G\gamma} = \frac{|\theta_x|}{4}(L_x + 2L_y), \tag{5.6}$$

where $L$ is the length of magnets and $\theta_x$ is the orbital bend corresponding to $\varphi_x$. It follows from these two equations that the spin tune is energy independent and the orbit excursion also remains constant, if the solenoidal fields are fixed and the dipole fields are ramped with energy during acceleration. It is easy to see that superconducting solenoids can be used in this insert since they do not have to be ramped with energy. The insert also does not introduce any coupling if the two solenoidal fields point in opposite directions.

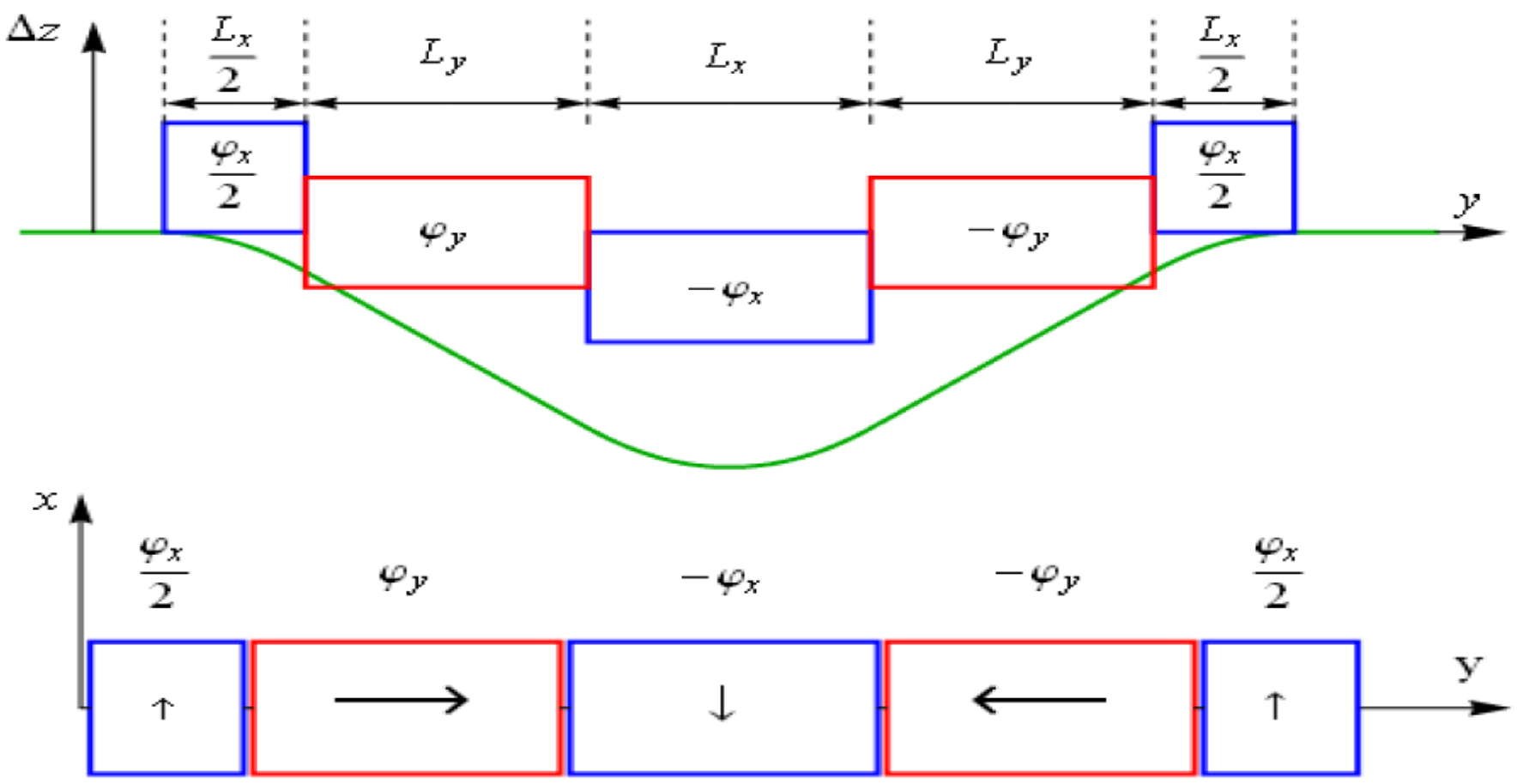

Figure 5.22: A schematic drawing of a magnetic insert which produces a net rotation of the spin by a small angle around the vertical axis.

# 6. Electron Cooling

Since there is no radiation damping of a medium energy ion beam in a storage ring, an efficient damping mechanism must be introduced to reduce and then preserve the emittance of the MEIC ion beam. The choice for the MEIC baseline design is conventional electron cooling [1,2]. While stochastic cooling [3] is another well developed technology, its cooling time is too long to be applicable to the MEIC.

New concepts have been proposed recently to cool hadron beams. The most notable one among them is coherent electron cooling [4,5]. It is an electron beam based cooling system in which the accompanying electron beam provides feedback, similar to stochastic cooling, with an extremely high frequency bandwidth (THz range) by use of the FEL amplification mechanism in an undulator. While this new concept, which requires very high quality, high charge electron bunches, holds promise for fast cooling rate of medium to high energy ion beams, it is yet to be demonstrated experimentally. Therefore, the MEIC baseline design does not invoke this method, although it could be considered after a proof-of-principle test [6] is successfully completed.

The present MEIC baseline has adopted a scheme of multi-phase cooling for supporting its high luminosities [1,2]. In this scheme, electron cooling will be utilized not only for assisting ion beam formation but also during collisions. The latter is particularly important for preserving the collider luminosity and its lifetime by suppressing IBS induced heating.

It should be noted that, while the electron cooling mechanism is well developed and experimentally tested with many successful low energy DC beam applications worldwide, the cooling parameters for the case of MEIC, in terms of the ion energy and bunched beam, are new and require R&D, particularly in accelerator technologies that provide a cooling electron beam.

The MEIC electron cooling demands a facility capable of providing a high energy, high current electron beam with a high bunch repetition rate and good quality (small emittance and energy spread). Such a beam will be generated in a photo-injector and accelerated by an SRF linac. Two advanced technologies, the energy recovery linac (ERL) and the compact circulator ring (CR), have been integrated into the conceptual design to meet the two most demanding technical challenges, namely, high beam power and adequate lifetime of the photo-cathode.

This chapter is devoted to the MEIC cooling design. We start with a presentation of the multi-phase cooling scheme in Section 6.1, followed by the design concept of the ERL circulator ring based cooling facility in Section 6.2. The last section discusses several key technologies for this facility, particularly a fast kicker for switching bunches into and out of the circulator ring.

## 6.1 Multi-Phase Electron Cooling

An ion beam in a storage ring typically has poor 6D emittance compared to a lepton beam. Two principal drivers for this are: 1) a large production emittance followed by a very slow and complicated process of beam formation and acceleration and 2) no natural (radiation) damping for reduction and preservation of the emittance. Space charge at low energy not only prevents accumulation of high intensity ion beams but also forces a large phase space footprint.

Without damping, the ion beam emittance can only grow under the influence of space charge, IBS, noise and other effects.

Electron cooling is a leading mechanism for reduction of the 6D emittance of a hadron beam in a storage ring. A cold electron beam accompanying a hot hadron beam, co-moving along a short straight section of a storage ring, acts like a thermal sink for the hadron beam via Coulomb collisions between electrons and hadrons. This mechanism has been widely used in low energy storage rings for production of high quality hadron (proton, antiproton or ion) beams. During the past decade, the first application in the medium energy range, namely, cooling of an 8 GeV antiproton beam, has been achieved in the Tevatron Recycler at Fermilab [7,8].

The MEIC design will utilize this traditional electron cooling for its ion beam cooling. The present baseline has adopted a scheme of staged cooling for producing and maintaining short bunch and low emittance ion beams. This new scheme utilizes electron cooling at four different stages of the ion beam life cycle for achieving distinct goals in beam formation and maintenance as described below.

***Step 1: Low Energy DC Cooling at the Pre-booster***

As described in Sections 5.2 and 5.5, the stacking of ions in the pre-booster, except in the cases of negative ions such as $H^-$ and $D^-$, requires electron cooling to achieve a quick reduction of the phase space footprint of the ions already in the buckets in order to allow injection of the next batch of ions from the ion linac. The pre-booster injection energies for $^3$He and lead ions, for example, are 190 MeV/n and 100 MeV/n respectively, and the corresponding energy range of the cooling electron beam is 56 keV to 103 keV. Considering the RF frequency in the pre-booster is below 1 MHz (ramping from 0.55 to 1 MHz), the ion beam is effectively a coasting beam. Therefore, a conventional DC electron cooler with beam energy from 56 keV to 103 keV will suffice to cool ion beams with species $^3$He to lead during their stacking in the pre-booster. Such a low energy DC electron cooler is now a mature technology. The cooling time is estimated in the range of 16 to 55 ms. Table 5.12 presents additional parameters for this cooling stage.

***Step 2: Cooling at the Injection Energy of the Collider Ring***

Electron cooling is applied right after an ion beam is injected into the collider ring but before boosting of its energy. Since the cooling time is proportional to the normalized emittance and approximately 5/2 power of the ion beam energy, it is advantageous to perform an electron cooling while the ion beam energy is still low. This pre-cooling could lead to a factor of 50 reduction of the cooling time for a proton beam at an injection energy of 20 GeV, compared to cooling at the top beam energy of 100 GeV. A similar reduction factor for the cooling time applies to other ion beams as well. Further, a reduction of the normalized emittance at this stage also will cut the cooling time at the next stage (top energy) significantly.

***Step 3: Cooling at the Top Energy of the Collider Ring***

This stage of cooling of the ion beam happens in the collider ring, after the ion beam is boosted to the collision energy and de-bunched to a coasting beam, during re-bunching of the ion beam, and continued after that. The ion beam is captured and bunched by SRF cavities with a 748.5 MHz frequency. The electron cooling conditions the ion beam 6D emittances to the design values and makes the ion beam ready for collision with the electron beam. The cooling time at this stage, due to an already significantly reduced emittance achieved by the pre-cooling at stage 2, is on the order of a few minutes.

***Step 4: Cooling During Collision***

It is critical to continue electron cooling of the ion beam during collisions in order to preserve the beam emittances. The main effect that could cause emittance degradation is multiple small-angle IBS while the single large-angle Coulomb collisions can cause particle loss (the Touschek effect). It is estimated that, with the MEIC design parameters, the IBS time is very short, in the range of a few minutes, as shown in Table 6.1. It means that, without cooling, the MEIC luminosity could drop very fast, due to rapid emittance growth. Therefore, this stage of continuous electron cooling is necessary for suppressing IBS.

The design parameters of electron cooling of a proton beam at the collider ring are given in the following table. They can be scaled to meet the requirement for the other ion beams. We assume that one electron cooler will be constructed and responsible for three stages of cooling in the collider ring. The first of the two data columns on the right is for cooling at the injection energy; at that stage the proton beam is still a coasted beam. The short (~2 cm RMS) electron bunches will cool a very small part of the coasted proton beam during each passage of the cooling channel, while the proton synchrotron motion will bring an averaging effect over the longitudinal direction. The second data column is for cooling in the final stage as well as during collisions.

Table 6.1: Electron cooling of proton beam in MEIC ion collider ring

| Energy (proton / electron) | GeV / MeV | 20 / 10.9 | 100 / 54 |
|---|---|---|---|
| Cooling channel length/circumference | m | 60 / 1350 | |
| Particles/bunch | $10^{10}$ | 0.417 / 1.25 | |
| Bunch frequency | MHz | ~ 1 / 748.5 | 748.5 |
| Beam current | A | 0.5 / 1.5 | |
| Energy spread | $10^{-4}$ | 10 / 3 | 5 / 3 |
| Ion bunch length | cm | coasted | coasted → 1 |
| Electron bunch length | cm | 2 | |
| Proton normalized emittance, horiz. /vert. | μm | 4 | 4 → 0.35/0.07 |
| Cooling time | min | 10 | ~ 0.4 |

***Cooling Against IBS and Extending Beam Lifetime***

Above the transition energy of the collider ring, energy exchanges between ions in IBS lead to growth of the horizontal emittance due to the energy-orbit coupling while the vertical emittance is determined by the *x-y* coupling. Multiple small angle IBS has a relatively large probability and is contributing to phase space equilibrium of an ion beam under cooling, while Touschek scatterings will limit the core and, hence, luminosity lifetime. Specifically, after the cooling starts, the ion bunch will shrink; its horizontal emittance will reach an equilibrium value determined by a balance between the multiple IBS and cooling. Following this stage, an interplay of Touschek scatterings and cooling of scattered particles determines the beam core (and luminosity) lifetime. At ion energies far above the transition value, the area of cooling beam should exceed that of the ion beam, in order to extend the ion core lifetime. To avoid the flip-flop instability of colliding beams [9], the vertical emittance of the ion beam, which is determined by the coupling and multiple IBS, should be not below the limit set by the beam-beam effect. Using this phenomenology, and a proper choice of the target luminosity lifetime, one can derive an optimum set of parameters for reaching a maximum average luminosity of a collider.

## 6.2 ERL Circulator-Ring Based Electron Cooler

Two electron cooling facilities are needed for implementing the multi-step electron cooling scheme in MEIC. The first one is a low energy electron cooler with a DC beam and is used in the pre-booster. Such a cooler is technically well developed and readily available. The second one is a medium energy electron cooler required in the ion collider ring, with machine parameters an order of magnitude beyond state-of-the-art; therefore, it demands new accelerator technologies to realize it. This rest of this chapter will focus on the conceptual design of this medium energy cooler for MEIC and development of the supporting technologies for this cooler design.

### 6.2.1 Design Choices

As required by the MEIC design, the medium energy electron cooler must deliver a cooling electron beam with a high bunch charge and a high repetition rate. The energy range of this cooler, up to 54 MeV, rules out the electrostatic accelerating apparatuses that are used for accelerating the electron beam in all existing low energy electron cooling facilities including the Fermilab cooler. Therefore, the MEIC cooler must rely on RF or SRF technology. Next, the requirement of a superior beam quality (small emittance and energy spread) in order to achieve high cooling efficiency suggests that a photo-cathode electron source with either an ultra DC voltage (as a DC gun) or SRF cavities (as an SRF gun) should be a choice for the electron gun of this cooler.

To ensure the success of a medium energy electron cooler for MEIC, two key accelerator technologies, namely, energy recovery linac (ERL) and circulator ring (CR), will be utilized to overcome the two technical challenges of the cooler design. The first challenge is the required beam power, up to 81 MW, for a 1.5 A electron beam with 55 MeV energy. Delivering such large power demands not only very high capital cost for hardware but also an unacceptablly high operational cost. Furthermore, safely dumping a used beam with such high power, which is about a hundred times of the CEBAF beam, is technically unfeasible. The ERL, a continually maturing technology, provides a perfect solution for overcoming this challenge.

The second challenge is the necessarily long lifetime of the photo-injector, in terms of the total charge extracted from a cathode, which greatly exceeds the present state-of-the-art. Drawing such high average current beam could cause a rapid reduction of the quantum efficiency of the cathode, hence leading to an unacceptable short operational time between services of the cathode surface. A compact circulator ring, in which the cooling electron beam will circulate up to 100 times while continuously cooling an ion beam, could lead to a reduction of average current of the electron beam from the cathode and ERL by a factor equal to the number of circulations, thus greatly extending the effective injector lifetime.

Use of a circulator ring also makes a great reduction of the DC and RF power non-recoverable in the ERL design—namely, the part of beam power associated with the DC or/and SRF cavities in the photo-cathode gun and in the injector—by reducing the current drawn from the photo-cathode and injector. This further reduces the operational cost of the cooler and the burden of dumping a used beam.

### 6.2.2 Design Concept

Figure 6.1 illustrates a generic electron cooler based on a magnetized photo-cathode gun, an ERL and a compact circulator ring [2]. A high charge electron bunch from an injector is accelerated in an SRF linac up to 55 MeV and then sent to a circulator ring with an optically matched cooling channel for the ion bunch. The bunch circulates a large number of revolutions (up to 100) in the circulator cooler ring before its quality degrades significantly by disruption by intra- and inter-beam scatterings, then returns to the same SRF linac for energy recovery. The recovered energy is used for accelerating a new electron bunch from the injector. The JLab FEL device has already demonstrated excellent energy recovery efficiency to nearly 100% [10].

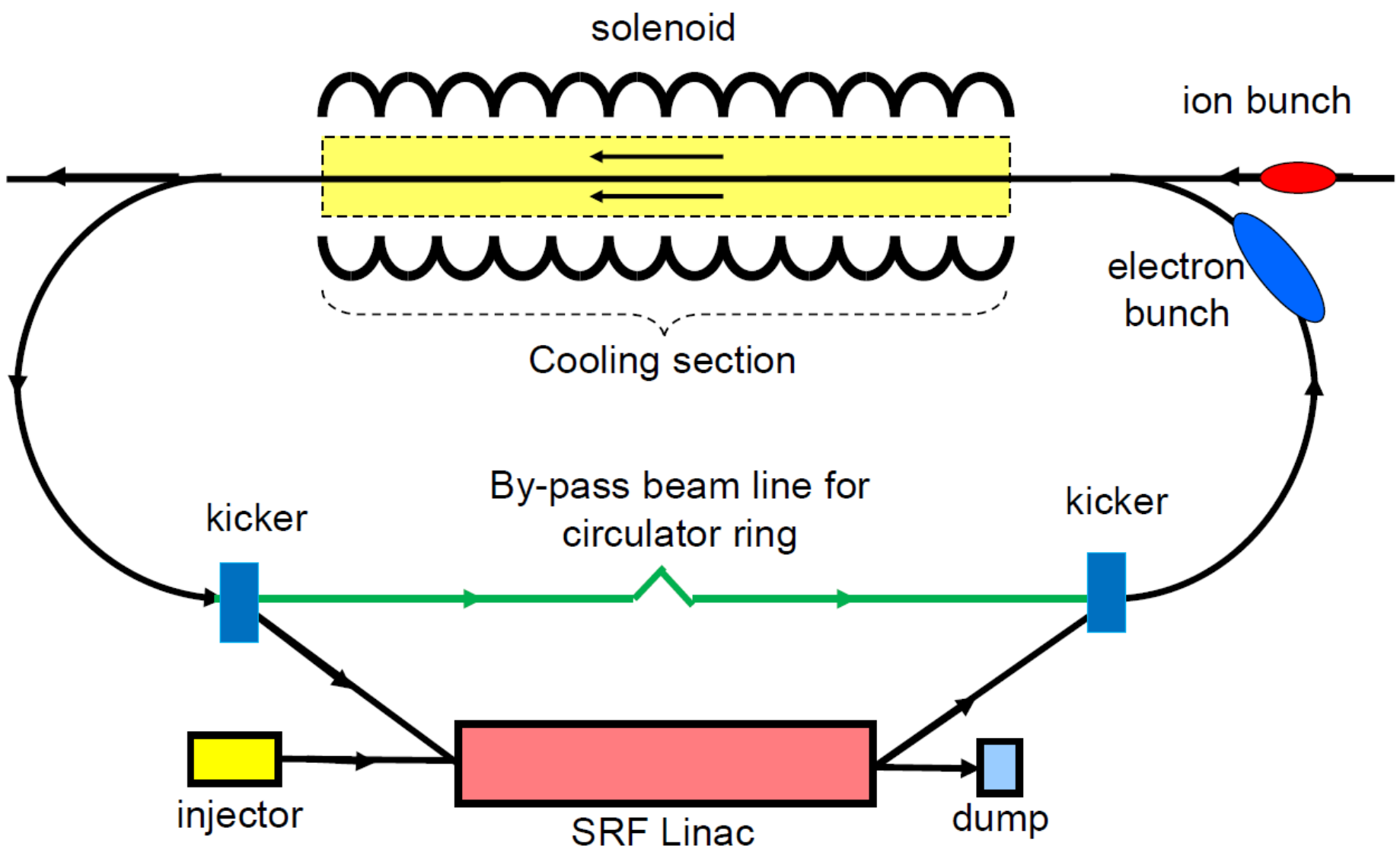

Figure 6.1: A schematic drawing of an ERL circulator cooling facility.

In the above schematic drawing, a straight beamline with an equal length of the cooling channel is needed to complete the circulator ring. A design optimization could be achieved by taking an advantage of the figure-8 shape of the collider ring: placing the circulator ring in the vertex of the figure-8, as illustrated in Figure 6.2, such that the electron cooling can happen in both long straight sections. With this optimization, the total length of cooling channel is doubled; so is the cooling rate. Table 6.2 presents the design parameters for this circulator cooler.

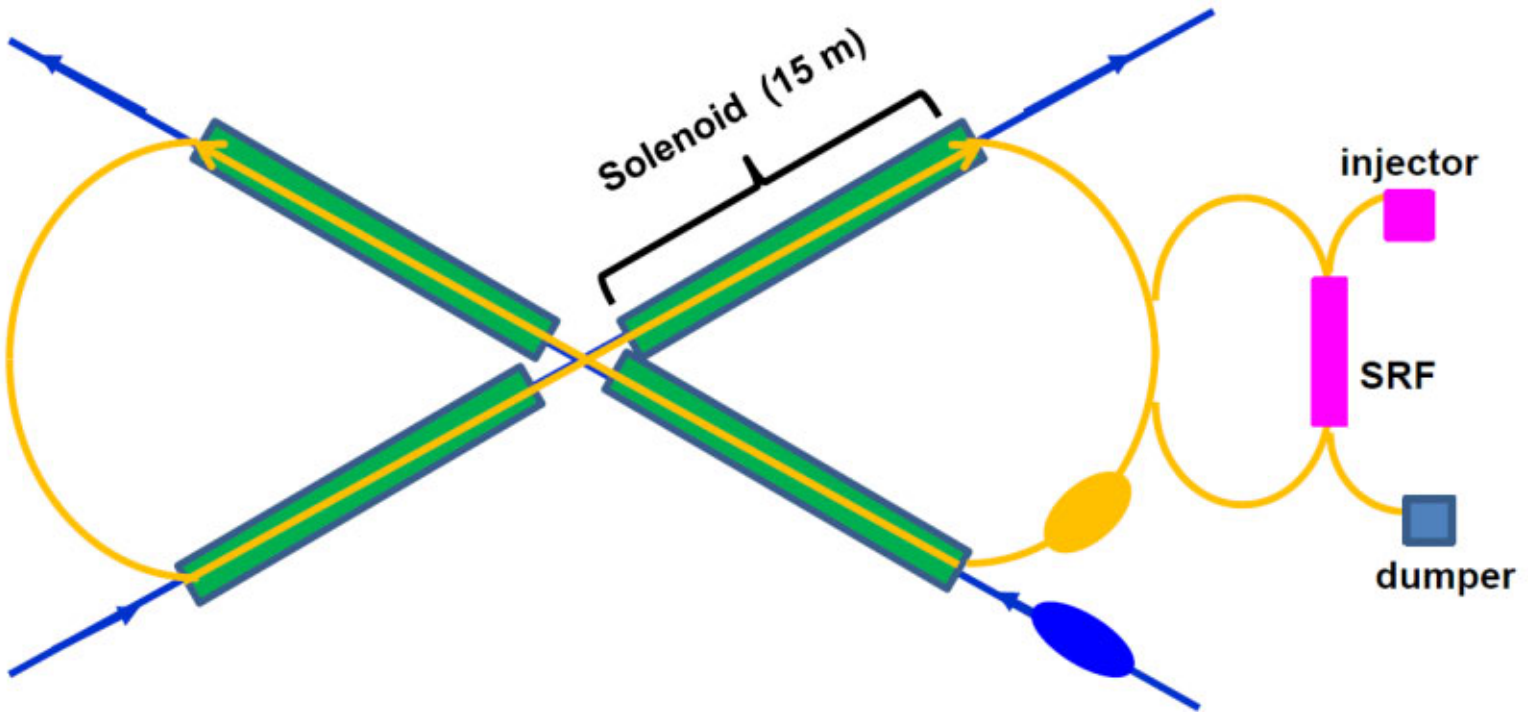

Figure 6.2: An electron cooler is placed in the center of the MEIC figure-8 collider ring.

Table 6.2: MEIC ERL circulator cooler design parameters

| | | | |
|---|---|---|---|
| Beam energy (*p/e*) | GeV/MeV | 30 / 15 | 150 / 75 |
| Length of cooling section | m | 2 × 30 | 2 × 30 |
| Particles per bunch (p/e) | $10^{10}$ | 0.4 / 1.25 | 0.4 / 1.25 |
| Number of circulations | | ~100 | ~100 |
| Average current in ERL | mA | 15 | 15 |
| Average in circulator ring | A | 1.5 | 1.5 |

` A linear optics design of the ERL circulator cooler ring for MEIC has been completed; its footprint is shown in Figure 6.3. On the right side is an SRF linac consisting of a single cryomodule. The figure-8 shape circulator ring on the left side provides space for two 30 m long cooling channels. A fast kicker at the middle of the ERL ring opposite the SRF linac is responsible for switching the electron bunches in and out of the circulator ring. The RMS bunch length of the electron beam should be 2 to 3 cm in the cooling channel; however, it has to be very short in the SRF linac in order to attain very small momentum spread (~$10^{-3}$). Change of the bunch length is achieved through the arc bending dipoles in the ERL ring with large energy chirp. To achieve this, an SRF cavity is placed on each side of the fast kicker for longitudinal matching (de-chirping and chirping), thus expanding the bunch length to a few cm for cooling in the circulator cooler and re-compressing it back to the original length before it returns to the SRF module for energy recovery.

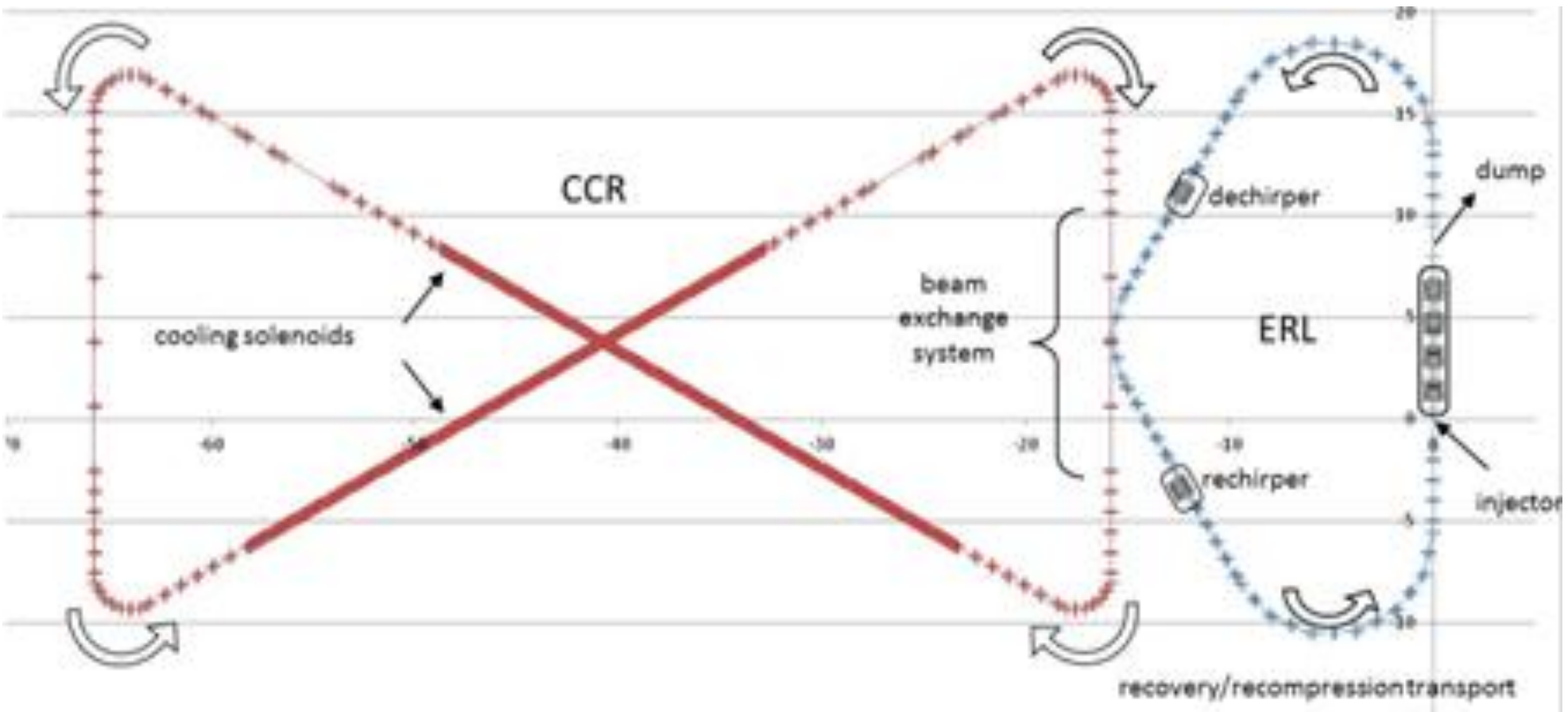

Figure 6.3: Footprint of the ERL circulator cooler ring.

## 6.3 Development of Supporting Technologies

### 6.3.1 Magnetized Beam

A magnetized electron beam could significantly improve the cooling efficiency [1,7,8,11]. There are several additional advantages for magnetizing the cooling beam in the MEIC design:

- It could suppress the non-linear Laslett detuning and provide stable transport of the beam over multiple revolutions in the circulator ring.
- It may be the only effective way to suppress the impact of the beam angular divergence caused by space-charge effects on the cooling rate.
- The strong solenoids in the cooling sections can drastically diminish the impact of the short-range beam misalignment on the cooling rates while the misalignment of the solenoids can be effectively controlled by beam position monitors and correcting loops.
- Large beam size in the arcs acquired by the canonical angular momentum dominated beam may alleviate the CSR instability.
- In the cases of cooling heavy ions, recombination must be taken into account. The transverse velocity spread of the electron beam should be large in order to maintain a sufficiently long ion beam lifetime. This requirement is, nevertheless, contrary to small electron beam emittance for achieving good cooling efficiency. High magnetization of the electron beam could provide a solution to this contradiction at the price of an acceptable reduction of the cooling rate.

Generating and transporting a magnetized beam is rather nontrivial for the MEIC cooler. For a low energy DC electron cooler, the whole beam line starting from the source to the dump can be wrapped entirely by long solenoids. It is not feasible to make a similar arrangement for an ERL circulator cooler with energies up to 55 MeV and a circulator ring with 100 m plus circumference. Our present design concept is to combine a magnetized photo-cathode/injector and a superconducting solenoid around the cooling section with a specially designed beam transport system.

***Magnetized Electron Gun and Injector***

A magnetized beam acquires an angular momentum at the exit of the solenoid around the cathode. Figure 6.4 illustrates a design concept of a magnetized photo-cathode DC injector. It consists of a high (350 to 500 kV) voltage DC gun immersed in a solenoid, a DC buncher cavity and one to several SRF cavities for bringing the electron energy to several MeV. Once generated in a solenoid, the electron beam is guided and the space charge force is counteracted by the solenoid field. There are two solenoids downstream of the gun for implementing the emittance compensation scheme. A set of quadrupoles after the SRF module will be used for matching the optics between the injector and the ERL.

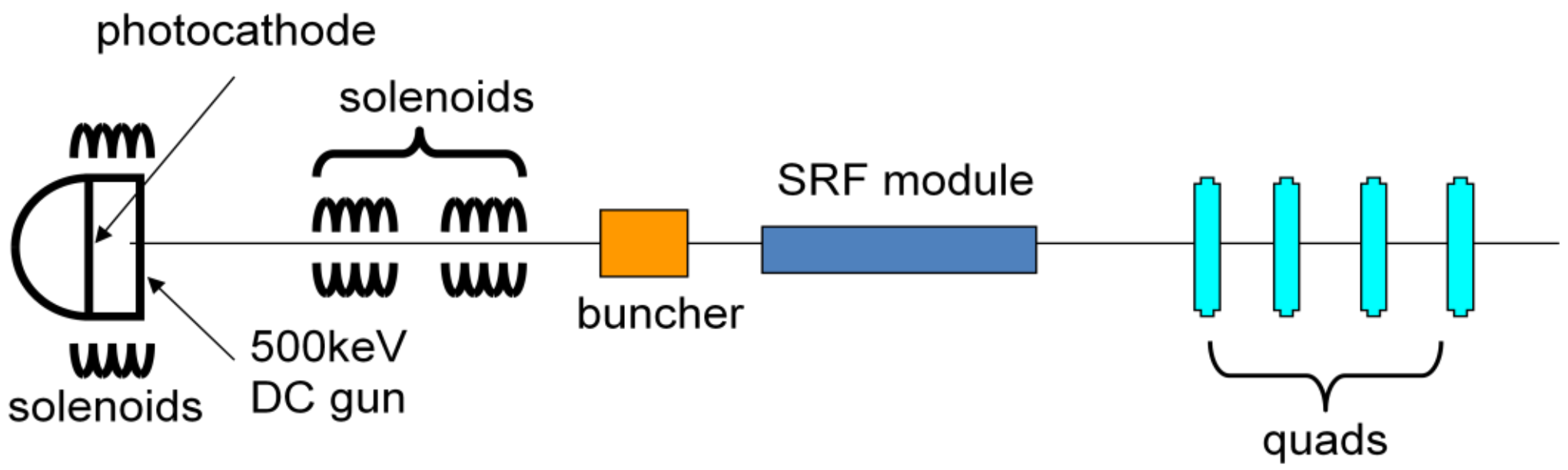

Figure 6.4: A schematic drawing of a magnetized DC electron injector.

We will also investigate both DC and normal conducting RF guns with thermionic cathode immersed in a solenoid field. Once generated in a solenoid, the electron beam is guided and the space charge force is counteracted by the solenoid field.

***Beam Transport System***

The round rotational beam will quickly expand. Here we have two options [8,12]. The first is to transport the rotated beam with axi-symmetric or conventional non-axi-symmetric quadrupole magnets and RF cavities with a goal to match the beam to the cooling section solenoid at its entrance. The second option has multiple steps. The angular momentum dominated beam at the exit of the gun is first converted to a flat beam with a beam adapter consisting of skew quadrupoles. After this conversion, the beam optics is decoupled and the flat beam is transported to the front of the cooling section. Once again with a beam adapter the flat beam is converted to a rotational round beam matched to the cooling section solenoid. We will study pros and cons of these two approaches.

### 6.3.2 Fast Beam Kicker

A key component of a circulator cooler is a fast kicker which switches electron bunches in and out of the circulator ring. The main parameters for this kicker are (1) 7.5 to 75 MHz repetition rates (assuming 10 to 100 revolutions in the ring); (2) about 1.3 ns rise/fall time; (3) 5 kV integrated kicking voltage. These parameters are beyond the state-of-the-art. Several ideas are currently under investigation and are discussed briefly below.

***Harmonic RF Kicker***

This approach focuses on generating the RF pulse structure that will be required to drive a physical kicker such as a cavity or a strip-line [13]. The pulse is essentially a superposition of a set of harmonic RF signals. Figure 6.5 illustrates a signal comprising the first 11 harmonic modes. The problem therefore reduces to how to sum harmonically related frequencies of equal amplitude. One solution is to use several narrow-band amplifiers, one for each frequency, and add them together at high power. This may not be efficient because the total power required in the final waveform is considerably less than the sum of the power in the harmonics because of the (desired) interference. The best solution is to create an amplifier with gain at a series of harmonically related frequencies.

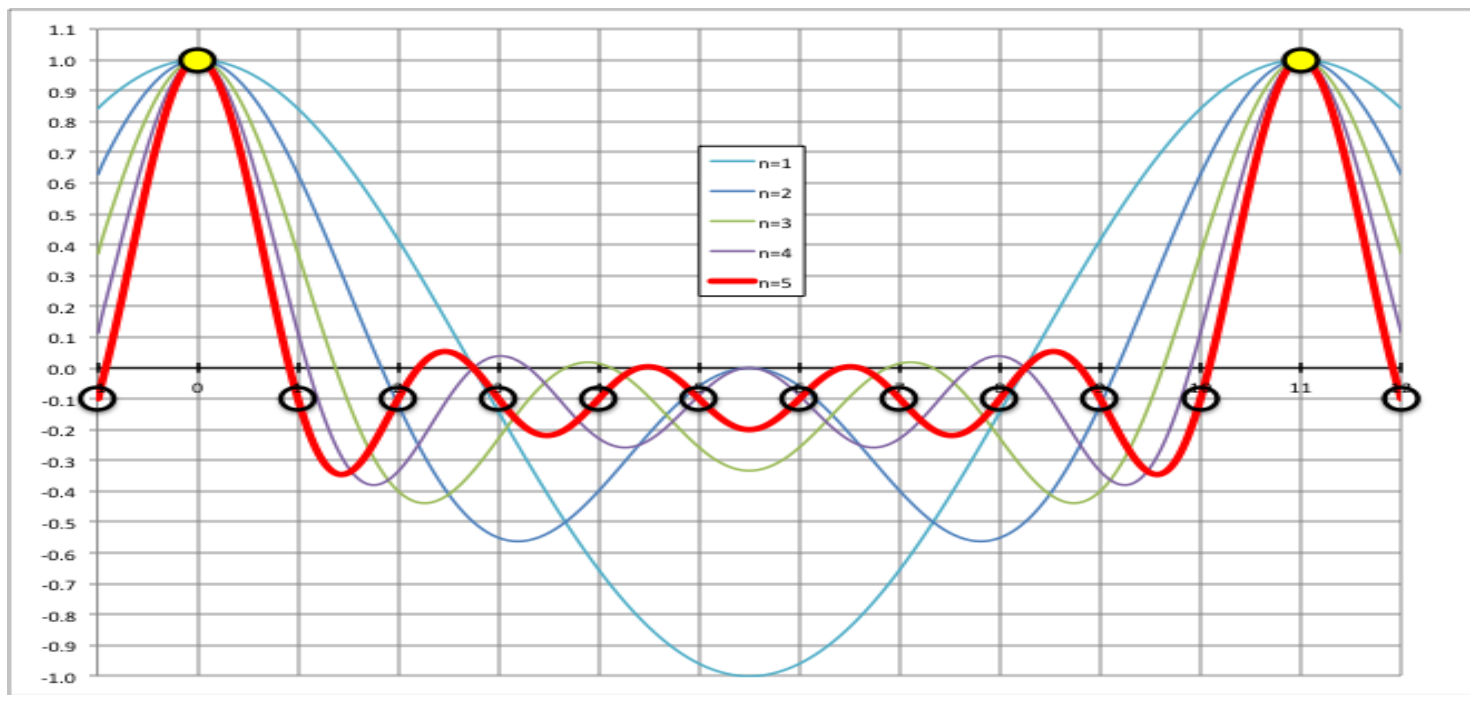

Figure 6.5: Illustration of a waveform for a harmonic kicker.

The usual way to create a kicker is to use either a cavity, which is tuned to the desired frequency, or a strip-line to create a traveling wave. The modes in a cavity are not usually harmonically related so it would be difficult to create the desired pulse structure in a cavity. The strip-line is usually terminated in a matched load to ensure rapid response; however this makes it very energy inefficient. An alternative is to use a resonant ring in which the power exiting the strip-line is brought back to the input, and the incoming power is fed into the ring via a directional coupler. In this case, the length of the resonant ring (or to be exact, the time taken to make the circuit) should equal the spacing between extracted bunches.

*Pulse-compression Technique*

Present researches involving *pulse compression techniques* may provide several potential options for producing short, medium-power RF pulses with 10 to 15 MHz repetition rate. Pulse compression employs a swept RF source in conjunction with a dispersive element, resulting in multiple wave-fronts piling up at the output to produce a very short, high peak power pulse [2,14]. Recent experiments using a helically corrugated waveguide as the dispersive element have achieved compression and power enhancement ratios of 12 or better, creating 2 ns pulses at a 12 MHz repetition rate, and having a peak power of 11 kW [2,14]. In these tests, a swept RF source provides a gated "chirp" signal amplified by a gated traveling-wave tube amplifier, which is coupled to the load via the dispersive helically corrugated waveguide. Figure 6.6 illustrates the technique, along with associated frequency spectra.

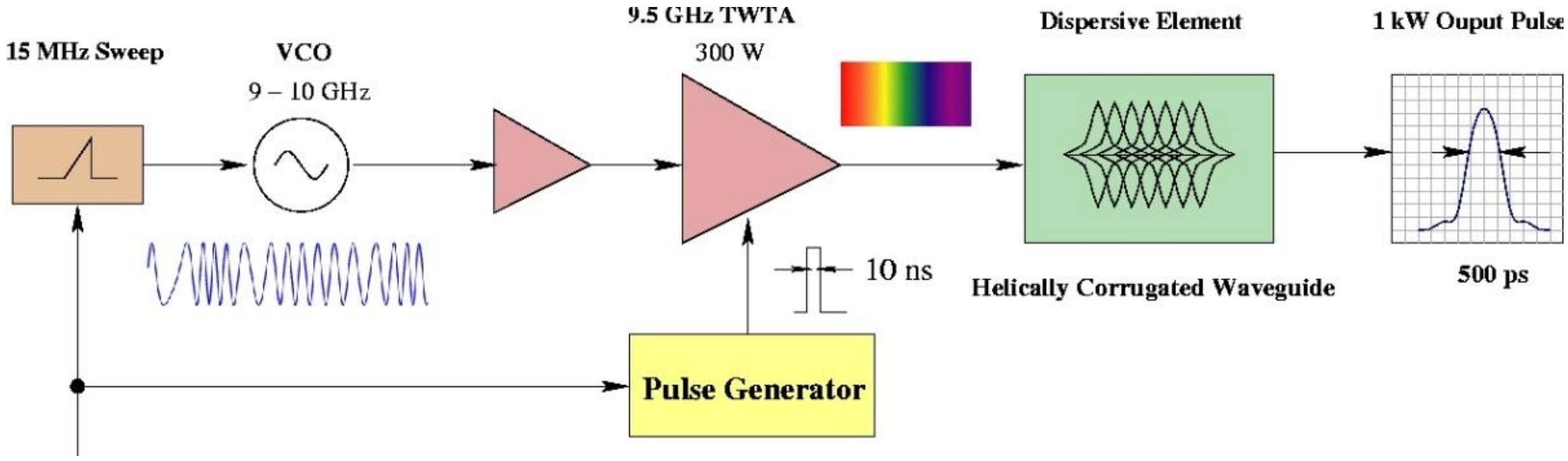

Figure 6.6: Schematic of pulse compression technique using helically corrugated waveguide. Output pulse envelope is the result of "chirping" the RF input, and letting all wave-fronts pile up at the output.

Although experiments have shown that it is possible to nearly realize the required pulse parameters with a single stage, it is likely that more stages (10–20) having reduced power will provide the required BdL (1.25 G m). This is especially attractive, since the required output power from each individual amplifier decreases quadratically with the total number of stages. Also, since the primary goal of the experimenters was high peak power, specifically for plasma physics and radar applications, it can be expected that a reduction in peak power will further reduce pulse width. Conceptually, the power electronics would reside above-ground to facilitate accessibility and repair. Due to its manageable size, the dispersive element can be located either with the electronics or at the beamline.

***Beam-Beam Kicker***

A rather innovative idea currently under investigation utilizes a non-relativistic sheet beam for providing transverse kicking to an electron bunch [2]. This beam-beam kicker idea was first proposed by Shiltsev [15] for two round Gaussian beams. Here we consider a case of flat beams as shown in Figure 6.7; the analytical treatments for these two cases are quite similar.

Since the target (cooling) beam is moving at the speed of light, it passes through the non-relativistic kicking flat beam in a period of time determined by the length of the kicker beam $l_k$. At a very close distance to the kicking beam, an electron in the target beam receives an instant angular kick determined by integration of the transverse force over that passing time

$$\delta\theta_y = \frac{2\pi N_k r_e}{\gamma \sigma_{xk}} \tag{6.1}$$

where $N_k$ and $\sigma_{xk}$ are number of electrons and horizontal RMS size of the kicker bunch and $r_e$ is the electron classical radius. Table 6.3 summarizes the design parameters for a beam-beam kicker for the MEIC circulator cooler design.

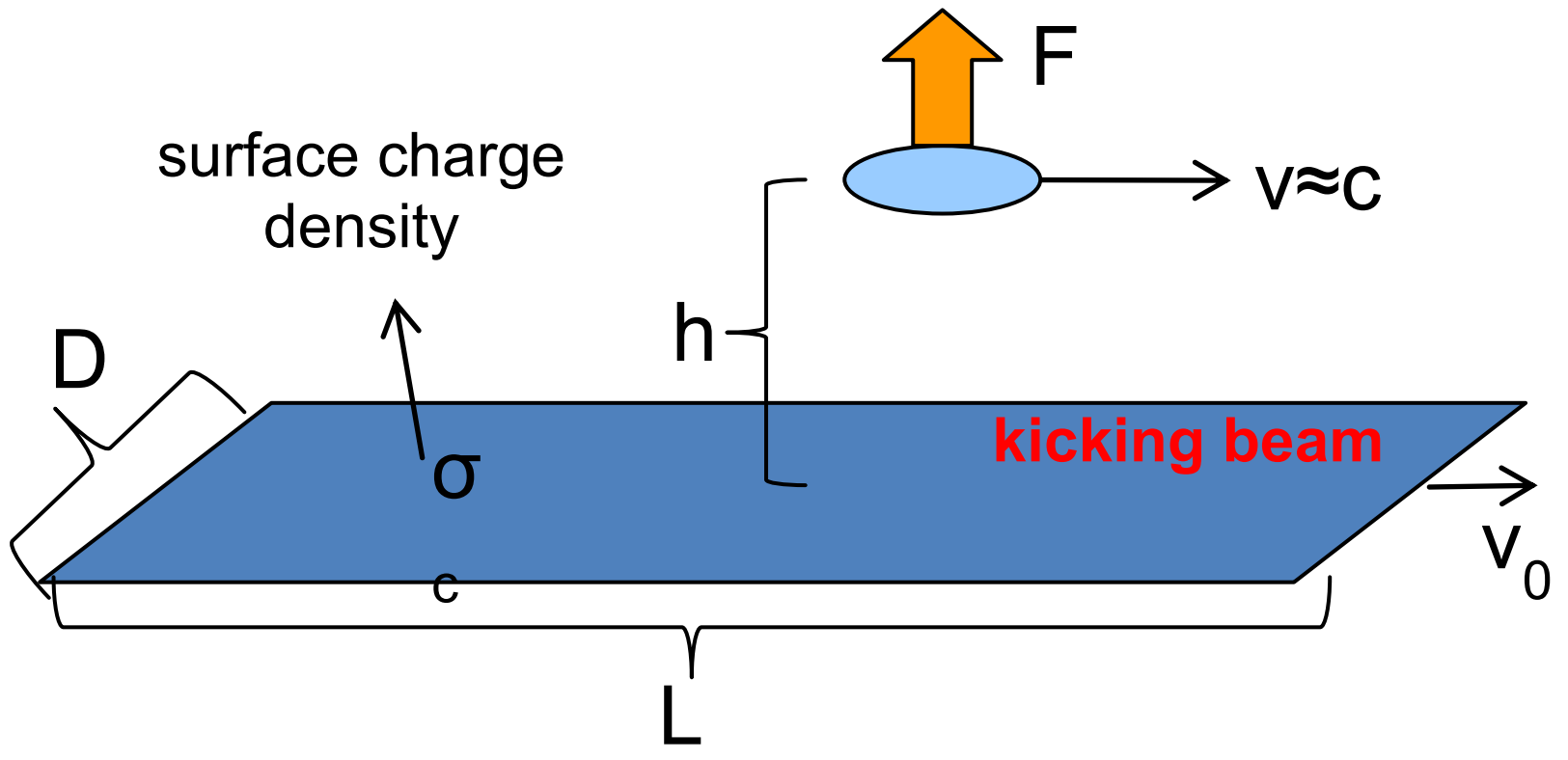

Figure 6.7: Schematic drawing of beam-beam fast kicker.

Table 6.3: Design parameters for a beam-beam kicker for MEIC

| | | |
|---|---|---|
| Circulating beam energy | MeV | 33 |
| Kicking beam energy | MeV | ~0.3 |
| Kicking repetition rate | MHz | 5 – 15 |
| Kicking angle | mrad | 0.2 |
| Kicking bunch length | cm | 15 – 50 |
| Kicking bunch width | cm | 0.5 |
| Kicking bunch charge | nC | 2 |

A flat kicker beam can be produced utilizing a grid-operated DC (thermionic) electron gun with a round magnetized cathode. While maintaining the beam in solenoid, one can impose a constant quadrupole field that causes beam shrinking in one plane while enlarging in the other plane due to the drift motion of particles. The process should be adiabatic relative to the particles' cyclotron motion in the solenoid [2]. The beam current density could be specifically profiled at the cathode to create uniform distribution in a homogenous field in a direction transverse to in the "plane" of flattened beam.

# 7. Interaction Regions

Presently, the MEIC is designed to accommodate two physics detectors for electron-proton or electron-ion collisions in the medium energy range—a primary one with full acceptance and excellent resolution, and a secondary, high-luminosity one with a less demanding specification [1], including a reduction of the magnetic free detector space. In this chapter we focus on the design of the interaction region (IR) for the primary detector. We start with a brief description of the full acceptance detector, followed by a discussion of the IR design considerations for supporting this detector. The layout of the IR, the detector integration, and a demonstration of full acceptance are presented in section 7.3. The linear optics of the IR presented in section 7.4 reflects a generic design incorporating features of both the primary and secondary detector IRs (the size of the former and the symmetry of the latter). This has been used for developing the scheme for chromatic compensation presented in section 7.5, and preliminary studies of momentum acceptance and dynamic aperture in section 7.6. In the next two sections, several special IR issues such as crab crossing, synchrotron radiation, and detector background are discussed. Finally, a preliminary simulation study of the beam-beam effect is presented in the last section.

## 7.1 MEIC Primary Detector

The two main requirements for a full-acceptance detector of MEIC integrated into the MEIC design are the following [1]:

1. *Hermeticity*: It allows for detection of scattered electrons, mesons, and baryons without holes in the acceptance, even in forward regions.
2. *High-luminosity capability*: It operates in a high-luminosity environment with moderate event multiplicities and acceptable background conditions.

Figure 7.1 illustrates the layout of this full-acceptance detector [1]. At center of this layout is a 5 m long SC solenoid extending 3 m from the interaction point (IP) on the side of the outgoing ions and 2 m on the opposite side. The maximum solenoid magnetic field will be between 2 and 4 T, adjustable independently of the beam energies in order to optimize detection for different processes. Further, the magnetic enclosure around the solenoid has to be compatible with the end-cap geometry.

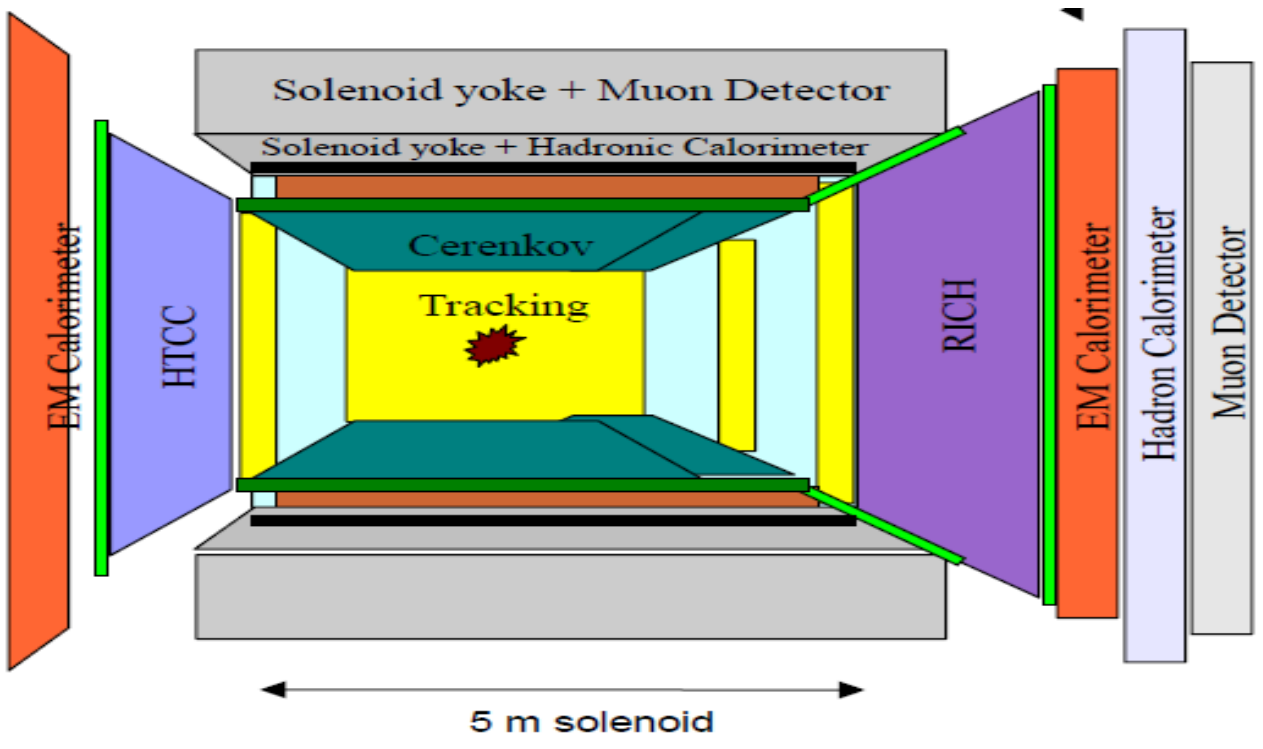

Figure 7.1: Schematic layout of the MEIC primary full-acceptance physics detector.

A true full-acceptance detector requires the capability for detecting small angle outgoing particles in addition to large-angle scatterings. The hardware for such a detector includes:

1. End-cap detector packages (approximately 2 m thickness).
2. Small-angle detectors and analyzing magnets on both sides of the FFQs.

As a consequence, a sufficiently large magnet-free space near the IP must be made available for these detector components. On the outgoing ion side, a 7 m space from the IP to the first final focusing quadrupole (FFQ) is needed, while on the other side, this space could be smaller (down to 3.5 m) as long as the ion FFQs do not interfere with the end-caps or the low-$Q^2$ tagger.

Detection of small-angle outgoing particles in MEIC falls into the following two categories:

***Outgoing Electrons***

Detection of the small-angle outgoing electrons that have no significant change of their angle (low-$Q^2$ tagger) after radiating a photon demands large-aperture electron FFQs followed by a horizontal dipole magnet for momentum analysis and a 5–10 m long drift space. The latter is needed to allow electrons that still carry a significant fraction of the beam momentum to be deflected well outside of the beam-stay-clear area where detectors can be placed. The dipole magnet should at least approximately track the electron beam energy. Optical elements that could interfere with functionality of the low-$Q^2$ tagger (e.g., horizontally focusing quadrupoles) should be avoided in the dipole/drift area.

***Outgoing Hadrons***

Detection of the small-angle hadrons on the outgoing ion side is done in two stages, and with the barrel and the end-cap detectors, a full angular coverage can be achieved in principle.

The first stage, extending an additional 2 m past the end-cap, is centered around a large acceptance dipole magnet of 2 T·m BDL in front of the ion FFQs for covering at least 3–5° to either side, depending on the azimuthal angle. This dipole will be about 1 m long, and followed by 1 m drift space and detectors. The dipole field strength can be scaled with the ion beam energy. If the electron FFQs are placed in front of the dipole, their positions, gradients, and lengths need to be optimized to minimize the maximum outer radius of the doublet/triplet. For instance, if the first electron FFQ has an outer radius of 10 cm and is placed at a distance of 3.5 m from the IP, the clearance in the beam plane (worst azimuthal case) will only be 1.2° on the side of the FFQ if the crossing angle is 50 mrad. This would be acceptable, but a larger crossing angle and smaller FFQs placed further away could significantly improve the acceptance.

The second stage is aimed at detecting the hadrons that pass through the apertures of the ion FFQs. This requires:

1. A large-aperture dipole with a 10 to 20 Tm BDL placed not too far downstream of the ion FFQs followed by at least 10 m of drift space for allowing scattered particles, ranging from significantly less than half to almost the full beam momentum, to move to outside the beam-stay-clear area. The dipole field may be adjusted to the ion beam energy.

2. The ion beam aperture should be at least 7 cm at the front face of the FFQs, and become progressively larger with increasing distance from the IP all the way until reaching the downstream detectors in order to allow detection of particles scattered between 0° and 0.5°. Neutrons, which travel in straight lines, require apertures of all magnets to scale with the distance from the IP to its far end.

Transporting charged particles and ion fragments with up to half of the beam momentum can be demanding. It could impose constraints not only on the apertures, but also the ion optics and the distance to the downstream dipole. Since not all particles that enter the FFQs will reach the detectors in the drift space following the downstream dipole, additional detectors will be placed between the FFQs and the dipole in order to achieve full acceptance.

## 7.2 Interaction Region Design Considerations

The interaction region, acting as an interface between the beam acceleration/transport system and the detector, is one of the most challenging parts of the MEIC design, and it largely determines the performance of the collider. To reach the high performance goals, the IR design should go through an optimization process that is subject to several important physics constraint conditions. Some of these factors are:

1. Beam transport and manipulation (linear and high orders).
2. Single particle beam dynamics, especially the impact on dynamic aperture.
3. Collective effects and instabilities (the leading one is the beam-beam effect).
4. Detector acceptance
5. Synchrotron radiation, residual gas and detector background.

In the following, we will show how these factors plus the main project goals are addressed while selecting the main design parameters and also the layout of the interaction region for the primary full-acceptance detector.

### *Choice of $\beta^*$*

To achieve a high luminosity, it is standard to de-amplify the transverse spot sizes of the colliding beams at the IP through a final focusing block (FFB) which is typically made of a doublet or triplet of strong magnetic quadrupoles. Such de-amplification is characterized by a $\beta^*$, the betatron function at the IP. From the beam spot size formula, $\sigma=(\beta\varepsilon)^{½}$, where ε is the geometric emittance, a small $\beta^*$ means a small spot size at the IP, as a result of a strong focusing, and requires strong quadrupole fields.

Choice of a small $\beta^*$ is constrained by aperture of the final focusing quadrupoles, chromaticity, nonlinear effects of magnets, and some additional factors. For the full-acceptance detector of MEIC, the magnet-free space for the ion beam must be at least 7 m. Therefore, to maintain large apertures of the FFQs and to satisfy other constraints, a 2 cm vertical $\beta^*$ has been chosen for the colliding beam. The maximum value of the betatron function in the FFB, given by a simple formula $\beta^{max} = f^2/\beta^*$, reaches 2450 m for ions and 625 m for electrons, assuming the focal length $f$ approximately equals 7 m and 3.5 m detector space respectively.

Further, since the bunch lengths for the both colliding electron and ion beams are designed to either equal or be less than 1 cm, it is clear that there will be no luminosity loss due to the hour-glass effect.

It is stated in section 3.3 that the maximum betatron value $\beta^{max}$ is to be 2.5 km, as a general design guideline. It is clear that if a detector space can be reduced, then $\beta^*$ can be reduced accordingly, leading to a further reduction of the beam spot size and a gain of luminosity. As an example, the detector space for the secondary MEIC detector, which is presently optimized for higher luminosity, can be reduced to 4.5 m, thus we can choose a much smaller $\beta^*$, 8 mm.

It should be noted that the ion beams of MEIC are not round as a results of a balance of electron cooling and (un-isotropic) intra-beam scattering; thus the ratio of the horizontal and vertical emittances depends on ion beam energies. For a 60 GeV proton beam, the aspect ratio is about 5, so following the B-factory designs [2,3], a 10 cm horizontal $\beta^*$ has been selected for an optimization of horizontal and vertical beam-beam parameters for both beams.

Lastly, in MEIC, an ion beam will collide with an electron beam of various energies and vice versa. Because electron beam emittance in a storage ring strongly depends on the electron energy and the ion beam emittance could also vary depending on the electron cooling, in order to always match the beam spots at the collision points, the value of $\beta^*$ must be adjusted accordingly for both beams over the energy ranges. The interaction region design must be able to accommodate such flexibility.

***Crab Crossing of Colliding Beams***

For a beam with a 748.5 MHz bunch repetition frequency, the bunch space is 40 cm. A fast beam separation near an IP is required in order to eliminate all harmful long range parasitic collisions. In the MEIC design, a crossing angle of 50 mrad has been chosen for the colliding beams at an IP to provide sufficiently fast beam separation after collisions as well as enabling detection of small angle outgoing particles. To avoid loss of luminosity associated with non-head-on collisions, following the success of KEK B-factory, a scheme of crab crossing [3] utilizing SRF crab cavities will be employed for restoring the head-on collisions [4,5].

***Organizing the Interaction Region***

Synchrotron radiation of the electron beam at or near the IR could cause serious background problems for the detector; thus bending of the electron beam in the IR has been minimized. Ideally, the electron beam should travel along a straight line after exiting the arc until reaching the IP. This is the exact reason that ion beam will go through a vertical chicane to be at the plane of the electron ring for horizontal crossing IPs, as it is mentioned in section 3.2. For the same reason, the electron beam line should be aligned with the detector solenoid. The strength of the last dipole(s) where the electron beam exits the arc should be reduced to minimize synchrotron radiation reaching the IP. Further, to reduce the random background from the interactions between the ion beam and residual gas inside the beam pipe, the section between the ion beam exit arc and the IP should be short and suitable for holding ultrahigh vacuum. As a corollary, the IP on each straight section should be located as close as possible to the arc where the ion beam comes.

***Beam-Stay-Clear***

In general, one would like to make magnet apertures and beam pipes as large as possible in order to maximize the flexibility of the accelerator design. Beam-stay-clear is affected by several design constraints such as dynamic aperture, magnet strength, distance between the IP and the front face of the first FFQ, beam crossing angle, etc. We use the MEIC nominal design parameters in Table 3.1 as a starting point for an estimation of the physical aperture imposed by the beam-stay-clear. For the primary full acceptance detector, the maximum RMS size of a 5 GeV electron beam at the FFQ is about 1.8 mm while the maximum RMS size at the ion FFQ is up to 3.7 mm due to a large 7 m detector space. Away from the IP, the beam sizes are smaller since the beta functions are of order of 10 to 20 m for electrons and ions. With these beam sizes, one possible configuration for the beam-stay-clear of the MEIC is:

- Near the IP: 2.2 cm ($12\sigma$) for the electron beam, 3.7 cm ($10\sigma$) for the proton beam.
- Away from the IP: 1.8 mm ($8\sigma$) for the electron beam, 3 mm ($8\sigma$) for the proton beam.

An additional aspect of the beam-stay-clear is how close the detectors can be moved to the beam without causing beam loss. The closer the detectors the better the detector coverage is. As discussed in the next section, it is advantageous to focus the ion beam downstream of the ion forward final focusing block so that a smaller opening is required for the beam in the surrounding detectors. In combination with a large dispersion at that point, this allows for detection of reaction particles produced with small momentum offsets and improves the detector momentum resolution. Due to a small beam size at that point, the beam pipe aperture is not an issue.

## 7.3 Interaction Region Layout and Detector Acceptance

Figure 7.2 shows the schematic layout of the IR for the full-acceptance detector. To minimize hadronic background from collision of ions with residual gas, the IP is located asymmetrically on the straight section, close to where the ion beam exits the arc. To reduce backgrounds from synchrotron radiation, the electron beam propagates in a straight line between the arcs, and is aligned with the axis of the detector solenoid. The ion beam crosses the electron beam at the IP in the horizontal plane at a 50 mrad angle. In addition to its benefits for the accelerator, this large crossing angle is also a key feature of the detector concept. A solenoid field does not provide momentum analysis for particles moving close to its axis, but the crossing angle moves the particles scattered at small angles away from the solenoid axis into a region with good tracking resolution. The spot of poor resolution on the axis is moved into the periphery off to one side, affecting only a small portion of the azimuthal coverage where high-resolution detection is complicated anyway by shadowing by the electron quadrupoles. For proper integration with the solenoid, the latter have to be either permanent (first pair) or superconducting. The impact on detection is minimized by adopting quadrupoles with a small outer diameter. The crossing angle also allows for the introduction of two spectrometer dipoles on the ion beamline. In addition to generating the large dispersion needed to detect high-momentum particles scattered at small angles, the dipoles correct for the crossing angle, returning the ion beam to nearly parallel with respect to the electron beamline, but with a separation of about 1 m, allowing for easy integration and providing ample space for detectors.

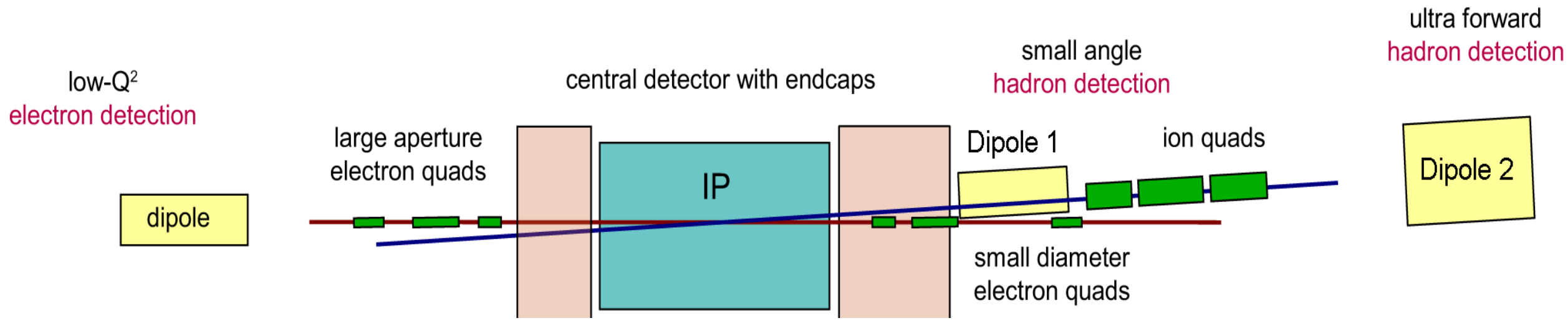

Figure 7.2: Layout of the interaction region for an MEIC primary detector.

Due to the large asymmetry in the electron (3–11 GeV) and ion beam energies (20-100 GeV for protons), the layout of the interaction region design is asymmetric. Much of the science of an electron-ion collider is associated with the detection of spectator quarks and nuclear fragments moving along the ion beam direction in coincidence with hadrons originating from the struck partons. Thus, special emphasis is placed on the ion beam line elements downstream of the IP in an effort to measure all the particles in the event with good resolution, including the high-momentum ones emerging at small angles. In order to accomplish this, the detection is performed in three stages. The first uses the endcap of the central detector. Due to the much higher momenta of the particles produced along the direction of the outgoing ion, the IP is located asymmetrically inside the detector solenoid (2 + 3 m) leaving more space for tracking there. As noted above, the tracking resolution at small angles is enhanced by the large beam crossing angle. The second detection stage covers an intermediate region of a few degrees around the ion beam, where the resolution is further improved by a 2 Tm dipole (Dipole 1 in Figure 7.2), acting in conjunction with the central solenoid (for which the maximum field is expected to be somewhere between 2 and 4 T). Making this dipole outbending (for positive particles) offers several benefits, including larger beamline separation, better acceptance for the zero-degree calorimeter (ZDC) used for neutral-particle detection, and better momentum acceptance for the large downstream Dipole 2, as the dispersion inside it becomes small. The third detection stage covers particles scattered within the first degree, which are detected after passing through the ion FFQs.

The layout of the full-acceptance interaction region up to the ion FFQs is given in Figure 7.3. Both the assumed beam-stay-clear area and a 0.5 degree cone are indicated along the ion beam direction. The electron FFQs are relatively small and moved in to a 3 m distance from the IP. Dipole 1 and the three ion FFQs are also indicated, with the first ion FFQ located at a distance of 7 m from the IP. As shown in Figure 7.4, a complete 3D model of the interaction region has also been implemented [6] using G4beamline/GEANT4 [7]. This has been used for comprehensive studies of the detector resolution and acceptance through particle tracking [6], validating the detection capabilities. Special attention was paid to sizes and positions of the detector region elements to avoid interference with each other and with the detector functionality. The IR optics design, described in the next section, was optimized for detection, but the optimization also reduces the associated accelerator challenges.

Perhaps the most challenging of the three detection stages is the last one, where very ambitious resolution and acceptance goals have been achieved simultaneously. Particular attention was thus paid to the simulations of particles detected downstream of the FFQs. The acceptance is on one hand limited by the magnet apertures and other possible obstructions, and on the other by the beam size and dispersion at the detection point. The former is important for the detection of particles with rigidities different from that of the beam (high-*t* recoil baryons, nuclear fragments with *Z/A* different from that of the beam, and tagged spectators), while the latter is essential for the detection of particles with a rigidity similar to that of the beam (low-*t* recoil baryons and nuclear fragments with *Z/A* similar to that of the beam). Achieving good resolution also requires a large dispersion in order to spatially separate particles with different momenta, and long magnet-free drift sections allowing precise angular measurements. In addition, it is important to minimize the number of magnetic elements to ensure that the transport of the scattered particles is well understood. It is also worth keeping in mind that a good intrinsic resolution for all detection stages has the advantage of reducing the requirements on the tracking detectors and their alignment, potentially offering considerable cost savings in instrumenting the detector. It also allows for the use of more radiation resistant technologies where needed.

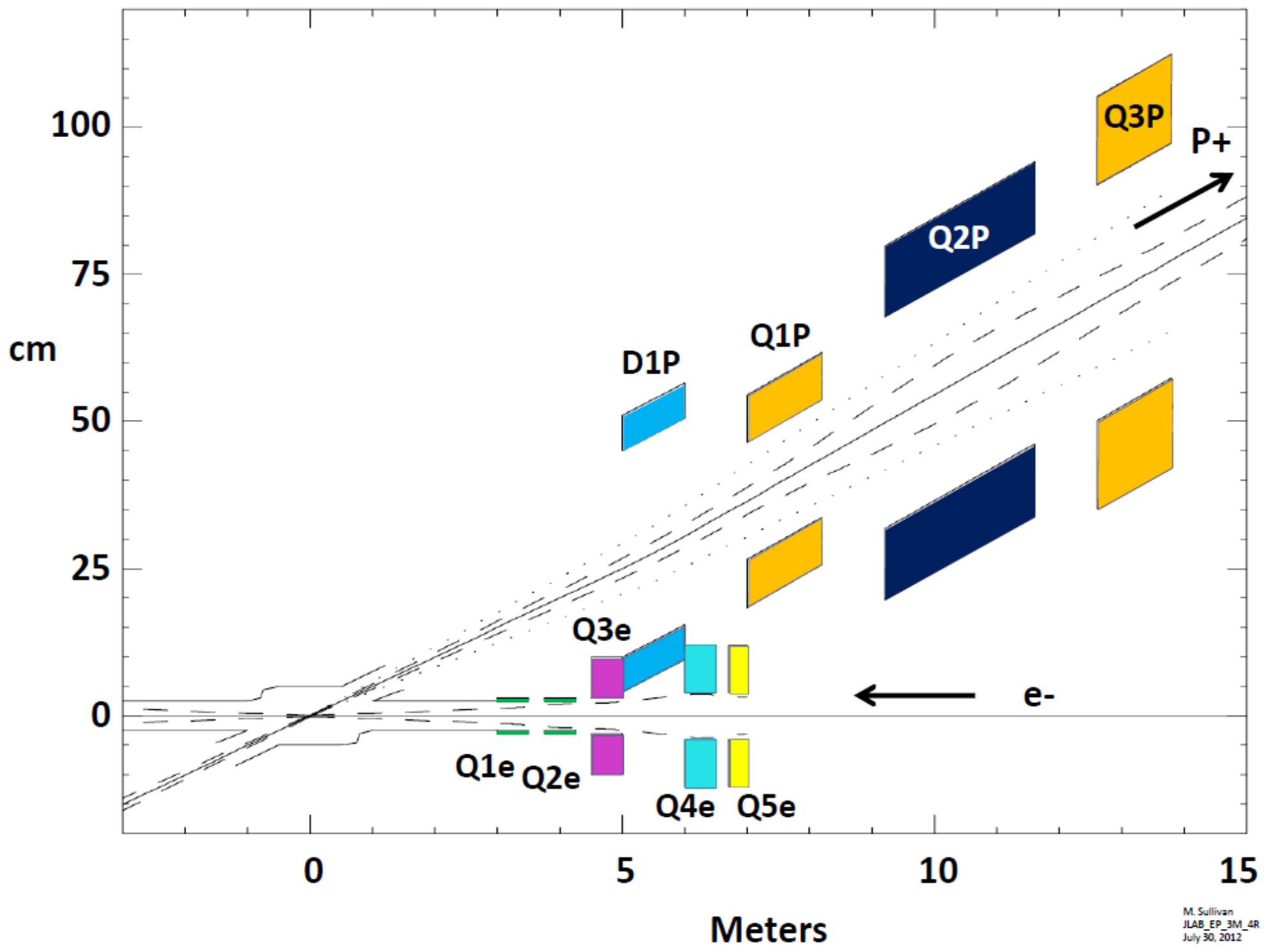

Figure 7.3: Layout of the asymmetric IR of the MEIC. The vertical scale is highly exaggerated, and the emphasis of the figure is on the outgoing ion beam direction elements. The beam-stay-clear area and a 0.5° detection cone are indicated. The detector solenoid is offset by 0.5 m in the direction of the outgoing electron beam, and extends 3 m (2 m) in the outgoing ion (electron) beam direction. The interleaved positions of the incident electron and outgoing ion beam elements are illustrated.

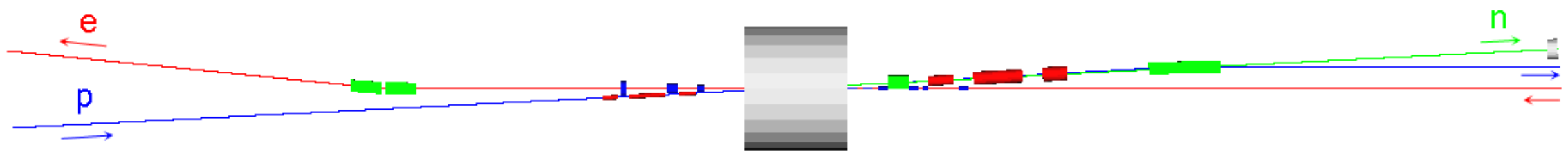

Figure 7.4: To-scale G4beamline/GEANT4 3D model of the MEIC's full-acceptance detector.

The acceptance of small-angle particles with rigidities different from that of the beam depends on the magnet apertures, which in turn are limited by the peak field at the maximum beam energy supported by the ring. Since the peak field is simply the product of the magnet gradient and the size of the aperture, achieving a good acceptance hinges on a linear optics design with low maximum quadrupole gradients, and benefits from a lower maximum energy. Figure 7.5 shows simulations where particles were tracked from the IP and through the FFQs for different initial scattering angles $\theta_x$ (with respect to the ion beam direction), and momentum offset $\Delta p/p$ (or equivalently $\Delta(m/q)$ for ion fragments). Both planes are shown for three choices of the quadrupole peak field, with the latter ranging from state-of-the-art (6 T) to extremely challenging (12 T). Here $\Delta p/p$ = -0.5 corresponds to, for instance, a tagged spectator proton from deuterium, while $\Delta p/p$= +0.5 would be a fairly exotic fragment with low charge. Particles reaching detectors beyond the ion FFQs are indicated in blue, while particles in red are detected in the intermediate stage, between the upstream 2 Tm dipole and the FFQs, where the resolution is not as good. The ultra-forward detection of the MEIC can thus achieve excellent acceptance (up to 0.5 degrees) with the most conservative choice of magnet field strengths, but the final choice can be adjusted depending on feasibility and science requirements. To maximize the acceptance, the large spectrometer dipole is located close to the FFQs, but leaving a few meters of space for detectors in between. The low gradients and large apertures also benefit the detection of neutral particles. With the 9 T peak field option, the ZDC covers a cone of 25 mrad with line-of-sight from the IP around zero degrees with respect to the ion beam.

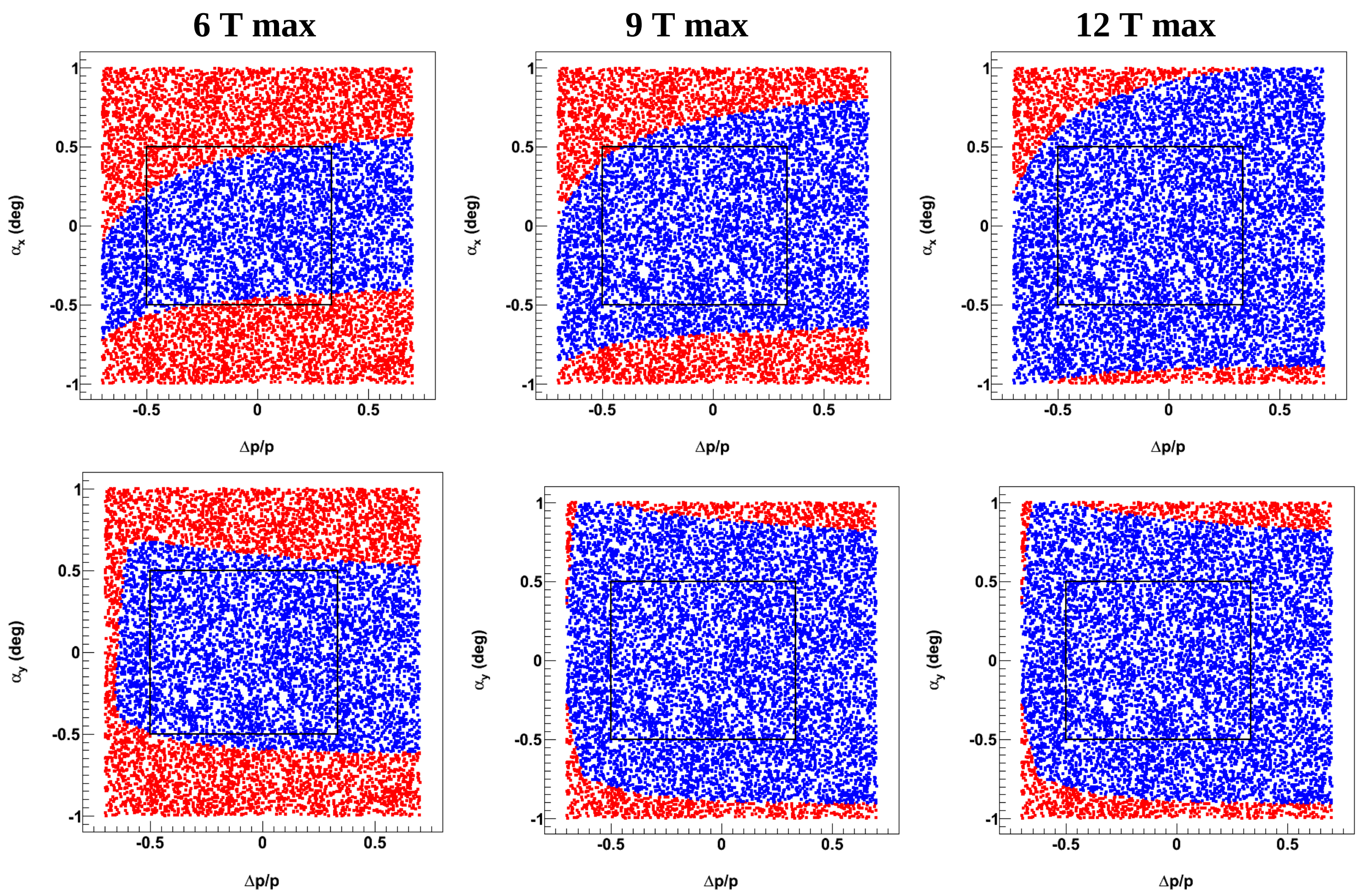

Figure 7.5: The forward acceptance of the MEIC detector for three choices of the peak field of the downstream ion FFQs in both the horizontal and vertical planes. Particles in red areas are detected prior to the FFQs. Particles in blue areas can be detected beyond the ion FFQs with improved resolutions.

The acceptance for particles with scattering angles and rigidities very close to that of the beam depends on the beam size (σ) and the dispersion at the detection point. The excellent performance of the MEIC is achieved by creating a focal point 16 m downstream of the 20 Tm dipole, where both the β function (related to the size of the beam) and the dispersion are about 1 m. As shown in Figure 7.6, if the detectors are placed at a distance of 10σ from the beam (marked beam-stay-clear in the figure) at a beam energy of 60 GeV, the acceptance for scattered particles covers momenta up to 99.5% of the beam energy for all angles, and down to 2–3 mrad for all momenta, or a combination of the two. Particles scattered at even smaller angles can be detected by supplementary detectors further downstream, but with degraded resolution. At higher energies, the σ of the ion beam may (or may not, depending on the cooling) somewhat become smaller, but due to trivial kinematics, the distribution of the scattered particles is always compressed much more rapidly as a function of ion energy. Thus, the best acceptance and resolution for particles with rigidities close to that of the beam are always reached at the lowest beam energies. Figure 7.6 also shows the detector resolutions at the focal point downstream of the 4 m long, 5 T Dipole 2 [6]. The resulting momentum resolution is limited only by the intrinsic beam momentum spread, and due to the long magnet-free drift space, the angular resolution is excellent.

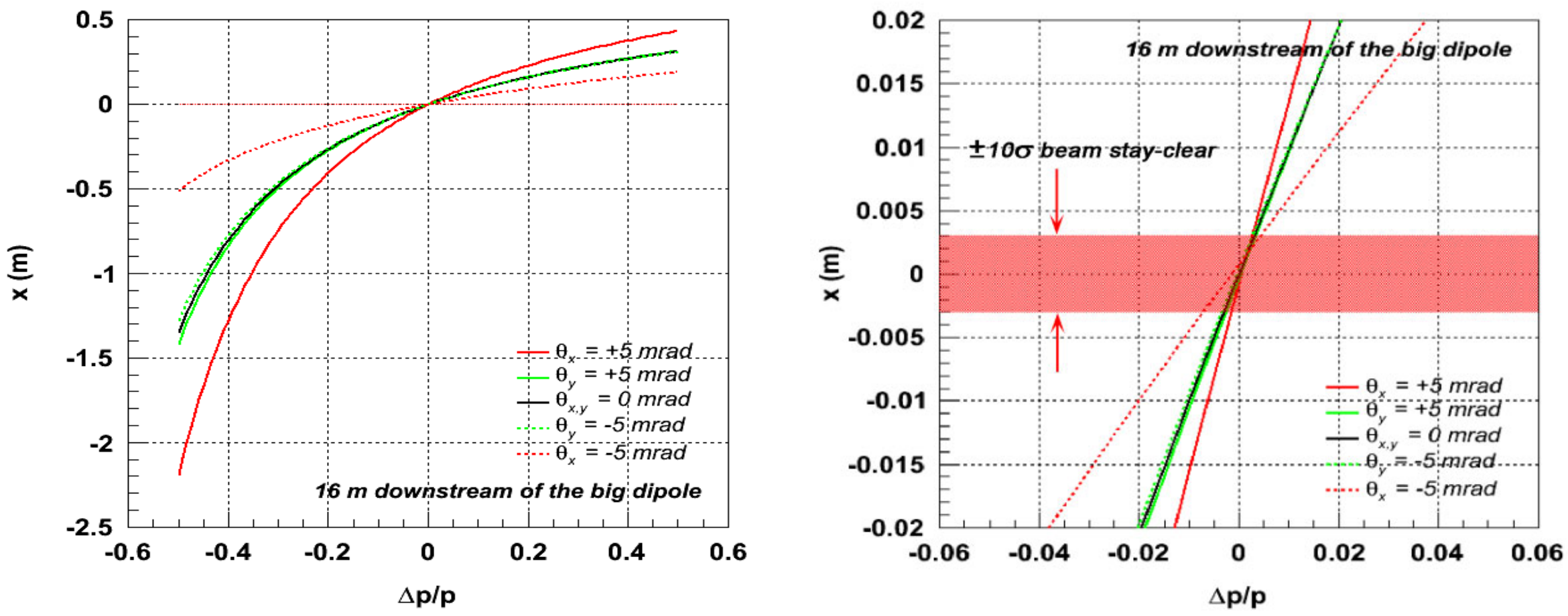

Figure 7.6: MEIC detector resolution at a focal point 16 m downstream of Dipole 2 for a characteristic set of initial angles. The right plot is an expanded version of the left figure. These plots show the ultra-forward detection capabilities of the MEIC, including for particles approaching the full ion beam rigidity.

On the outgoing electron side, a conceptually similar but much less demanding layout line allows tagging of low-$Q^2$ electrons and performing high-resolution momentum analysis. Initially designed to use large-aperture quadrupoles covering a degree, this angle may be reduced somewhat in the future in favor of stronger integration with the ion quadrupoles to simplify the arrangement. However, the less demanding requirements and more compact layout have already allowed moving the ion quadrupoles on the incoming side very close to the detector endcap. While this requires higher gradients (since the focal length is shorter), the peak fields remain moderate as the incoming beam does not demand large apertures. Combined with the asymmetric location of the IP within the central detector (a thickness of 2 m is assumed for each endcap), the IR naturally lends itself to an asymmetric optics design requiring large values of β only on the outgoing ion side. Thus, a fully-integrated detector/interaction region design has been established in the MEIC.

## 7.4 Linear Optics Design

In this section, we present the linear optics design of the IR. From the beam dynamics point-of-view, one key task in IR design is to compensate the large chromatic effect caused by the strong final focus while preserving an adequate dynamic aperture. For the case of MEIC, given the design parameter of ultra small $\beta^*$, the chromatic effect is an order of magnitude beyond state-of-the-art, especially for ion beams; therefore the task of chromatic compensation is particularly challenging such that it becomes the dominating issue in the optics design. On the other hand, the IR design must satisfy the detector requirements outlined in section 7.1, and particularly, it must support the design concept of a full-acceptance detector presented in the previous section. Earlier work on this subject can be found in [8]

In principle, the IR design must be optimized simultaneously for both design goals discussed above. However, since the chromaticity compensation scheme adapted for MEIC and presented in the next section is largely independent of a particular final focus design, the two design optimizations—namely, the final focus for best detector integration and the chromaticity

compensation— can be carried out in parallel, as long as a final integration is performed. This is the exact approach we have taken.

Below we will present a final focus optics that has been optimized for the full acceptance detector, which utilizes a novel asymmetric design. We will then describe the IR optics with chromaticity compensation. The second IR optics, which was built up around a conventional symmetric IR arrangement, serves as a middle step for demonstrating powerful and effective compensation of chromaticity by the adapted scheme. Both optics designs have been developed independently; integration of them should be straightforward, and will be optimized in the next MEIC design iteration. To illustrate the integration, as a proof-of-concept, we will show a linear optics of the full ion collider ring with a "drop-in" of the asymmetric IR associated with the full acceptance detector.

It should be noted that the vertical chicanes at the two long straights of the figure-8 rings through which the ion beams will travel to the plane of the electron collider ring are not included in the ion ring optics presented in this report. Design of such vertical chicanes is straightforward, with one additional design goal of complete suppression of the vertical dispersion in order to avoid its propagation around the ring and potentially causing various problems at the IPs. Such a design is well known and is used at different existing machines, and will be incorporated into the next version of the IR design.

***Detector-Optimized Optics***

Figure 7.7 (left) shows the detector-optimized ion IR optics design with the magnet parameters listed in Table 7.1. The detector solenoid is not taken into account in Figure 7.7; however, it is included in the detector region model used for the tracking studies presented in section 7.3. Note that the optics in Figure 7.7 (left) is asymmetric, with the upstream final focusing elements placed much closer to the IP than these on the downstream. This should be beneficial in reducing both the maximum betatron functions on the upstream side as shown in Figure 7.7 (left) and also the contribution to the chromaticity. The downstream final focusing block was designed such that the apertures of its quadrupoles become progressively larger (while remaining uniform within each quadrupole) with increasing distance from the IP to maximize its acceptance to the forward-scattered hadrons. The ion beam is focused downstream of the forward final focus. Having a small beam size at the focal point allows one to place the detectors closer to the beam center. In combination with ~1 m dispersion at that point, this allows detection of particles with small momentum offsets $\Delta p/p$.

Similarly, the optics of the detector-optimized IR for electrons is shown in Figure 7.7 (right). The magnet parameters are listed in Table 7.2. An important additional design feature of this IR is introduction of two permanent magnetic quadrupoles upstream of and very close to the IP to maximize the detector acceptance by reducing the solid angle blocked by the final focusing quadrupoles. Change of their focusing strengths with energy can be compensated by adjusting the upstream electric quadrupoles.

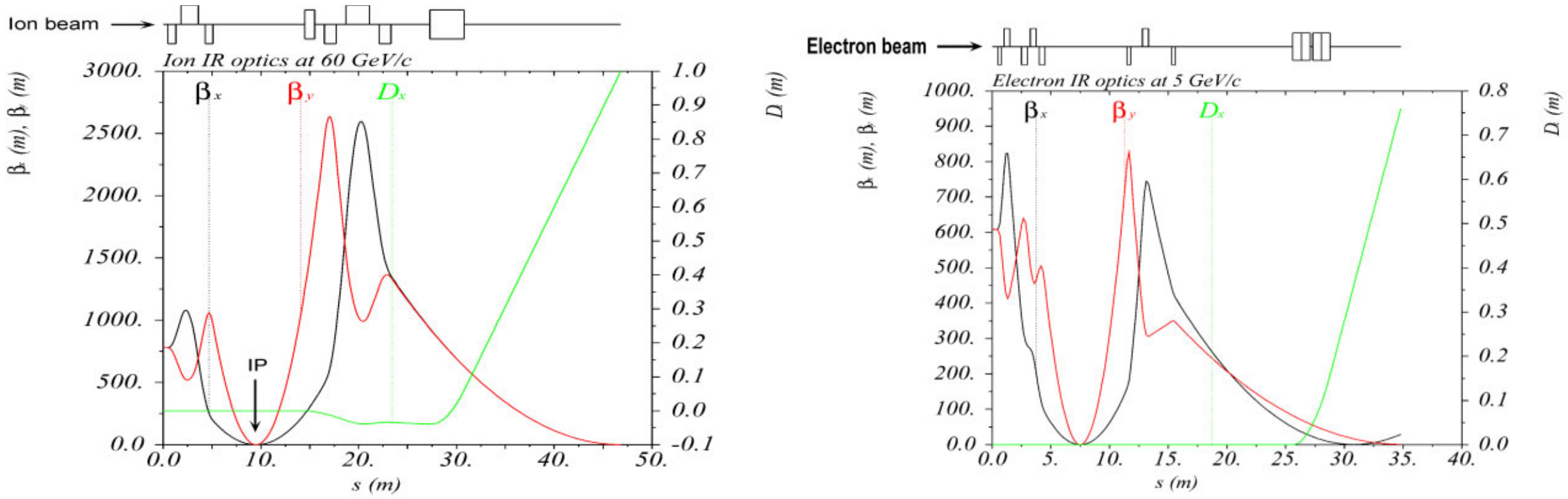

Figure 7.7: Optics of the MEIC interaction region for ions (left) and electrons (right).

Table 7.1: Ion detector region magnet parameters

| Magnet | $B_{\text{pole-tip}}$* | $\partial B_y/\partial x$* | Length | $R_{\text{inner}}$ | $L$ to IP** |
|---|---|---|---|---|---|
| | T* | T/m | m | cm | M |
| Upstream ion FFB | | | | | |
| Quad 1u | 5.4 | 180.5 | 0.8 | 3 | 4.4 |
| Quad 2u | 3.4 | 85.7 | 1.8 | 4 | 5.9 |
| Quad 3u | 2.5 | 63.0 | 0.8 | 4 | 8.2 |
| Downstream ion FFB | | | | | |
| Dipole 1 | 2 | N/A | 1 | 20 | 5 |
| Quad 1d | 9 | 89.0 | 1.2 | 10.1 | 7 |
| Quad 2d | 9 | 51.1 | 2.4 | 17.6 | 9.2 |
| Quad 3d | 7 | 35.7 | 1.2 | 19.6 | 12.6 |
| Dipole 2 | 6 | N/A | 3.5 | 30 | 17.8 |

* Maximum values corresponding to 100 GeV protons
** Distance from the IP to the magnet side facing the IP

Table 7.2: Electron detector region magnet parameters

| Magnet | $B_{pole\text{-}tip}$* | $\partial B_y/\partial x$* | Length | $R_{inner}$ | $L$ to IP** |
|---|---|---|---|---|---|
| | T | T/m | M | Cm | M |
| Upstream electron FFB | | | | | |
| Quad 1u (perm.) | 0.3 | 15 | 0.5 | 2 | 3 |
| Quad 2u (perm.) | 0.3 | 15 | 0.5 | 2 | 3.75 |
| Quad 3u | 0.7 | 34.0 | 0.5 | 2 | 4.5 |
| Quad 4u | 1.8 | 45.6 | 0.5 | 4 | 6 |
| Quad 5u | 1.5 | 38.0 | 0.3 | 4 | 6.7 |
| Downstream electron FFB | | | | | |
| Quad 1d | 6 | 67.1 | 0.3 | 9 | 4 |
| Quad 2d | 6 | 37.3 | 0.5 | 16 | 5.3 |
| Quad 3d | 3 | 10.2 | 0.3 | 30 | 7.8 |
| Dipole 1 | 1.2 | N/A | 1.5 | 20 | 18.1 |
| Dipole 2 | 1.2 | N/A | 1.5 | 20 | 19.8 |

* Maximum values corresponding to 11 GeV electrons
** Distance from the IP to the magnet side facing the IP

As was mentioned above, combining the detector-optimized optics with the chromaticity compensation scheme and integrating them into the basic collider ring lattices described in sections 4.3 and 5.7 is rather straightforward. As a proof-of-concept, the following two figures illustrate how this can be done for the ion collider ring. Figure 7.8 shows a layout of the ion collider with the final focusing blocks "dropped" in, while Figure 7.9 presents the complete optics of the whole collider ring. Compared to the symmetric design below, the horizontal and vertical natural chromaticities of this design are significantly less, reduced from -278 and -268 to -207 and -191, respectively. Figures 7.8 and 7.9 assume two identical interaction regions shown in Figure 7.7 (left), but in the future we plan to optimize each interaction region individually for specific experimental needs. Such optimization will be done in the next MEIC design iteration.

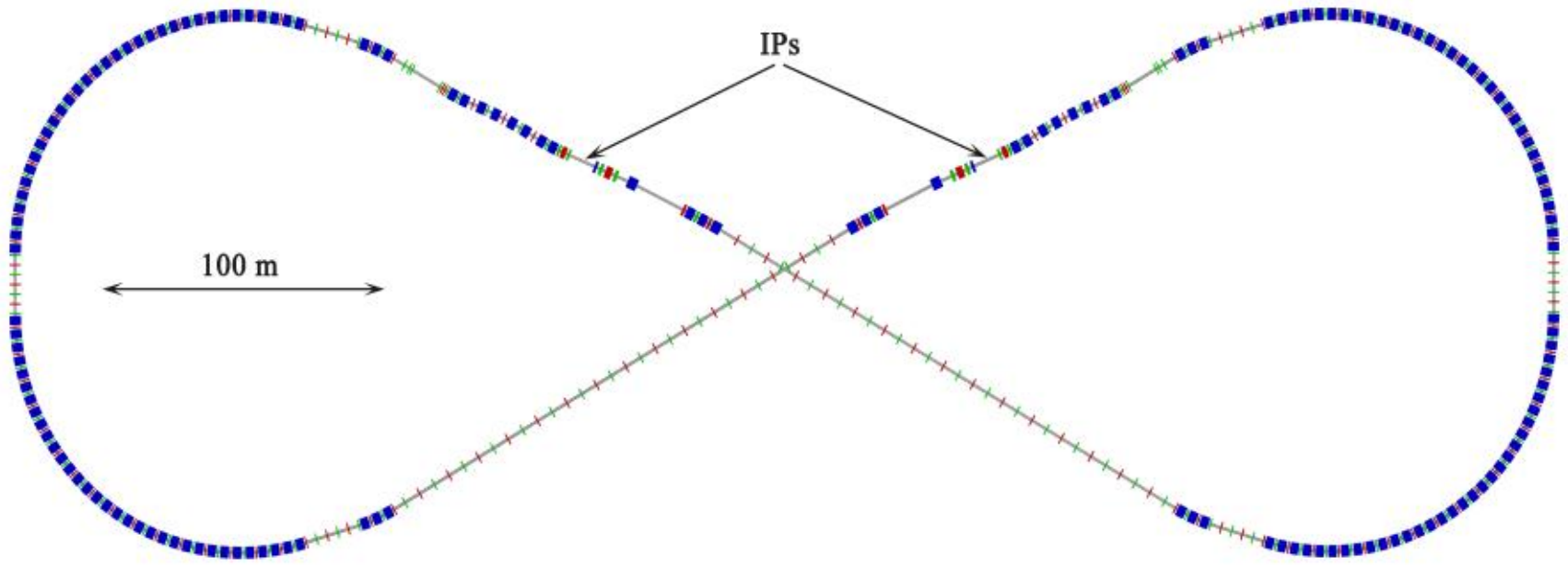

Figure 7.8: Layout of the MEIC ion collider ring incorporating two detector-optimized IRs.

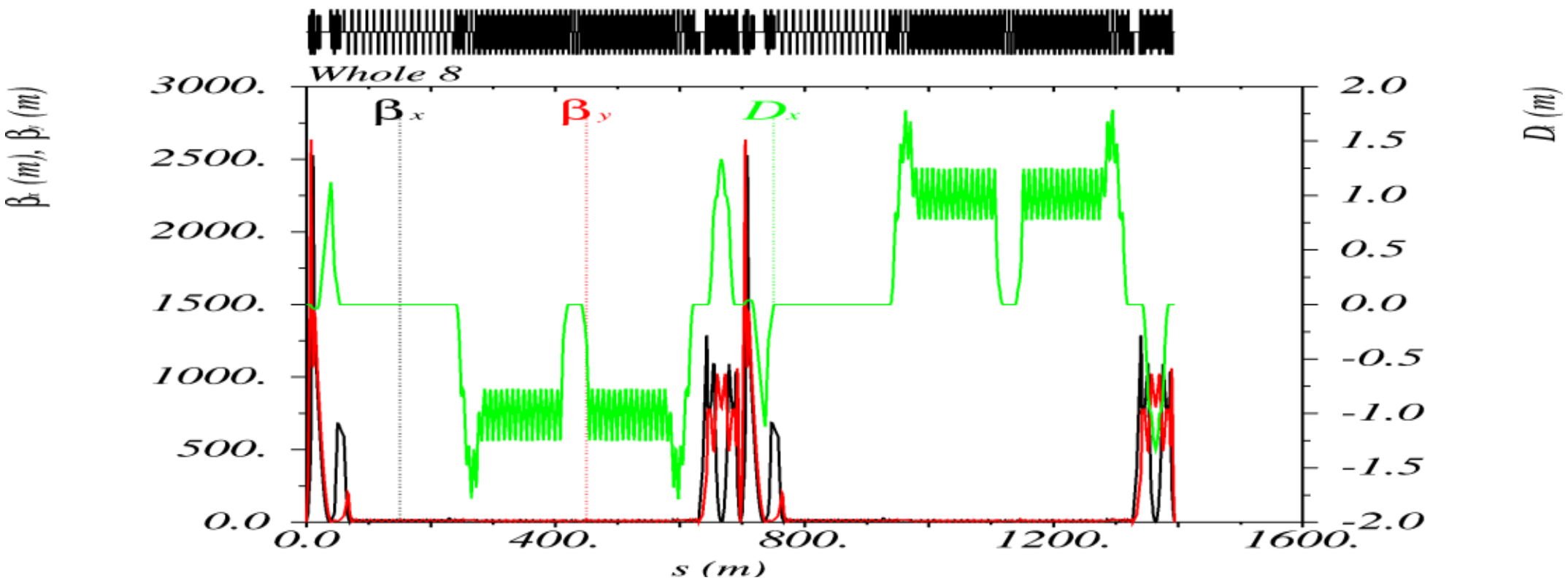

Figure 7.9: Optics of the ion collider ring incorporating two detector-optimized IRs.

### *Optics with Chromaticity Compensation*

We next present an IR linear optics optimized for chromaticity compensation. The IR is arranged symmetrically with a detector space of ±7 m for ions and ±3.5 m for electrons, resulting in a larger chromaticity compared to the asymmetric IR discussed above.

The MEIC IR consists of three types of function blocks [9] as illustrated in Figure 7.10: a final focusing block (FFB), a chromaticity compensation block (CCB) and a beam extension section (BES), distributed from the IP toward the arc. If the IR is designed in mirror symmetry about the IP, then there are three identical blocks on the other side of the IP. It is obvious that a strong focusing at an IP is unavoidably accompanied by a large beam extension [10]. The size of the required beam extension is determined (inverse linearly) by the focal length of the FFB, which approximately equals the dimension of the magnet-free detector space. The CCB is required for a local compensation of the large chromatic effect caused by the strong final focusing at the IP. The chromatic correction is the subject of the next section.

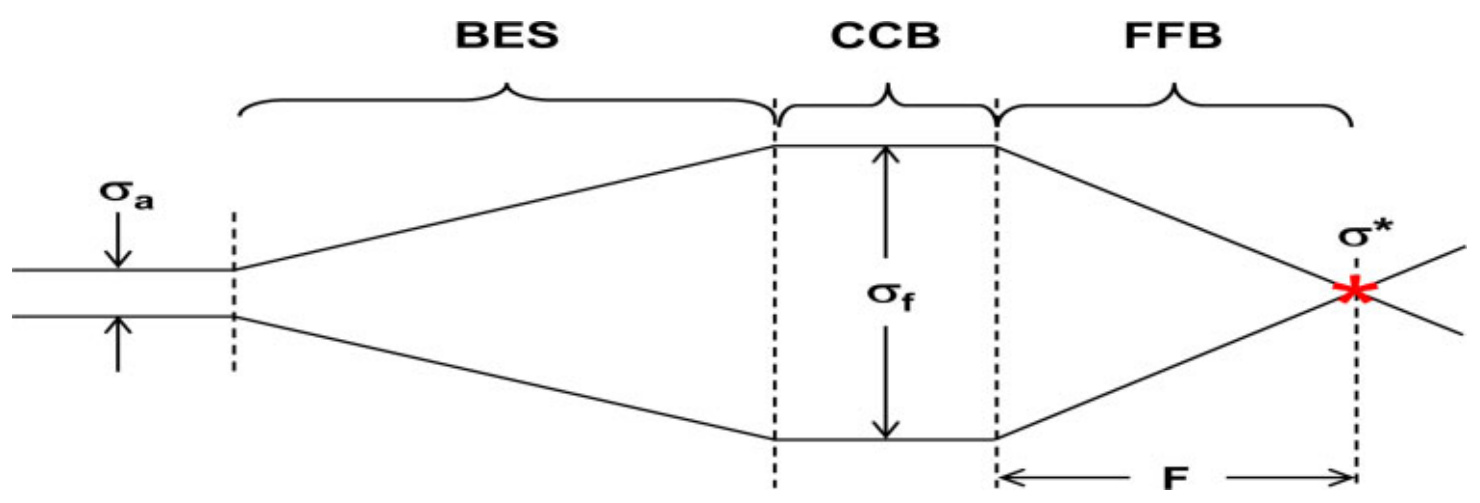

Figure 7.10: A layout of an interaction region.

The horizontal and vertical beta functions at the IP ($\beta_x$* and $\beta_y$*) are 10 and 2 cm. A quadrupole doublet was used in this study for the electron FFB. The BES for the electron beam is made of six quadrupoles for matching optics of the CCB to optics either at the end of an arc or the normal lattice of the long straight of the figure-8, thus requiring a substantial longitudinal space. The optical functions for the FFB, CCB and BES are shown in Figure 7.11 using MAD-X [11]. The quadrupole parameters are adjusted so that the beta functions have roughly equal values at the exit from the FFB and the alpha functions are zero in the middle of the CCB. The complete electron optics in the IR is shown in Figure 7.12.

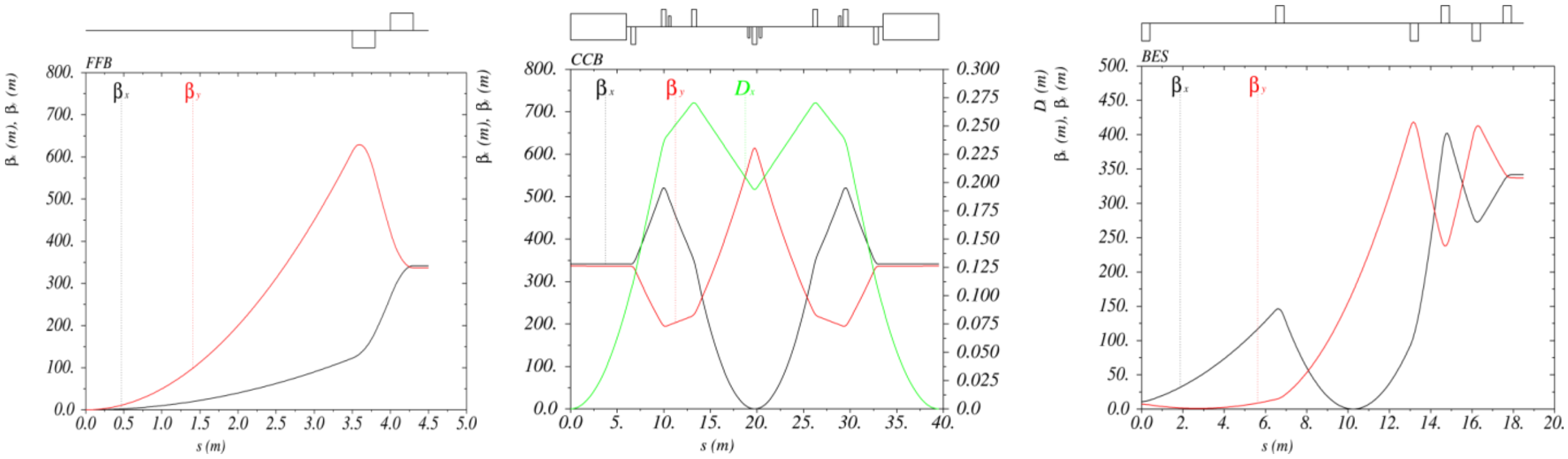

Figure 7.11: Optical function of the electron beam at final focusing block (left), chromaticity compensation block (middle) and beam extension section (right).

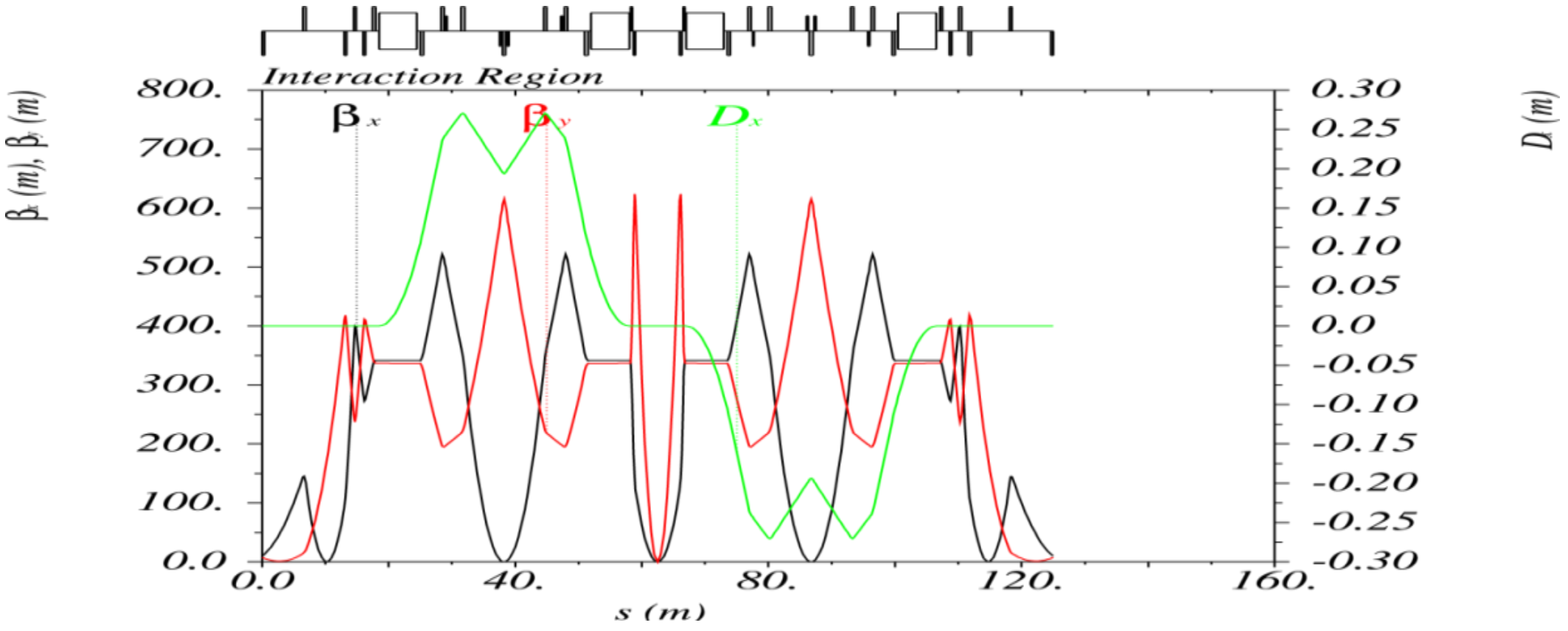

Figure 7.12: Complete optics of the electron interaction region.

A similar design has also been followed for ion beam linear optics. Figure 7.13 and 7.14 show the optical functions for the ion CCB, FFB and BES individually, and also over the complete IR, respectively. The linear optics design of the ion IR is very similar to the electron IR; nevertheless, the magnet-free distance space is 7 m, double that for the electron beam. Since the beta functions grow roughly as a square of that distance, the maximum beta functions are a factor of four greater than those of electrons.

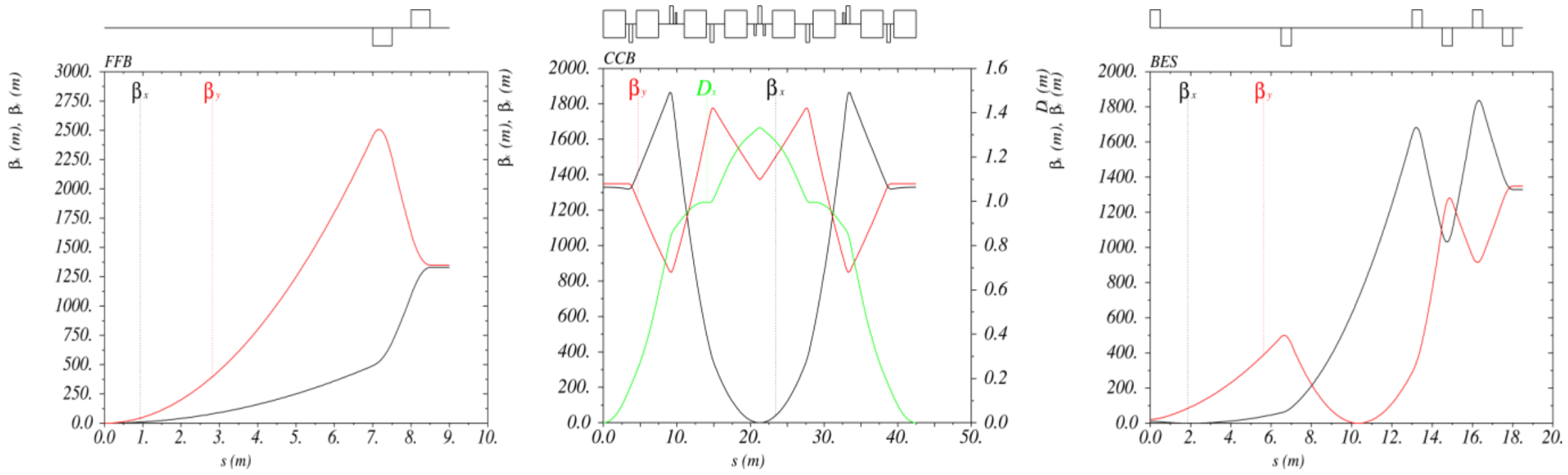

Figure 7.13: Optical functions of the ion beam at final focusing block (left), chromaticity compensation block (middle) and beam extension section (right).

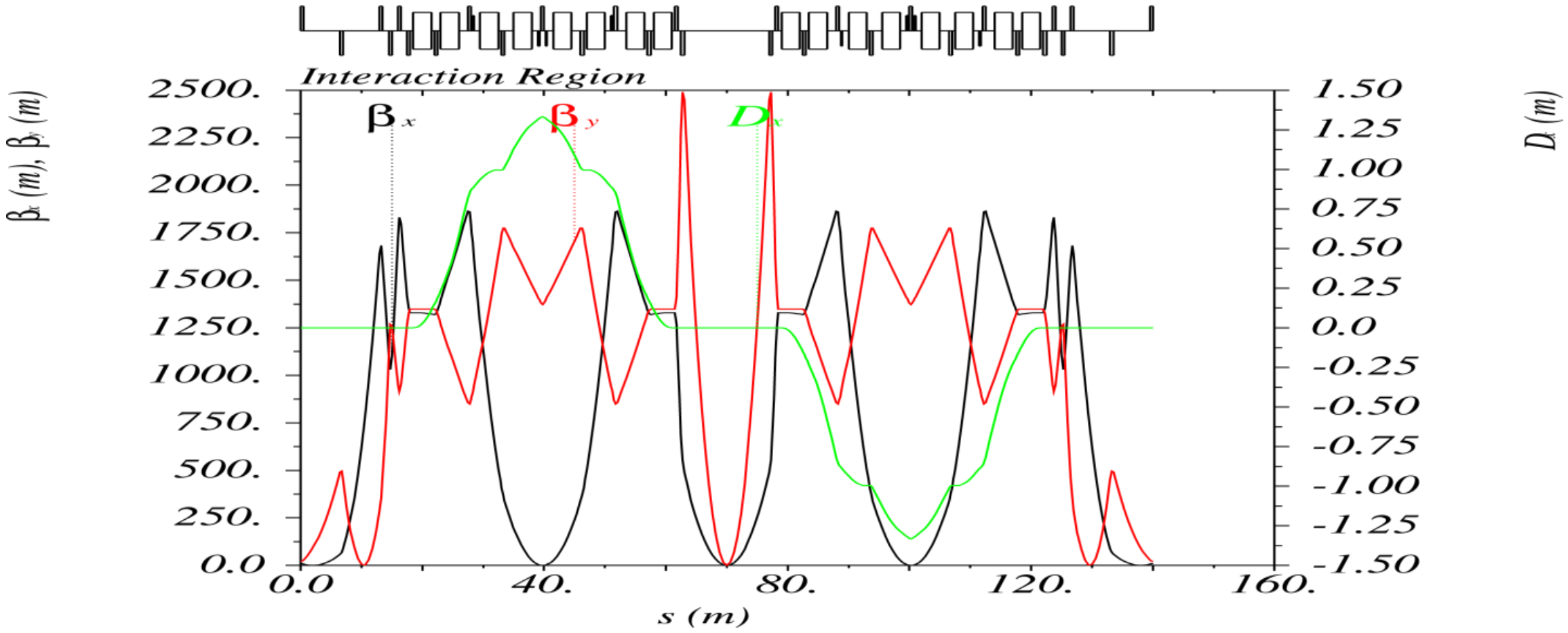

Figure 7.14: Complete optics of the ion interaction region.

It should be noted that the ion beam emittance after injection is about ten times larger than the design emittance before cooling. If the ion collider ring has the collider-mode optics at injection, the ion beam size becomes unacceptably large at FFQs, causing serious particle loss. Therefore, a scheme of varying the maximum beta functions in the IR is needed in order to first reduce the beta maximum to a proper value at injection and during acceleration, then restore it to the full value for the collider operation. The present IR design allows a straightforward implementation of such a “beta squeeze” scheme. Since the dispersion is suppressed at entering and exiting the CCB and the slopes of the beta functions are almost zero, the CCB’s optics is compatible with any initial beta function values without requiring any adjustment as long as the initial alpha functions are close to zero. Thus, the beta functions inside the IR can be controlled by adjusting their values at the end of the BES. This requires only changing the optics of the BES with the rest of the ring’s optics intact.

## 7.4 Chromaticity Compensation

The effective focal length of an FFB depends on the particle momentum [10]. The problem with such a correlation is two-fold. Firstly, it induces a large chromatic betatron tune spread, thus limiting the momentum aperture of the collider ring. Secondly, it causes chromatic beam smear at the IP, which can significantly increase the beam spot size, resulting in severe luminosity loss. Quadrupoles in the ring also contribute to the chromatic effect; however, for the MEIC design, these contributions are much smaller compared to that of the low beta insertion of the interaction region.

The standard approach for chromatic correction is utilizing families of sextupoles in large dispersion regions either in the arc (as a global correction scheme [12]) or near the collision point (as a local correction scheme [13]). The MEIC baseline has adopted a local correction scheme as its primary approach [9] and placing sextupoles in a specially designated area, i.e., chromaticity compensation block, as shown in Figure 7.15, near an IP and one on each side. Correction of the natural chromaticity is shown in the following equation:

$$C_{x,z} = -\frac{1}{2\pi}\oint ds\, \beta_{x,z}(s)\left[K_{x,z}(s) \pm S(s)D(s)\right] \tag{7.1}$$

where $K_{x,z}$ and $S$ are strengths of the quadrupoles and sextupoles and $D$ is the horizontal dispersion. However, the nonlinear sextupole fields generate 2$^{nd}$ and higher-order aberrations which may require higher magnetic moments to correct both the non-linear phase advance and the reduction of dynamic aperture [14].

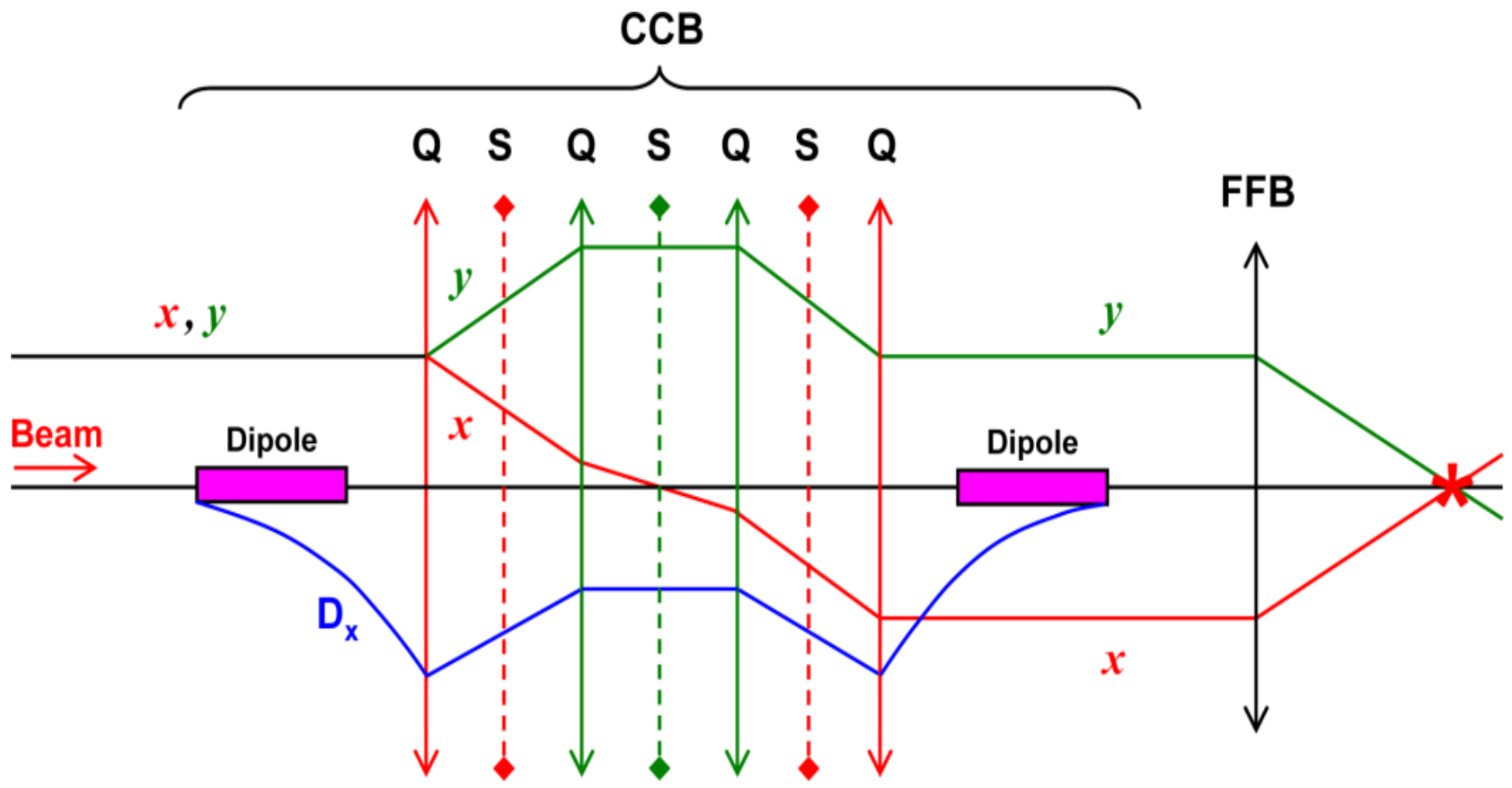

Figure 7.15: A conceptual drawing of a chromaticity compensation block with an odd-symmetry horizontal betatron trajectory and an even-symmetry vertical betatron trajectory and dispersion.

In the current optics design, the interaction point is dispersion free for both collider rings; thus, in order to correct the natural chromaticity, horizontal dispersion is generated locally inside the CCB by introducing a pair of horizontal bending dipoles as shown in Figure 7.15.

Another important feature of the CCB design is symmetric or anti-symmetric beam orbital motion and dispersion function with respect to the center of the CCB, also shown in Figure 7.15, achieved by a symmetric arrangement of dipoles and quadrupoles in the CCB. It can be shown [9] that such arrangement allows simultaneous compensation of the 1$^{st}$ order chromaticity and chromatic beam smear at the collision point without inducing significant 2nd order aberrations and therefore largely preserving the ring's dynamic aperture.

The two identical bends in Figure 7.15 are responsible for generating and then suppressing the dispersion in the CCB. In the electron ring, the maximum bending fields and therefore the maximum dispersion are limited by the emittance degradation impact of the bending magnets. In the ion ring, on the other hand, it is advantageous to have strong bends to produce large dispersion, which reduces the required sextupole fields and thus minimizes their nonlinear effects. These contradictory bend requirements complicate geometric matching of the electron and ion IRs. The solution for this problem involves use of alternating bends in the ion CCB. Further, the electron and ion CCBs located on the opposite sides of the IP have their bends reversed. The trajectories of the two beams [9] shown in Figure 7.16 are adjusted to produce a 50 mrad crab crossing angle at the IP. The optical functions of the electron and ion CCB are shown in Figures 7.11 and 7.13, respectively.

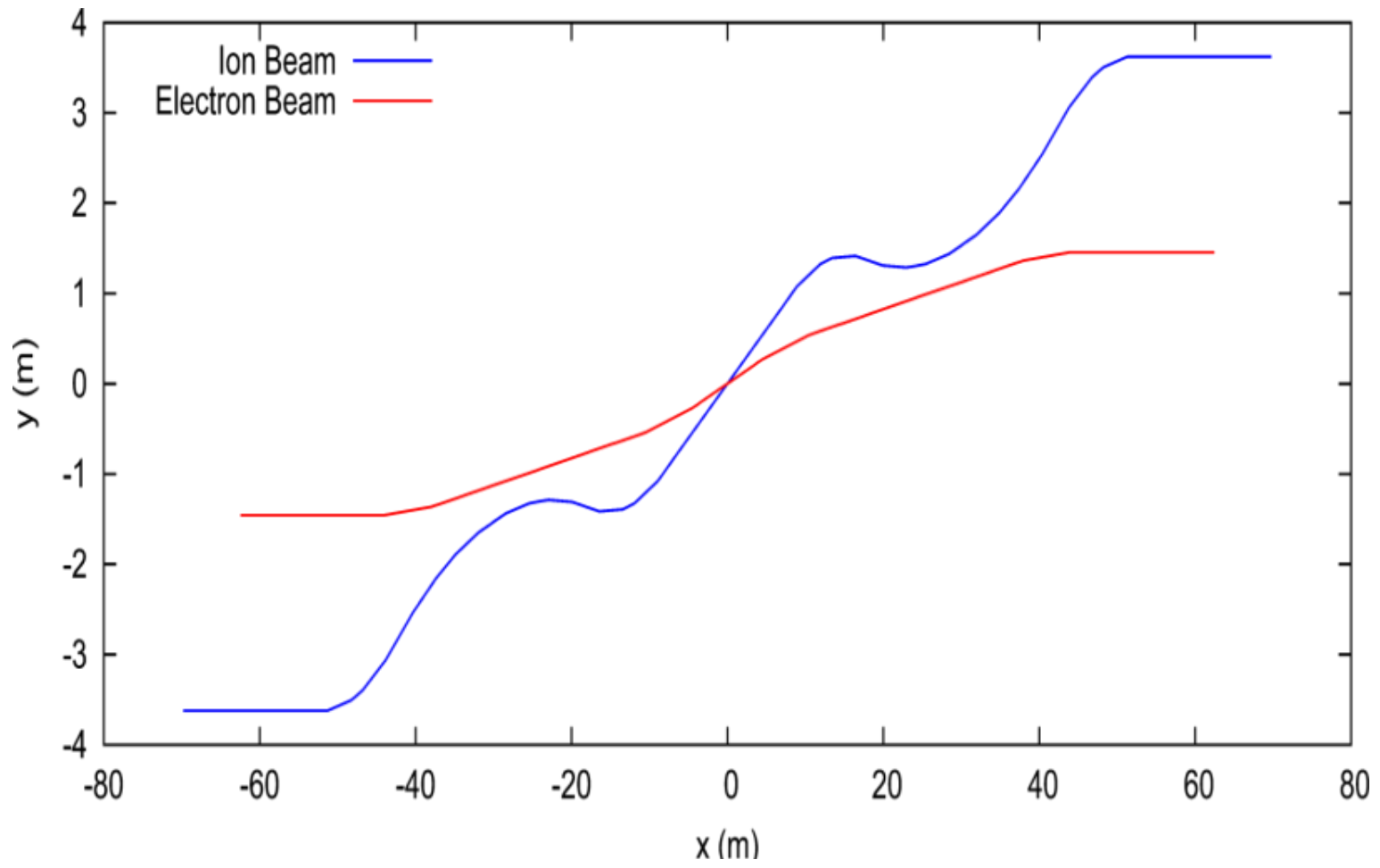

Figure 7.16: Electron and ion beam trajectories in an MEIC interaction region.

In accordance with this compensation scheme [9], two pairs of sextupoles are inserted in one CCB, and their positions, shown by the shorter bars in Figures 7.11 and 7.13, are placed at the points where the dispersion is near maximum and there is a large difference between the horizontal and vertical beta functions. Strengths of the sextupoles are adjusted to cancel the horizontal and vertical linear chromaticities [11]. The chromatic tune dependence before and after the sextupole compensation is shown for the electron and ion collider rings in Figure 7.17.

For the MEIC electron ring, after the sextupole compensation, change of the betatron tunes is less than 0.02 within a momentum range of $\pm 1.5\times 10^{-3}$, corresponding to roughly ±2 RMS momentum spreads. For the ion collider ring, the momentum acceptance is about ±10 and ±7 times of momentum spread for the same 0.02 horizontal and vertical tune shifts respectively.

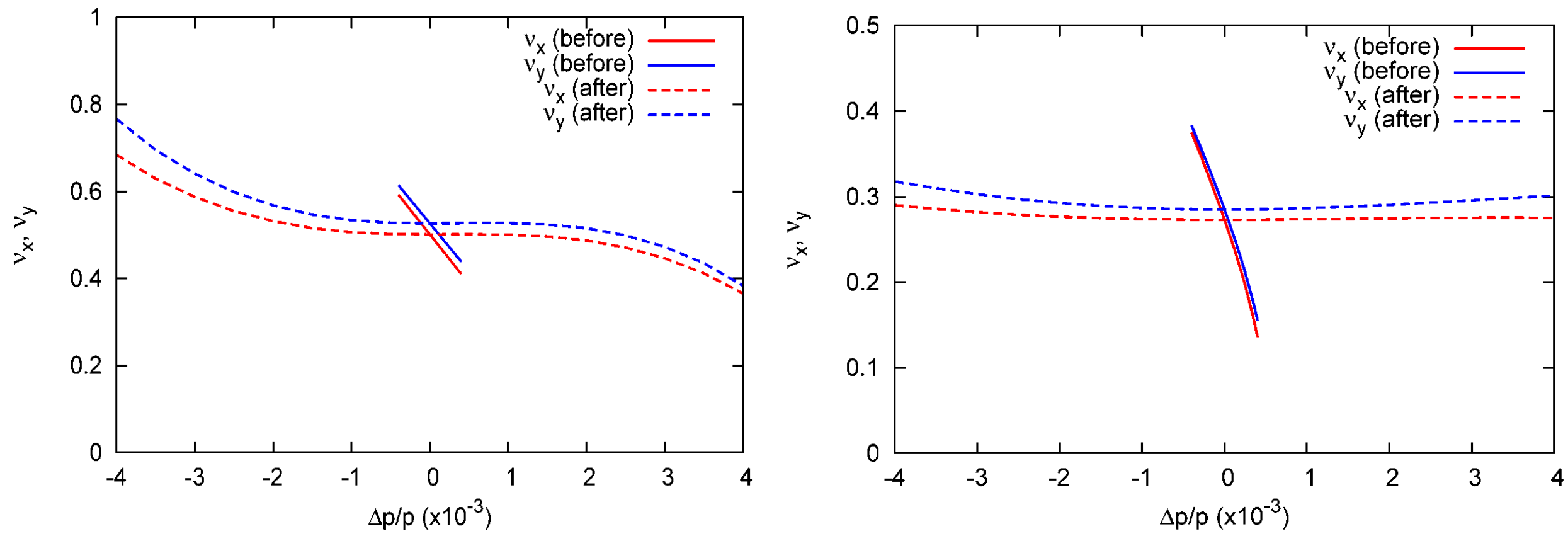

Figure 7.17: Chromatic dependence of the fractional betatron tunes of the electron collider ring (left) and ion collider ring (right) before and after the sextupole compensation

## 7.5 Momentum Acceptance and Dynamic Aperture

Optimizing the momentum acceptance and dynamic aperture is more challenging for the ion collider ring than for the electron one due to its larger natural chromaticities and maximum beam size. Therefore, below we focus first on evaluation and then on some very preliminary nonlinear optimization of the ion collider ring. The ring lattice used for tracking includes two interaction regions shown in Figure 7.14 incorporated into the two straight sections of the ring's basic lattice described in section 5.7. Components such as detector solenoids, spin control devices, vertical chicanes, injection/extraction, etc. were not included in the lattice. The beam-beam effect was not taken into account. The ring was simulated [15] using *Elegant* [16] particle tracking code. All ring components were modeled as canonical kick elements with exact Hamiltonians retaining all orders in momentum offset. The magnet fields were approximated as hard-edge and the lattice was assumed perfect, i.e. containing no alignment or field errors. Addressing the question of error tolerances is going to be our next step.

The momentum acceptance can be demonstrated by a frequency map [17] from particle tracking simulations. At the 60 GeV design energy of the MEIC ion ring, the maximum horizontal RMS beam size is 3.2 mm due to the large 7 m detector space, and the vertical rms beam size is 1.6 mm, which makes it challenging to obtain a large horizontal dynamic aperture. Therefore, the frequency map is computed in the ($x$-$\Delta p/p$) space from tracking particles for 2000 turns using *Elegant* and plotted in Figure 7.18. The particles in this simulation were launched parallel to the beam axis at a point at the entrance into one of the final focusing blocks where the horizontal RMS beam size is about 2.7 mm. The color reflects the tune change in terms of the tune diffusion defined as $d=log(\Delta \nu_x^2+\Delta \nu_y^2)$ where $\Delta \nu_{x,y}$ is the tune change from the first to second half of the simulation in the horizontal and vertical planes, respectively. The diffusion index is an excellent criterion for long term stability [18] and enables us to see the nonlinear behavior. The more negative the diffusion, the smaller the nonlinear effects are.

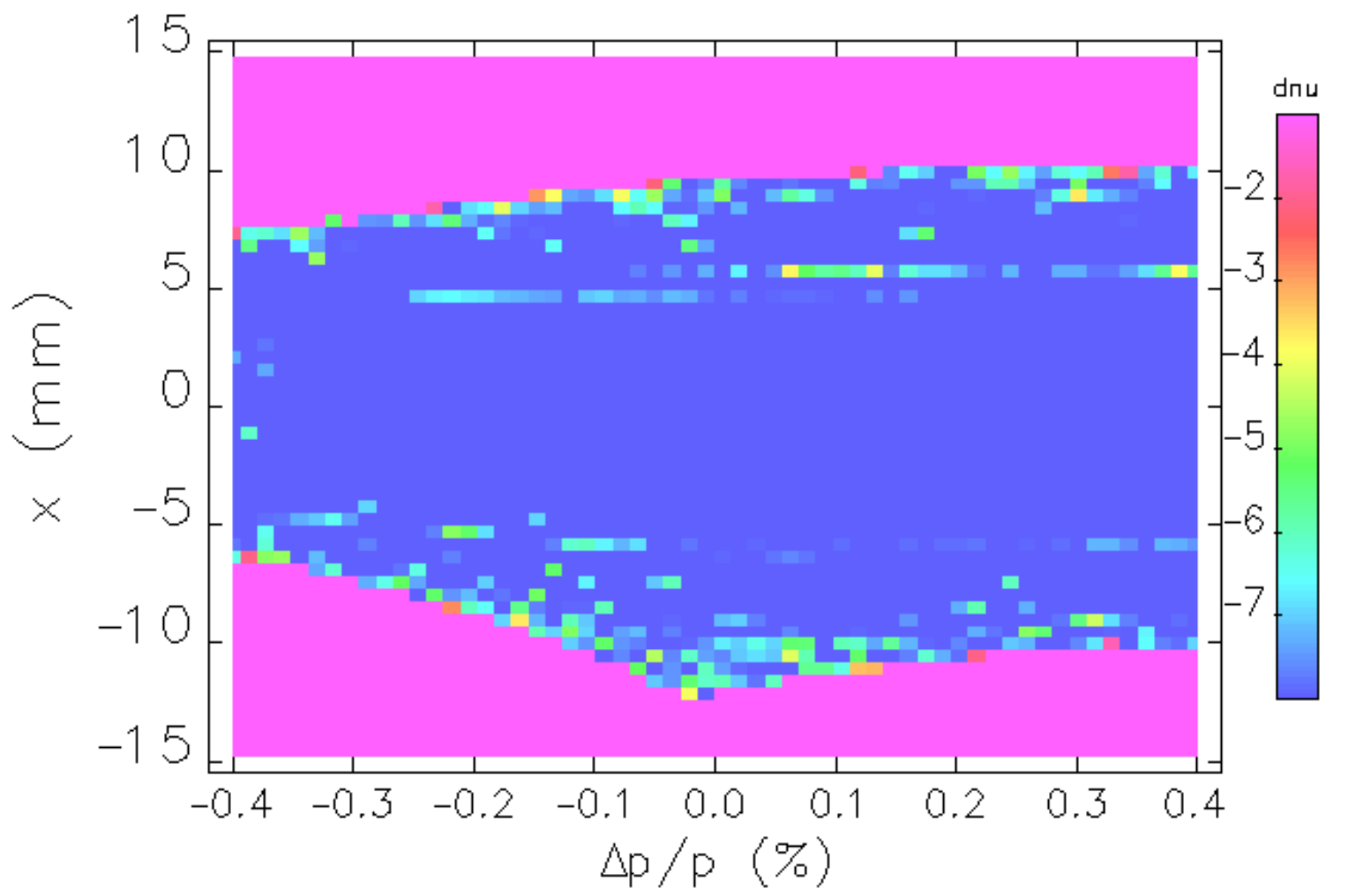

Figure 7.18: Frequency map in the ($x$-Δ$p/p$) space. The color reflects the tune change in terms of the tune diffusion defined as $d=\log(\Delta\nu_x^2+\Delta\nu_y^2)$.

Two conclusions can be drawn immediately from the frequency map: 1) the momentum acceptance can easily reach $\Delta p/p$ of ±0.4% with only linear chromaticity compensation, 2) the uniform distribution of color means that there are no strong resonant perturbations in the particle tracking. The simulation was terminated at $\Delta p/p$=±0.4%, which is about 14 times $\sigma_\delta$, the RMS momentum spread of about 0.03%, expected after cooling, and is conventionally considered to be sufficient to demonstrate an adequate momentum acceptance. Thus, the present chromaticity compensation scheme has resulted in an excellent momentum acceptance, without needing further compensation of the 2$^{nd}$ order chromaticities.

Figure 7.18 also indicates that the horizontal dynamic aperture size is about ±10 mm, which is reasonable given the large compensated values of the natural chromaticities and the fact that there was no nonlinear optimization after the linear chromaticity compensation. However, due to the large beam extension required to achieve the ambitiously small β* values, clearly, further optimization using more sextupole and even octupole families is required.

Figure 7.19 is the tune footprint derived from the same particle tracking simulations for the frequency map in Figure 7.18. The color for the tune footprint reflects the tune change in terms of the diffusion index. The lines stand for the betatron resonances up to 3$^{rd}$ order; higher-order resonances are not shown in this plot. Since the frequency map is calculated for particles with initial coordinates in ($x$-$\Delta p/p$) space, the vertical tune variation arises only from the chromatic tune dependence, which is around 0.03 within Δ$p/p$ of ±0.4% as shown in Figure 7.18. The horizontal tune variation is about 0.14, which is significantly larger than the chromatic tune change of 0.02 corresponding to Δ$p/p$ of ±0.4% in Figure 7.18. This indicates that, after the chromaticity compensation, the horizontal betatron motion produces a substantial tune change, which can drive the particles close to or into a resonance mode and cause particle loss. The tune footprint can instantly reveal the movement of the particle tunes, which allows us to understand the tune trend due to particle motion and/or adjust the design tunes to avoid approaching or crossing some strong resonances.

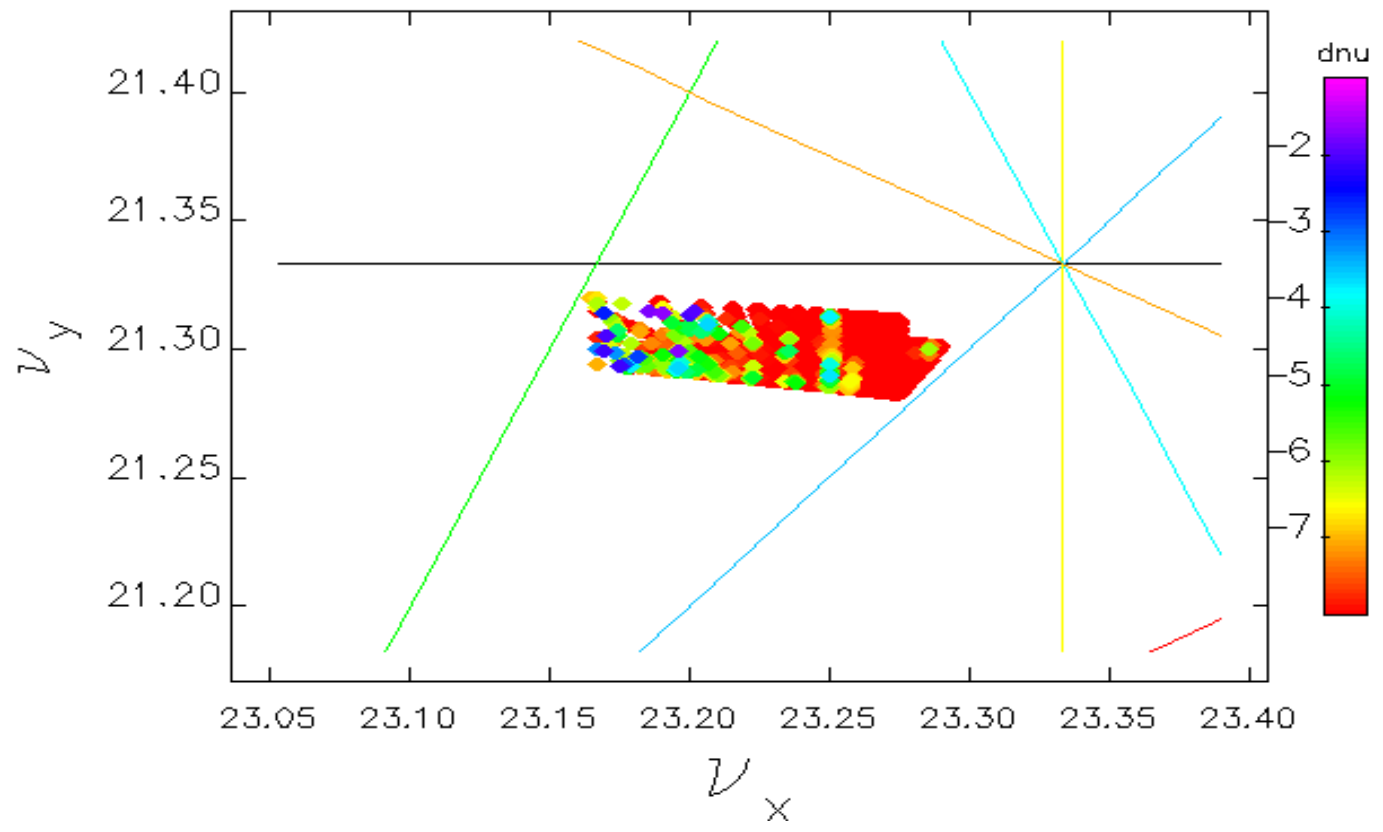

Figure 7.19: The MEIC ion collider ring tune footprint derived from a tracking simulation. The color reflects the tune change in terms of the tune diffusion defined as $d=\log(\Delta\nu_x^2+\Delta\nu_y^2)$.

The dynamic aperture in the transverse directions of the ion ring has been explored by tracking particles for 1000 turns with increasing initial horizontal ($x$) and vertical ($y$) amplitudes until the boundary between survival and loss is found. Such a bound is illustrated by a red curve in Figure 7.20. It can be seen that the dynamic aperture for the on-momentum particle at the entrance into the final focusing block can reach about 10 mm ($4\sigma_x$) horizontally and 18 mm ($\sim 15\sigma_y$) vertically. Simulations also show that the particles having large initial amplitudes experience stronger nonlinear fields in the sextupoles, resulting in the third-order aberration, namely, amplitude dependent tune shift.

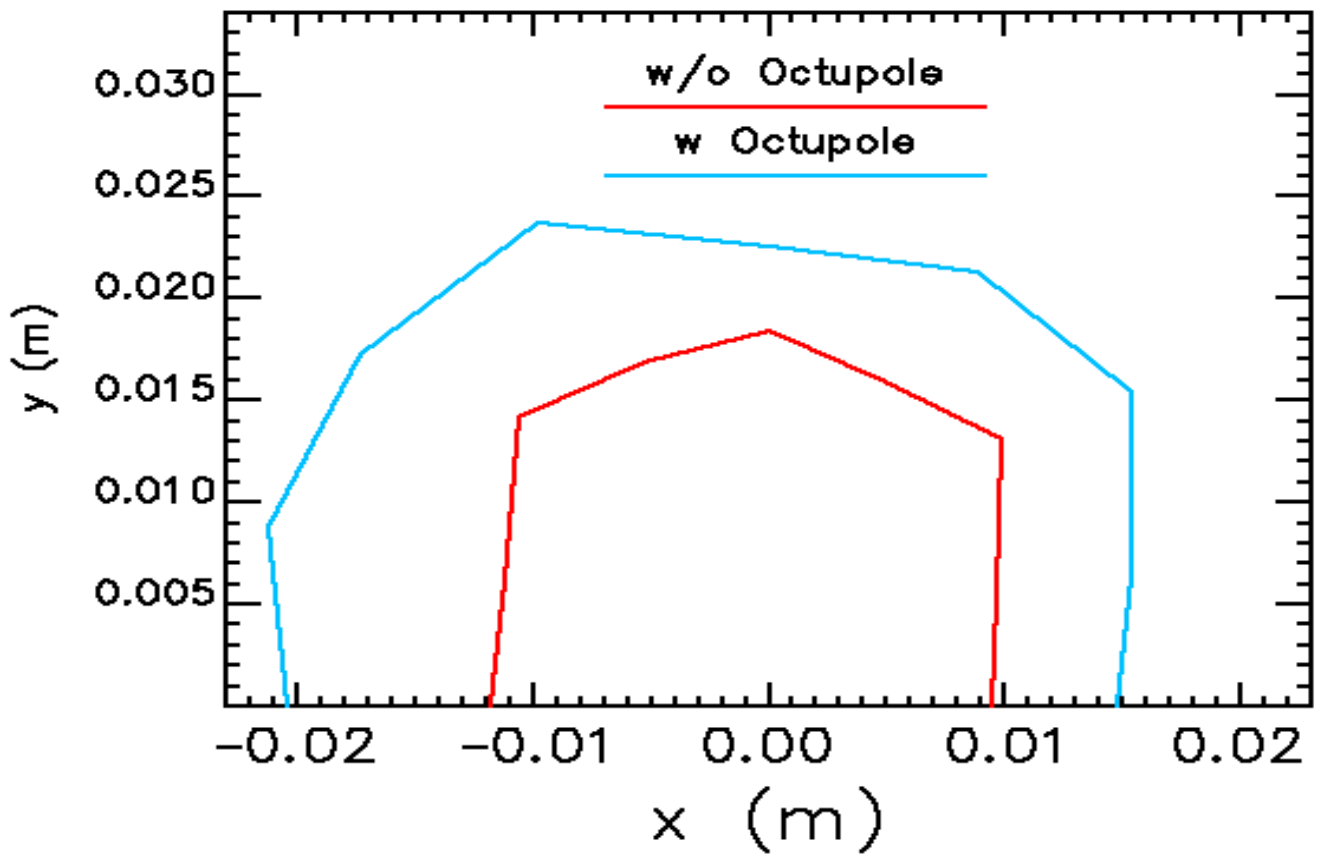

Figure 7.20: Dynamic aperture in the ($y$-$x$) space for the MEIC ion ring without (red) and with (blue) octupole minimization of the 1st order amplitude dependent tune shift.

As seen from the upper two plots in Figure 7.21, the horizontal and vertical tunes change from the design values by 0.08 (red line in the left plot) and 0.04 (blue line in the right plot), respectively, within the amplitude range of about ±10 mm. A straightforward approach to compensate the amplitude-dependent tune shift caused by the sextupoles is to introduce octupoles. Since the momentum acceptance is large enough after the chromatic correction using the CCB's, families of octupoles are placed in dispersion-free regions, leaving the chromatic correction unaffected. Besides, to reduce the octupole strengths, they should be placed at large beta-function points. After the compensation of the 1st order amplitude dependent tune shifts $d\nu_{x,y}/dJ_{x,y}$, the horizontal and vertical tune changes are now 0.04 and 0.03, respectively, within the amplitude range of over ±15 mm, as given in the lower two plots in Figure 7.21, and the dynamic aperture is increased as shown by the blue line in Figure 7.20, compared to the red line without the compensation of the 1st order amplitude-dependent tune shift.

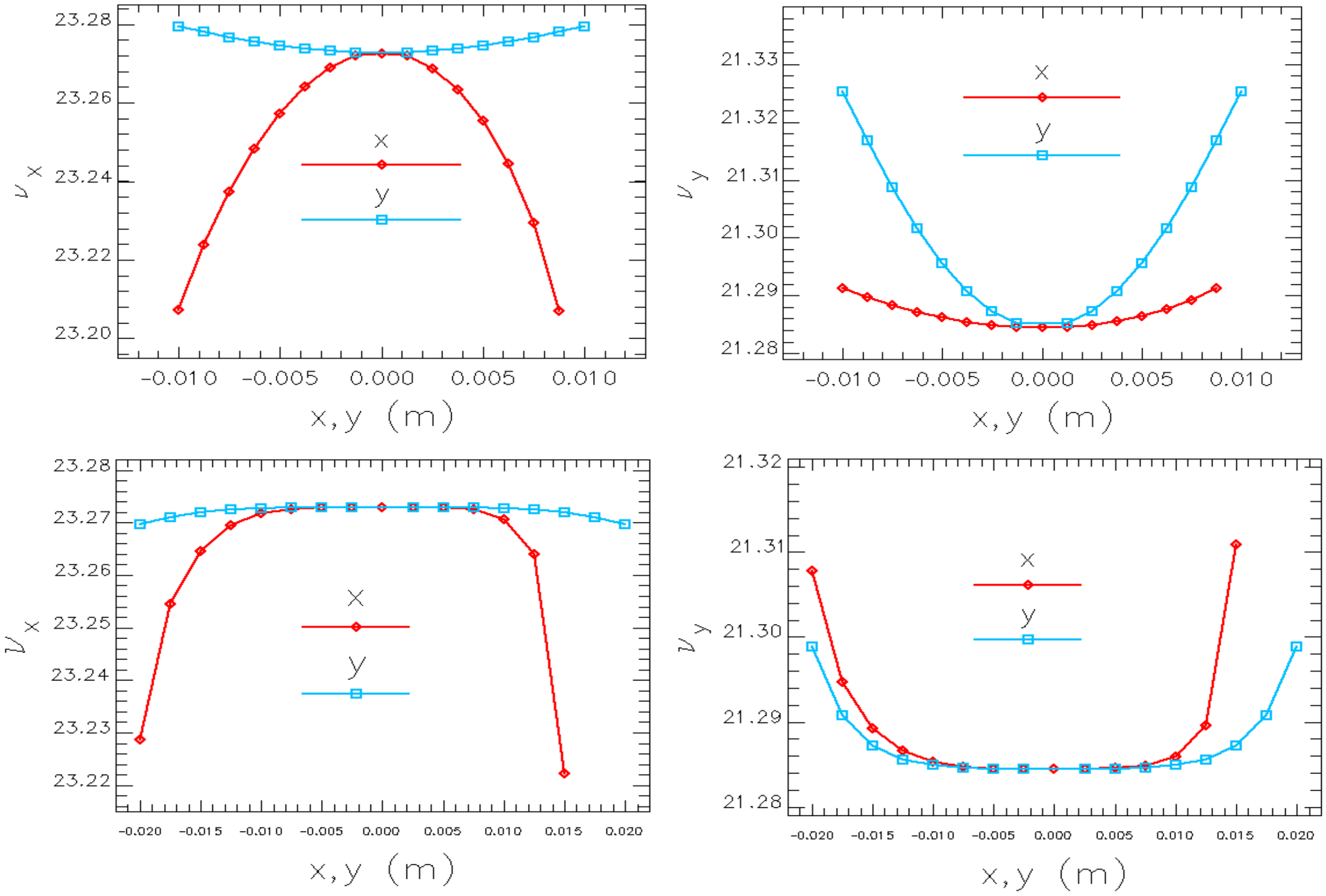

Figure 7.21: The horizontal (left) and vertical (right) tune variation as a function of initial transverse amplitudes *x* and *y* before (upper) and after (lower) minimizing the 1st order amplitude dependent tune shift.

The limitation for increasing the dynamic aperture comes from the effect of the 2nd order amplitude-dependent tune shift, which dominates in the tune behavior after squeezing the 1st order too much. This provides a guideline that the dynamic aperture should be further optimized by considering the tune shift from the 1st, 2nd and even higher order amplitude-dependent effects. These effects can be compensated using multiple sextupole and octupole families [18,19]. The initial nonlinear optimization results presented above look rather encouraging and make us confident that the design parameters can be achieved. We plan to use a systematic approach as much as possible in combination with advanced numerical techniques such as a genetic optimization algorithm. We will also specify a quantitative criterion for the dynamic aperture to provide the necessary luminosity lifetime (e.g., considering the balance of IBS, multiple and single scattering (Touschek) and electron cooling) and determine required parameter tolerances.

Neither the linear nor the nonlinear lattice design of a collider ring is complete without analysis of tolerances to field errors and element misalignments. Three critical aspects of this issue include control of field errors and magnet misalignment in the IR, control and correction of the reference orbit position in the CCB and control and correction of collisions (positions of the focal points of two beams, especially in the transverse plane). All of these aspects will be investigated in simulations based on particle tracking as the next task. Accordingly, an orbit correction system will be developed and the technical and engineering specifications of the magnets will be provided.

## 7.6 Crab Crossing

The MEIC IR design has adapted nonzero crossing angle for a rapid separation of colliding beams near the IP in order to eliminate numerous harmful parasitic collisions under an ultrahigh (750 MHz) bunch repetition frequency. To prevent loss of luminosity, a scheme of using transverse crab kicking to tilt the bunch has also been adopted in the baseline design for restoring the head-on collision condition [4,5]. This scheme is illustrated in Figure 7.22. The $e^+$-$e^-$ collider in the KEK B-Factory was first to successfully implement this scheme for achieving the world's highest collider luminosity [20].

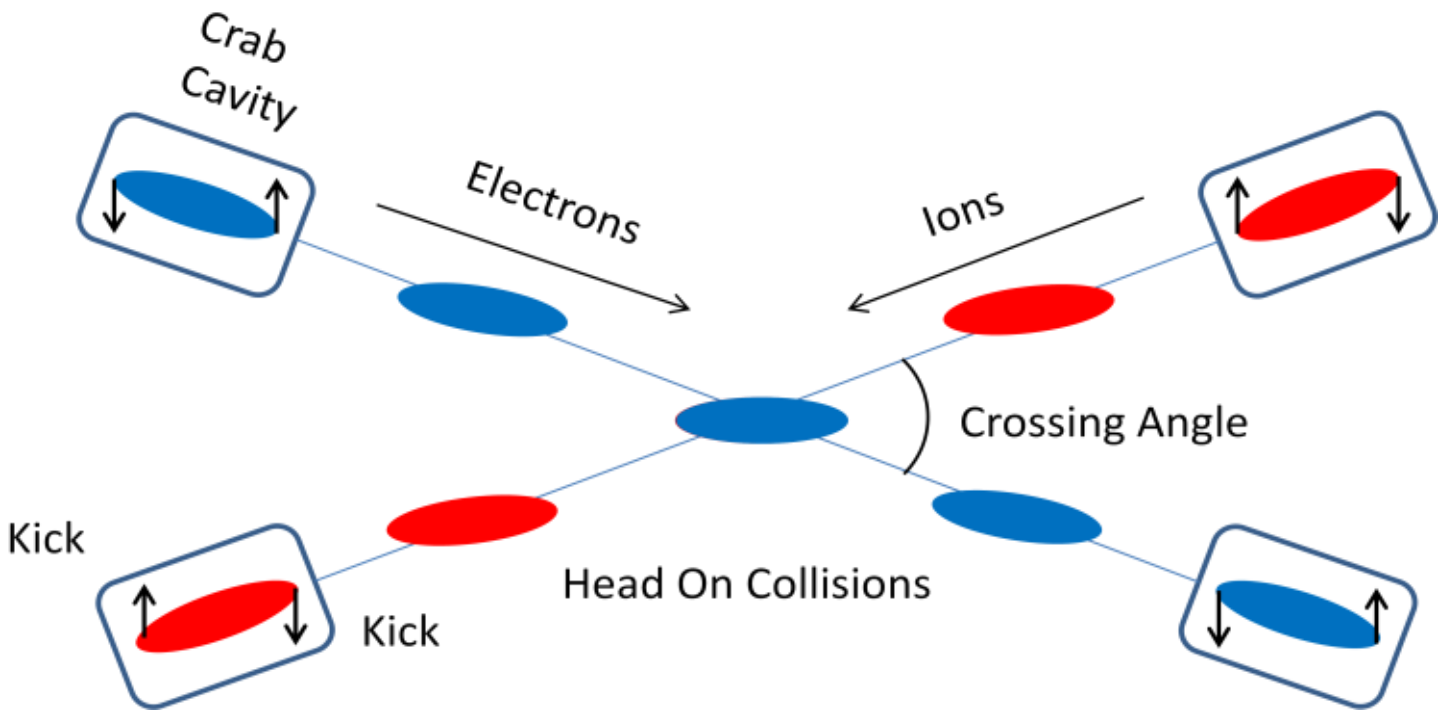

Figure 7.22: Schematic drawing of crab crossing scheme.

The transverse crabbing kick is usually provided by an SRF crab cavity with the integrated kicking voltage given by the following formula

$$V = \frac{cE_b \tan \varphi_{crab}}{2\pi f_{rf} \sqrt{\beta_x^* \beta_x^c}} \tag{7.2}$$

where $E_b$ is the beam energy, $\varphi_{crab}$ is the crab crossing angle, and $\beta^c{}_x$ is the value of the betatron function at the location of the crab cavity. With a 50 mrad crab crossing angle for MEIC, the voltage for a 5 GeV electron beam is about 1.35 MV, nearly identical to that of the KEK-B Factory; however, the required voltage for a 60 GeV proton beam is nearly six times larger. The additional design parameters for the crab crossing in MEIC are summarized in Table 7.3.

Table 7.3: MEIC crab crossing design parameters

| | | Electron | Proton |
|---|---|---|---|
| Energy | GeV | 5 | 60 |
| Bunch frequency | MHz | 748.5 | |
| Crab crossing angle | mrad | 50 | |
| Betatron function at IP | cm | 10 | |
| Betatron function at crab cavity | m | 300 | 1400 |
| Integrated kicking voltage | MV | 1.35 | 8 |

It should be noted that, while the KEK B-Factory adopted a global compensation scheme [20,21] such that only one cavity is installed in each collider ring, MEIC has a local scheme [4,5,21] in which two identical crab cavities, one for crabbing and the other for restoration, will be installed symmetrically on both sides of the IP, as shown in Figure 7.23. This configuration confines the beam gymnastics only in the IR. The crab cavities are placed in the expanded beam regions next to the FFB to ensure a $\pi/2$ betatron phase advance between the cavity and IP, as well as to minimize the required integrated crab kicking voltage.

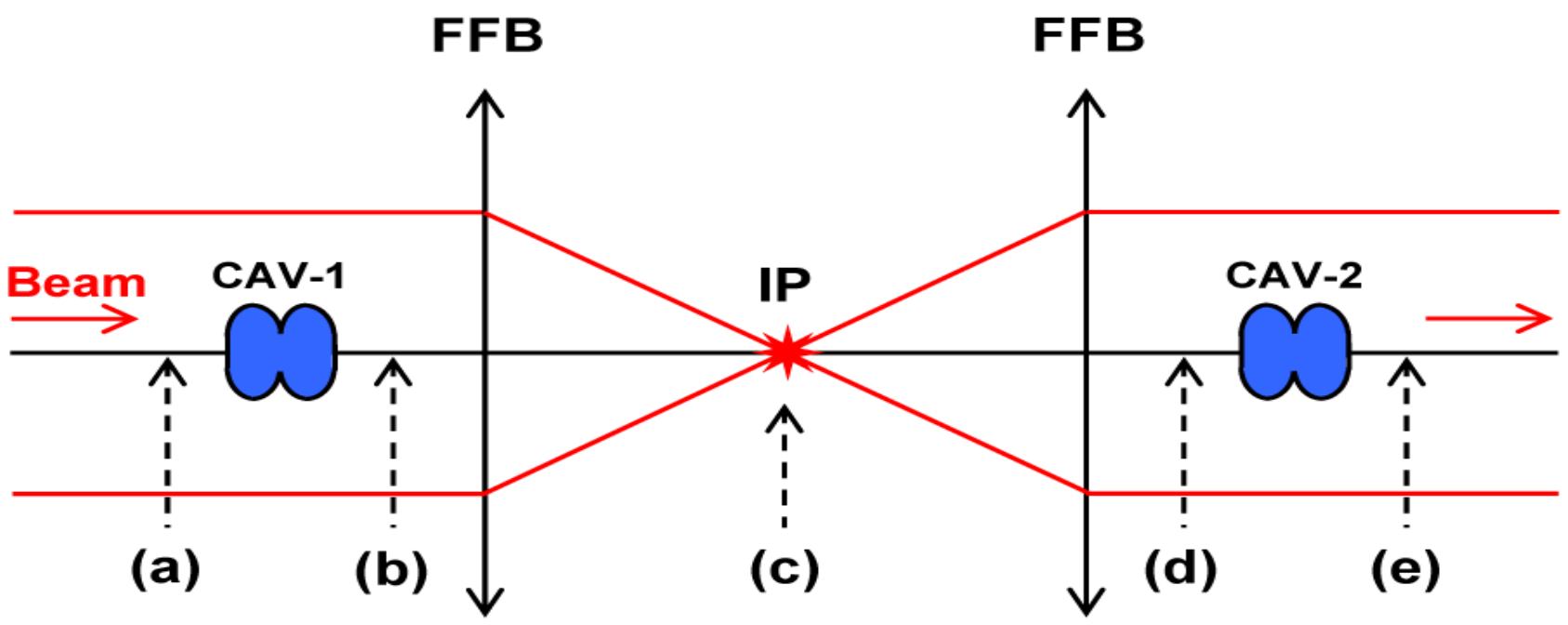

Figure 7.23: Crab cavity integrated into the MEIC interaction region

A novel type of SRF deflecting/crabbing cavity has been developed recently at Jefferson Lab and Old Dominion University for the MEIC crab crossing application [22,23]. It is a TEM-type parallel-bar cavity with cylindrical outer conductor and trapezoidal shaped loading elements, as shown in Figure 7.24 (left and center). Table 7.4 summarizes its design parameters. Such an SRF cavity can deliver high net deflection via the transverse electric field (Figure 7.24 (right)) to the particles on the beam axis while keeping lower and well balanced peak surface fields with higher shunt impedance. Another attractive feature is that this geometry has no lower-order mode and the nearest higher-order mode is far removed from the fundamental deflecting mode.

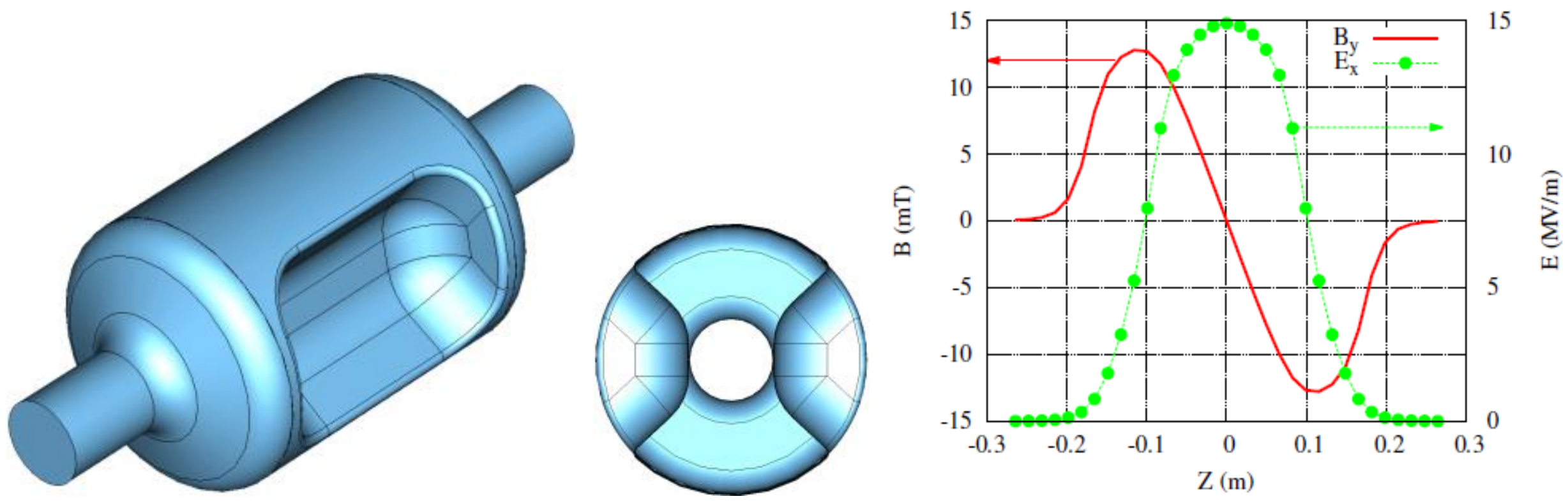

Figure 7.24: MEIC crab cavity design (left and middle) and transverse kicking fields (right).

Table 7.4 MEIC SRF crab cavity (normalized) design parameters

| | | |
|---|---|---|
| Frequency of π mode | MHz | 750.0 |
| λ/2 of π mode | mm | 199.9 |
| Frequency of 0 mode | MHz | 1314.4 |
| Frequency of near neighbour mode | MHz | 1143.1 |
| Cavity length | mm | 300.0 |
| Cavity diameter | mm | 193.0 |
| Bars length | mm | 185.0 |
| Bars inner height | mm | 57.5 |
| Angle | Deg | 36.2 |
| Aperture diameter | mm | 60.0 |
| Deflecting voltage ($V_T^*$) | MV | 0.2 |
| Peak electric field ($E_P^*$) | MV/m | 4.95 |
| Peak magnetic field ($B_P^*$) | mT | 8.74 |
| $B_P^* / E_P^*$ | mT/(MV/m) | 1.77 |
| Energy content ($U^*$) | J | 0.056 |
| Geometrical factor | Ω | 136.9 |
| $[R/Q]_T$ | Ω | 152.9 |
| $R_T R_S$ | $\Omega^2$ | $2.1\times10^4$ |

At $E_T^*$ = 1 MV/m

Tracking simulations to demonstrate the crab crossing concept were performed [24] in a single pass of a 5 GeV electron bunch with the design phase space parameters through the MEIC IR. A set of plots in Figure 7.25 illustrate evolution of the electron bunch phase space distribution at various locations in the IR: (a) it passes through a crab cavity where it receives the initial crabbing kick (b), it is then focused at the interaction point (c) by the FFB, it next goes through a second symmetrically located FFB (d), and finally it passes through the second crab cavity (e), which compensates the crabbing kick of the first cavity. The top row in Figure 7.25 shows the radial vs. longitudinal (*x-z*) particle positions in the bunch, and the bottom row shows the horizontal angles vs. longitudinal positions (*x'-z*) at the same locations; blue and red colors correspond to the cases that the crab cavities are turned off and on, respectively. These plots clearly show that the bunch is tilted at the IP and then restored to the original orientation after the second crab cavity. There is also no visible distortion of the phase space distribution of the bunch after crabbing and its compensation. Studies of long-term beam stability with crab crossing over multiple passes are underway.

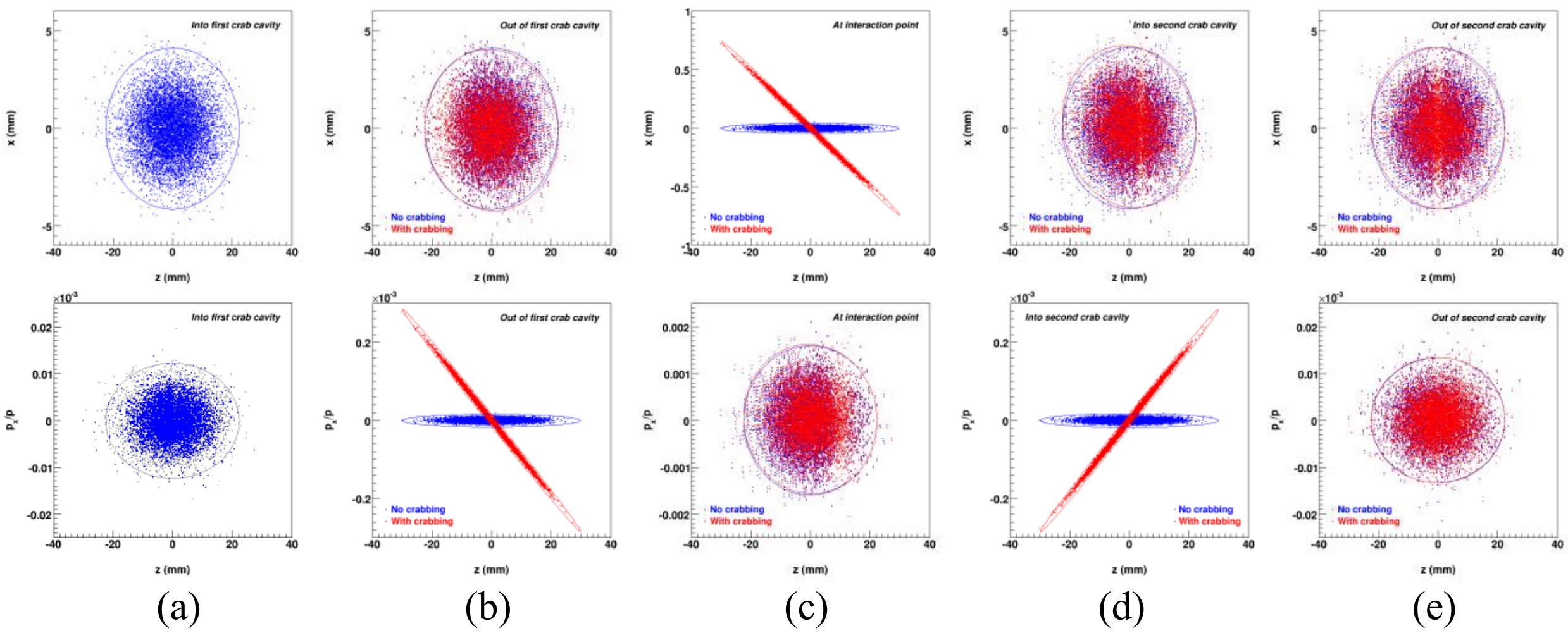

Figure 7.25: Evolution of an electron bunch phase space distribution during crabbing and its compensation. The top row shows particle radial and longitudinal positions before (a) and after (b) the first crab cavity, at the IP (c) and then before (d) and after (e) the second cavity, while the bottom row shows particle horizontal angles and longitudinal positions at the same locations. The blue and red colors compare the phase space dynamics with and without crabbing, respectively.

## 7.7 Synchrotron Radiation and Detector Background

Synchrotron radiation (SR) is exclusively generated by the electron beam in MEIC. The sources most important for the IR are the final dipole bending magnet(s) upstream of the collision and the quadrupole magnets between the last dipole bend and the IP. Figure 7.26 is a simple plan view drawing of a generic beam pipe for the collision point area with a greatly exaggerated scale in the horizontal plane. The electron beam coming in from the right and the proton beam from the left form a 50 mrad crossing angle. The figure shows four surfaces (labeled A-D) that are portions of the beam pipe that intercept SR from the last four upstream quadrupoles. Because of the desire to have a large beam aperture for the outgoing electron beam there are essentially no downstream beam pipe surfaces that intercept SR. Table 7.5 shows SR background rates for photons scattering from the labeled surfaces and striking the central detector beam pipe. The model uses a non-gaussian beam tail distribution that populates the large transverse beam sigma region with beam particles. The tail distribution is illustrated in Figure 7.27 and was used in the PEP-II design with reasonable agreement between the predicted and observed background rate for SR in the BaBar detector [25]. Table 7.5 represents a very preliminary investigation for MEIC; further work is needed to assure a comprehensive understanding and control of the SR backgrounds in the detector.

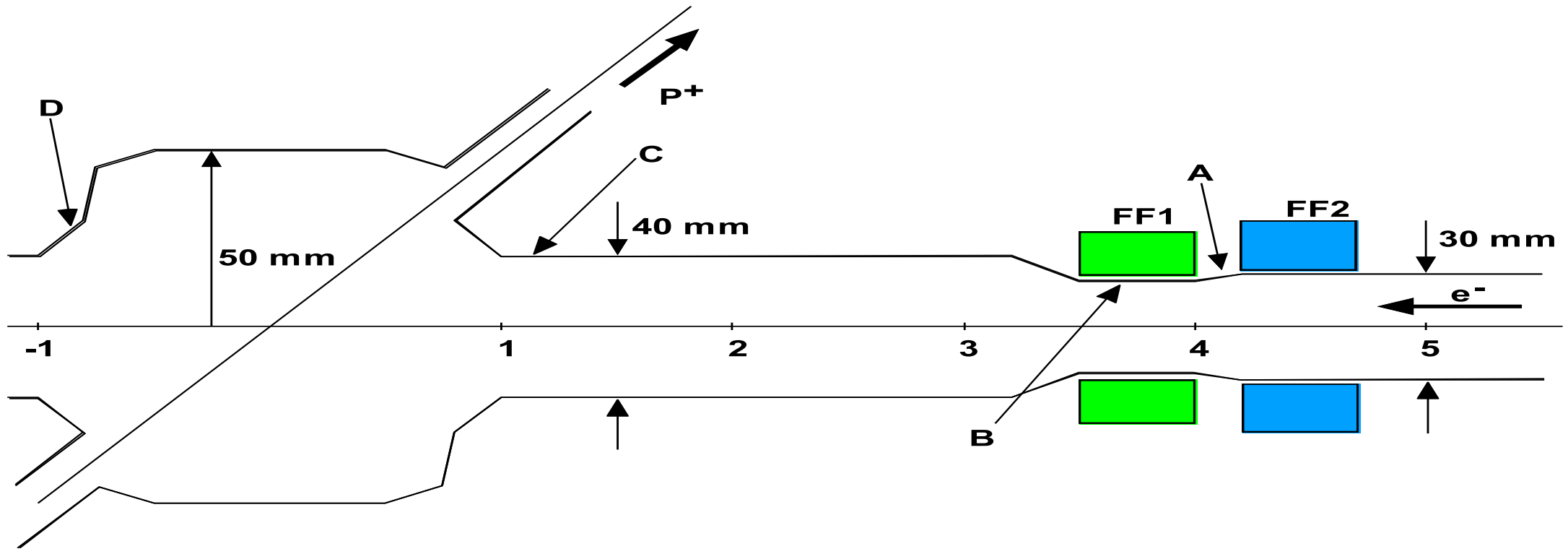

Figure 7.26: A schematic drawing of beam pipe in an IP. The electron beam enters from the right and the ion beam from the left. There are four surfaces pointed out in the figure labeled A-D. The colored boxes portray the last two quadrupoles of the electron beam.

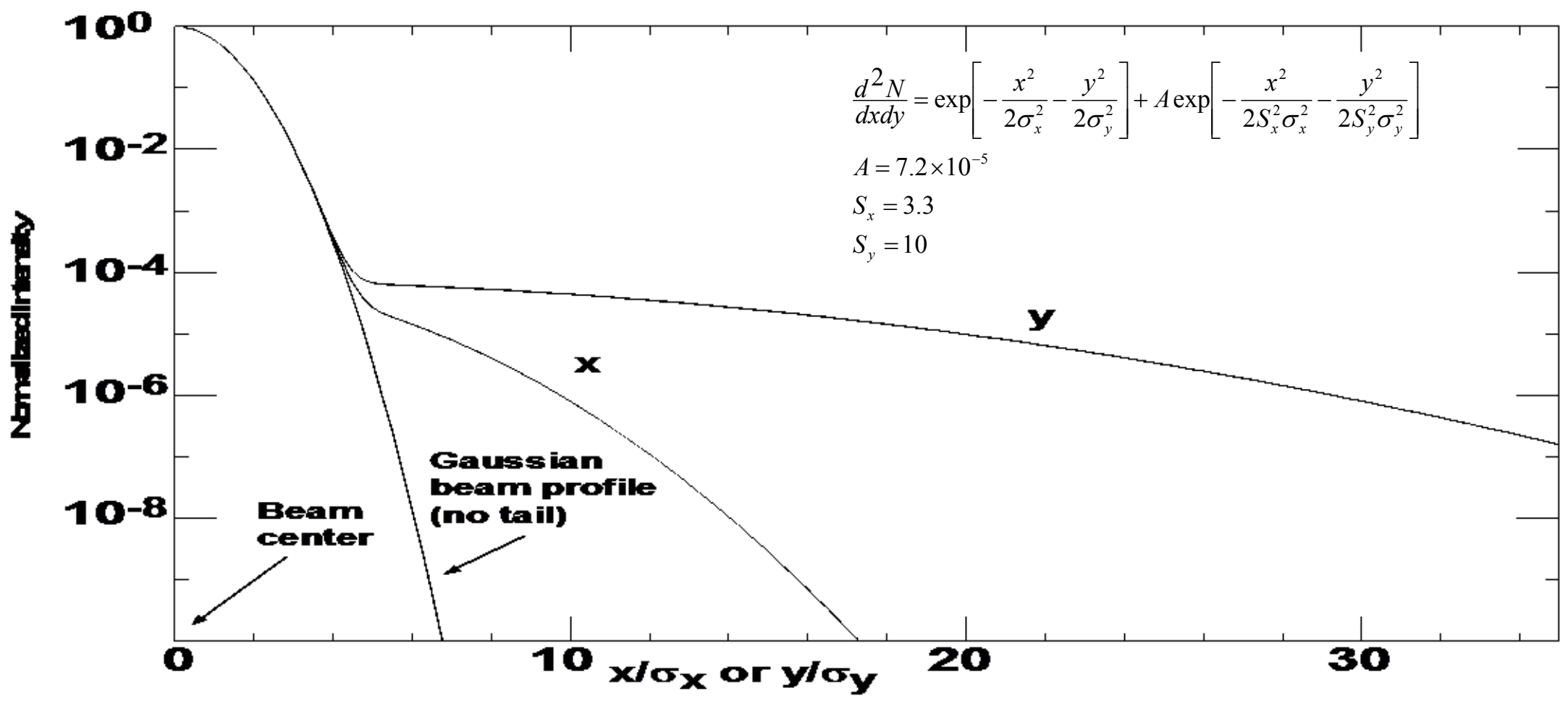

Figure 7.27: Transverse beam profile used to model the SR backgrounds in the IR.

Table 7.5: Estimated SR photon rates at the selected surfaces in beam pipe near an IP

| Energy (GeV) | 11 | | 5 | |
|---|---|---|---|---|
| Current (A) | 2.32 | | 2.32 | |
| Emittance (nm-rad) | 1.0 / 0.2 | | 5.5 / 01.0 | |
| Surface* | Photons >10 keV per beam bunch incident on surf. | Scattered to the central 5 cm beam pipe | Photons >10 keV per beam bunch in on surf. | Scattered to the central 5 cm beam pipe |
| A | 240 | None | $1.8\times10^5$ | None |
| B | 3080 | None | $6.4\times10^5$ | None |
| C | $4.6\times10^4$ | 2 | $9.0\times10^4$ | 4 |
| D | $8.5\times10^5$ # | 38 | $1.6\times10^5$ | 7 |

* Surface A–D are pointed out in Figure 7.25.
# This photon rate corresponds to 2.5 W of incident power.

The photon rates seen in the above table all come from particles out at large transverse positions. This means the background rate in the detector from SR is sensitive to the large beam sigma particle distribution. The particle density in the large sigma region is determined by beam-beam interactions, Coulomb scatterings and Touschek lifetime effects. All of these can be modeled. However, the effect which is most difficult to model correctly is the beam-beam effect. Collimation of the electron beam to limit the particle density in the high sigma region can be very helpful as long as the beam lifetime is not severely impacted. Collimators will be needed to help suppress other background sources from Beam-gas Bremsstrahlung (BGB) as well as the aforementioned Coulomb and Touschek scattering sources. The numbers in the table do not have any information about the photons from the last bending magnets. The majority of these photons should strike the beam pipe upstream of the surface labeled A as well as on surface A. Since surface A has no view to the central beam pipe there are no photons that can scatter from this surface and cause background in the detector. However, there is significant SR power that comes from the last few bending magnets and this power needs to be carefully tracked to make sure the beam pipe has adequate cooling where the radiation strikes the inside surface

Good vacuum pumping also decreases the background rate from the above mentioned sources of Coulomb scattering and BGB. An important part of the design is to have a very good vacuum right after the last bend magnet and up to the collision point in order to minimize background effects from the beam particles that have scattered enough in this region to be steered into the detector by the final focus elements. In general, good vacuum is required all around the ring in order to minimize BGB rates and Coulomb scattering rates. A large portion of the off-energy beam particles created upstream of the last bend magnets before the IP will be swept out by these magnets and strike the beam pipe before getting into the IR area. These regions can then be shielded in order to minimize backgrounds in the detector.

One of the discoveries of the B-factories concerning backgrounds is the generation of a significant neutron background. This background arose because of the high beam currents and due to the presence of bending magnets that were too close to the IP. In the case of the PEP-II B-factory the very strong bending magnets that brought the beams into and out of a head-on collision also created a significant neutron background from the sweeping out of the off-energy beam particles due to the collisions. The KEKB B-factory also had strong bending magnets (in the form of off-axis quadrupoles) on the downstream side of the IP and these too caused a neutron background in the detector, although this background was significantly lower than that of PEP-II because the bending magnets were further downstream from the collision point.

The MEIC design has no close bending elements for either beam and should therefore not have serious trouble with neutron backgrounds. However, many of the very low angle detectors for the electron beam may experience shower debris and consequent neutron backgrounds beyond the first bending magnet from the collision point.

## 7.9 Beam-Beam Interactions

The beam-beam interaction is one of the most important effects that limit the luminosity of MEIC. Investigating this effect becomes a critical part of feasibility study of this conceptual design. Preliminary simulations have been performed [26,27] recently with attention being focused on details of disturbance of colliding beams by nonlinear beam-beam kicks in a strong-strong regime within the current computing capability. The beam transport in the collider rings was idealized by a set of linear maps betweens IPs, thus effectively omitting the single nonlinear or collective effects in the rings. However, synchrotron radiation damping and associated quantum fluctuations of electrons were still taken into account. It should be acknowledged that the results of such simulations should not be used for predicting long-term (minutes or longer) beam and luminosity behavior.

The code utilized in this simulation study is BeamBeam3D [28], a self-consistent code developed at LBNL which solves the Poisson equation for electromagnetic fields over a 2D mesh for a number of longitudinal slices using the standard particle-in-cell method. The colliding bunches are modeled by groups of macro particles interacting with each other through nonlinear beam-beam kicks.

We first simulated a single IP case using nominal design parameters (Table 3.1) for collisions of 5 GeV electrons and 60 GeV protons with a synch-betatron tune working point selected empirically [26]. Figure 7.28 shows evolutions of the MEIC luminosity, and vertical emittances of the electron and proton beams over the turns of the collider ring. We then performed a luminosity scan, as shown in Figure 7.29, against the current of one of the two colliding beams in order to explore both limits of design parameters and threshold values for the onset of coherent beam-beam instabilities.

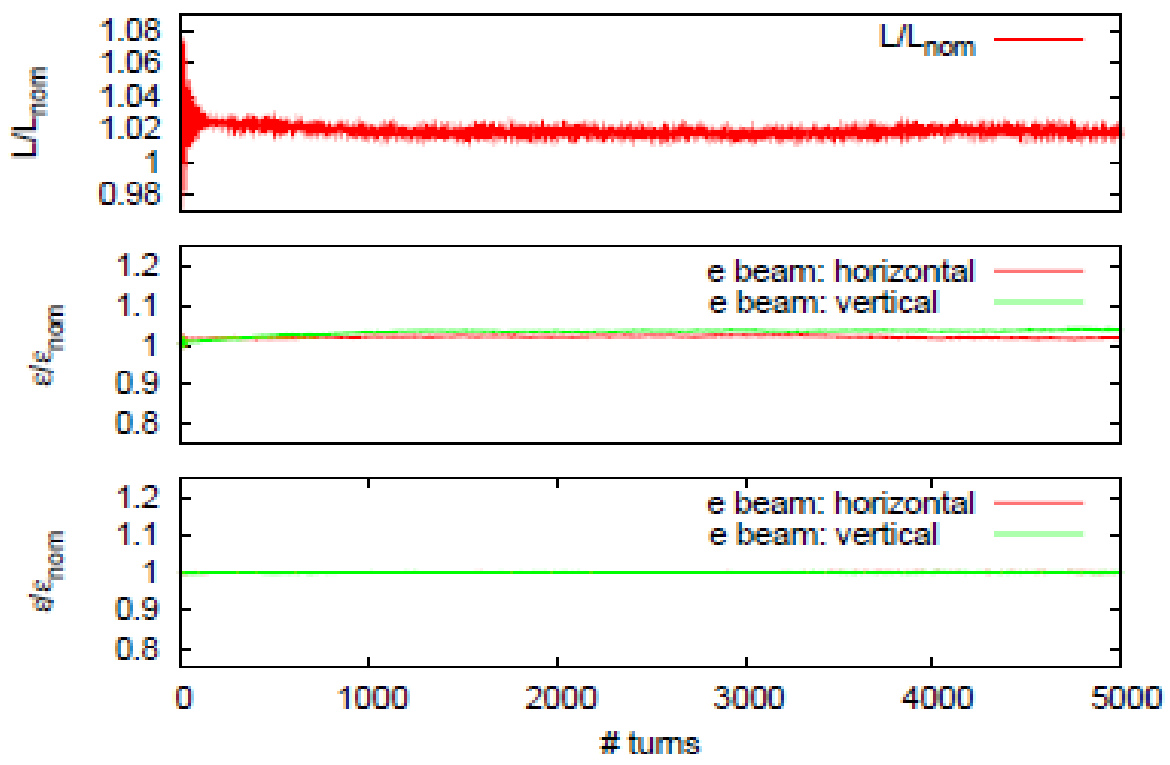

Figure 7.28: Evolution of MEIC luminosity (top) and emittance of electron beam (middle) and proton beam (bottom), all normalized to their respective design values.

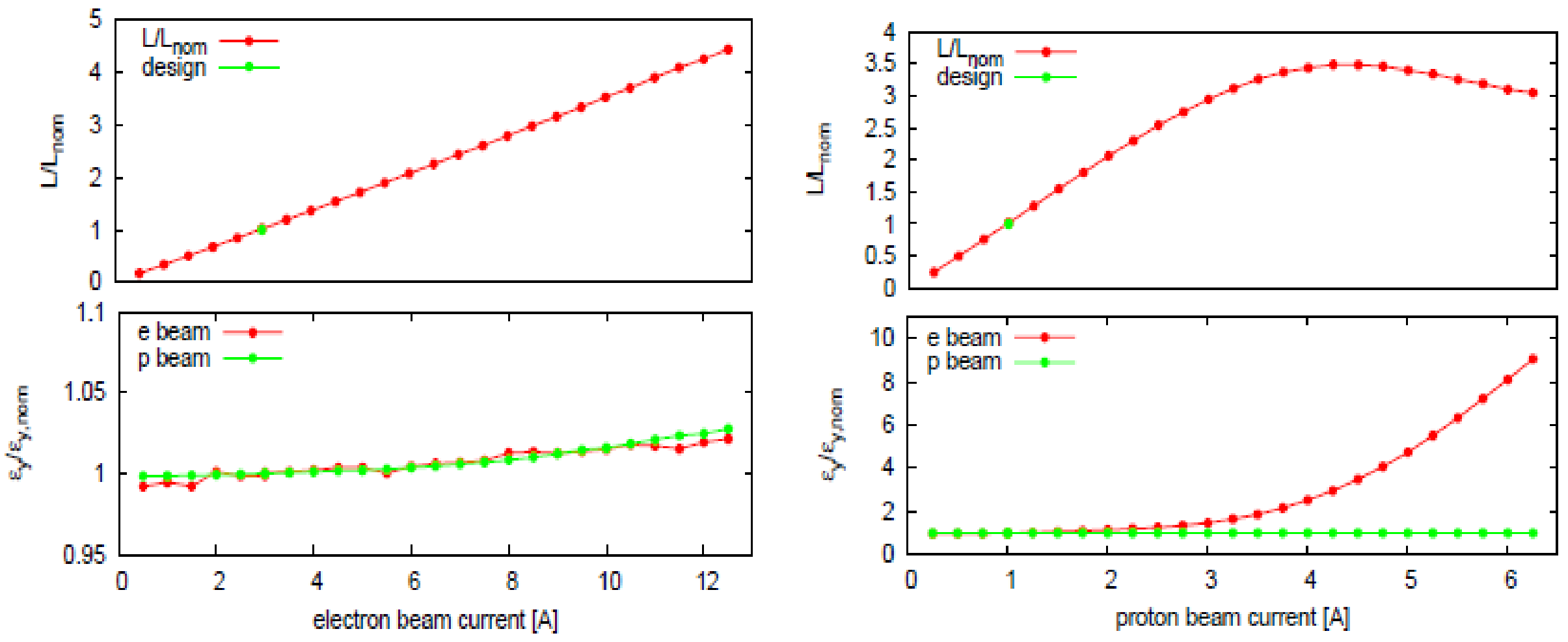

Figure 7.29: MEIC luminosity (top) and beam emittances (bottom), normalized to their design values, as current of the electron beam (left) and proton beam (right) is increased.

We summarize the main findings of this simulation study briefly below:

1. With a fair working point, MEIC luminosity for the 60×5 $GeV^2$ *e-p* collisions has almost no loss due to the beam-beam effects. However, a very slow decay of luminosity was also observed with a time scale much longer than the synchrotron radiation damping and far beyond the scope of this simulation study. This slow drift may be cured by the electron cooling.

2. The MEIC luminosity increases almost linearly as the current of each colliding beam is increased, one at a time. When a current is increased far beyond the design values of either proton or electron beams, nonlinear beam-beam interactions start to

   dominate, causing a notable slowdown of luminosity increase, and eventually breakup of the beams. This fact means that MEIC design parameters are still in the linear region when only the beam-beam interaction has been considered. It would be expected that this result will change when additional beam effects, particularly the nonlinear beam dynamics in the collider ring, are included in the simulation.

3. No coherent beam-beam effects have been observed by this preliminary simulation.

# 8. Outlook

Now that we have completed this baseline design, the focus of the MEIC machine studies is on two important aspects: design optimization and accelerator technology R&D.

There are multiple goals for the MEIC design optimization: best serving the science program, which itself is under continuously active development; extending the reach and enhancing performance of this collider; improving robustness and facility operability; reducing the technological challenges, uncertainties and R&D costs; and minimizing project cost and completion time. We foresee a process of design iterations informed by a steady progression of machine studies, closely coupled to development of both the science cases and accelerator technologies.

The MEIC design relies on adapting several new concepts/schemes and integrating an array of cutting-edge accelerator technologies for delivering high performance to meet scientific needs. A successful demonstration of these new concepts/schemes and development of various key supporting technologies are critical for proof-of-principle of the MEIC design approach. A group of critical R&D issues have been identified to address potential performance risks. In this chapter we give a brief discussion of each of these R&D issues, in an order roughly of their importance, and for some of them, we also outline the R&D plan in some detail.

***Electron Cooling in the Collider Ring***

Electron cooling in the MEIC collider ring is a most critical part of the multi-phase cooling design concept for achieving a high luminosity in MEIC. The main challenge is achieving the desired high cooling efficiency at medium energy. A key goal for R&D is a proof-of-principle demonstration of the design concept of an ERL-Circulator Cooler.

Conventional electron cooling has yet to be applied to bunched ion beams or at ion energies of the order of 100 GeV/u. Although initial estimates indicate that sufficient cooling rates can be obtained, much more detailed studies are necessary to ensure that unexpected effects will not reduce efficiency. We plan to carry out simulation studies of electron cooling of medium energy ion beams since it is not feasible to conduct direct full energy experimental measurements because of costs. At the first stage we limit the scope of the study to “pure” electron cooling; that is, we assume the cooling electron bunch is fresh from the linac with every interaction with an ion bunch. At the second stage, the coupling of the electron beam dynamics to the ion beam will also be included by tracking the electron bunches over multiple turns in the circulator cooler and through the energy recovery linac. Such extended simulations likely require significant computing resources and also additional software enhancements.

A proof-of-principle experiment [1] has been proposed recently to demonstrate this cooler concept. The Jefferson Lab FEL [2] is selected as the test facility for this experiment since it can produce a high quality electron beam with an energy range and bunch repetition rate relevant to the proposed MEIC cooler. It allows maximum reuse of existing hardware, dramatically reducing the capital cost of this experiment. As shown in Figure 8.1, the presence of the two parallel IR and UV beam lines in the Jefferson Lab FEL facility provides an opportunity for implementation of a compact circulator ring with minimum additional beam line. The purpose of this experiment is to demonstrate circulation of an electron beam in a circulator ring while the beam quality is satisfactorily preserved. The more turns that the beam can take while preserving its quality, the easier electron source development will be. Specifically, we will (1) demonstrate a scheme for bunch exchange between the ERL and the circulator ring; (2) develop and test supporting technologies such as a fast kickers; (3) test bunch length change and longitudinal phase matching between the ERL and the circulator ring; and (4) study beam dynamics and collective effects in the circulator ring.

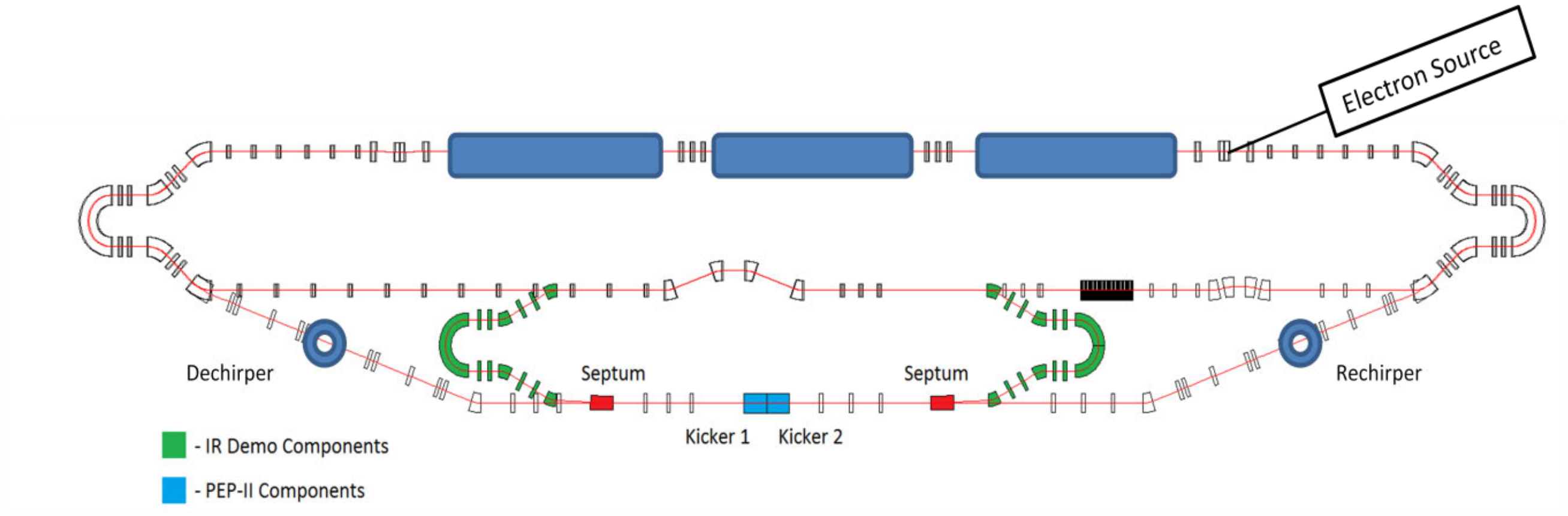

Figure 8.1: A test facility for an ERL circulator electron cooler based on Jefferson Lab FEL.

To support the ERL circulator cooler design concept, two key accelerator technologies, namely, a magnetized electron beam provided by a photo-cathode gun and an ultrafast kicker, must be successfully developed. Without a magnetized beam, the cooling ability of the electron beam will be far from optimal; and without the fast kicker, the compact circulator ring will not be able to refresh the cooling beam effectively. Preliminary conceptual development has already been started and was presented in section 6.3, and will be further developed at highest priority. An alternative technology based on high bunch charge and a high current thermionic gun is also under active evaluation for driving the ERL cooler without a compact circulator ring, and it provides a backup approach.

***Space-Charge Dominated Ion Beams in the Pre-booster***

Limitations on current and emittances from the space charge of the hadron beam in the booster synchrotrons or low energy storage rings are important issues that must be addressed in the final design of the facility. In the case of the MEIC pre-booster, the 6D phase space emittance of an ion beam after accumulation to a high current (up to 0.5 A) should be small enough to provide the necessary conditions to attain the required short (10 min or less) electron cooling time in the collider ring at the injection energy (i.e., the second phase of electron cooling as described in section 6.1). To reach these beam current and emittance goals, low energy DC electron cooling is proposed as an alternative to stripping injection for assisting accumulation of positive ion beams (including both polarized light ions and heavy ions) and for reduction of the phase space footprints, as discussed in section 5.2. This cooling-assisted beam accumulation scheme has been successfully implemented earlier at IUCF [3] but for relatively low circulating proton currents. Our estimates show [4] that the desired current, quality and stability of the accumulated ion beam can be achieved by use of conventional and tested techniques. To validate these estimations and also to explore the limits on the current and emittance of the accumulated ion beam, we include the following topics in the planned R&D:

- Conceptual study, design and modeling of the low energy electron cooling in the pre-booster for stacking the positive ion beams.
- Development of *circular optics* [5] and its implementation in the MEIC pre-booster to realize the *matched electron cooling* [6] as a way to diminish the space charge impact and attain a low 6D phase space emittance of the hadron beam.
- Conceptual study and design implementation of the painting technique for stripping injection of polarized light ion beams in the pre-booster synchrotron with circular optics.
- Development of a *barrier-bucket* RF technique [7] to be used in the pre-booster for stacking and acceleration as an alternative to the cavity-based RF system.

***Beam Synchronization***

As discussed in detail in section 5.7, the present preferred scheme for solving the challenging beam synchronization problem is based on varying the number of ion bunches in the ion collider ring with changes in the circumference of the electron ring and the RF frequency. The requirement of frequency variation of the SRF system in the ion collider ring requires careful mechanical design to ensure reproducibility and limited tuner stresses.

The SRF cavities used in the electron and ion collider rings can be separated into two categories. One is the single-cell, heavily HOM damped, high gradient SRF cavity like the CESR cavity at Cornell University for the electron collider ring. Another is the multi-cell, moderated HOM damping, high gradient SRF cavity like Jefferson Lab's FEL high current cavity (748.5 MHz) for the ion collider ring.

The frequency detuning of the SRF cavity necessary for stable ion beam re-bunching and also for reducing the klystron power requirement is small (~5 kHz). Another frequency detuning to match the reduced harmonic number (from 3378 to 3370 for the ion collision energy from 13.52 to 100 GeV/u) and continue increased $\beta$ ($\Delta\gamma\sim2$) for a fixed 1350 m path length is also relatively small (~1 kHz). So the detuning amount for the multi-cell cavity using a mechanical tuner on the stiff cavity wall presents no technical challenge. However, the detuning amount of 0.015% for the single-cell SRF cavity used in the electron collider ring is slightly larger compared to traditional the tuning method on normal conducting cavities.

Frequency deviations in excess of this amount are routine in normal conducting cavities where plunger tuners are used to change the volume of the cavity in a high magnetic field region, hence tuning the inductance of the cavity over a large range (e.g. the PEP-II specification was ± 340 kHz out of 476 MHz or ±0.07 %; the as-built range was ± 500 kHz or ± 0.1%). Such plungers have been proposed for SRF cavities, but care must be taken to avoid multipacting or particulate trapping in the gaps, to ensure adequate cooling and to engineer sufficient range of motion at cold temperatures over many cycles. SRF cavities normally require a much smaller tuning range because of the naturally small bandwidth. Tuning is generally achieved by small elastic distortions of the cell shape, usually in the capacitive region of the cavity (changing the iris to iris distance). The range of such motion is limited by the mechanical properties of the cavity cell, which is usually designed to be stiff. However, it is easy to conceive of cavity profiles that could have a much greater range of motion without exceeding mechanical stress limits. Such shapes would rely on the stiffness of the external tuner to resist pressure fluctuations and vacuum and Lorentz forces. Either inductive or capacitive tuning could therefore be employed to achieve the desired large tuning range for MEIC in an SRF cavity. Development and verification of a robust scheme will be a goal of the MEIC R&D program.

***Beam-Beam Interactions***

The nonlinear beam-beam force can cause strong perturbations to the particle dynamics, which lead to an emittance growth and a reduction of the collider luminosity. It may also affect dynamic aperture and beam lifetime. The initial phase of simulation studies on MEIC beam-beam effects is briefly summarized in section 7.9, and we are now ready to investigate many more subtle effects. These studies are important because the application of many advanced design features pushes the MEIC design parameters, and thus the beam-beam effect as well, to new regimes. For example, the low $\beta^*$ in MEIC is beyond the state-of-the-art for an ion collider, which can create new challenges for the stability of the ion beam when both nonlinear lattice and beam-beam effects are included. As a design feature, the large synchrotron tune of 0.045 in MEIC, which actually exceeds the total beam-beam tune shift of ions, is an order of magnitude higher than the values used for existing ion colliders. While such a large synchrotron tune brings a fast averaging effect over the particles within a bunch, thus helping to ease the beam-beam induced high order resonance, it can potentially cause a synchro-betatron resonance [8] in the single particle optics, which may be further amplified by the integration of a crab crossing collision scheme to the IR [9,10]. In the case of multiple IP operations, an idea of equal-phase-distance has been proposed such that the phase differences between any adjacent IP are designed to be identical; thus a system with multiple IPs is effectively the same as a system with a single IP as far as the beam-beam tune-shift is concerned. All the above design features and potential issues they might bring need thorough studies. We plan to investigate the following topics of beam-beam effects for the MEIC baseline design using frequency map method [11] and other techniques:

- Combined effects of beam-beam interactions with lattice nonlinearities.
- Interplay of synchro-betatron resonances and beam-beam interactions.
- Interplay of crab crossing and beam-beam interactions.
- The effect of multiple bunches and multiple IPs on incoherent beam-beam interactions.

- Coherent dipole, quadrupole, and synchrobetatron instabilities for single and coupled bunch beam-beam interaction.
- Dynamic beta effects [12] near the half-integer working point.
- Weak-strong beam-beam modeling for long term beam stability and lifetime.
- Coupling of beam-beam effects with other collective effects, such as geometrical impedance/wakefield or resistive wall, in the storage rings.
- Study the concept of phase equal-distance for the multiple IPs.

***Chromaticity Compensation and Dynamic Aperture***

Choosing a very small β* for achieving high luminosity while reserving a large magnet-free detector space for full detector coverage poses significant challenges to the IR design. Conceptually, the main challenge is to compensate the large chromatic effects caused by the strong final focus in an IR while preserving an adequately large dynamic aperture. The IR design approach of MEIC [13] provides an excellent momentum acceptance after only compensating for linear chromaticities, as presented in section 7.5 and 7.6. The dynamic aperture is also reasonably large, considering the large values of the natural chromaticity that have been successfully compensated. Due to the large beam extension required to accommodate small β*, further design optimization is required, likely requiring more sextupole families and even octupole families. The initial nonlinear optimization results presented above look rather encouraging and provide some confidence that the design parameters can be achieved [14,15].

It is important to specify a quantitative criterion for the momentum acceptance and dynamic aperture required for the necessary luminosity lifetime. IBS can cause particles to undergo momentum changes such that they may be lost due to having too large betatron amplitude or momentum deviation. On the other hand, the radiation damping of electron beam and electron cooling of ion beams may help to bring these particles back to the beam cores. Therefore a beam lifetime depends on the balance of multiple and single (Touschek) IBS and electron cooling for ion beams and the balance of Touschek effect and radiation damping for electron beams, both strongly affected by the achieved momentum acceptance and dynamic aperture. The chromaticity compensation scheme has to be optimized for the electron beam to minimize its emittance degradation and synchrotron radiation.

The initial dynamic aperture studies presented in chapter 7 do not include any error effects. Therefore, an important next step is to evaluate the error sensitivity of the IR design and demonstrate that the required error tolerances can be met using realistic technologies and optimized lattice concepts. It is clear that, in addition to the IR, other machine elements such as beamline magnets in arcs and straights, and particularly special machine insertions such as electron spin rotators, Siberian snakes for ion beams and crab cavities, etc., are contributing to the chromaticity and affecting dynamic apertures, though much less than the final focusing quadrupoles in the IR. The future studies will constitute a thorough investigation of the fully integrated system.

***Beam Polarization***

As presented in section 5.9, the unique figure-8 geometry has been adopted for all the MEIC ion rings. This design choice greatly benefits the preservation of polarization of all the interested light ions (including deuterons) during their staged acceleration and final storage in the MEIC ion complex. We should not expect any fundamental problem that prevents MEIC from achieving the high polarization goal demanded by the science program for polarized light ions. Further, while the ion collider ring needs full or partial Siberian snakes [15] or other additional special magnetic insertions [16] (see section 5.9.5) to fulfill the task of arrangement of polarization at IPs, these magnetic spin rotators are all currently available and tested in other facilities, notably RHIC at BNL [17,18]. Therefore, the focus of this R&D area will be, through spin tracking simulations, (1) a demonstration of high beam polarization in the collider rings; and (2) a proof of the advantages of a figure-8 shaped ring, compared to conventional rings, in preserving ion polarization during acceleration.

The electron beam injected into the MEIC collider ring from CEBAF is essentially fully polarized (over 80%). The electron polarization is maintained by the Sokolov-Ternov effect [19] if the electron spin direction is arranged properly (i.e., anti-parallel to the dipole magnetic field) in the arc of the collider ring. However, the spread of spin directions inside the magnets, particularly those inside universal spin rotators due to betatron oscillations and energy spread, can cause serious depolarization [20]. Spin matching [21], organizing the beam optics in a special way to cancel these depolarization effects, is critical for preserving a high electron polarization, and therefore will be the focus of future spin tracking based R&D activities.

Other important aspects related to electron polarization include the development of spin flipping techniques [22,23]. The electron spins that are in a "disadvantaged" state, namely, the spins that are parallel to the magnetic fields of arc dipoles, will be depolarized by the Sokolov-Ternov effect. Such a reverse effect becomes very severe when the beam energy is high. Simulation studies have been planned to explore both the decay rate of the electron polarization and also mitigation schemes such as frequent replacement of the electron beam in the collider ring.

***Electron Clouds***

Clouds of low energy (few eV) electrons [24] are generated in the environment of a positively charged particle beam by synchrotron radiation, residual gas ionization or particle losses, and subsequently increased by the surface secondary electron emission process. For the case of MEIC, the electron clouds will likely arise in the ion rings (due to residual gas ionization and particle losses) and also in the lepton ring (predominately by synchrotron radiation) if a positron beam is stored.

The electron clouds are drawn into the path of a positively charged beam as observed in several lepton and hadron colliders [25]. They can introduce nonlinear focusing and amplitude dependent tune shifts that contribute to emittance blowup, as well as coupled transverse motion from one bunch to the next along a train that can lead to dipole instabilities [11]. This effect can constrain the MEIC design parameters such as the beam current, bunch length and spacing, and vacuum chamber aperture.

Initial simulation studies [26,27] in dipole magnet regions of the MEIC ion collider ring have explored the parameter space and the related variation of the electron cloud build-up. A saturated cloud density of $1.54 \times 10^{12}$ e/m$^3$ (equivalent to a line density of 0.7 nC/m) is achieved after a fairly large number (1500) of consecutive bunches, after which space charge forces limit further emission of secondary electrons from the chamber walls.

Understanding the electron cloud effect in MEIC is critical for gaining confidence that the ring design will deliver the requisite high luminosity. An R&D program has been initiated to methodically evaluate the cloud build-up and the induced instability by simulation for MEIC. We will characterize the electron cloud effect in various magnetic regions of the MEIC rings and for different values of the secondary electron yield. By using a realistic MEIC ring lattice as input for the simulations, we will determine the thresholds for the single-bunch instability, emittance (head tail) blow-up and coupled bunch instabilities as well as emittance dilution below these instability thresholds. The combination between beam instability and build-up simulations will allow determining the maximum allowed secondary electron yield at the surface.

If the electron cloud evaluation shows significant detrimental effects in MEIC, then mitigation will be required. Several techniques appear promising, including coatings with low secondary-electron yield, grooved chambers, and the use of clearing electrodes [28]. We will determine the efficacy of potential mitigations as well as their impact on the beam impedance. A recommendation for mitigations will be given for the various regions in the MEIC rings.

***Ion Sources***

The design concept of a universal $H^-/D^-$/He ion source design [29] has been proposed recently. It combines the most advanced developments in the field of polarized ion sources to provide high-current high-brightness ion beams with greater than 90% polarization and improved lifetime, reliability and power efficiency. The main innovation of this approach is the strong suppression of parasitic generation of unpolarized H-/D-. The new source design will be based on the atomic beam polarized ion source (ABPIS) with resonant charge-exchange ionization of neutral atoms by negative and positive ions generated by the interaction of plasma with a "cesiated" surface. The ABPIS design will be improved by using new materials, an optimized magnetic focusing system, and novel designs for the dissociator, plasma generator and a surface-plasma ionizer to provide greater beam polarization.

# Acknowledgement

This work was supported by the U.S. Department of Energy, Office of Nuclear Physics, under Contract No. DE-AC05-06OR23177, DE-AC02-06CH11357, DE-AC05-060R23177, and DE-SC0005823.